\documentclass[a4paper,fleqn,usenatbib]{mnras}

\usepackage{newtxtext,newtxmath}

\usepackage[T1]{fontenc}
\usepackage{ae,aecompl}

\usepackage{graphicx}	
\usepackage{amsmath}	
\usepackage{amssymb}	

\usepackage{threeparttable} 
\usepackage{pdflscape} 
\usepackage{longtable} 

\usepackage[normalem]{ulem}

\newcommand{\rk}[1]{#1}

\title[Rotation of Cometary Nuclei]{Rotation of Cometary Nuclei: New Lightcurves and an Update of the Ensemble Properties of Jupiter-Family Comets}

\author[R. Kokotanekova et al.]{R. Kokotanekova$^{1,2}$\thanks{E-mail: kokotanekova@mps.mpg.de}, C. Snodgrass$^{2}$, P. Lacerda$^{3}$, S. F. Green$^{2}$, S. C. Lowry$^{4}$, 
\newauthor Y. R. Fern\'andez$^{5}$, C. Tubiana$^{1}$, A. Fitzsimmons$^{3}$ and H. H. Hsieh$^{6,7}$
\\
$^{1}$Max Planck Institute for Solar System Research,
              Justus-von-Liebig-Weg 3, 37077, G\"ottingen, Germany\\
$^{2}$Planetary and Space Sciences, School of Physical Sciences, The Open University, Milton Keynes, MK7 6AA, UK\\
$^{3}$Astrophysics Research Centre, Queen's University Belfast, Belfast BT7 1NN, UK\\
$^{4}$Centre for Astrophysics and Planetary Science, School of Physical Sciences (SEPnet), The University of Kent, Canterbury, CT2 7NH, UK \\
$^{5}$Dept. of Physics, Univ. of Central Florida, 4000 Central Florida Blvd., Orlando, FL 32816-2385, USA \\
$^{6}$Planetary Science Institute, 1700 E. Ft. Lowell Road, Suite 106, Tucson, AZ 85719, USA \\
$^{7}$Institute of Astronomy and Astrophysics, Academia Sinica, P.O. Box 23-141, Taipei 10617, Taiwan
}

\date{Accepted XXX. Received YYY; in original form ZZZ}

\pubyear{2017}

\begin{document}
\label{firstpage}
\pagerange{\pageref{firstpage}--\pageref{lastpage}}
\maketitle


\begin{abstract}

We report new lightcurves and phase functions for nine Jupiter-family comets (JFCs). They were observed in the period 2004-2015 with various ground telescopes as part of the Survey of Ensemble Physical Properties of Cometary Nuclei (SEPPCoN) as well as during devoted observing campaigns. 
We add to this a review of the properties of 35 JFCs with previously published rotation properties. 

	The photometric time-series were \rk{obtained in Bessel R, Harris R and SDSS r' filters and were} absolutely calibrated using stars from the Pan-STARRS survey. This specially-developed method allowed us to combine data sets taken at different epochs and instruments with \rk{absolute-calibration uncertainty} down to 0.02 mag. We used the resulting time series to improve the rotation periods for comets 14P/Wolf, 47P/Ashbrook-Jackson, 94P/Russell, and 110P/Hartley 3 and to determine the rotation rates of comets 93P/Lovas and 162P/Siding-Spring for the first time. In addition to this, we determined the phase functions for seven of the examined comets and derived geometric albedos for eight of them. 

    We confirm the known cut-off in bulk densities at $\sim$0.6 g $\mathrm{cm^{-3}}$ if JFCs are strengthless. Using the model of \cite{Davidsson2001} for prolate ellipsoids with typical density and elongations, we conclude that none of the known JFCs require tensile strength larger than 10-25 Pa to remain stable against rotational instabilities. We find evidence for an increasing linear phase function coefficient with increasing \rk{geometric} albedo. The median linear phase function coefficient for JFCs is \rk{0.046} mag/deg and the median \rk{geometric} albedo is 4.2 per cent.

\end{abstract}

\begin{keywords}
comets: general -- comets: individual
\end{keywords}


\section{Introduction}
\label{sec:Intro}

Comets are believed to preserve pristine material from the epoch of planet formation. Therefore, their properties have often been studied in the search for constraints on the conditions during solar system formation. A milestone in cometary exploration came from the European Space Agency's Rosetta mission which followed comet 67P/Churyumov-Gerasimenko through its perihelion passage in the period 2014-2016. The successful rendezvous of Rosetta with comet 67P/C-G has provided a unique ensemble of comprehensive observations which are set to fully characterise this comet. However, in order to interpret the detailed measurements from Rosetta, as well as those for other comets visited by spacecraft, we need to consider them in the context of other known comets. 

One fundamental technique to derive the physical properties of comet nuclei is through their rotational lightcurves. Lightcurves can be used to extract the individual objects' spin rates and axis ratios, which in turn can be used to constrain important properties of the comets, e.g. collisional history, density, tensile strength. Additionally, knowing the lightcurve brightness variation of JFCs can significantly improve the results of optical studies of JFC colour and size distributions. Despite being such a rich source of information, just a small fraction of JFCs have well-studied lightcurves.

There are two main techniques to derive rotational lightcurves from telescope observations: (1) photometric time-series of bare nuclei and (2) periodic variability of coma structures of active comets \citep[for an overview see][]{Samarasinha2004}. 
The former relies on the direct detection of the nucleus signal, and is expected to produce more precise results \citep{Samarasinha2004}. In order to detect the nucleus brightness variation directly, the comets need to be observed at large heliocentric distances when they are inactive. Observing the comets when they are weakly active can also allow reliable lightcurve derivations, but only in the cases when the nucleus signal dominates over the coma contribution. It is also possible to derive the nucleus rotation rate of active comet nuclei, provided that they are observed with sufficient spatial resolution to distinguish the nucleus signal from that of the coma. Such observations have been performed with the Hubble Space Telescope \citep[HST; see][]{Lamy2004}. 

Additionally, comet rotations can be studied during spacecraft flybys. Such missions have allowed the rotational properties of three comets to be studied in greater detail: 9P \citep[][and references therein]{Chesley2013}, 103P \citep[][and references therein]{Belton2013}, and 67P \citep{Jorda2016}. The ground- and space-based telescope techniques for period derivations usually do not account for Sun-comet-Earth geometry changes and therefore produce synodic rotation periods, while the spacecraft observations allow the sidereal spin periods to be derived \citep[e.g. see][]{Samarasinha2004}.  \rk{However, the difference between the synodic and sidereal rotation periods is usually small ($\sim$0.001 hour) even for near-Earth asteroids \citep[e.g.][]{Pravec1996}, so the synodic rotation periods are good approximations for the spin rates.}

The photometric observations used to study the rotational properties of comets can also be a valuable source of information about the comets' surface properties.  The lightcurve-resolved photometry allows a precise determination of the nucleus absolute magnitude. Combined with thermal infrared data, the photometric magnitude can be used to determine the geometric albedo \rk{(hereafter, albedo)} of the nucleus. In some cases when the comets have been observed at multiple epochs, the photometric data can be used to derive the phase functions of the nuclei. Albedos and phase functions provide us with the opportunity to characterise the surfaces of comets \cite[e.g][]{Li2013,Fornasier2015a,	
Ciarniello2015} and to compare the icy populations in the Solar system \citep[e.g.][]{Belskaya2008,Masoumzadeh2017}.

To distinguish between the various small body populations in the Solar system, we use the Tisserand parameter with respect to Jupiter:  

\begin{equation}
T_J = \frac{a_j}{a} + 2 \ \sqrt[]{\frac{a(1-e^2)}{a_J}} \cos(i), 
\end{equation}
where $e$, $i$, and $a$ are the eccentricity, inclination and semi-major axis of the orbit of the object, and $a_J$ is the semi-major axis of Jupiter's orbit ($a_J$ is approximately 5.2 au). The Tisserand parameter is a useful characteristic of the orbits of minor planets since it remains approximately constant for any object even after perturbations by Jupiter. \citep{Levison1996}. For the purposes of this paper, we consider JFCs to be objects with 2 $\leq$ $T_\mathrm{J}$ $\leq$ 3 and periodic comet designations. \rk{According to the distinction in \cite{Levison1996}, all objects with $T_\mathrm{J}$ > 2 are classified as ecliptic comets. Thus, the class of ecliptic comets includes objects with $T_\mathrm{J}$ > 3 such as 2P/Encke, active asteroids and active Centaurs. Since active asteroids and Centaurs are believed to have different physical properties from JFCs, we focus the analysis only on objects with 2 $\leq$ $T_\mathrm{J}$ $\leq$ 3.} We, however, include 2P/Encke at $T_\mathrm{J}$=3.025 as it is possible for comets of JFC origin to achieve $T_\mathrm{J}$ of slightly above $T_\mathrm{J}$ = 3 following terrestrial planet interactions \citep[e.g.][]{Levison2006}.

Previous extensive overviews of the known JFCs surface and rotation properties were published by \cite{Lamy2004}, \cite{Samarasinha2004},  \cite{Snodgrass2006}, and \cite{Lowry2008}. In this work we study the lightcurves and the surface properties of nine JFCs and compare them with a broad sample of JFCs with known rotational properties. This updated sample contains a collection of 37 well-studied comets, and allows us to investigate the population properties of JFCs for the first time after NASA's Deep Impact and EPOXI and ESA's Rosetta missions.

In Section \ref{sec:review} we review the studies of all comets, which to our knowledge have period determinations published after the last review by \cite{Snodgrass2008}. This section includes Table \ref{tab:review} which contains the nucleus properties of the whole sample of JFCs used in this work. We describe the observations and the method for precise absolute calibration of multi-epoch time-series observations in Section \ref{sec:methods}. In Section \ref{sec:results} we present the results from the observations of nine JFCs. After adding our comets to the rest of the JFCs with known rotational properties, we study the cumulative properties of JFCs in Section \ref{sec:discussion}. Finally, Section \ref{sec:conclusions} contains a summary of our results.

\section{Overview of the known JFC rotation properties}
\label{sec:review}
The aim of this study is to combine the newly obtained nuclei properties with those from previous works in order to analyse the bulk properties of the expanded sample of JFCs. Previously, the collective rotational properties of JFCs were studied by \cite{Lamy2004}, \cite{Samarasinha2004} and \cite{Snodgrass2006}. We expand their samples to include the cometary nuclei whose rotations were derived since then, and complement them with the newly obtained results from this work. Table \ref{tab:review} contains the properties of all considered comets together with the sources of all known parameters. However, the sections below focus in detail only on the comets with updates since the reviews in \citep{Lamy2004} and \cite{Snodgrass2006}, including the unpublished HST results quoted in \citet{Lamy2004} that were revised by \citet{Lamy2011}.\\

In addition to the rotational properties, we also review below the published size and shape estimates of the considered comets. While photometric lightcurves can be used to determine nucleus shapes, they do not provide absolute sizes. For those comets visited by spacecraft, the dimensions of shape models are directly measured. Radar data also provide absolute sizes for the comets with close approaches to the Earth. Combined thermal infrared and optical (reflected sunlight) data allow the albedo and cross-sectional area of the body (and hence an effective radius) to be determined. For those objects with only photometric data, the nucleus size can be estimated by assuming a geometric albedo of typically 4\%. The most recent reviews of comet sizes from visible photometry and thermal IR Spitzer photometry are given by  \cite{Snodgrass2011} and \cite{Fernandez2013} respectively.

	\subsection{Jupiter-family comets with recently updated rotation rates}
        \subsubsection{2P/Encke}

        Comet 2P/Encke is among the comets with the shortest known orbital periods, 3.3 years, which has allowed different observers to study its properties over multiple apparitions. Its relatively small heliocentric distance at aphelion of 4.1 au allows the comet to stay mildly active at almost all times, which has hindered the direct observation of the comet's nucleus. Nevertheless, 2P is one of the best-studied JFCs, having well-constrained spin rate, rotation changes, colour, albedo and phase function. All of the earlier works leading to today's relatively good understanding of 2P are thoroughly described in \cite{Lamy2004} and \cite{Lowry2007}. Newer papers have added spectroscopy of the nucleus \citep{Tubiana2015} and a study of the aphelion activity of this comet (Kelley et al., in prep.). Here, we provide an outline of the most important results on the nucleus shape and rotation rate.
        
        The earliest attempts to determine the rotational lightcurve of 2P came from \cite{Jewitt1987}. Their time-series optical photometry suggested a most-likely period of 22.43 $\pm$ 0.08 hours. A later study by \cite{Luu1990a} led to a best-fit period of 15.08 $\pm$ 0.08 hours, although both studies note that alternative periods were also consistent with their data. \cite{Fernandez2000} used thermal infrared time series data to confirm the 15.08 hour period. A large data set of observations between July 2001 and September 2002 when 2P was close to perihelion was used by \cite{Fernandez2005} to determine that the comet's synodic period was either 11.079 $\pm$ 0.009 hours or 22.158  $\pm$ 0.012 hours.  \cite{Fernandez2005} also discussed that these periods are not compatible with the spin rates found by  \cite{Jewitt1987} and  \cite{Luu1990a}. 
        
        \cite{Belton2005} compiled the available optical and infrared photometry and reached the conclusion that the nucleus of 2P is in a complex or excited rotation state. According to this analysis, the nucleus precesses about the total angular momentum vector with a period 11.8 hours and oscillates around the long axis with period 47.8 hours.
        
        \cite{Lowry2007} added new optical data sets collected in October 2002, just a few weeks apart from some of the observations in \cite{Fernandez2005}. This allowed \cite{Lowry2007} to combine data from the two studies and to derive an effective radius 3.95 $\pm$ 0.06 km, an axis ratio of 1.44 $\pm$ 0.06 and a rotation period of 11.083 $\pm$ 0.003 hours.
        
        2P was later observed during the following aphelion, and the lightcurves obtained suggested that the spin period increases by  $\sim$ 4 minutes per orbit  \citep{Mueller2008,Samarasinha2013}.

	 The early nucleus size estimates of $\leq$ 2.9 km  \rk{\citep[][we use effective radius to characterise the nucleus size hereafter]{Campins1988}} and $\mathrm{2.8 \leq r_{eff} \leq 6.4}$ km 
 \citep{Jewitt1987,Luu1990a} were confirmed by the later estimate of 2.4 $\pm$ 0.3 km by \cite{Fernandez2000} . Comet 2P was also observed with radar during two apparitions \citep{Kamoun1982,Harmon2005}. The data from \cite{Harmon2005} confirmed a period of $\sim$ 11 hours and excluded the longer periods of $\sim$ 15 and $\sim$ 22 hours. \cite{Harmon2005} combined the radar data with previous infrared observations and obtained a solution for 2P's shape with an effective radius of 2.42 km and an axis ratio of 2.6.  

	\cite{Fernandez2000} also managed to obtain the phase function of 2P with phase coefficient 0.06 mag $\mathrm{degree^{-1}}$ (in the range between 0 and 106 degrees) as well as a relatively high visual geometric albedo of 5 $\pm$ 2 \%.

        \subsubsection{9P/Tempel 1}
        9P/Tempel 1 was the target for two NASA missions: Deep Impact and Stardust-NExT. It was also extensively observed from ground during the supporting campaigns \citep{Meech2005,Meech2011a}. 
        
        Multiple authors studied the size, shape and rotation rate of 9P before the Deep Impact flyby \citep[e.g.][]{Weissman1999,Lowry1999,Lowry2001a,Lamy2001,Fernandez2003}. A detailed overview of their contributions can be found in \cite{Lamy2004}. 
        
        The two flybys provided sufficient information to determine the size of the nucleus with good precision. The mean radius of the shape model after the Deep Impact flyby was estimated as 3.0 $\pm$ 0.1 km, with axes of 7.6 and 4.9 km, and an axis ratio $a/b$ = 1.55 \citep{AHearn2005}. \cite{Thomas2013} combined the data sets from the two spacecraft and obtained a radius of 2.83 $\pm$ 0.1 km.  They reported a shape model with radii between 2.10 and 3.97 km, which gives an axis ratio $a/b$ = 1.89.
        
        The two flybys combined with the ground observing campaigns gave an insight into  the rotation of 9P. \cite{Belton2011} analysed multiple available data sets and determined that 9P had the following sidereal rotation periods: 41.335 $\pm$ 0.005 h before the 2000 perihelion passage; 41.055 $\pm$ 0.003 h between the perihelion passages in 2000 and 2005; 40.783 $\pm$ 0.006 h from the Deep Impact photometry slightly before the 2005 perihelion passage, and 40.827 $\pm$ 0.002 h in the period 2006-2010. \cite{Chesley2013} updated their work and concluded that 9P/Tempel 1 spun up by either 12 or 17 minutes during perihelion passage in 2000 and by 13.49 $\pm$ 0.01 minutes during the perihelion passage in 2005.

        \subsubsection{10P/Tempel 2}
        
        10P/Tempel 2 is one of the largest known JFCs. It is also known to be only weakly active at perihelion. The combination of these two factors has allowed its nucleus to be observed with very small coma contribution both at aphelion and perihelion, making 10P one of the best-studied comets. 
    
   A series of works have determined that 10P has a spheroidal shape with dimensions a=8-8.15 km and b=c=4-4.3 km (axis ratio of 1.9), albedo $\mathrm{A_{R}}$ = 2.4 $\pm$ 0.5\% and rotation period about 9 hours \citep{Sekanina1987,AHearn1989,Jewitt1989}. A detailed summary of the works which have estimated the size of the nucleus of 10P can be found in \cite{Lamy2004}. \cite{Lamy2009} used HST photometry to determine a nucleus radius of 5.98 $\pm$ 0.04 km.

   	10P is one of the first comets observed to change its spin rate on orbital timescales. It is progressively slowing down by $\sim$ 16 s per perihelion passage \citep{Mueller1996,Knight2011,Knight2012}. The most recent analysis by \cite{Schleicher2013} led to the conclusions that 10P has a prograde rotation with a period of 8.948 $\pm$ 0.001 hours, and that the rate of spin down has decreased over time, most likely in accordance with the known decrease in water production by the comet since 1988.

        \subsubsection{19P/Borrelly}
        The nucleus of comet 19P/Borelly was studied using HST images by \cite{Lamy1998}. Their analysis  suggested a rotation rate of  25.0 $\pm$ 0.5  hours and  dimensions of  4.4 $\pm$ 0.3 km $\times$ 1.8 $\pm$ 0.15 km, assuming an albedo  of 4\%. The comet was observed during five nights in July/August 2000 at the CTIO-1.5 m telescope \citep{Mueller2002}. These data yielded a lightcurve with period 26.0 $\pm$ 1 hours and a large lightcurve variation - between 0.84 mag and 1.0 mag. 
             
        On September 22, 2001, just eight days after 19P passed perihelion, the NASA-JPL Deep Space 1 Mission had a flyby of the comet \citep{Soderblom2002}. Using the encounter images, \cite{Buratti2004} determined that the nucleus has a radius of 2.5 $\pm$ 0.1 km and axes 4.0 $\pm$ 0.1 km and 1.58 $\pm$ 0.06 km. Dividing these two values yields an axis ratio $a/b$ = 2.53 $\pm$ 0.12.

        HST/STIS observations were conducted in parallel to the Deep Space 1 encounter \citep{Weaver2003}. They could not be used to derive an independent measure of the nucleus rotation rate but were in agreement with the previous period  measurement from \cite{Lamy1998}.
\cite{Mueller2002} collected all available ground-based data from 2000 and the HST data from 2001 and improved the period by one order of magnitude. They narrowed down the possible periods to three values P = 1.088 $\pm$ 0.003 days, P = 1.108 $\pm$ 0.002 days, and P = 1.135 $\pm$ 0.003 days, which were consistent with the initial period of P = 1.08 $\pm$ 0.04 days from \cite{Mueller2002} \citep{Mueller2010}. These authors continued studying the comet with observations from the SOAR telescope in Chile in September/October 2014 \citep{Mueller2015}. These new data were used in an attempt to choose between the three possible rotation periods as well as to look for activity-induced spin changes of the nucleus during the two apparitions since the last observations. The most likely period was 1.209 days (29.016 hours) but 1.187 days (28.488 hours) could not be excluded \citep{Mueller2015}. The newly derived period suggested that the rotation of 19P slows down by approximately 20 minutes per orbit \citep{Mueller2015}.

        \subsubsection{61P/Shajn-Schaldach} 
        
        \cite{Lowry2003b} used snapshot observations of the nucleus of 61P (in non-photometric conditions) to determine a radius of 0.92 $\pm$ 0.24 km. \cite{Lamy2011} observed the comet at heliocentric distance 2.96 au (inbound) and determined a mean nucleus radius of 0.61 $\pm$ 0.03 km and axis ratio $a/b$ $\geq$ 1.3. Their partial rotational lightcurve suggested a few possible periods, but the shortest one of them, 4.9 $\pm$ 0.2 hours was considered as most likely \citep{Lamy2011}.

        \subsubsection{67P/Churyumov-Gerasimenko}
        
       Comet 67P/Churyumov-Gerasimenko was selected as the backup target for the Rosetta mission after the \rk{2003} launch of the mission had to be postponed due to a failure of the Ariane rocket \citep{Glassmeier2007}. The comet was observed in detail during only two apparitions before the rendezvous in August 2014. 
       
       The rotation period of 67P was first constrained to $\sim$12 hours by Hubble Space Telescope observations in March 2003, soon after its perihelion passage in September 2002 \citep{Lamy2006}. After the comet moved to greater heliocentric distances and its activity was quenched, it was possible to directly observe the nucleus from ground and to determine the spin rate with greater precision. \cite{Lowry2012a} combined all available ground observations \citep{Lowry2006,Tubiana2008,Tubiana2011} and determined the sidereal rotation period of the nucleus to be $P = 12.76137 \pm 0.00006$ hours. \cite{Mottola2014} revised the period  before the second perihelion passage in 2009, and set it to $P = 12.76129 \pm 0.00005$ hours.

	The next period determination was done with measurements from the Rosetta camera OSIRIS in March 2014 \citep{Mottola2014}. The new period of the comet was determined as $P = 12.4043 \pm 0.0007$ hours and suggested that the nucleus had spun up  by 1285 s \citep[$\sim$ 21 minutes;][]{Mottola2014}.
    
    OSIRIS continued monitoring the temporal evolution of the rotation rate of 67P throughout the extent of the mission \citep{Jorda2016}. The perihelion measurements of the orientation of the comet's rotational axis determined an excited rotational state with period of 11.5 $\pm$ 0.5 days and an amplitude of 0.15 $\pm$ 0.03\textsuperscript{$\circ$} \citep{Jorda2016}. They determined a rotation period of 12.4041 $\pm$ 0.0001 h, which stayed constant from early July 2014 until the end of October 2014. After that, the rotation rate slowly increased to 12.4304 h until 19 May 2015, when it started dropping to reach 12.305 h just before perihelion on August 10, 2015 \citep{Jorda2016}. 
    
    According to the Rosetta measurements made available by ESA\footnote{\url{http://sci.esa.int/rosetta/58367-comet-rotation-period/}}, the rotation rate continued decreasing until February 2016, and at the end of the mission, the sidereal period of 67P was 12.055 hours (ESA provided no uncertainty on this value). These measurements imply that 67P spun up by 1257 s ($\sim$ 21 minutes) during its latest perihelion passage (2014-2016). This period change is similar to the change of 1285 s measured by \cite{Mottola2014}, which suggests that the comet spins up with a rate of approximately 21 minutes per orbit.

    The overall spin evolution of 67P is in very close agreement with the activity model of \cite{Keller2015}. According to their analysis, the sign of the rotation period change is determined by the nucleus shape, while the magnitude of the change is controlled by the activity of the comet.

Rosetta measured the precise dimensions of the bilobate nucleus of 67P \citep{Sierks2015}. The overall dimensions along the principal axes are (4.34 $\pm$ 0.02) $\times$ (2.60 $\pm$ 0.02) $\times$  (2.12 $\pm$ 0.06) km, with the two lobes being 4.10 $\times$ 3.52 $\times$ 1.63 km and 2.50 $\times$ 2.14 $\times$ 1.64 km \citep{Jorda2016}. Using the longest and the shortest axes of the comet, we calculated an axis ratio $a/b$ = 2.05 $\pm$ 0.06. 

The mean radius derived from the shape model of 67P is 1.743 $\pm$ 0.007 km. The area equivalent radius and the volume equivalent radius are 1.93 $\pm$ 0.05 km and 1.649 $\pm$ 0.007 km, respectively \citep{Jorda2016}.

        \subsubsection{73P/Schwassmann-Wachmann 3}
        \label{sec:rev_73P}
        
        Comet 73P/Schwassmann-Wachmann had a strong outburst in September 1995  \citep{Crovisier1995} which was accompanied by a split-up into at least four pieces \citep{Bohnhardt1995,Scotti1996}. The remnants of the 73P nucleus were detected during the subsequent apparitions. The largest one of them is fragment C, which was estimated to have a radius of 0.5 km \citep{Toth2005,Toth2006,Nolan2006}. 
        
        In 2006, the comet approached Earth to less than 1 au and provided an excellent opportunity for different observers to study the lightcurve of fragment C. \cite{Toth2005} and \cite{Toth2006} used HST data to determine the dimensions of fragment C. Assuming an albedo of 0.04 and a linear phase coefficient of 0.04 mag $\mathrm{deg^{-1}}$ for the R-band, they obtained an effective radius of 0.41 $\pm$ 0.02 km. The derived lightcurve suggested an elongated body with axes 0.57 $\pm$ 0.08 km and 0.31 $\pm$ 0.02 km, which results in a minimum axis ratio $a/b$ $\geq$ 1.8 $\pm$ 0.3 \citep{Toth2006}. 
        
        \cite{Drahus2010} collected all of the reported lightcurves \citep{Farnham2001,Toth2006,Storm2006,Nolan2006}, and added a further estimate of the spin rate using variations in the production rates of the HCN molecule from sub-mm observations. Their analysis showed that 73P-C had a stable rotation during the 21-day observing campaign in May 2006 and narrowed down the possible periods to 3.392 h, 3.349 h, or 3.019 h. Since  none of these values could be excluded, \cite{Drahus2010} concluded that the rotation period of 73P-C was between 3.0 and 3.4 hours during the duration of their observing campaign.  This is the fastest known rotation period of a JFC and its stability against rotational splitting suggests that 73P-C has a bulk tensile strength of  at least  14-45 Pa \citep{Drahus2010}, or that it has a higher than expected density (see Section \ref{sec:disc-density}). Given that 73P has previously split, and continues to fragment \citep{Williams2017}, it is most likely at the very limit of stability.
        
        \subsubsection{76P/West-Kohoutek-Ikemura} 
        
        \cite{Tancredi2000} observed the nucleus of 76P and estimated a radius of 1.3 km. However, the authors note that the collected photometric measurements of the nucleus brightness had a large scatter which makes the radius value uncertain.  \cite{Lamy2011} obtained a  partial lightcurve of the comet with most likely period of 6.6 $\pm$ 1.0 hours and brightness variation of 0.56 mag which corresponds to an axis ratio $a/b$ $\geq$ 1.45.  They estimated the nucleus radius to be 0.31	$\pm$ 0.01 km \citep{Lamy2011}.

        \subsubsection{81P/Wild 2} 
        
        Comet 81P/Wild 2 was the primary target of the sample-return mission Stardust. The observations of 81P before 2004 provided an estimate of its size \citep[summarised in][]{Lamy2004}. During the Stardust flyby in January 2004, the instruments on board revealed the shape of the nucleus as well as great details from the surface. \cite{Duxbury2004} used the obtained images to model the nucleus as a triaxial ellipsoid with radii 1.65 $\times$ 2.00 $\times$ 2.75 km $\pm$ 0.05 km, while the model of \citet{Sekanina2004} provided an effective radius of 1.98 km. 
        
        The rotation rate of the comet remained unknown until 81P was observed  at perigee in March/April 2010 \citep{Mueller2010a}. Their narrow-band filter photometry revealed a periodic variation in the CN features of the coma with a period of 13.5 $\pm$ 0.1 hours. 

        \subsubsection{82P/Gehrels 3} 
		The radius of 82P was estimated to be ${R_\mathrm{eff}}$ $<$ 3.0 km \citep{Licandro2000} or ${R_\mathrm{eff}}$ = 2.0 km \citep{Tancredi2000}. However, 82P shows signs of activity all along its orbit \citep[e.g.][]{Licandro2000}, and these values are therefore most likely influenced by the presence of coma. 
        
        \cite{Lamy2011} obtained a partial lightcurve with a  rotation period P = 24 $\pm$ 5 hours. However, the lightcurve is poorly sampled and this result most likely corresponds to a lower limit of the comet's rotation period \citep{Lamy2011}. The authors used the same data set to derive a mean radius ${R_\mathrm{eff}}$ = 0.59 $\pm$ 0.04 km and axis ratio $a/b$ $\geq$ 1.59.

        \subsubsection{87P/Bus} 
        The attempts to determine the size of the nucleus  of 87P resulted in the following upper limits:  $\mathrm{r_{n}}$ $\leq$ 0.8 km  \citep{Lowry2001a}, $\mathrm{r_{n}}$ $\leq$ 0.6 km \citep{Lowry2003b} and $\mathrm{r_{n}}$ $<$ 3.14-3.42 \citep{Meech2004a}. 

        \cite{Lamy2011} analysed a partial HST lightcurve of 87P and determined a most likely period of 32 $\pm$ 9 hours, a mean radius of 0.26 $\pm$ 0.01 km and an axis ratio $a/b$ $\geq$ 2.2.

        \subsubsection{103P/Hartley 2} 
        103P/Hartley 2 was extensively studied during the EPOXI flyby on 4 November 2010, and has been the target of multiple ground observations due to its favourable observing geometry during close approaches to Earth.  The first determinations of its radius $\mathrm{r_{n}}$ = 0.58 km came from \cite{Jorda2000} but was later revised to $\mathrm{r_{n}}$ = 0.71 $\pm$ 0.13 km \citep{Groussin2004}. This result was consistent with the upper limits set by  \cite{Licandro2000}, \cite{Lowry2003b}, \cite{Lowry2001a} and \citet{Snodgrass2008}. In preparation for the EPOXI mission \cite{Lisse2009} used Spitzer to measure an effective radius of 0.57 $\pm$ 0.08 km. This value was practically the same as the mean radius of 0.580 $\pm$ 0.018 km measured with the in situ instruments of EPOXI \citep{Thomas2013a}. The shape model presented in \cite{Thomas2013a} results in an estimated diameter range for the nucleus of 0.69 - 2.33 km. We divided the two extreme diameter values to obtain an axis ratio $a/b$ = 3.38. 

	The rotation period of 103P was studied in detail using the EPOXI data as well as the  extensive support observations from ground. It was established that the nucleus is  slowing down during the perihelion passage  and that it  is in a non-principal axis rotation  \citep{AHearn2011a,Belton2013,Drahus2011a,Harmon2011,Jehin2010,Knight2011,Knight2015b,Meech2011,Samarasinha2010,Samarasinha2011,Samarasinha2012}. The EPOXI lightcurve suggested several periodicities ranging from 17 to 90 hours \citep{AHearn2011a,Belton2013}, which were used to understand the complex rotation of the nucleus \citep{AHearn2011a,Belton2013,Samarasinha2012}.  The ground observations between  April 2009 and December 2010 monitored the change in the strongest periodicity of $\sim$ 18 hours, which corresponds to the precession of the long axis of the nucleus around the angular momentum vector \citep{Meech2011}. Over the period covered by the campaign, the rotation rate increased by $\sim$ 2 hours, from 16.4 $\pm$ 0.1 hours \citep{Meech2009,Meech2011} to  18.4 $\pm$ 0.3 or 19 hours \citep{Jehin2010}.

        \subsubsection{147P/Kushida-Muramatsu} 
        
        147P is among the smallest known JFC nuclei. \rk{Regarding the orbit class of this comet, \cite{Ohtsuka2008} showed that 147P is a quasi-Hilda comet, which underwent a temporary satellite capture by Jupiter between 1949 and 1961.} \cite{Tancredi2000} reported a nucleus radius of 2.3 km but noted that the measurement is uncertain.  \cite{Lowry2003b} reported $\mathrm{r_{n}}$ $\leq$ 2.0 km after a non-detection at heliocentric distance of  4.11 au. \cite{Lamy2011} derived a complete but poorly sampled lightcurve, which suggested that the rotation period  of 147P was either 
10.5 $\pm$ 1 hours or 4.8 $\pm$ 0.2 hours, where the former period is slightly favoured by the obtained periodogram. They estimate a radius of 0.21 $\pm$ 0.02 km and an axis ratio $a/b$ $\geq$ 1.53.

        \subsubsection{169P/NEAT}

        Comet 169P/NEAT was discovered as asteroid 2002 EX12 by the NEAT survey in 2002. Later it was designated as 169P/NEAT due to the detection of cometary activity \citep{Warner2005}. Due to its albedo of 0.03 $\pm$ 0.01 \citep{DeMeo2008} and its weak activity level, 169P is considered to be a transition object on its way to becoming a dormant comet. 
        
        \cite{Warner2006} reported the first rotational lightcurve of 169P with a double-peaked period 8.369 $\pm$ 0.05 hours and peak-to-peak amplitude $\Delta$m = 0.60 $\pm$ 0.02 mag. \rk{Later, \cite{Kasuga2010} observed the comet with a much larger (1.85-m) telescope and separated the nucleus brightness from the slight coma contribution.Therefore their derived lightcurve period of 8.4096 $\pm$ 0.0012 hours, photometric range $\Delta$m = 0.29 $\pm$ 0.02 mag and consequent effective radius of 2.3 $\pm$ 0.4 km are more reliable measures of the nucleus properties. However, the presence of coma during the observations done by \cite{Warner2006} would suppress the lightcurve amplitude. Therefore the higher amplitude measured by \cite{Warner2006} must instead be the result of a more elongated shape, measured at a different aspect than \cite{Kasuga2010}, unless the coma is highly variable on a timescale shorter than the spin period. However, due to the weak levels of activity present in this comet, this level of variability is unrealistic and we adopt the larger implied axis ratio limit from the \cite{Warner2006} data.} 
        
        \citet{Fernandez2013} determined an effective radius of $\mathrm{2.48^{+0.13}_{-0.14}}$ km for 169P using Spitzer mid-infrared data.
    

        \subsubsection{209P/LINEAR}
        
        \cite{Hergenrother2014} observed 209P and found its rotation rate to be either 10.93 or 21.86 hours. In May 2014, the comet had an exceptionally close approach to Earth (0.6 AU) which provided an opportunity for detailed studies of its intrinsically faint nucleus. \cite{Howell2014} used the Arecibo and Goldstone planetary radar systems to directly measure the nucleus to be 3.9 $\times$ 2.7 $\times$ 2.6 km in size, and calculated an effective radius of $\sim$ 1.53 km. These observations ruled out the longer period by \cite{Hergenrother2014} since the measured rotational velocities were too fast for the longer period.  
        
        \cite{Schleicher2016} also observed 209P during its perigee in May 2014. They used images obtained mainly with the 4.3 m Discovery Channel Telescope to study the coma and the nucleus of the comet. They used a small aperture with fixed projected size of 312 km, minimising the coma contribution so that the estimated nucleus fraction of the obtained light was 52-69 percent \citep{Schleicher2016}. Their lightcurve was consistent with the two periods from  \cite{Hergenrother2014}. However, \cite{Schleicher2016} preferred the shorter value, 10.93 hours, since it also agreed with the radar observations. \cite{Schleicher2016} reported that their lightcurve had a different shape than the one in \cite{Hergenrother2014}. Additionally, they measured variation of 0.6-0.7 mag, which is larger than the prediction of 0.4 mag based on the radar measurements. These differences can be explained by a possible interplay between shape and viewing geometry as well as albedo effects \citep{Schleicher2016}. Despite these discrepancies, all three investigations agree on the spin period of 10.93 hours.

        \subsubsection{260P/McNaught}
        
        260P was discovered in 2012, and the most reliable estimate of its effective radius to date is $1.54^{+0.09}_{-0.08}$ km \citep{Fernandez2013}. Its rotational characteristics were studied by \cite{Manzini2014} with ground photometric observations while the comet was around perihelion in 2012 and 2013. \cite{Manzini2014} used coma structures to constrain the pole orientation of the comet, but they were unable to use the coma morphology to derive a rotational period. Instead, the comet's lightcurve was obtained by measuring the coma brightness with apertures larger than the seeing disc but small enough to include only contribution from the coma at a distance up to 2000 -- 2500 km from the surface \citep{Manzini2014}. The resulting lightcurve had a variation of 0.07 mag and could be phased with a few possible periods, best summarised as 8.16 $\pm$ 0.24 hours.
        
While the method used in \cite{Manzini2014} has been used successfully to derive other rotations periods of comets with weak jet activity \citep[e.g.][]{Reyniers2009}, we regard the results on 260P with caution. It is very likely that the coma contribution in the selected apertures dilutes the received nucleus signal and dampens the possible variation caused by rotation. Therefore the limit on the nucleus elongation derived from the brightness variation is a weak constraint on the nucleus shape.

        \subsubsection{322P/SOHO 1} 
        \label{sec:rev_322P}
        Comet 332P/SOHO 1 was discovered by \emph{SOHO} as C/1999 R1, but after it was identified again in the \emph{SOHO} fields during the following apparitions \citep{Hoenig2005}, it became the first \emph{SOHO}-discovered comet with conclusive orbital periodicity. The observations of 322P during four consecutive apparitions displayed no clear signatures of a coma or tail and showed a nearly identical asymmetrical heliocentric lightcurve, implying repeated activity at similar levels each orbit \citep{Lamy2013}. 
        
        Despite its comet-like orbit with Tisserand parameter with respect to Jupiter of 2.3, the unusual properties of 322P suggest that it has asteroidal rather than cometary origin \citep{Knight2016}. Their optical lightcurve indicates a fast rotation rate of 2.8 $\pm$ 0.3 hr and photometric range of $\ga$ 0.3 mag. These figures imply a density of $>$ 1000 kg $\mathrm{m^{-3}}$, which strengthens the argument for asteroidal origin \citep{Knight2016}. This density is significantly higher than the typical values of other known comets but is typical for asteroids (see Section \ref{sec:disc-density}). Additionally, the colour of 322P is indicative of V- and Q-type asteroids, and its albedo (estimated to be between 0.09 and 0.42) is higher than the albedos measured for any other comet \citep{Knight2016}. These, together with the very low activity of the nucleus, indicate the possibility that 322P is an asteroid which  becomes active when very close to the Sun. However, since no other comet nucleus has been studied so close to the Sun, it is not excluded that it has a cometary origin, but  proximity to the Sun has changed the properties of its surface \citep{Knight2016}.

	\subsection{Comets with new rotation rates derived in this work}

    	\subsubsection{14P/Wolf}

         The first attempt to find the size of the nucleus of comet 14P/Wolf resulted in an effective radius of 1.3 km \citep{Tancredi2000}. However, the authors classified the estimate as poor due to the large scatter in the  data points. \cite{Lowry2003b} determined a radius of 2.3 km using snapshots of  the comet  at large heliocentric distance (3.98 au). The most recent value for the comet effective radius is 2.95 $\pm$ 0.19 km, obtained within the SEPPCoN survey \citep{Fernandez2013}. SEPPCoN used Spitzer infra-red photometry to measure sizes, and should be more reliable than visible photometry from earlier ground-based surveys.

\cite{Snodgrass2005} obtained time-series of the bare nucleus of 14P on 20 and 21 January 2004 with the New Technology Telescope (NTT) in La Silla. The observations showed a clear brightness variation of the nucleus with a period of 7.53 $\pm$ 0.10 hours. The peak-to-peak variation of the lightcurve was 0.55 $\pm$ 0.05 mag, which corresponds to an axis ratio $a/b$ $\geq$ 1.7 $\pm$ 0.1. The mean absolute magnitude of the time series was 22.281 $\pm$ 0.007, which suggested an effective radius 3.16 $\pm$ 0.01, assuming an albedo of 4\% \citep{Snodgrass2005}. 

In Section \ref{sec_res_14P} we provide the results from our lightcurve analysis. We combined the re-analysed data from 2004 with a SEPPCoN dataset from 2007 in order to improve the lightcurve of the comet and to derive its phase function.

        \subsubsection{47P/Ashbrook-Jackson} 
        \label{sec:past_47P}
        
       The early estimates of the nucleus size of 47P from photometric observations close to aphelion determined an effective radius $R_{\mathrm{eff}}$ = 3.0 km \cite{Licandro2000} and $R_{\mathrm{eff}}$ = 2.9 km \citep{Tancredi2000}. \cite{Snodgrass2006} and \cite{Snodgrass2008} observed the nucleus in 2005 and 2006 at large heliocentric distance close to aphelion and estimated $R_{\mathrm{eff}}$ = 2.96 $\pm$ 0.05 km. However, their photometric comet profiles showed signatures of activity, and therefore this estimate was considered an upper limit of the nucleus size. \cite{Lamy2011} used HST observations of the active nucleus of 47P to determine a mean effective radius of 2.86 $\pm$ 0.08 km. The most recent effective radius measurement of $\mathrm{3.11^{+0.20}_{- 0.21}}$ km was obtained within the SEPPCoN survey \citep{Fernandez2013}. 
        
        \cite{Lamy2011} derived a partial lightcurve with multiple possible periods. Analysing the periodogram, they suggested that the rotation period of the comet is $\geq$ 16 $\pm$ 8 hours. Both \cite{Snodgrass2008} and \cite{Lamy2011} attempted to constrain the phase function of 47P by combining all mentioned photometric observations. While the analysis of \cite{Snodgrass2008} clearly suggested a linear phase function with a slope $\mathrm{\beta}$ = 0.083 mag/deg , \cite{Lamy2011} showed that a less steep phase function similar to that of 19P/Borelly \citep[0.072 $\pm$ 0.020 \%;][]{Li2007a} is also possible. 

In Section \ref{sec_res_47P}, we show the result from our analysis of the data from \cite{Snodgrass2008} complemented by a new data set obtained in 2015. We determined the lightcurve and the phase function of 47P, but the derived results need to be considered with caution since the comet was active during both observing runs.

    	\subsubsection{93P/Lovas}

        Comet 93P/Lovas was one of the targets of the SEPPCoN survey. Its effective radius  $R_{\mathrm{eff}}$ = 2.59$\pm$ 0.26 km  was derived from Spitzer thermal emission observations \citep{Fernandez2013}. 

Our optical time-series observations are presented in Section \ref{sec_res_93P}. Despite the weak activity detected on the frames, we attempted to constrain the comet's rotation lightcurve.

    	\subsubsection{94P/Russell 4}

        \cite{Tancredi2000} tried to estimate the effective radius of 94P. However, at the time of the observations, the comet exhibited slight activity and the absolute magnitude measurements of the nucleus had large scatter. Therefore \cite{Tancredi2000} considered their effective radius estimate of 1.9 km as uncertain and estimated the error bars of the measurement to be between $\pm$ 0.6 and $\pm$ 1 mag. 
        
        \cite{Snodgrass2008} observed the comet during  four nights in July 2005  at heliocentric distance 4.14 au, outbound. The analysis pointed to a nucleus with effective radius of 2.62 $\pm$ 0.02 km and a lightcurve with period $\sim$ 33 hours \citep{Snodgrass2008}. The peak-to-peak variation of the lightcurve was 1.2 $\pm$ 0.2 mag, implying axis ratio $a/b$ $\geq$ 3.0 $\pm$ 0.5. Their nucleus size estimate $R_{\mathrm{eff}}$ = 2.62 $\pm$ 0.02 km is in a good agreement with the SEPPCoN Spitzer data from \cite{Fernandez2013}, who reported an effective radius of $2.27^{+0.13}_{-0.15}$ km. 
         
         In Section \ref{sec_res_94P}, we present two additional data sets from 2007 and 2009 with time-series photometry of 94P. They allowed us to determine the rotational lightcurve and the phase function of the comet.

    	\subsubsection{110P/Hartley 3}
        \label{sec:rev_110P}
        
        110P/Hartley 3 was observed with HST on November 24 2000 at heliocentric distance of 2.58 au, inbound \citep{Lamy2011}. The data  yielded an estimate of the effective radius of the nucleus $R_{\mathrm{eff}}$ = 2.15 $\pm$ 0.04 km and a lightcurve with period 9.4 $\pm$ 1 hours. The peak-to-peak amplitude of the obtained lightcurve was 0.4 mag, which suggested an axis ratio $a/b$ $\geq$ 1.30. 
        
        In Section \ref{sec_res_110P}, we analyse a further data set from 2012 which our team had obtained in order to derive the comet's phase function. We used the data to derive a precise phase function of the comet as well as to constrain better the lightcurve of 110P.

    	\subsubsection{123P/West-Hartley}

        \cite{Tancredi2000} estimated a radius of 2.2 km for the nucleus of comet 123P/West-Hartley. However, the authors consider this result as very uncertain as the individual photometric measurements of the comet nucleus displayed a large scatter. The SEPPCoN mid-infrared observations of 123P yielded an effective radius of 2.18 $\pm$ 0.23 km \citep{Fernandez2013}. 
        
        In Section \ref{sec_res_123P} we present the results from our analysis of a SEPPCoN data set from three observing nights in 2007. The comet was very faint ($m_{\mathrm{r}}$ = 23.3 $\pm$ 0.1 mag) and weakly active during the observations, which significantly obstructed the lightcurve analysis.

    	\subsubsection{137P/Shoemaker-Levy 2}
        
        \cite{Licandro2000} observed 137P at heliocentric distance 4.24 AU and determined an effective radius of 4.2 km  and a brightness variation of 0.4 mag. As described in \cite{Licandro2000}, their observations suffered from different technical problems, and therefore this result is uncertain. \cite{Lowry2003b} obtained a radius $\leq$ 3.4 km from observations of the still active nucleus of 137P at heliocentric distance 2.29 au. \cite{Tancredi2000} observed the comet at 5 au from the sun and estimated the effective nucleus radius to be 2.9 km. Finally, \cite{Fernandez2013} targeted the comet as part of SEPPCoN and measured an effective radius of $4.04^{+0.31}_{-0.32}$ km. 

        \cite{Snodgrass2006} obtained time-series photometry from one night on NTT/EMMI in La Silla. The data did not show brightness variation within the 3 hours of the observations and could not be used to determine the rotation rate of the nucleus. However, \cite{Snodgrass2006} used these frames to estimate the nucleus radius as 3.58 $\pm$ 0.05 km. We added 2 further nights of time-series obtained within SEPPCoN to the one night reported in \cite{Snodgrass2006} and we used the combined data set in an attempt to characterise the phase function and the rotational properties of the comet (Section \ref{sec_res_137P}).

        \subsubsection{149P/Mueller 4}
        
        149P/Mueller was among the SEPPCoN targets. The Spitzer observations revealed a nucleus with an effective radius of $1.42^{+0.09}_{-0.10}$ km \citep{Fernandez2013}. To our knowledge, no previous lightcurves of this comet are available. 
        
        In Section \ref{sec_res_149P}, we present an analysis of the optical observations taken as part of SEPPCoN. We use the data to derive the phase function of the comet and to place constraints on its shape and albedo.

    	\subsubsection{162P/Siding Spring}
        \label{sec:rev_162P}
         
        Comet 162P was discovered as asteroid 2004 TU12 but was later identified as a comet since it shows weak intermittent activity \citep[][and references therein]{Campins2006}. 
        
        \cite{Fernandez2006} analysed its thermal emission from NASA's Infrared Telescope Facility in December 2004 during the same apparition. Their measurements suggested a remarkably large nucleus with an effective radius of 6.0 $\pm$ 0.8 km \citep{Fernandez2006}. 162P was also observed within SEPPCoN. The Spitzer mid-infrared observations from 2007 provided a more precise estimate of the effective radius, $R_{\mathrm{eff}}=7.03^{+0.47}_{-0.48}$ km \citep{Fernandez2013}.  
        
        There are no published rotational lightcurves  of the nucleus of 162P to our knowledge. However, there is a well-sampled lightcurve with period $P_{\mathrm{rot}}$ $\sim$ 33 hours by the amateur observatory La Ca\~{n}ada{\footnote {\url{http://www.lacanada.es/Docs/162P.htm}}}. Those data were taken in November 2004, just a month after the discovery of the comet.
        
       In Section \ref{sec_res_162P}, we analyse two time-series data sets from 2007 and 2012. These data allow us to derive the phase function of 162P and to estimate its rotation period at two different epochs.

	\subsection{Other objects}
    \label{sec:rev_others}
        
    There are a number of objects which are not comets but  have been observed as active during multiple orbits, and therefore have been given periodic-comet designations. These objects are either Centaurs or active asteroids, and can be distinguished from JFCs dynamically using the Tisserand parameter with respect to Jupiter. 
    
    While JFCs have $ 2 \leq T_J \leq 3$, Centaurs have a Jovian Tisserand's parameter above 3.05 and semi-major axes between these of Jupiter and Neptune. The list of Centaurs with known activity includes 29P/Schwassmann-Wachmann 1, 39P/Oterma, 95P/Chiron, 165P/Linear, and 174P/Echeclus \citep[see][]{Jewitt2009a}. JFCs are likely to have originally been Centaur objects as both are believed to have evolved from the scattered disk in the Kuipter belt inwards towards the inner Solar system \citep[e.g.][]{Duncan2004,Volk2008}. However, the known active Centaurs are larger than JFCs and show mass loss at heliocentric distances larger than 5 au where water sublimation cannot be the major driving mechanism for the observed activity. This suggests that Centaurs are shaped by different processes and must be studied as a separate population.

    Active asteroids have semi-major axes   
$ a < a_J$ and  $T_J > 3.08$ \citep[see][]{Jewitt2015b}. Despite showing evidence for mass loss, these objects have typical asteroid-like characteristics such as orbital dynamics, colours, and albedos \citep[for a review, see][]{Jewitt2015b}.  Active asteroids must therefore also be considered as a separate population from JFCs, and we do not include them when considering the ensemble properties of JFCs (in Section \ref{sec:discussion}).


\begin{table*}

\caption{Summary of the properties of the comets with published rotation rates and the comets studied in this work}
\begin{tabular}{lllllllll}

\hline
Comet     & R\_eff (km) & Ref. R\_eff & $\Delta$m & Ref. $\Delta$m  & $a/b$  & Ref. $a/b$ & P\_rot & Ref. P\_rot (hr) \\
\hline
2P			& 3.95 $\pm$ 0.06		&(1)		& 0.4 $\pm$ 0.04	& (1)		& $\geq$ 1.44 $\pm$ 0.06	& (1)		& 11.0830 $\pm$	0.0030	&(1)	\\															
6P			& $\mathrm{2.23^{+0.13}_{-0.15}}$	&(2)	& 0.082	$\pm$ 0.016	& (3)	& $\geq$ 1.08				& -\textsuperscript{a}						& 6.67 $\pm$ 0.03		&(3)	\\																			
7P			& 2.64	$\pm$ 0.17		&(2)	& 0.30 $\pm$ 0.05	& (4)	& $\geq$ 1.3 $\pm$ 0.1		& (4)	& $\mathrm{7.9^{+1.6}_{-1.1}}$	&(4)	\\
9P			& 2.83	$\pm$ 0.1		& (5)		& 0.6 $\pm$ 0.2		& (6)	& 1.89\textsuperscript{b}					 & (5) 	& 41.335 $\pm$ 0.005\textsuperscript{c}  & (7)	\\	
10P			& 5.98	$\pm$ 0.04		& (8)		& 0.7				& (9)		& $\geq$ 1.9				& (9)		& 8.948	$\pm$ 0.001		& (10)	\\
14P			& 2.95	$\pm$ 0.19		& (2)						& 0.37 $\pm$ 0.05	& (*)					& $\geq$ 1.41 $\pm$ 0.06			& (*)		& 9.02 $\pm$ 0.01								& (*) 	 \\	
17P			& 1.62	$\pm$ 0.01		& (11)	& 0.30 $\pm$ 0.05	& (11)	& $\geq$ 1.3 $\pm$ 0.1 & (11)	& 7.2/8.6/10.3/12.8	& (11)	\\	
19P			& 2.5	$\pm$ 0.1		& (12)	& 0.84-1.00 		& (13)	& 2.53 $\pm$ 0.12\textsuperscript{b}	 & (12)		& 26.0 $\pm$ 1.0		& (13) \\ 
21P			& 1.0					& (14)						& 0.43				& (15)				& $\geq$ 1.5								& (15)			& 9.50 $\pm$ 0.2								& (16) 				\\ 
22P			& 2.15 $\pm$ 0.17		& (2)	& 0.55 $\pm$ 0.07	& (17)	 	& $\geq$ 1.66 $\pm$ 0.11 & (17)	& 12.30 $\pm$ 0.8		& (17) \\	
28P			& 10.7 $\pm$ 0.7		& (18)		& 0.45 $\pm$  0.07	& (19)	& $\geq$ 1.51 $\pm$ 0.07 & (19)	& 12.75 $\pm$ 0.03	& (19) \\ 
31P			& $\mathrm{1.65^{+0.11}_{-0.12}}$		& (2)	& 0.5 $\pm$  0.1	& (20)	 	& $\geq$ 1.6 $\pm$ 0.15	 & (20)	& 5.58 $\pm$ 0.03		& (20) \\ 
36P			& 2.55 $\pm$ 0.01		& (21)	& 0.7 $\pm$ 0.1		& (21)	& $\geq$ 1.9 $\pm$ 0.1	 & (21)	& $\sim$ 40			& (21) \\ 
46P			& 0.56 $\pm$ 0.04		& (22)	& 0.38				& (22)	& $\geq$ 1.4 $\pm$ 0.1	 & (22) & 6.00	$\pm$ 0.3		&(23) \\ 
47P			& $\mathrm{3.11^{+0.20}_{-0.21}}$ 	& (2)						& 0.33 $\pm$ 0.06	& (*)							& $\geq$ 1.36 $\pm$ 0.07					& (*)						& 15.6 $\pm$ 0.1 								& (*)					\\ 
48P			&  $\mathrm{2.97^{+0.19}_{-0.20}}$		& (2)	& 0.32 $\pm$ 0.05	& (24)		& $\geq$ 1.34 $\pm$ 0.06	 &(24)		& 29.00	$\pm$ 0.04		& (24) \\ 
49P		 	& 4.24 $\pm$ 0.2		& (18,25,26) &0.5 &(25)		& $\geq$ 1.63 $\pm$ 0.07	 & (25)	& 13.47	$\pm$ 0.017		& (25) \\ 
61P			& 0.61 $\pm$ 0.03		& (27)		& 0.26				& (27)		& 	$\geq$ 1.3		& (27)			& 4.9 $\pm$ 0.2			& (27) \\ 
67P 		& 1.649 $\pm$ 0.007		& (28)		& 0.4 $\pm$ 0.07	& (29)				& 2.05 $\pm$ 0.06\textsuperscript{b}	& (28)				& 12.055 $\pm$ 0.001							& ESA/Rosetta 						\\ 
73P 		& 0.41	$\pm$ 0.02		& (30)		& - & 	-									& $\geq$ 1.8 $\pm$ 0.3	 &(30)			& 3.0 - 3.4			& (31) \\ 
76P			& 0.31	$\pm$ 0.01		& (27)		& 0.56				& (27)		& $\geq$ 1.45			 & (27)			& 6.6 $\pm$ 1.0			& (27) \\ 
81P			& 1.98 $\pm$ 0.05					& (32)						&	-				& 	-								& 1.67 $\pm$ 0.04 							& (33)			& 13.5 $\pm$ 0.1								& (34)				\\ 
82P 		& 0.59 $\pm$ 0.04		& (27)		& 0.58				& (27) 		& $\geq$ 1.59			 & (27)			& $\geq$ 24 $\pm$ 5		& (27) \\ 
87P			& 0.26 $\pm$ 0.01		& (27)		& 0.94				& (27)		& $\geq$ 2.2			 & (27)			& 32 $\pm$ 9			& (27) \\ 
92P			& 2.08 $\pm$ 0.01		& (4)	& 0.6 $\pm$ 0.05	& (4)	& $\geq$ 1.7  $\pm$ 0.1	 & (4)		& 6.22 $\pm$ 0.05		& (4) \\ 
93P			& 2.59 $\pm$ 0.26					& (2)						& 0.21 $\pm$ 0.05	& (*)							& $\geq$ 1.21 $\pm$ 0.06				 			& (*)						& $\mathrm{18.2^{+1.5}_{-15}}$					& (*)							\\  
94P 		& $\mathrm{2.27^{+0.13}_{-0.15}}$ 	& (2)						& 1.11 $\pm$ 0.09		& (*)							& $\geq$ 2.8 $\pm$ 0.2				 				& (*)						& 20.70 $\pm$ 0.07								& (*)							\\  
103P		& 0.58	$\pm$ 0.018		& (35)	& --				& --					& 3.38\textsuperscript{b}				 & (35)  & 16.4 $\pm$ 0.1		& (36) \\
110P		& 2.31 $\pm$ 0.03						& (*)							& 0.20 $\pm$ 0.03	& (*)							& $\geq$ 1.20 $\pm$ 0.03				 			& (*)						& 10.153 $\pm$ 0.001 							& (*)							\\ 
121P		& $\mathrm{3.87^{+0.26}_{-0.21}}$  & (2)	& 0.15 $\pm$ 0.03	& (21)	& $\geq$ 1.15 $\pm$ 0.03 & (21)		& $\mathrm{10^{+8}_{-2}}$ 			& (21) \\ 
123P		& 2.18 $\pm$ 0.23				   	& (2)						& 0.5 $\pm$ 0.1		& (*)							& 1.6 $\pm$ 0.1				 				& (*)						& --											& -- 								\\  
137P		& $\mathrm{4.04^{+0.31}_{-0.32}}$ 	& (2)						& 0.18 $\pm$ 0.05	& (*)							& 1.18 $\pm$ 0.05				 			& (*)						& --										&  							\\  
143P		& $\mathrm{4.79^{+0.32}_{-0.33}}$		& (2)	& 0.45 $\pm$ 0.05	& (37)		& $\geq$ 1.49 $\pm$ 0.05 & (18) & 17.21	$\pm$ 0.1		& (37) \\ 
147P		& 0.21 $\pm$ 0.02		& (27)		& 0.40				& (27)		& $\geq$ 1.53			 & (27)			& 10.5 $\pm$ 1 / 4.8 $\pm$ 0.2		& (27) \\ 
149P		& $\mathrm{1.42^{+0.09}_{-0.10}}$ 	& (2)						& 0.11 $\pm$ 0.04	& (*)							& 1.11 $\pm$ 0.04				 			& (*)						& --											& --								\\  
162P		& $\mathrm{7.03^{+0.47}_{-0.48}}$ 	& (2)  					& 0.59 $\pm$ 0.04	& (*)							& $\geq$ 1.72 $\pm$ 0.06				 			& (*)						& 32.853 $\pm$ 0.002							& (*)							\\  
169P		& $\mathrm{2.48^{+0.13}_{-0.14}}$ 		& (2)	& 0.60 $\pm$ 0.02	& (38)		& $\geq$ 1.74 $\pm$ 0.03		 & -\textsuperscript{a}		& 8.4096 $\pm$ 0.0012	& (39) \\ 
209P		& $\sim$ 1.53			& (40)		& 0.4 - 0.7			& (40,41)	& $\geq$ 1.55 & (40)		& 10.93	$\pm$ 0.020		& (40,41) \\ 
260P		& $\mathrm{1.54^{+0.09}_{-0.08}}$ 		& (2)	& 0.07				& (42)	& $\geq$ 1.07			 & 		-\textsuperscript{a}			& 8.16 $\pm$ 0.24		& (42) \\ 
322P		& 0.150 - 0.320			& (43)		& $\geq$ 0.3		& (43)		& $\geq$ 1.3			 & (43)		& 2.8 $\pm$ 0.3			& (43) \\
\hline

\multicolumn{9}{l}{%
\begin{minipage}{\textwidth}~\\
    \textsuperscript{a} Calculated with Eq. \ref{eq:elong} using the brightness variation $\Delta$m. \\
    \textsuperscript{b} The exact shape model was derived by spacecraft observations in the cited paper. The provided axis ratio is obtained by dividing the highest shape model radius to the lowest one. \\
	\textsuperscript{c} The comet is known to increase its period and this is the minimum known value measured with sufficient precision. \\
    * Results derived in this work. 
    
    References:  1  \cite{Lowry2007}; 2  \cite{Fernandez2013}; 3  \cite{Gutierrez2003}; ; 4  \cite{Snodgrass2005}; 5  \cite{Thomas2013}; 6  \cite{Fernandez2003}; 7  \cite{Belton2011}; 8  \cite{Lamy2009}; 9  \cite{Jewitt1989}; 10 \cite{Schleicher2013}; 11 \cite{Snodgrass2006}; 12 \cite{Buratti2004}; 13 \cite{Mueller2002}; 14 \cite{Tancredi2000}; 15 \cite{Mueller1992}; 16 \cite{Leibowitz1986}; 17 \cite{Lowry2003}; 18 \cite{Lamy2004}; 19 \cite{Delahodde2001}; 20 \cite{Luu1992}; 21 \cite{Snodgrass2008}; 22 \cite{Boehnhardt2002}; 23 \cite{Lamy1998a}; 24 \cite{Jewitt2004}; 25 \cite{Millis1988}; 26 \cite{Campins1995}; 27 \cite{Lamy2011}; 28 \cite{Jorda2016}; 29 \cite{Tubiana2008}; 30 \cite{Toth2006}; 31 \cite{Drahus2010}; 32 \cite{Sekanina2004}; 33 \cite{Duxbury2004}; 34 \cite{Mueller2010a}; 35 \cite{Thomas2013a}; 36 \cite{Meech2009}; 37 \cite{Jewitt2003}; 38 \cite{Warner2006}; 39 \cite{Kasuga2010}; 40 \cite{Howell2014}; 41 \cite{Schleicher2016}; 42 \cite{Manzini2014}; 43 \cite{Knight2016}
    
\end{minipage}%
}\\
\label{tab:review}
\end{tabular}
\end{table*}

\section{Observations and Data Analysis}
\label{sec:methods}

\subsection{Data collection}

The main goal of this paper is to expand the sample of JFCs with known rotational  properties, in an attempt to define better constraints on their bulk properties. Below we present the optical lightcurves of nine JFC nuclei which were observed in the period 2004-2015 (Table \ref{table_all_obs}). 

Most of the data come from SEPPCoN (Survey of Ensemble Physical Properties of Cometary Nuclei). SEPPCoN surveyed over 100 comets between 2006 and 2013 in order to determine distributions of the radius, geometric albedo, thermal inertia, colours, and axis ratio of the JFC nuclei \citep{Fernandez2013}. The survey combined mid-infrared measurements from the Spitzer Space Telescope with quasi-simultaneous ground-based visible light observations from 2-8m telescopes. For a small subset of the SEPPCoN targets, the optical data sets included long time-series observations aimed at detecting the rotational variation of bare nuclei around aphelion. Here, we present the lightcurves of eight of those comets. The remaining comets had time series which were not sufficient to measure reliable brightness variations. They will be included in a further publication which will focus on the sizes, albedos and phase curves of all observed comets.

For some of the SEPPCoN comets presented below, we were also able to retrieve archival time-series from other programmes. For 14P and 94P, this included already published data from previous studies \citep{Snodgrass2005,Snodgrass2006}. These archival data sets could be consolidated with the newly obtained data, since all observations were from the same aphelion passages. All observations were analysed with our newly developed method which ensured that the combined time series from all different epochs were consistent. Combining all available data allowed us to derive more accurate lightcurves and phase functions for these two comets. 

Comet 47P was also part of SEPPCoN although it was at an unfavourable orbital configuration during the ground observing campaign. We managed to collect time series of the comet later, in 2015, when 47P was observed as a backup target of the ESO large program 194.C-0207. These data were combined with an archival data set from 2005 \citep{Snodgrass2008}.

Another major source of time-series data were the ESO observing programmes P87.C-107 and P89.C-0372. Those campaigns, led by our team, aimed to follow the same comets over an extended period in order to provide a good phase-function sampling. Despite having a different observing strategy, those datasets were suitable for the extraction of rotational lightcurves. They provided short-time series of comets 110P and 162P over the course of a few months. Although the data came from different epochs and geometries, they could be linked together owing to our specially-developed procedure for absolute photometric calibration described in section \ref{sec:data_analysis}.

\begin{table*}
\centering
\caption{Summary of all analysed observations.}
\label{table_all_obs}
\begin{tabular}{llllllllll}
\hline
Comet & UT date & $R_{\mathrm{h}}$ {[}au{]}\textsuperscript{a} & $\Delta$ {[}au{]} & $\alpha$ {[}deg.{]} & Filter & Number & Exposure time (s) & Instrument & Proposal ID \\
\hline
14P   & 2004-01-20 & 5.51\textsuperscript{O}  & 4.96 & 8.96  & R      & 29     & 220               & NTT-EMMI   & 072.C-0233(A)  \\	
      & 2004-01-21 & 5.51\textsuperscript{O}  & 4.95 & 8.87  & R      & 29     & 220               & NTT-EMMI   & 072.C-0233(A)  \\	
      & 2007-05-14 & 4.36\textsuperscript{I}  & 3.43 & 6.05  & R      & 6      & 60                & NTT-EMMI   & 079.C-0297(A)  \\	
      & 2007-05-18 & 4.35\textsuperscript{I}  & 3.41 & 5.79  & R      & 18     & 70                & WHT-PFIP   & W/2007A/20         \\	
      & 2007-05-19 & 4.34\textsuperscript{I}  & 3.41 & 5.75  & R      & 29     & 70                & WHT-PFIP   & W/2007A/20         \\	
47P   & 2005-03-05 & 5.42\textsuperscript{I}  & 4.47 & 3.49  & R      & 20     & 85                & NTT-EMMI   & 074.C-0125(A)  \\	
      & 2005-03-06 & 5.42\textsuperscript{I}  & 4.47 & 3.30  & R      & 34     & 85                & NTT-EMMI   & 074.C-0125(A)  \\	
      & 2006-06-01 & 4.96\textsuperscript{I}  & 4.23 & 8.87  & R*     & 4      & 300               & VLT-FORS2  & 077.C-0609(B)  \\	
      & 2015-04-19 & 4.55\textsuperscript{I}  & 3.64 & 5.77  & r'     & 5      & 100               & NTT-EFOSC2 & 194.C-0207(C)  \\	
      & 2015-04-21 & 4.55\textsuperscript{I}  & 3.62 & 5.40  & r'     & 7      & 150               & NTT-EFOSC2 & 194.C-0207(C)  \\	
      & 2015-04-22 & 4.55\textsuperscript{I}  & 3.61 & 5.22  & r'     & 19     & 17x80 , 2x100     & NTT-EFOSC2 & 194.C-0207(C)  \\	
      & 2015-04-23 & 4.54\textsuperscript{I}  & 3.60 & 5.04  & r'     & 21     & 20x80 , 1x120     & NTT-EFOSC2 & 194.C-0207(C)  \\	
      & 2015-04-24 & 4.54\textsuperscript{I}  & 3.60 & 4.86  & r'     & 29     & 26x80 , 3x120     & NTT-EFOSC2 & 194.C-0207(C)  \\	
93P   & 2009-01-21 & 3.79\textsuperscript{O}  & 3.25 & 13.40 & R      & 4      & 150               & WHT-PFIP   & W/2008B/23         \\	
      & 2009-01-22 & 3.80\textsuperscript{O}  & 3.24 & 13.30 & R      & 2      & 250               & VLT-FORS2  & 082.C-0517(B)  \\	
      & 2009-01-24 & 3.81\textsuperscript{O}  & 3.22 & 13.00 & R      & 8      & 250               & VLT-FORS2  & 082.C-0517(B)  \\	
      & 2009-01-27 & 3.83\textsuperscript{O}  & 3.20 & 12.50 & R      & 18     & 120               & NTT-EFOSC2 & 082.C-0517(A)  \\	
      & 2009-01-28 & 3.83\textsuperscript{O}  & 3.19 & 12.30 & R      & 29     & 120               & NTT-EFOSC2 & 082.C-0517(A)  \\	
      & 2009-01-29 & 3.84\textsuperscript{O}  & 3.19 & 12.20 & R      & 16     & 120               & NTT-EFOSC2 & 082.C-0517(A)  \\  
94P   & 2005-07-04 & 4.14\textsuperscript{O}  & 3.19 & 5.62  & r'      & 7      & 75                & INT-WFC    & I/2005A/11         \\	
      & 2005-07-05 & 4.14\textsuperscript{O}  & 3.18 & 5.37  & r'      & 17     & 75                & INT-WFC    & I/2005A/11         \\	
      & 2005-07-06 & 4.14\textsuperscript{O}  & 3.18 & 5.13  & r'      & 17     & 75                & INT-WFC    & I/2005A/11         \\	
      & 2005-07-07 & 4.15\textsuperscript{O}  & 3.18 & 4.88  & r'      & 15     & 75                & INT-WFC    & I/2005A/11         \\	
      & 2007-07-17 & 4.68\textsuperscript{I}  & 4.38 & 12.30 & R      & 1      & 750               & NTT-EMMI   & 079.C-0297(B)  \\	
      & 2007-07-18 & 4.68\textsuperscript{I}  & 4.36 & 12.30 & R      & 4      & 340               & NTT-EMMI   & 079.C-0297(B)  \\	
      & 2007-07-19 & 4.68\textsuperscript{I}  & 4.35 & 12.20 & R      & 6      & 360               & NTT-EMMI   & 079.C-0297(B)  \\	
      & 2007-07-20 & 4.68\textsuperscript{I}  & 4.33 & 12.20 & R      & 8      & 400               & NTT-EMMI   & 079.C-0297(B)  \\	
      & 2009-01-22 & 3.41\textsuperscript{I}  & 3.12 & 16.60 & R      & 6      & 120               & WHT-PFIP   & W/2008B/23         \\	
      & 2009-01-27 & 3.39\textsuperscript{I}  & 3.18 & 16.80 & R      & 6      & 100               & NTT-EFOSC2 & 082.C-0517(A)  \\	
      & 2009-01-28 & 3.39\textsuperscript{I}  & 3.19 & 16.90 & R      & 8      & 100               & NTT-EFOSC2 & 082.C-0517(A)  \\	
      & 2009-01-29 & 3.39\textsuperscript{I}  & 3.21 & 16.90 & R      & 8      & 100               & NTT-EFOSC2 & 082.C-0517(A)  \\	
110P  & 2012-06-17 & 4.51\textsuperscript{I}  & 3.73 & 9.22  & R      & 26     & 160               & NTT-EFOSC2 & 089.C-0372(A)  \\	
      & 2012-06-18 & 4.51\textsuperscript{I}  & 3.72 & 9.06  & R      & 42     & 10x250, 32x180    & NTT-EFOSC2 & 089.C-0372(A)  \\	
      & 2012-06-22 & 4.50\textsuperscript{I}  & 3.67 & 8.37  & R*     & 22     & 21x70, 1x40       & VLT-FORS2  & 089.C-0372(B)  \\	
      & 2012-06-24 & 4.50\textsuperscript{I}  & 3.65 & 8.01  & R*     & 28     & 70                & VLT-FORS2  & 089.C-0372(B)  \\	
      & 2012-07-12 & 4.47\textsuperscript{I}  & 3.50 & 4.23  & R*     & 25     & 70                & VLT-FORS2  & 089.C-0372(B)  \\	
      & 2012-07-15 & 4.47\textsuperscript{I}  & 3.48 & 3.54  & R*     & 18     & 70                & VLT-FORS2  & 089.C-0372(B)  \\	
      & 2012-07-26 & 4.45\textsuperscript{I}  & 3.44 & 1.28  & R*     & 13     & 70                & VLT-FORS2  & 089.C-0372(B)  \\	
      & 2012-08-19 & 4.41\textsuperscript{I}  & 3.47 & 5.49  & R*     & 11     & 70                & VLT-FORS2  & 089.C-0372(B)  \\	
123P  & 2007-07-17 & 5.57\textsuperscript{O}  & 4.77 & 6.92  & R      & 14     & 150               & NTT-EMMI   & 079.C-0297(B)  \\	
      & 2007-07-18 & 5.57\textsuperscript{O}  & 4.76 & 6.79  & R      & 23     & 110               & NTT-EMMI   & 079.C-0297(B)  \\	
      & 2007-07-20 & 5.57\textsuperscript{O}  & 4.74 & 6.53  & R      & 18     & 200               & NTT-EMMI   & 079.C-0297(B)  \\	
137P  & 2005-03-06 & 6.95\textsuperscript{I}  & 6.17 & 5.36  & R      & 18     & 140               & NTT-EMMI   & 074.C-0125(A)  \\	
      & 2007-05-13 & 5.26\textsuperscript{I}  & 4.25 & 0.83  & R      & 26     & 1x14, 1x30, 24x75 & NTT-EMMI   & 079.C-0297(A)  \\	
      & 2007-05-14 & 5.25\textsuperscript{I}  & 4.24 & 0.62  & R      & 31     & 1x15, 30x75       & NTT-EMMI   & 079.C-0297(A)  \\	
149P  & 2009-01-21 & 3.56\textsuperscript{I}  & 2.69 & 8.41  & R      & 8      & 60                & WHT-PFIP   & W/2008B/23         \\	
      & 2009-01-22 & 3.56\textsuperscript{I}  & 2.69 & 8.57  & R*     & 21     & 3x130, 18x80      & VLT-FORS2  & 082.C-0517(B)  \\	
      & 2009-01-23 & 3.56\textsuperscript{I}  & 2.69 & 8.73  & R*     & 19     & 4x110, 15x80      & VLT-FORS2  & 082.C-0517(B)  \\	
      & 2009-01-24 & 3.55\textsuperscript{I}  & 2.69 & 8.90  & R*     & 34     & 80                & VLT-FORS2  & 082.C-0517(B)  \\	
      & 2009-01-27 & 3.54\textsuperscript{I}  & 2.70 & 9.42  & R      & 16     & 60                & NTT-EFOSC2 & 082.C-0517(A)  \\	
      & 2009-01-28 & 3.54\textsuperscript{I}  & 2.70 & 9.61  & R      & 14     & 60                & NTT-EFOSC2 & 082.C-0517(A)  \\	
      & 2009-01-29 & 3.54\textsuperscript{I}  & 2.70 & 9.79  & R      & 36     & 60                & NTT-EFOSC2 & 082.C-0517(A)  \\	
162P  & 2007-05-17 & 4.86\textsuperscript{O}  & 4.03 & 7.51  & R      & 13     & 90                & WHT-PFIP   & W/2007A/20         \\	
      & 2007-05-18 & 4.86\textsuperscript{O}  & 4.04 & 7.69  & R      & 13     & 3x90, 10x110      & WHT-PFIP   & W/2007A/20         \\	
      & 2007-05-19 & 4.86\textsuperscript{O}  & 4.05 & 7.86  & R      & 12     & 90                & WHT-PFIP   & W/2007A/20         \\	
      & 2012-04-23 & 4.73\textsuperscript{O}  & 3.79 & 4.68  & R*     & 30     & 60                & VLT-FORS2  & 089.C-0372(B)  \\	
      & 2012-05-24 & 4.77\textsuperscript{O}  & 4.12 & 10.02 & R*     & 5      & 60                & VLT-FORS2  & 089.C-0372(B)  \\	
      & 2012-06-14 & 4.80\textsuperscript{O}  & 4.44 & 11.84 & R      & 18     & 180               & NTT-EFOSC2 & 089.C-0372(A)  \\	
      & 2012-06-17 & 4.80\textsuperscript{O}  & 4.49 & 11.97 & R      & 13     & 300               & NTT-EFOSC2 & 089.C-0372(A)  \\	
      & 2012-06-23 & 4.81\textsuperscript{O}  & 4.59 & 12.14 & R*     & 29     & 60                & VLT-FORS2  & 089.C-0372(B) 	\\        \hline
\end{tabular}
\begin{minipage}{\textwidth}
	\textsuperscript{a} Superscripts I and O indicate whether the comet is inbound (pre-perihelion) or outbound (post-perihelion). \\
	\textsuperscript{*} ESO R\_SPECIAL+76 filter with effective wavelength 655 nm and and FWHM 165.0 nm. \\
\end{minipage}
\end{table*}

\subsection{Instruments}

The lightcurve data analysed in this paper were obtained from five different instruments on four telescopes (see Table \ref{table_all_obs}). 

Comets 14P, 47P, 94P, 123P and 137P were observed using the red arm of the EMMI instrument which was mounted at the f/11 \rk{Nasmyth}-B focus of the 3.6m New Technology Telescope (NTT) at the European Southern Observatory's (ESO) La Silla site. 
The red arm of EMMI was equipped with a mosaic of two MIT/LL 2048 $\times$ 4096 CCDs. The observations were done in 2 $\times$ 2 binning mode which gave a pixel scale of 0.332 arcsec pixel\textsuperscript{-1}. The effective size of the field of view was 9.1 $\times$ 9.9 arcmin\textsuperscript{2}. All images presented here were taken with the Bessel R filter. 

EFOSC2 replaced EMMI at the \rk{Nasmyth} focus of the NTT in 2008 \citep{Buzzoni1984,Snodgrass2008a}. The effective field of view of EFOSC2 is  4.1 $\times$ 4.1 arcmin\textsuperscript{2}. It contains a LORAL 2048 $\times$ 2048 CCD which was used in a 2 $\times$ 2 binning mode with an effective pixel scale of 0.24 arcsec pixel\textsuperscript{-1}. The observations of comets 93P, 94P, 110P, 149P and 162P were taken through a Bessel R filter, while 47P was observed with an SDSS r' filter. 

Some of the data for the lightcurves of 93P, 110P, 149P and 162P were obtained with the visual and near-UV FOcal Reducer and low-dispersion Spectrograph (FORS2) instrument at ESO's 8.2 m Very Large Telescope (VLT) on Cerro Paranal, Chile \citep{Appenzeller1998}. The detector of FORS2 consists of a mosaic of two 2k $\times$ 4k MIT CCDs. The pixel scale at the default readout mode used (2 $\times$ 2 pixel binning) is 0.25 arcsec pixel\textsuperscript{-1}. The field of view of the instrument is 6.8 $\times$ 6.8 arcmin\textsuperscript{2}.

Comets 14P, 93P, 149P and 162P were observed with the 4.2m William Herschel Telescope (WHT) at the Roque de Los Muchachos observatory on the island of La Palma, Spain. The observations were done using  the Prime Focus Imaging Platform (PFIP) which contains an optical mosaic of two EEV 2k $\times$ 4k CCDs. The total field of view of the instrument is 16.2 $\times$ 16.2 arcmin\textsuperscript{2} with a gap of 9 arcsec between the two chips. Both chips were used in an unbinned mode with a pixel scale of  0.24 arcsec pixel\textsuperscript{-1}. All observations were done using CCD2, as it has fewer bad pixels and defective columns than CCD1. The filter used for the observations was Harris R with a central wavelength 640.8 nm.

Finally, the re-analysed dataset from \cite{Snodgrass2006}, used to obtain the lightcurve of 94P, was taken using the 2.5m Isaac Newton Telescope (INT) at the Roque de Los Muchachos observatory. The Wide Field Camera (WFC), mounted at the primary focus of INT, was used for the observations. The WFC is a mosaic of four thinned EEV 2048 $\times$ 4096 pixel CCDs. Only CCD3 was used for collecting the 94P time series. It has an effective field of view of 11.5 $\times$ 23 arcmin\textsuperscript{2} and the pixel scale of the instrument is 0.33 arcsec pixel\textsuperscript{-1}. All observations were done through an SDSS r' filter.

\subsection{Data reduction}

To ensure compatibility, the same reduction routine was followed consistently for each individual dataset. We performed the data reduction using standard IRAF tasks \citep{Tody1986,Tody1993} implemented on PyRAF{\footnote{\url{http://www.stsci.edu/institute/software_hardware/pyraf}}}.
A master bias frame for each night was created by using 9-19 individual bias frames. The master bias frame was then subtracted from each frame. If at least five twilight sky flats for the corresponding night were taken, the normalised sky flats were median combined. Since all used instruments have demonstrated stable night-to-night flat fields, in some cases the same flat field was used for more than one night. This was done only when there were no sky flats available for some of the nights within the same run. In the cases when no sky flats were obtained within 2 nights of the observations, dome flats were used. All science images were flat-field corrected by division to the median-combined flat field of the corresponding night. The R-band images affected by fringing were corrected using the IRAF script provided by \cite{Snodgrass.2013}.

\subsection{Data analysis} 
\label{sec:data_analysis}

In an attempt to expand the sample of comets with known rotation rates, we had to analyse archival data sets taken during different observing runs which belong to different scientific programs. This posed the challenge of combining data from different instruments and different observing geometries. In order to be able to reconcile all observations, we developed a robust method for absolute photometric calibration which uses the Pan-STARRS1 (PS1) survey \citep{Chambers2016}. The main advantage of this method is that on each frame the comet is compared to numerous neighbouring stars with precisely measured PS1 magnitudes. This provides the opportunity to calibrate absolutely the comet's magnitude even in non-photometric conditions, \rk{and allows absolute photometric calibration with uncertainties as low as 0.02 mag. }

\subsubsection{Selecting comparison stars}

The first step of our photometric calibration procedure was to identify comparison stars on the science frames. For each observing night, the comet brightness variation was determined with respect to a number of rigorously selected neighbouring stars. The selected stars had to be present on all comet frames for the corresponding night, so that we could measure the comet variation with respect to each comparison star throughout the night. We ensured that no stars located in bad sections of the CCDs were used. In order to avoid vignetting effects, all stars close to the edges of the frames were excluded, taking care that the specific limits of each instrument were respected. 

All stars used were taken from the Pan-STARRS PS1 Data Release 1\footnote{\url{http://panstarrs.stsci.edu}} (DR1) archive which was publicly released on 16 December 2016 \citep[and references therein]{Kaiser2002,Kaiser2010,Chambers2016}. PS1 used a 1.4 Gigapixel camera mounted on a 1.8 metre telescope to complete a 3$\pi$ steradian survey of the sky in five broadband filters ($\mathrm{g_{P1}}$, $\mathrm{r_{P1}}$, $\mathrm{i_{P1}}$, $\mathrm{z_{P1}}$, $\mathrm{y_{P1}}$). The PS1 filter system is slightly different from SDSS, and the magnitudes from the two systems can be converted using the equations presented in \cite{Tonry2012}.

The catalogue stars were matched to the objects on our science frames after the WCS system of each frame was fixed using WCSTOOLS\footnote{\url{http://tdc-www.harvard.edu/wcstools/}}. This was done in order to maximise the number of PS1 stars identified on the frames. 

The survey provides positions and magnitudes of both stars and extended objects. To distinguish between them, we followed the PS1 DR1 guidelines for star-galaxy separation. A careful comparison of the PSF of the selected PS1 objects identified on FORS2 images confirmed that indeed the selected catalogue objects corresponded to objects with stellar profiles on the frames. This study of the 8.2m VLT telescope data allowed excellent identification of non-stellar profiles and gave us confidence that very few galaxies should be contaminating our selected comparison stars. Even if some galaxies were left in the list of selected catalogue objects, their influence would become negligible due to the large total number of comparison stars per frame (typically $>$ 20). 

To ensure that the photometric calibration is dominated by good comparison stars, we applied two additional criteria for selecting PS1 stars. We removed PS1 entries with uncertainties in the $\mathrm{r_{P1}}$, magnitude larger than 0.08 mag and used stars with colours $\mathrm{g_{P1}}$-$\mathrm{r_{P1}}$ $<$ 1.5 mag.

\subsubsection{Photometry}

To measure the frame magnitudes of the comet and the selected comparison stars, we performed circular aperture photometry. All measurements were done using the IRAF packages DIGIPHOT and APPHOT \citep{Davis1999}. 

The observations were taken with telescope tracking at sidereal rate. Exposure times were generally short enough so that the apparent motion of the comet would be less than 0.5''-0.6'' and the comet would thus remain within the seeing disk. The few frames which did not fulfil this criterion were excluded from the analysis below. Having stellar profiles for both the comet and the background comparison stars guaranteed that the adopted circular aperture photometry procedures allowed direct comparison with the catalogue magnitudes of the stars.

The aperture radius used to measure the brightness of the comet nucleus was set equal (within the nearest integer pixel) to the full width at half maximum (FWHM) of the stellar point spread function (PSF) for each frame. This approach was previously found to be optimal for maximising the signal-to-noise ratio \citep[S/N; e.g.][]{Howell1989}. This was also beneficial for slightly more crowded sky fields, as it decreased the probability that light from neighbouring stars influences the measured brightness.

To find the FWHM of the stellar PSF on each frame, we used the IRAF routine PSFMEASURE. The value for each frame was determined using the median of the measured FWHM of the best fit Gaussian profile to each of the selected comparison stars. 

The motion of the comet on the sky over the course of the observing night can be non-linear. Therefore instead of using the position of the comet predicted from its ephemeris, we determined the centre of the comet on each frame \rk{interactively} using the IRAF task IMEXAMINE. 

The main purpose of the analysis is to derive the brightness variation of the comet during the individual nights, and subsequently to combine all data points into a common lightcurve. This is best achieved by first deriving a differential lightcurve of the comet with respect to the comparison stars for each night. Then, the lightcurves from the separate nights can be calibrated absolutely by shifting all points by a factor derived from the absolute calibration of just one reference frame for each night. Taking the differential magnitude of the comet rather than absolutely calibrating each frame is a better approach since the brightness variation within each night is independent of the absolute calibration uncertainty.

The differential photometry was implemented as follows. First, once the magnitudes of the comet and the stars were determined, we calculated the differences between the comet magnitude and each star, $i$ ($\Delta m_{\textrm{comet,i}} = m_{\textrm{comet}} - m_{\textrm{i}}$). We also determined the difference in brightness between each star and the brightest non-saturated star ($\Delta m_{\ast\textrm{,i}} = m_{i} - m_{\ast}$). The brightest star was selected because it had the highest S/N.  Then, we scaled the difference of the comet and each star with $\Delta m_{\ast\textrm{,i}}$ ($\Delta m_{\textrm{frame,i}} = \Delta m_{\textrm{comet,i}} - \Delta m_{\ast\textrm{,i}}$). Finally, the differential photometry magnitude of the comet with respect to the brightest star, $m_{\textrm{comet,diff}}$, was calculated as the median of $\Delta m_{\textrm{frame,i}}$. Its uncertainty was estimated from the median absolute deviation of $\Delta m_{\textrm{frame,i}}$.

\subsubsection{Absolute calibration}

A key aspect of our method is the absolute calibration of comet magnitudes using stars from the PS1 catalogue. This procedure allows us to combine data from different observing runs with smaller systematic uncertainties than traditional absolute calibration methods (e.g. using Landolt stars).  

In order to convert the relative magnitudes of the comet to standard magnitudes, we need to derive a correction factor for each night. There are two main factors we need to take into account while deriving the conversion: 1) the colour term of the instrument set up (CCD chip and filter) with respect to the star catalogue (PS1), and 2) the zero point for each night.

\begin{figure}
    \centering
   \includegraphics[width=0.48\textwidth]{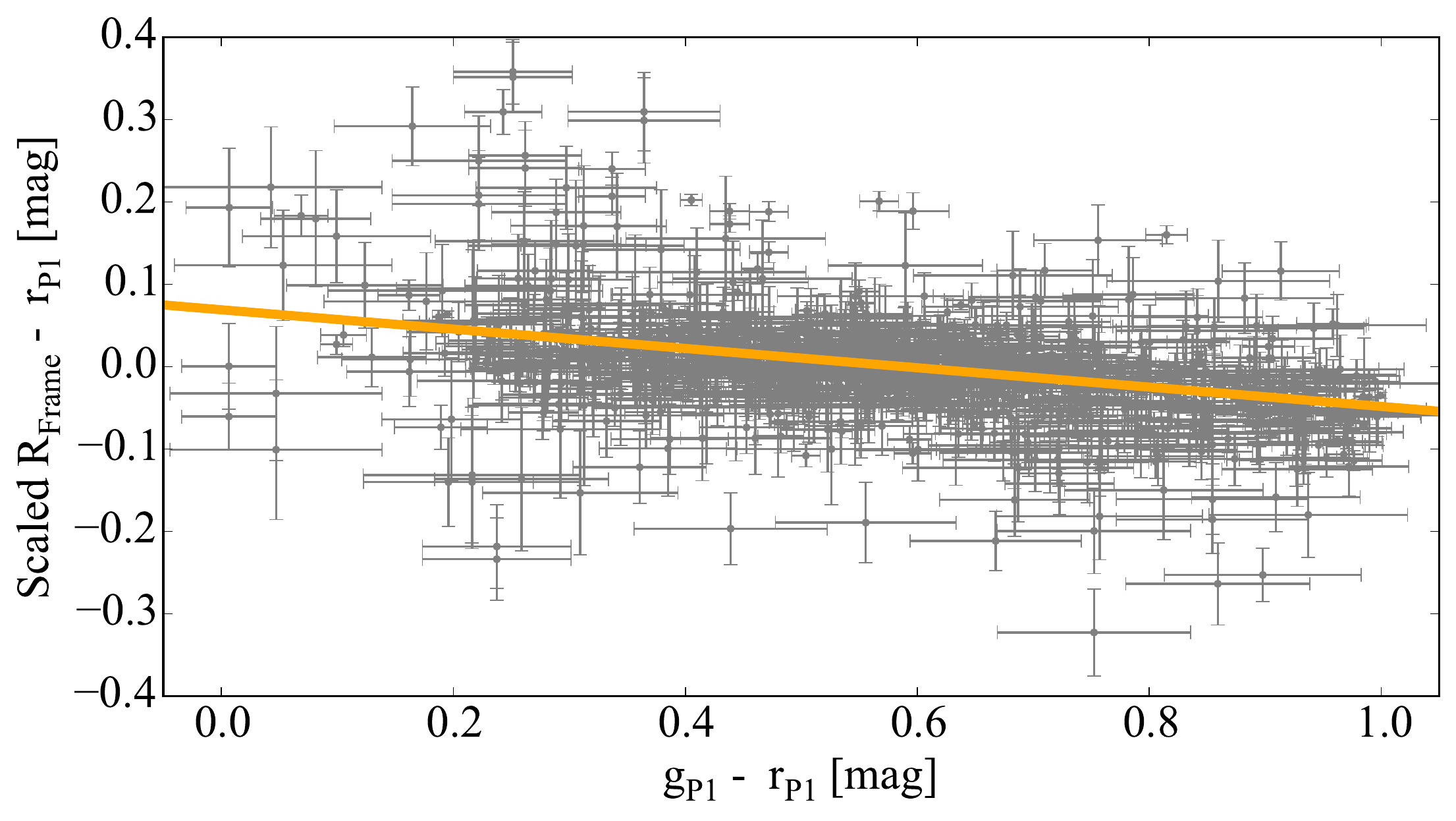}
   \caption{Colour term of the red arm of NTT-EMMI used with a Bessel R filter. The scaled difference of the measured R magnitudes and the PS1 $\mathrm{r_{P1}}$ magnitudes of the comparison stars from all used datasets are plotted against their PS1 ($\mathrm{g_{P1}}$-$\mathrm{r_{P1}}$) colour indices. The orange line indicates the best linear fit to all points and its slope corresponds to the colour term. The colour term is used to correct the magnitude of the comparison stars before finding the zero point of each frame taken with that instrument configuration. }
    \label{Colour_term_emmi}%
    \end{figure}

The colour term for each of the set ups was determined from comparison between the frame magnitudes and the PS1 magnitudes of 500-1500 stars in total. For each observing night, we chose the frame with the best seeing as a reference frame. The frame magnitudes of the comparison stars on the reference frame ($\mathrm{R_{frame}}$) were then compared to the corresponding PS1 $\mathrm{r_{P1}}$ and $\mathrm{g_{P1}}$ magnitudes. After PS1 stars with extreme colour indices ($\mathrm{g_{P1}}$ - $\mathrm{r_{P1}}$ $>$ 1.5 mag) were excluded, the differences $\mathrm{R_{frame}}$ - $\mathrm{r_{P1}}$ were plotted versus the colour indices of the stars. All points were scaled so that the median of $\mathrm{R_{frame}}$ - $\mathrm{r_{P1}}$ was brought to 0 mag. After this was done for all observed fields, all points were combined into a common plot such as Fig. \ref{Colour_term_emmi}. The colour term of the instrument was determined by taking the slope of the best fitting linear function. The derived colour indices of each instrument and their uncertainties are presented in Table \ref{table_color_ind}. 

\begin{table}
\centering
\caption{Derived colour terms for all instruments used in this work }
\label{table_color_ind}
\begin{tabular}{l|l|l|l|l}
\hline
\textbf{Instrument} & \textbf{Filter} & \textbf{c}\textsuperscript{a} & \textbf{$\sigma_\mathrm{c}$}\textsuperscript{b} & \textbf{($\mathrm{g_{P1}}$- $\mathrm{r_{P1}}$) range}\textsuperscript{c} \\
\hline
NTT-EMMI          & R               & -0.117     & 0.005               & 0.0 - 1.0              \\
NTT-EFOSC         & R               & -0.158     & 0.012               & 0.4 - 1.5*              \\
NTT-EFOSC         & r'              & -0.194     & 0.005               & 0.0 - 1.5              \\
VLT-FORS2         & R**       & -0.071     & 0.006               & 0.0 - 1.0              \\
WHT-PFIP          & R               & -0.100     & 0.008               & 0.0 - 1.0              \\
INT-WFC           & r'              & -0.007     & 0.002               & 0.0 - 1.5  \\ 
\hline
\end{tabular}
\begin{minipage}{0.48\textwidth}
	\textsuperscript{a} Colour term c derived from comparison with PS1 star magnitudes in $\mathrm{r_{P1}}$ and $\mathrm{g_{P1}}$ \\
    \textsuperscript{b} Uncertainty in the colour term \\
    \textsuperscript{c} Range of the PS1 $\mathrm{g_{P1}}$- $\mathrm{r_{P1}}$ colour indices of the used stars. Colour indices $<$ 1 for Johnson-Cousins R filters and $<$ 1.5 for SDSS r filters\\
    \textsuperscript{*} This range was selected due to an insufficient number of stars with $\mathrm{g_{P1}}$- $\mathrm{r_{P1}}$$<$ 0.4 in the observations used in this paper. \\
    \textsuperscript{**} ESO R\_SPECIAL+76 filter with effective wavelength 655 nm and and FWHM 165.0 nm. 

\end{minipage}

\end{table}

In order to use the colour term, we need to know the comet's colour index. The surface colours of JFCs are relatively well constrained with average colour indices (V - R) = 0.50 $\pm$ 0.03 and (B - V) = 0.87 $\pm$ 0.05 \citep{Lamy2009}. The (V - R) colour index can be converted to SDSS filter system: (g' - r') = 0.67 $\pm$ 0.06 using the relations in \cite{Jester2005} and to ($\mathrm{g_{P1}}$- $\mathrm{r_{P1}}$) = 0.58 $\pm$ 0.06 in the PS1 system \citep{Tonry2012}. Since no further colour information was available for most comets, this colour index was used for the absolute calibration throughout the analysis. 
 
The next step was to find the zero point of the reference frame from the difference between the colour-corrected frame magnitudes and the corresponding PS1 $\mathrm{r_{P1}}$ magnitudes. With the colour term and the zero point of the reference frame at hand, we converted the comet's magnitude to the PS1 system. Once the comet magnitude on the reference frame was converted to PS1 $\mathrm{r_{P1}}$ magnitudes, we shifted all the relative magnitudes to produce an absolutely-calibrated lightcurve of the comet for each night.

\subsubsection{Observing geometry correction}

The absolutely calibrated lightcurves from each night had to be corrected for viewing geometry effects. Firstly, we corrected each time series for light-travel time, converting ``observation times'' to ``times when the light left the nucleus''. The next step was to convert the absolutely calibrated frame magnitudes, $m_{\mathrm{r}}$, to absolute magnitudes, $m_{\mathrm{r}}(1,1,0)$. The comet magnitude, $m_{\mathrm{r}}$, depends on the observing geometry of the comet. It is given by :
\begin{equation}
m_{\mathrm{r}} = H_{\mathrm{r}} + 5\log(R_{\mathrm{h}}\Delta) + \beta\alpha,
\end{equation}
where $H_{\mathrm{r}}=m_{\mathrm{r}}(1,1,0)$ is the hypothetical absolute magnitude of the comet nucleus measured at an imaginary point at heliocentric distance  $R_{\mathrm{h}}$ = 1 au; geocentric distance $\Delta$ = 1 au and phase angle $\alpha=0^\circ$. This equation is valid for objects whose phase functions don't show an opposition surge and can be described by a linear fit with slope $\beta$. In the case of JFCs, a linear model with $\beta$ of 0.035 mag/deg is generally accepted \citep[e.g][]{Lowry2001a,Snodgrass2005}. 

For most comets we have data from different epochs, which could be used to derive a phase function slope $\beta$ independently. In all other cases, where the observations covered phase angle ranges smaller than $2^\circ$, we used $\beta$ of 0.04 mag/deg to find the absolute magnitude of the nucleus $H_{\mathrm{r}}$. For such single-run observations, we used the frame magnitude $m_{\mathrm{r}}$, rather than $H_{\mathrm{r}}$, to derive the lightcurves.

\subsubsection{Checking for activity}
\label{sec:activity_check}

To determine whether the comets were active at the time of the observations, we compared the average comet PSF profile to that of a star. We first median-combined all sky-subtracted images for the night to produce a deep image of the background stars without cosmic rays and the moving comet. We then scaled this image and subtracted it from each comet frame in order to remove the background stars. Next, we centred each difference frame on the comet and combined all frames using a median filter, removing all cosmic rays. Finally, the measured comet profile on the combined frame was compared to the PSF of a bright star measured on the combined star field image. 

In some cases described in detail below, the comet profile was noticeably different from that of the comparison stars (see. Sections \ref{sec_res_47P} and \ref{sec_res_93P}). This was interpreted as a strong indication of activity around the time of the observations. Nevertheless, we attempted to use these datasets to estimate the rotation rates and the properties of the nuclei. However, the derived results need to be interpreted with caution.

\subsubsection{\rk{Period search}}

We used the Lomb-Scargle method \citep[LS;][]{Lomb1976,Scargle1982} to detect periodicities in the brightness variation of the observed nuclei. LS is among the most widely used methods for finding periods in unevenly-sampled time series. We ran the python \texttt{gatspy}\footnote{\url{http://www.astroml.org/gatspy/}} \texttt{LombScargleFast} implementation of LS \citep{VanderPlas2015} to look for periods between 3 and 40 hours. In the cases where the nightly brightness variations suggested slower rotation, we extended the range to cover larger periods. Since we sampled a large range of possible periods, we computed the periodogram with the option of \texttt{LombScargleFast} to automatically determine the period grid. This guaranteed that the longer periods are as well-sampled as the shorter ones.  

We assumed that the brightness variation of the comets is a result of their shape rather than surface albedo variations. As the lightcurves of elongated bodies have two minima and two maxima per rotation cycle, we focused our search on double-peaked lightcurves. Experience shows that Lomb-Scargle periodograms preferentially fit single-peaked lightcurves. Therefore, we interpreted the derived peaks in the periodograms as half the rotation period of comets. 

The LS  periods were cross-checked using two other methods for detecting periods of unevenly spaced samples: phase dispersion minimization\footnote{\url{https://github.com/sczesla/PyAstronomy}}  \cite[PDM;][]{Stellingwerf1978} and string-length minimization \cite[SLM;][]{Dworetsky1983}. For all comets below, the three methods detected the same set of possible periods and showed general agreement.  Therefore, for simplicity, we have chosen to show only the LS periodograms.

\subsubsection{Nucleus size and shape and density estimates}
\label{sec:method-equations}

We used the lightcurves we derived to set constraints on the sizes, shapes and albedos of the observed nuclei. The mean apparent magnitude of the comet ($\overline{m_{\mathrm{r}}}$) and the mean absolute magnitude ($\overline{H_{\mathrm{r}}}$) were calculated as the arithmetic mean of all magnitudes $m_{\mathrm{r}}$ and $H_{\mathrm{r}}$. The uncertainty we report corresponds to the median of the uncertainties of all individual points. The mean absolute magnitude can be converted to an average radius for the nucleus in kilometres using: 
\begin{equation}
\label{eq:radius}
r_{\mathrm{N}} = (k \, / \ {\sqrt[]{A_{\mathrm{r}}}}) \times 10^{0.2(m_{\sun} - \overline{H_{\mathrm{r}}})},
\end{equation}
where k = 1.496 $\times$ $10^{8}$ km is the conversion factor between au and km; $A_{\mathrm{r}}$ is the geometric albedo of the comet  and $m_{\sun} = -27.08$ mag is the apparent magnitude of the Sun, both in PS1 $r_{\mathrm{P1}}$-band. We used the commonly assumed geometric albedo value for comets of $A_{\mathrm{r}}$ = 0.04.

The reported uncertainties on the radii are based only on the photometric uncertainty. They do not account for the uncertainties introduced by the albedo and the phase function slope. The albedos of JFCs are between 2-7 percent (see Table \ref{tab_albedo_phase}), which is within a factor of 2 of the commonly assumed value of 4 percent. Therefore, the radius estimate can vary with maximum $\sqrt{2}$ from the reported value. Since we observed all comets in a narrow phase angle range (typically $<$ 10 deg), the influence of the phase function uncertainty is also small. In the worst case, if the phase function slope varies with up to 0.08 mag/deg, the absolute magnitude of the comet will vary with 0.8 mag, and the estimated radius will be within a factor of 1.5 from the estimated value.   

Eight of the comets have SEPPCoN thermal measurements of the radii. We can use our absolute magnitudes $H_{\mathrm{r}}$ and the SEPPCoN effective radii $R_{\mathrm{eff}}$ to derive their geometric albedos using:

\begin{equation}
\label{eq:albedo}
A_{\mathrm{r}} = (k^2 \, / \ R_{\mathrm{eff}}^2 ) \times 10^{0.4(m_{\sun} - \overline{H_{\mathrm{r}}})}.
\end{equation}

The peak-to-peak variation $\Delta H_{\mathrm{r}}$ can also be used to set a lower limit on the elongation of the comet nucleus. We determined $\Delta H_{\mathrm{r}}$ by taking the observed range of magnitudes of the corresponding dataset. If the nucleus is modelled as a prolate ellipsoid with semi-axes $a$,$b$ and $c$, where $b = c$ and $a > b$, the axis ratio $a/b$ can be determined by
\begin{equation}
\label{eq:elong}
\frac{a}{b} \geq 10^{0.4\Delta H_{\mathrm{r}}} .
\end{equation}
Since we do not know the orientation of the rotational axis of any of the considered nuclei, we can only measure the projection of the axis ratio onto the plane of the sky. Therefore, Eq. \ref{eq:elong} provides only a lower limit of the elongation. 

We can also place a lower limit on the bulk density of the comets by combining the derived rotation periods ($P_{\mathrm{rot}}$) in hours and axis ratios ($a/b$). For a strengthless body, the nucleus density ($D_{\mathrm{N}}$) must be sufficient to prevent rotational break up due to centrifugal forces. In units of  $\mathrm{g \ cm^{-3}}$ this constraint can be approximated to:
\begin{equation}
\label{eq:density}
D_{\mathrm{N}} \geq \frac{10.9}{P_{\mathrm{rot}}^2}\frac{a}{b},
\end{equation}
where the period is given in  hours  \citep{Pravec2000}.

\subsubsection{Monte Carlo method}
\label{sec:MC}

Determining the uncertainty in the lightcurve period is a challenging and often neglected task. In this work, that problem is often additionally complicated by the large time span between the different observations, which leads to aliases in the periodograms. Additionally, as is shown for the individual comets below, sometimes more than one period seems to characterise the variation well, and it is not possible to decide on the most likely spin rate. In such cases, providing an uncertainty in the determined period based just on the information on the periodogram (e.g. FWHM of the highest peak) can be misleading.

Moreover, it is not clear to what extent the detected periods are influenced by the intrinsic uncertainties of the comet magnitudes. Two main effects are at play when considering what might dominate the uncertainties of the available time series. Firstly, the data from the different nights are linked using absolute calibration. In some cases the sky area under consideration has few stars, which increases the absolute calibration uncertainty. Second, when we combine observations from two different observing runs, the applied phase angle correction determines the relative difference between the comet magnitudes from the different epochs. This effect is hard to quantify, unless the influence of the different possible phase function correction parameters is explored.

In an attempt to account for these effects, we adopt a Monte Carlo method which allows us to retrieve better-validated values for the phase function coefficients and the rotation periods for the comets. The Monte Carlo method consists of the following steps:
\begin{enumerate}
\item Each magnitude from the time series of the comet is replaced by another randomly selected value. The new magnitude is selected from a normal distribution with mean equal to the original magnitude value and standard deviation equal to the uncertainty of the magnitude. The result is a clone $i$ of the original time series, where the times and observing geometries are the same as the original time series, but the magnitudes were varied within the uncertainty space.

\item The clone magnitudes are used to find the best fitting linear phase function coefficient $\beta_{\textrm{i}}$. 

\item The clone data set is corrected for the phase function by converting from $m(1,1,\alpha)$ to $m(1,1,0)$ using the derived $\beta_{\textrm{i}}$.

\item The Lomb-Scargle period search routine is run on the clone magnitudes $m(1,1,0)$ to determine the best-fitting period $P_{i}$.

\item This procedure is repeated for $i=1,2,\dots,5000$.

\item To determine the phase function coefficient, we plot the histogram of the determined $\beta_{\textrm{i}}$ and fit a gaussian probability density function to it. In the final results, we report the best fit for the phase function coefficient to be the mean of the distribution, while its uncertainty is taken to be equal to the central 3$\sigma$ range of the distribution. 

\item To determine the most likely rotation period, we plot the histogram of the derived $P_{i}$ and fit a gaussian probability density function to it. As a final result we report the period of the comet as equal to the mean of the distribution, and an uncertainty equal to the central 3$\sigma$ range of the fitted probability density function. 

\end{enumerate}

In all cases the distribution of the derived $\beta_{\textrm{i}}$ can be described well by a normal distribution. However, for some comets the $\mathrm{P_{i}}$ distributions are more irregular. In the cases when the distribution is irregular, we take the highest peak as the most-likely period candidate, but we carefully explore the alternatives in the analysis.


\section{Time Series Photometry Results}
\label{sec:results}
\subsection{14P/Wolf}
\label{sec_res_14P}

The lightcurve of comet 14P/Wolf was first determined from 2 observing nights close to aphelion in 2004 by \cite{Snodgrass2005}. Our team observed 14P as part of SEPPCoN once more in 2007 during the same aphelion passage. We analysed both datasets with our method for absolute photometric calibration and combined them in order to constrain better the comet's rotational period. 

\begin{figure}
    \centering
   \includegraphics[width=0.48\textwidth]{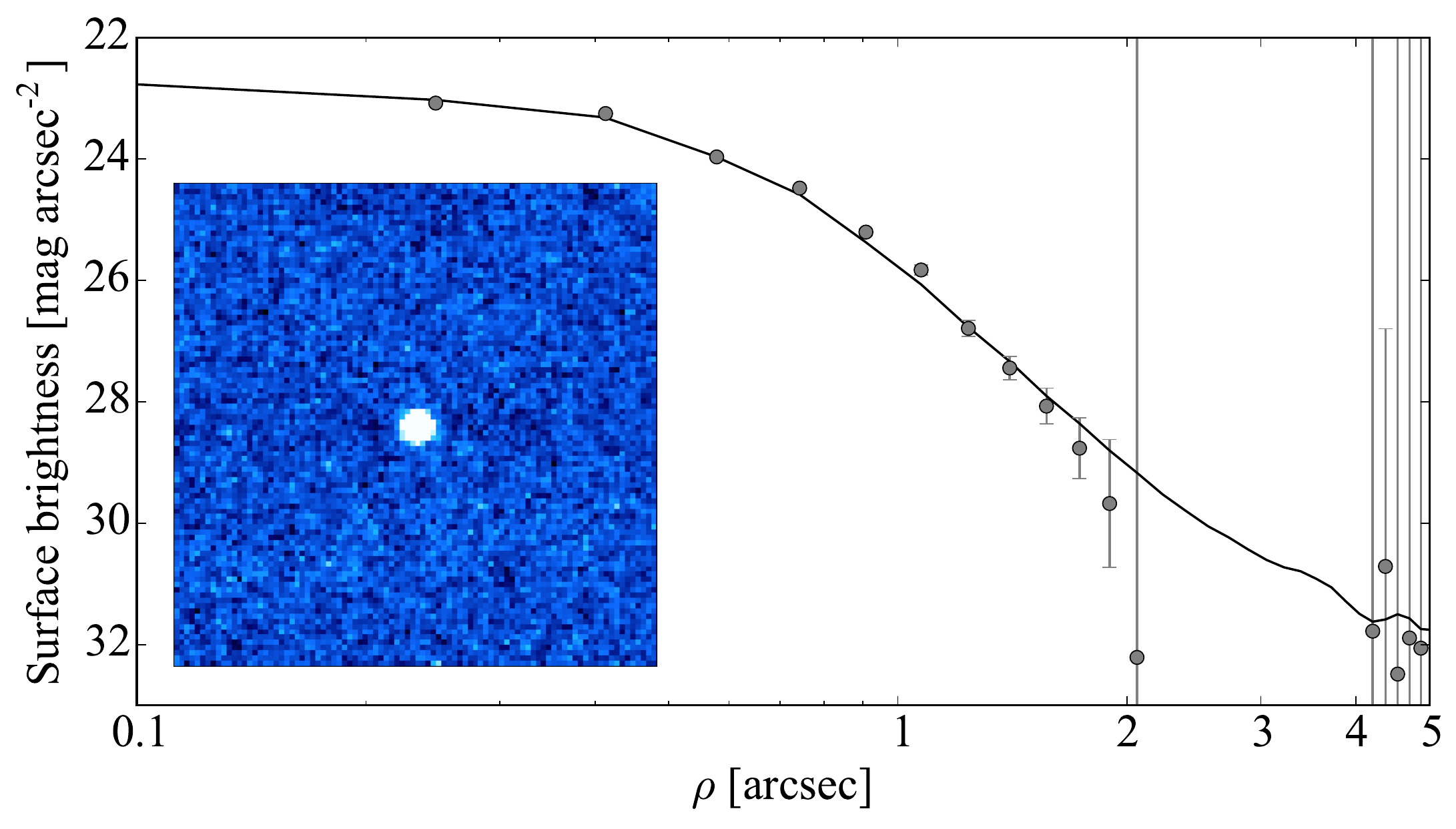}
   \caption{Surface brightness profile of 14P from the 2004 data set. The lower panel shows a 30 $\times$ 30 arcseconds composite image of 14P made up of 29 $\times$ 220 s exposures taken on 21 January 2004. The frames are added in a method which removes cosmic rays, the background sky and fixed objects. The comet appears stellar and no signatures of activity can be recognised. The surface brightness of the comet is plotted as a function of radius $\rho$ from the centre of the comet. The profile matches the scaled stellar PSF (solid line), indicating that the comet appears as a point source and is therefore considered to be inactive.}
    \label{14P_2004_PSF}%
    \end{figure}

We used the procedure described in Section \ref{sec:activity_check} to check whether 14P was active during the observations in 2004. The comet appears stellar in the co-added comet composite image and its surface brightness profile is indistinguishable from that of the comparison star (Fig. \ref{14P_2004_PSF}). This confirms the conclusion of \cite{Snodgrass2005} that 14P was not active during the observations in 2004. 

\begin{figure}
    \centering
   \includegraphics[width=0.48\textwidth]{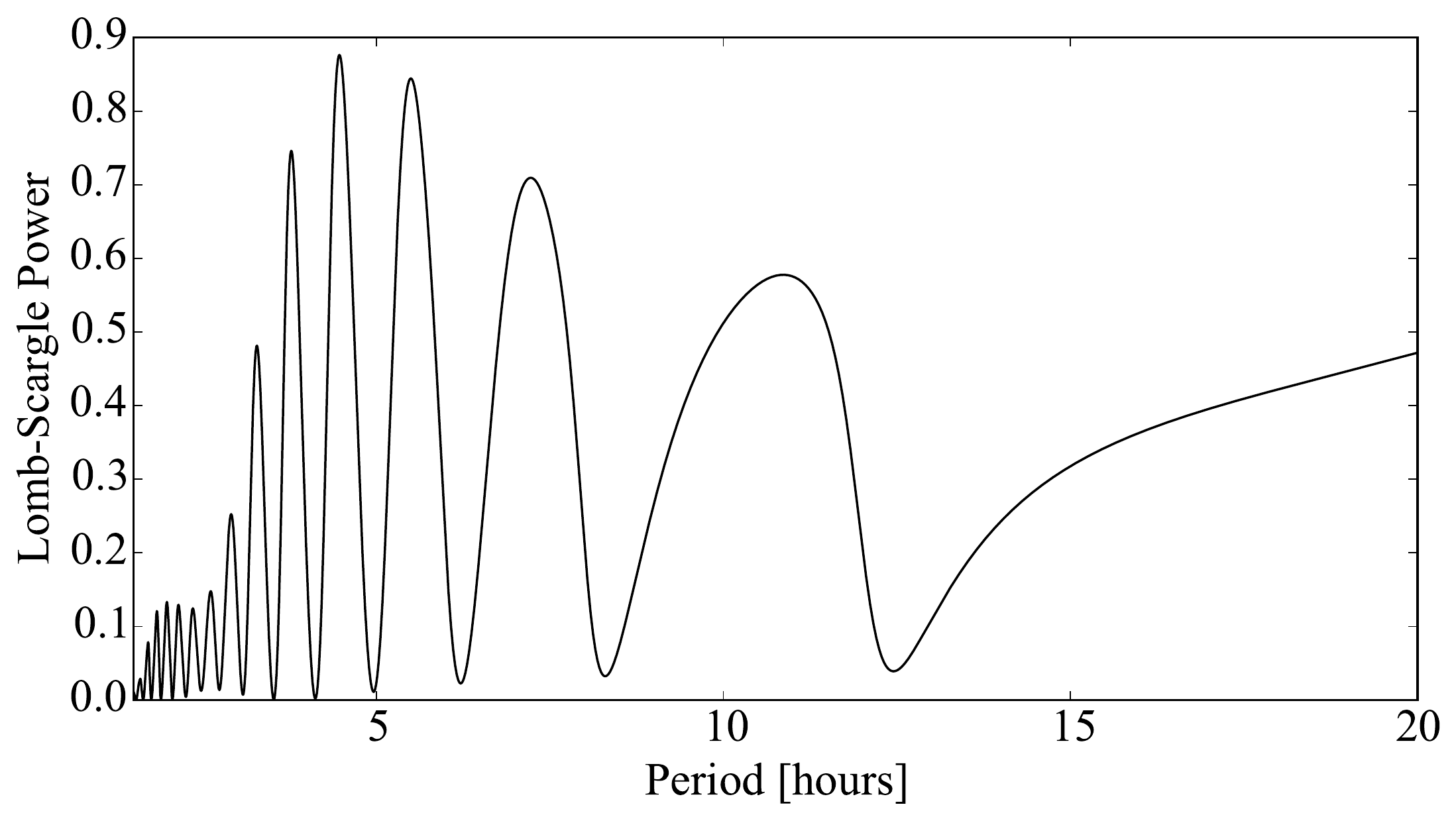}
   \caption{Lomb-Scargle periodogram for 14P from the dataset collected in 2004. The plot shows the LS power versus period. The highest peak occurs at 4.46 hours, which corresponds to the most likely period $P_{\mathrm{rot}}$ = 8.93 hours.}
    \label{14P_2004_LS}%
    \end{figure}

\begin{figure}
    \centering
   \includegraphics[width=0.48\textwidth]{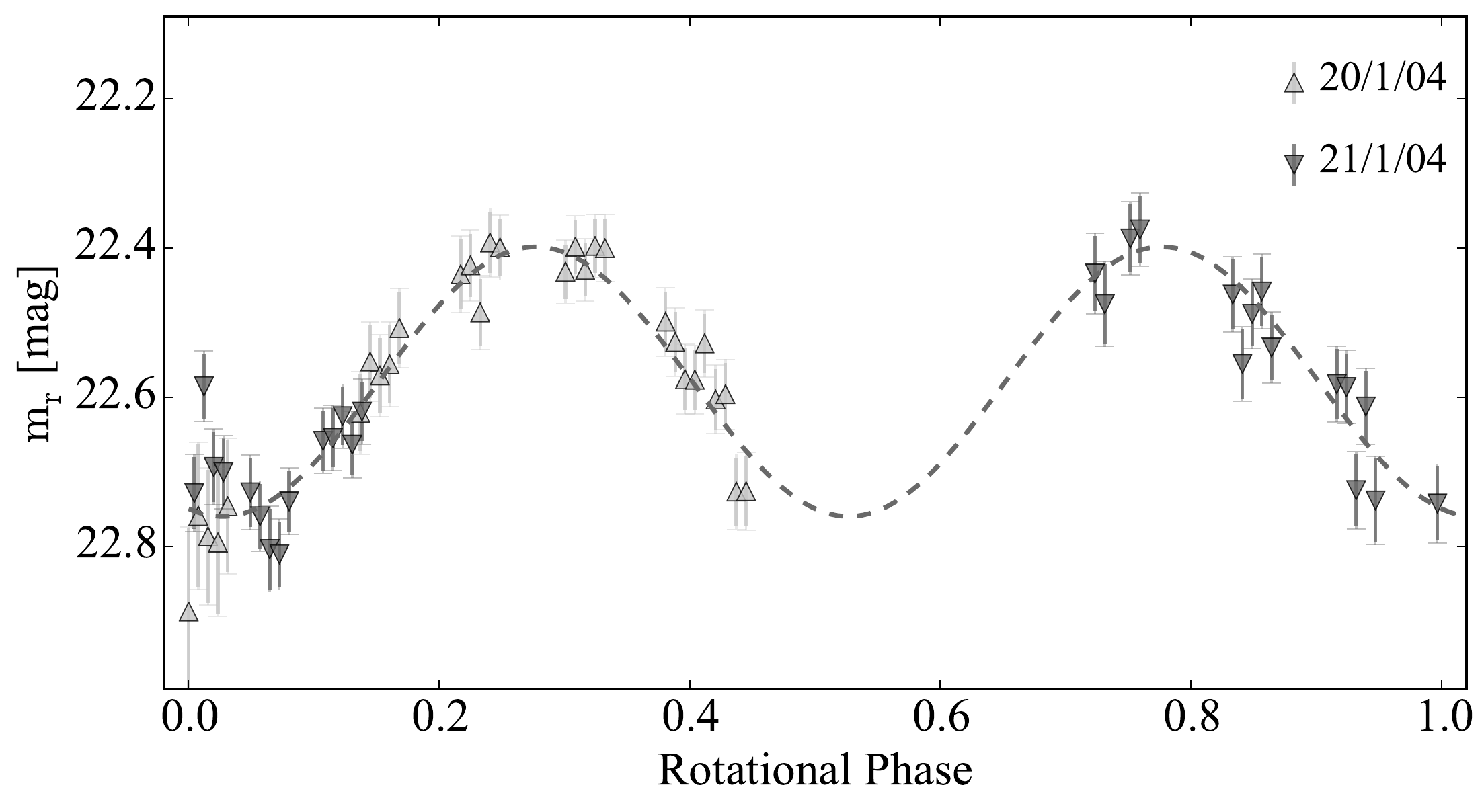}
   \caption{Rotational lightcurve of 14P with the data from 2004. The lightcurve is folded with the LS best period of 8.93 hours.}
    \label{14P_2004}%
    \end{figure}

\begin{figure}
    \centering
   \includegraphics[width=0.48\textwidth]{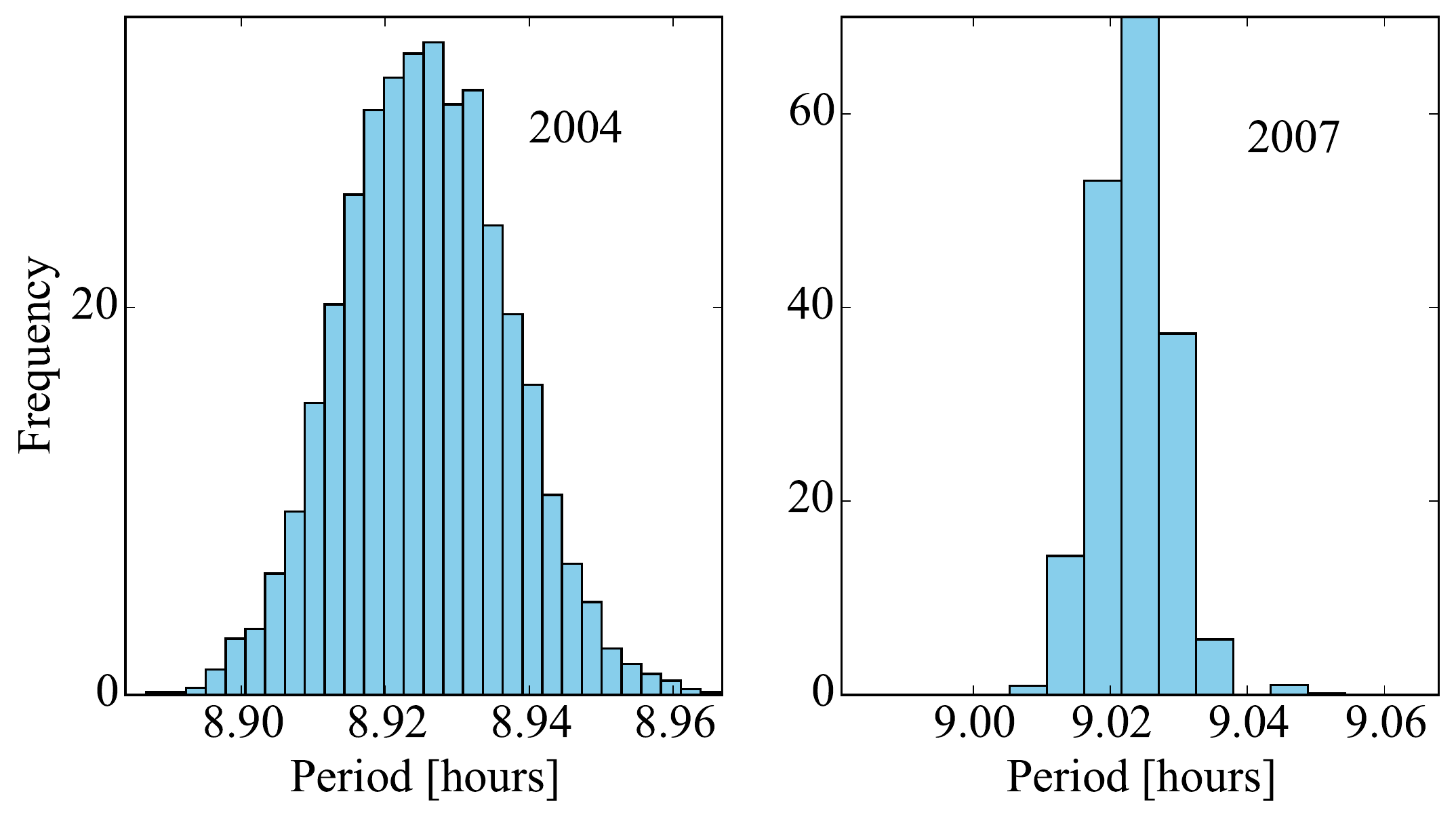}
   \caption{Results from the Monte Carlo simulations used to determine the rotation period of 14P from the datasets in 2004 (left) and 2007 (right). The resulting rotation periods for 2004 and 2007 are 8.93 $\pm$ 0.04 and 9.02 $\pm$ 0.04 respectively.}
    \label{14P_P_MC}%
    \end{figure}

Figure \ref{14P_2004_LS} shows the Lomb-Scargle periodogram for the 2004 observations of 14P. The highest peak is at $P_{\mathrm{fit}}$ = 4.46 hours, corresponding to a rotation period $P_{\mathrm{rot}}$ = 8.93 hours (Fig. \ref{14P_2004}). Using the Monte Carlo method without phase function correction, we determined that the best-fitting rotation period is $P_{\mathrm{rot}}$ = 8.93 $\pm$ 0.04 hours (Fig. \ref{14P_P_MC}).

Using the same dataset, \cite{Snodgrass2005} identified 7.53 $\pm$ 0.10 hours as the most likely rotation period of 14P. That period corresponds to the third highest peak in our periodogram and results in an unusual asymmetric lightcurve. The difference in the periods likely originates from the different methods for night-to-night calibration adopted in the two works. While \cite{Snodgrass2011} used Landolt star calibration, here we applied our newly developed method for absolute calibration with PS1, which allows precise absolute calibration independent of the changing observing conditions during the night. Thus, by re-analysing the data from 2004 with our method, we improved the period determination of 14P.

The lightcurve of 14P in 2004 phased with $P_{\mathrm{rot}}$ = 8.93 $\pm$ 0.04 hours has  a peak-to-peak brightness variation of $\Delta m_{\mathrm{r}}$ = 0.36 $\pm$ 0.05 mag, which corresponds to axis ratio $a/b$ $\geq$ 1.39 $\pm$ 0.06. From Eq. \ref{eq:density} we estimated a minimum nucleus density of 0.19 $\pm$ 0.04 $\mathrm{g \ cm^{-3}}$. 

\begin{figure}
    \centering
   \includegraphics[width=0.48\textwidth]{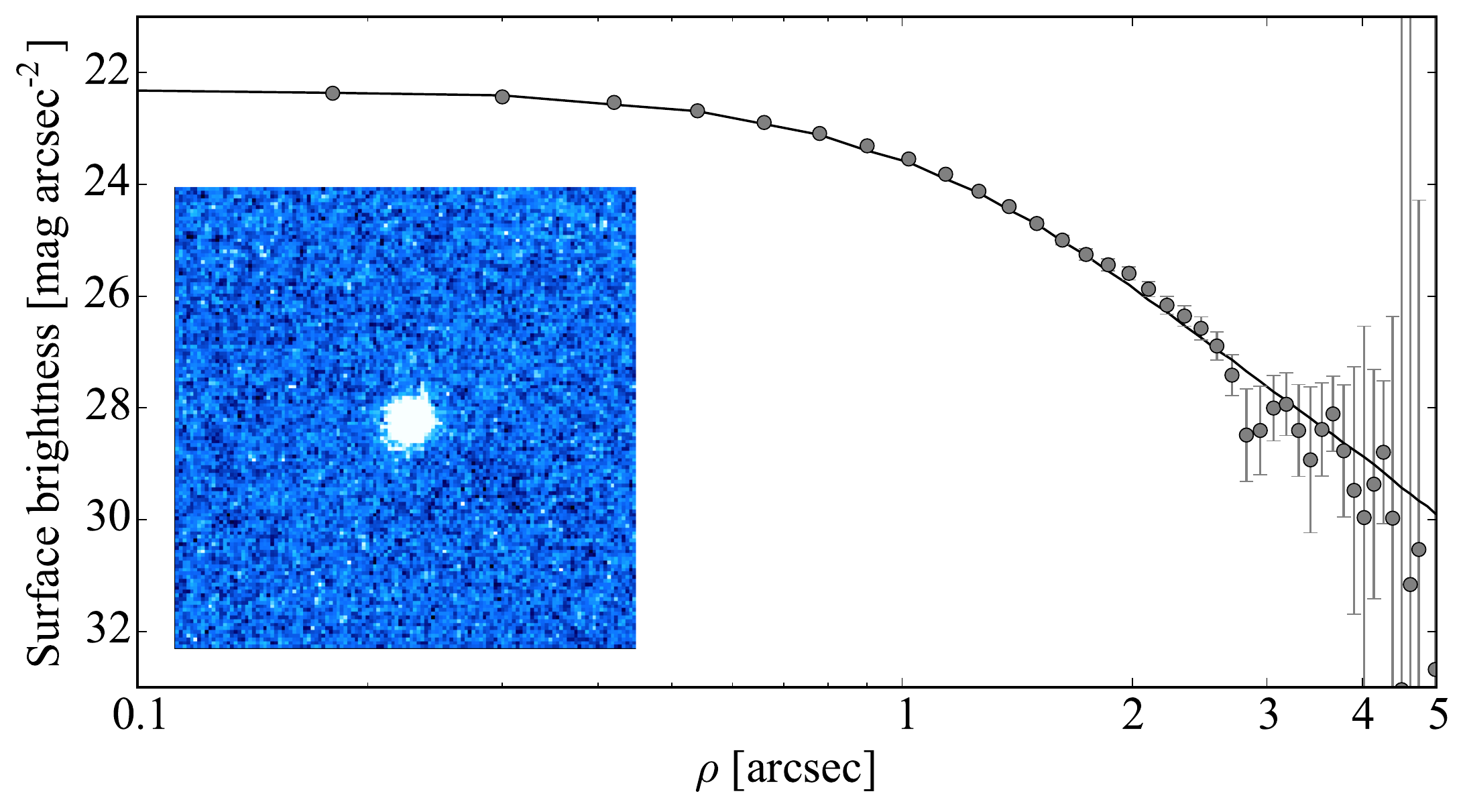}
   \caption{Same as Fig. \ref{14P_2004_PSF}, for 14P on 18 May 2007. The co-added composite image of 14P was made up of 18 $\times$ 70 s exposures. The stellar appearance on the composite image and the surface brightness profile of the comet suggest that 14P was inactive during the observations in 2007.}
    \label{14P_2007_PSF}%
    \end{figure}

Next, we analysed the observations from 2007. The comet appears stellar on the composite images and its surface brightness profile does not deviate from that of the comparison star (Fig. \ref{14P_2007_PSF}). We can therefore assume that 14P was inactive at the time of the observations. 

\begin{figure}
    \centering
   \includegraphics[width=0.48\textwidth]{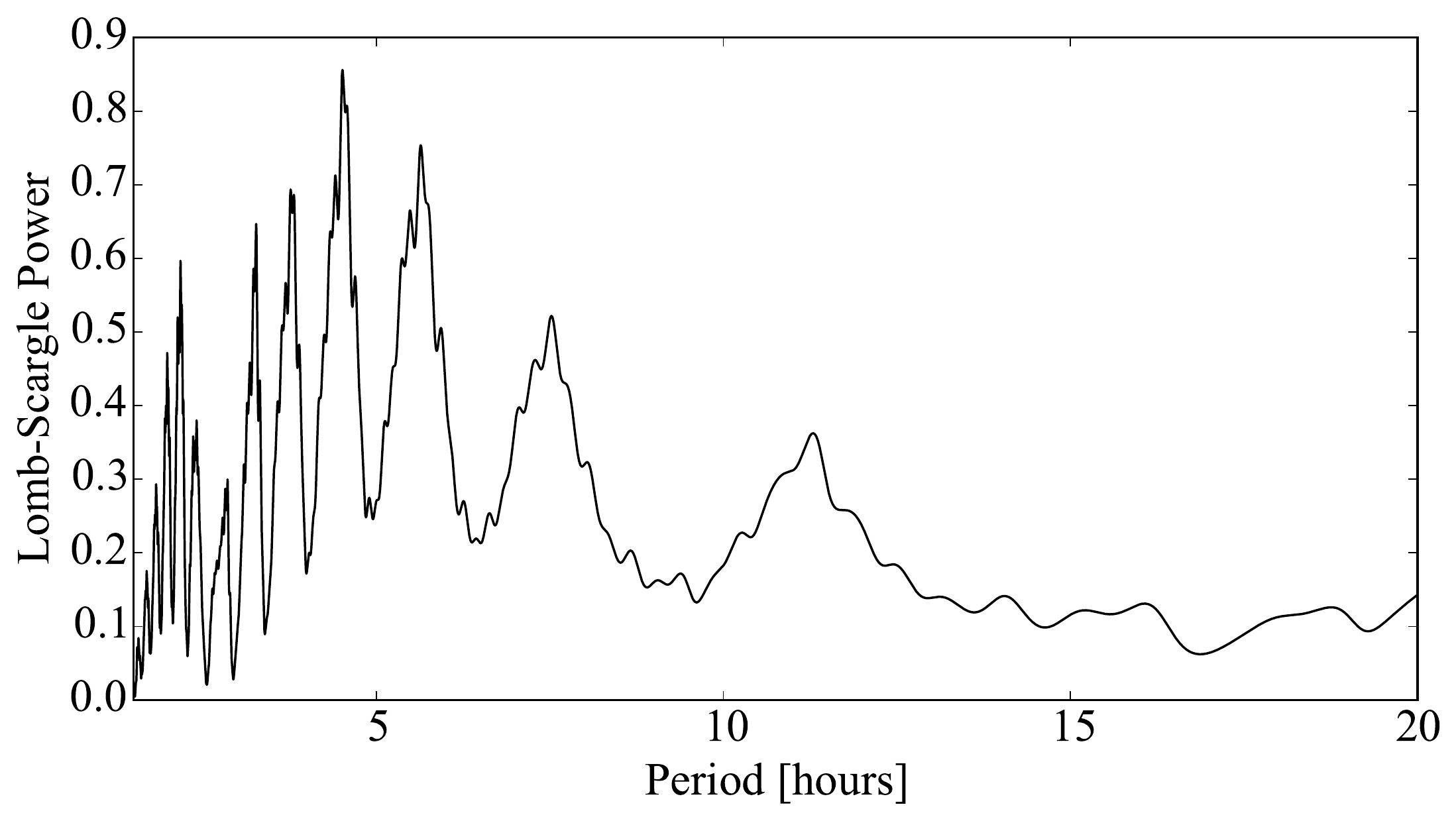}
   \caption{Lomb-Scargle periodogram for 14P with the dataset from 2007. The highest peak corresponds a period $P_{\mathrm{rot}}$ = 9.02 hours.}
    \label{14P_2007_LS}%
    \end{figure}

\begin{figure}
    \centering
   \includegraphics[width=0.48\textwidth]{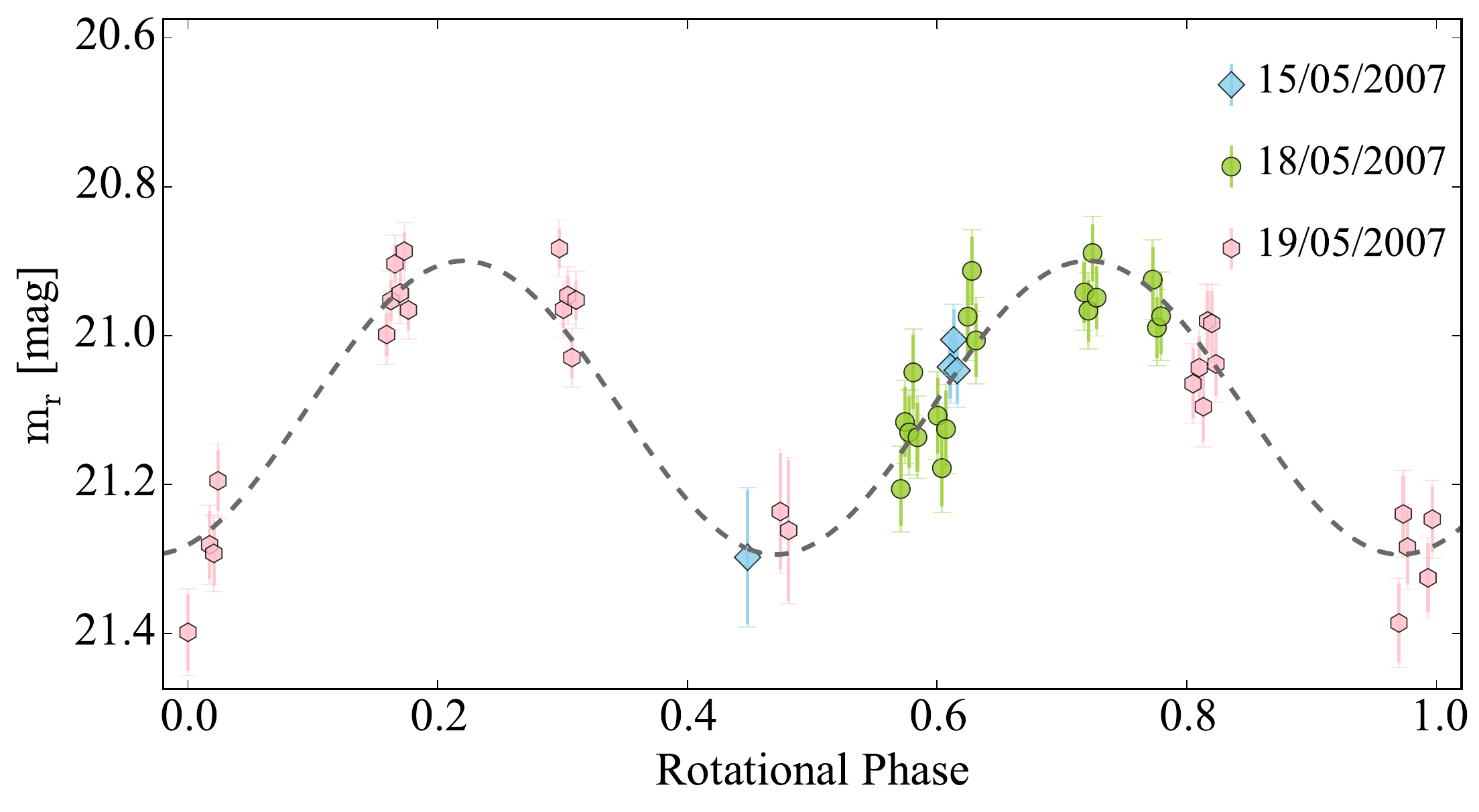}
   \caption{Rotational lightcurve of 14P with the data from 2007. The lightcurve is folded with period 9.02 hours.}
    \label{14P_2007}%
    \end{figure}

The highest peak of the Lomb-Scargle periodogram for the 2007 observations is at $P_{\mathrm{fit}}$ = 4.51 hours corresponding to a rotation period $P_{\mathrm{rot}}$ = 9.02 hours. (Fig. \ref{14P_2007_LS}). We used the Monte Carlo approach without geometric corrections to determine a rotation rate  $P_{\mathrm{rot}}$ = 9.02 $\pm$ 0.04 hours (right panel on Fig. \ref{14P_P_MC}). The lightcurve phased with the identified period (Fig. \ref{14P_2007}) has a peak-to-peak variation  $\Delta m_{\mathrm{r}}$ = 0.39 $\pm$ 0.05 mag corresponding to $a/b$ $\geq$ 1.43 $\pm$ 0.07 and $D_{\mathrm{N}}\geq 0.19 \pm 0.04$ g cm$^{-3}$.

The periods from 2004 and 2007, around the same aphelion passage, are compatible within the uncertainties. Furthermore, the fact that the comet was inactive at both epochs suggests that 14P probably remained inactive around aphelion and a period change due to outgassing is unlikely to have occurred. Since we have no knowledge about the comet spin axis orientation, it is not possible to exclude the possibility that the viewing geometry changed between the two epochs. However, both individual lightcurves have the same peak-to-peak brightness variation (within the corresponding uncertainties), and therefore we can assume that the change in geometry did not influence the observed lightcurve. With these assumptions at hand, we proceeded to combine the two datasets in order to determine a phase function and a common rotation period. 

\begin{figure}
    \centering
   \includegraphics[width=0.48\textwidth]{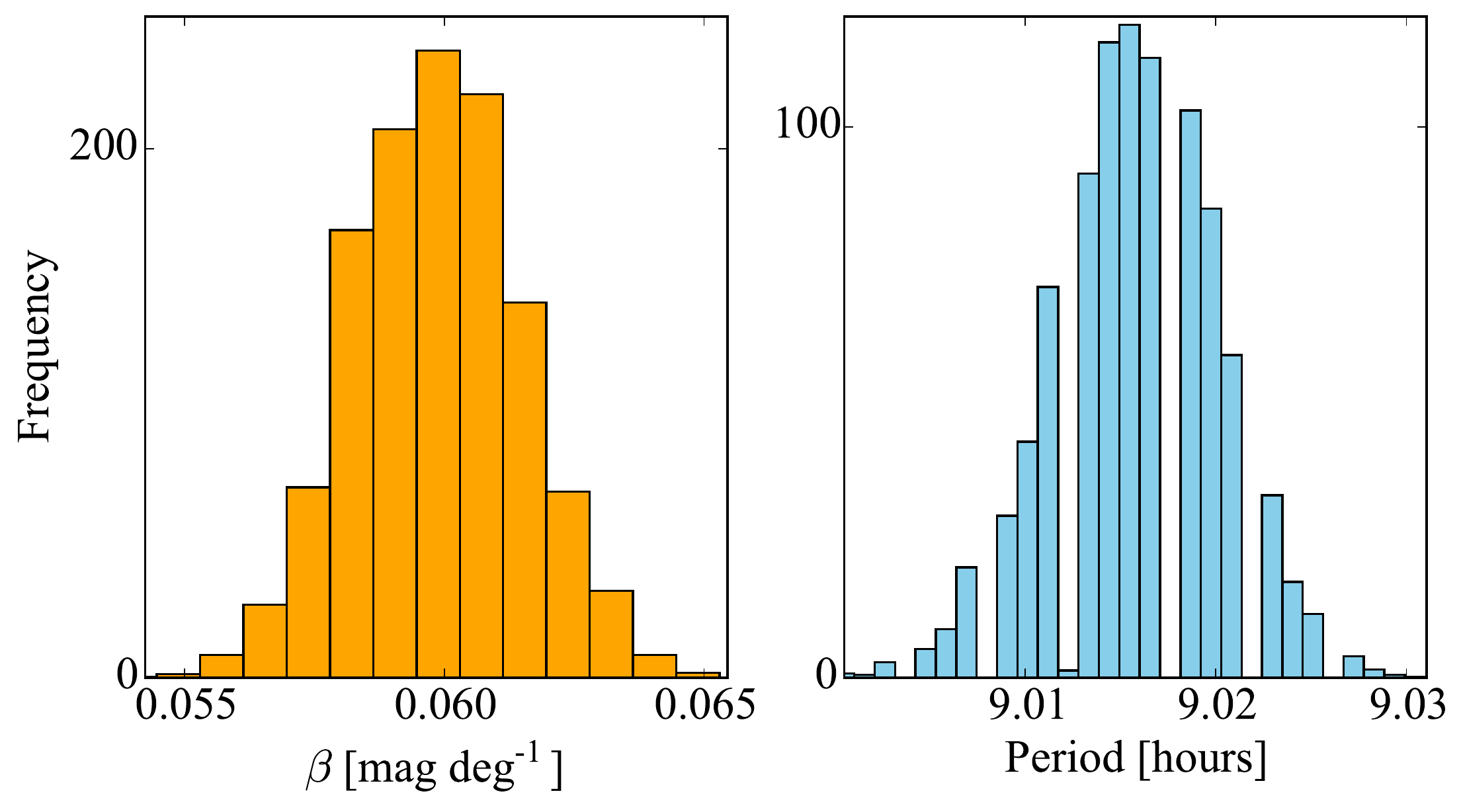}
   \caption{Monte Carlo simulation results for the phase function and the rotation period of 14P for the combined dataset from 2004 and 2007. The determined linear phase function slope is $\beta$ = 0.060 $\pm$ 0.005 (left) and the rotation period is   $P_{\mathrm{rot}}$ = 9.02 $\pm$ 0.01 hours (right).  }
    \label{14P_BOTH_MC}%
    \end{figure}

\begin{figure}
    \centering
   \includegraphics[width=0.48\textwidth]{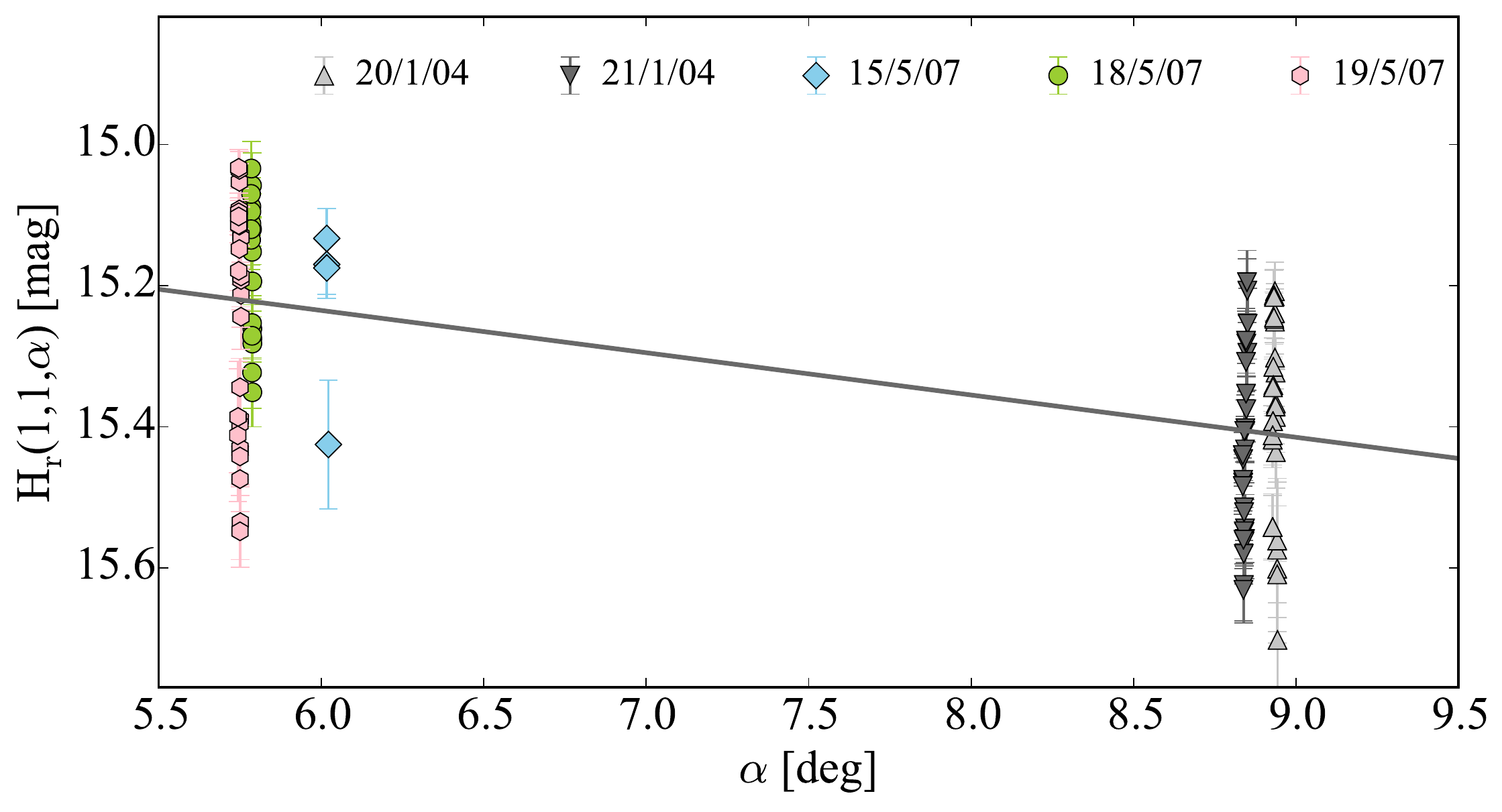}
   \caption{Phase function of comet 14P. The absolutely calibrated comet magnitudes corrected for heliocentric and geocentric distance are plotted versus phase angle $\alpha$. The linear phase function with the best-fitting slope $\beta$ = 0.060 $\pm$ 0.005 mag deg\textsuperscript{-1} is plotted as a solid line.}
    \label{14P_BOTH_PHASE}%
    \end{figure}

\begin{figure}
    \centering
   \includegraphics[width=0.48\textwidth]{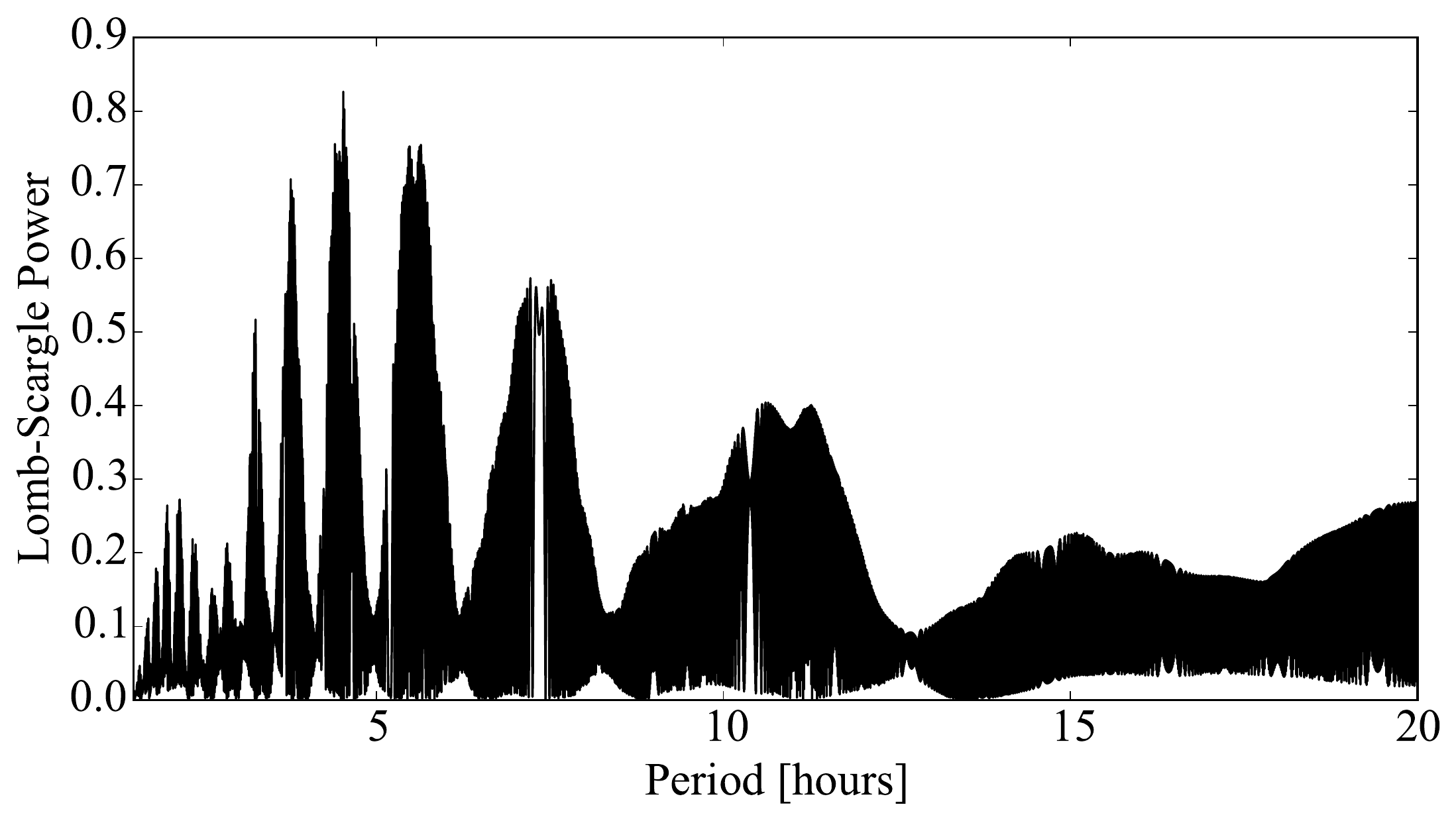}
   \caption{Lomb-Scargle periodogram of 14P with the combined datasets from 2004 and 2007. The highest peak corresponds to the most likely period $P_{\mathrm{rot}}$ = 9.02 hours. The periodogram is very densely populated with peaks from the aliases which are present due to the large time span between the two observing runs. }
    \label{14P_BOTH_LS}%
    \end{figure}

We ran the Monte Carlo simulation on the combined dataset and determined a phase function slope $\beta$ = 0.060 $\pm$ 0.005 mag/deg and period $P_{\mathrm{rot}}$ = 9.02 $\pm$ 0.01 hours (Fig. \ref{14P_BOTH_MC}, \ref{14P_BOTH_PHASE}). The Lomb-Scargle periodogram of the combined datasets (Fig. \ref{14P_BOTH_LS}) has a pronounced peak at $P_{\mathrm{fit}}$ = 4.51 hours which corresponds to the best period from the Monte Carlo simulation. 

\begin{figure}
    \centering
   \includegraphics[width=0.48\textwidth]{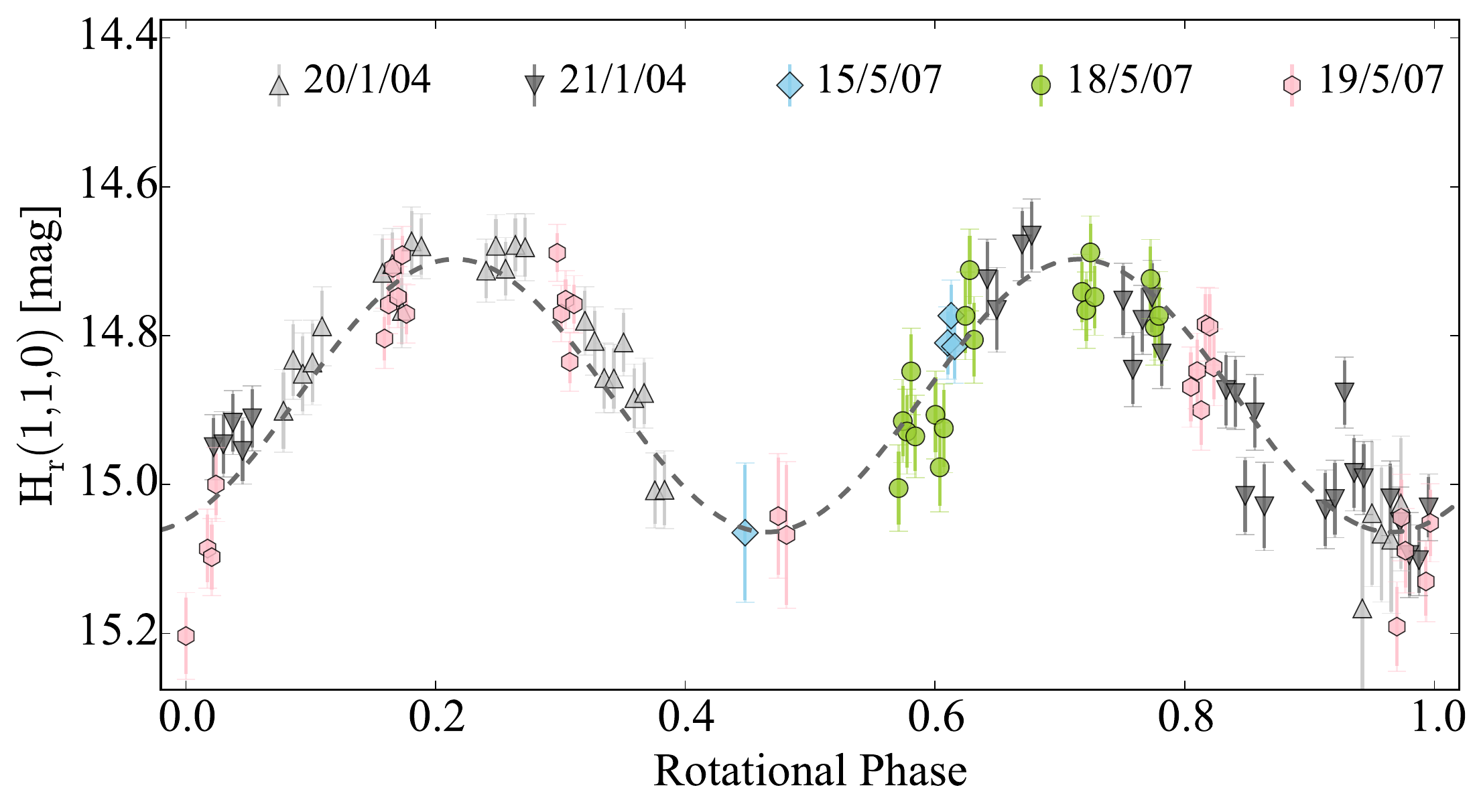}
   \caption{Rotational lightcurve of 14P/Wolf with the data from 2004 and 2007. The lightcurve is folded with period 9.02 hours.}
    \label{14P_BOTH}%
    \end{figure}

The lightcurve phased with the best period $P_{\mathrm{rot}}$ = 9.02 (Fig. \ref{14P_BOTH}) has a range $\Delta H_{\mathrm{r}}$ = 0.37 $\pm$ 0.05 mag corresponding to $a/b$ $\geq$ 1.41 $\pm$ 0.06 and $D_{\mathrm{N}} \geq 0.19 \pm 0.03$ g cm$^{-3}$. The mean absolute magnitude was $H_{\mathrm{r}}$(1,1,0) = 14.87 $\pm$ 0.05 mag. Using eq. \ref{eq:albedo} and the radius from \cite{Fernandez2013}, we estimated the comet's albedo to be $A_{\mathrm{r}}$ = 4.3$\pm$0.6\%. 

\subsection{47P/Ashbrook-Jackson} 
\label{sec_res_47P}
        
    The first attempt to determine the rotation rate of 47P was made by \cite{Snodgrass2006} using data from two observing nights in 2005. However, the resulting time series were not sufficient to choose between four possible periods: 11.2, 15.5, 21.6 and 44 hours. Moreover, as discussed in Section \ref{sec:past_47P}, the attempts to determine the comet's phase function have also remained unconsolidated \citep{Snodgrass2008,Lamy2011}. 
    
    In order to address these inconsistencies, we obtained new time-series observations of the comet in April 2015. The new data were taken at a different apparition than those from 2005, and could not be used to look for a common period without introducing further uncertainties. Nevertheless, the two datasets could still be combined for an attempt to derive the phase function of the nucleus.    
        
\begin{figure}
    \centering
   \includegraphics[width=0.48\textwidth]{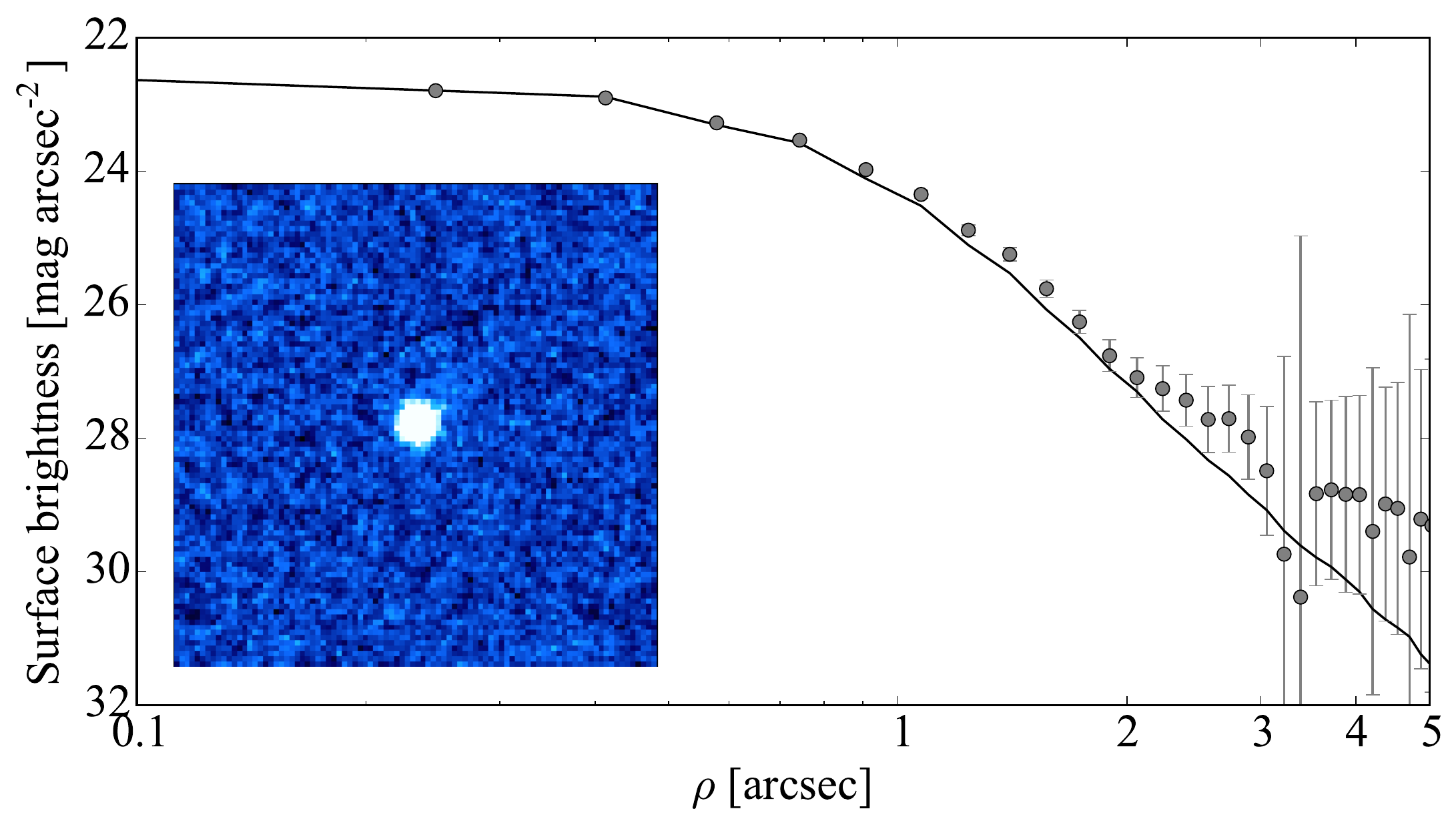}
   \caption{Same as Fig. \ref{14P_2004_PSF}, for 47P on 6 March 2005. The co-added composite image of 47P is made up of 27 $\times$ 85 s exposures. The surface brightness profile of the comet slightly deviates from the stellar one beyond 2 arcseconds, which suggests that the comet was weakly active during the time of the observations.}
    \label{47P_2005_PSF}%
    \end{figure}

   	Upon re-analysing the 2005 data set, we found that 47P was faintly active during the observing run. However, the inner surface brightness profile of the coma matched that of the comparison star well, suggesting that the activity was clearly weak (Fig. \ref{47P_2005_PSF}). 
    
\begin{figure}
    \centering
   \includegraphics[width=0.48\textwidth]{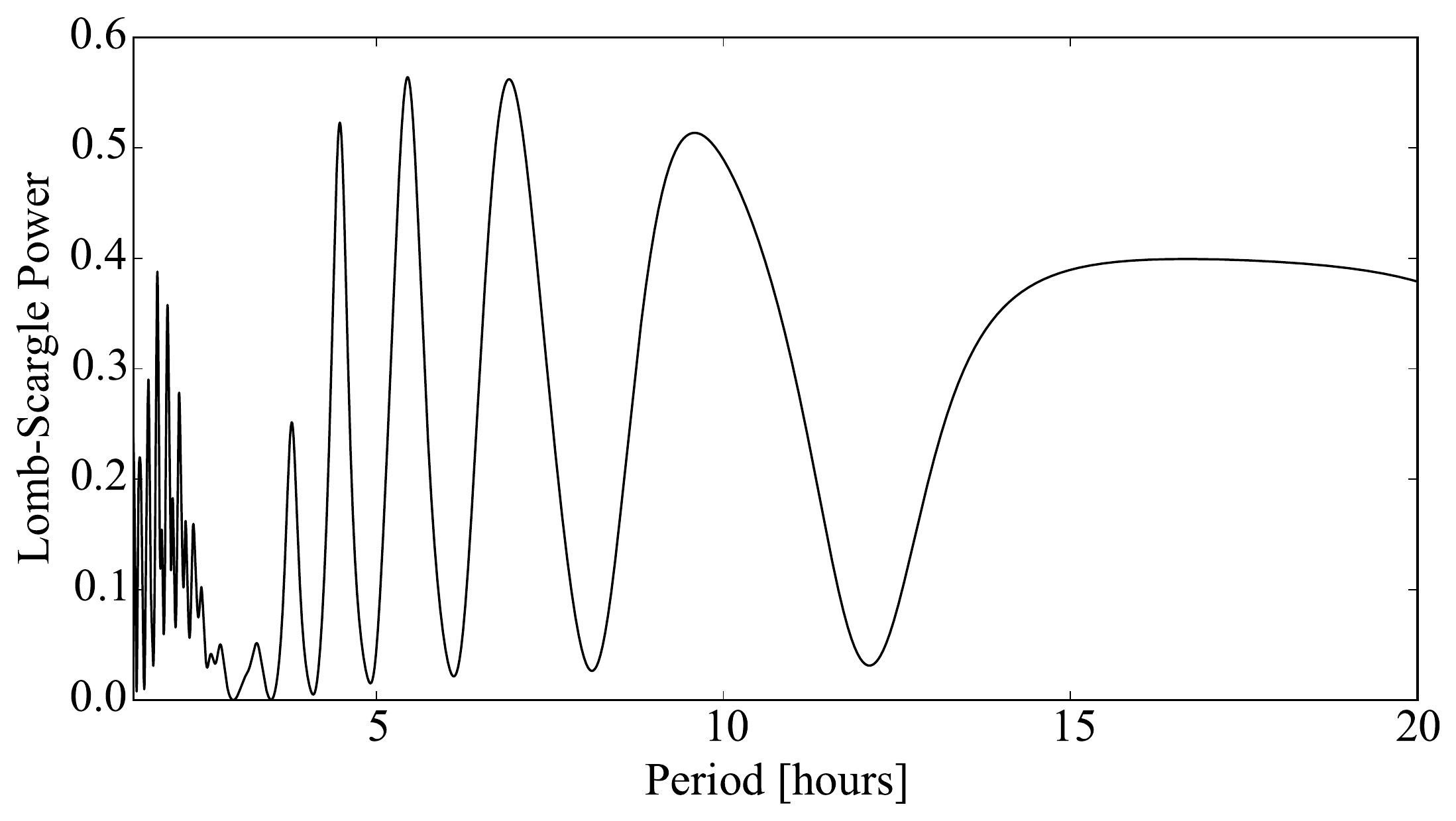}
   \caption{Lomb-Scargle periodogram of 47P with the data from 2005. The two highest peaks correspond to $P_{\mathrm{rot,1}}$ = 10.8 hours and $P_{\mathrm{rot,2}}$ = 14.1 hours.}
    \label{47P_2005_LS}%
    \end{figure}

\begin{figure}
    \centering
   \includegraphics[width=0.48\textwidth]{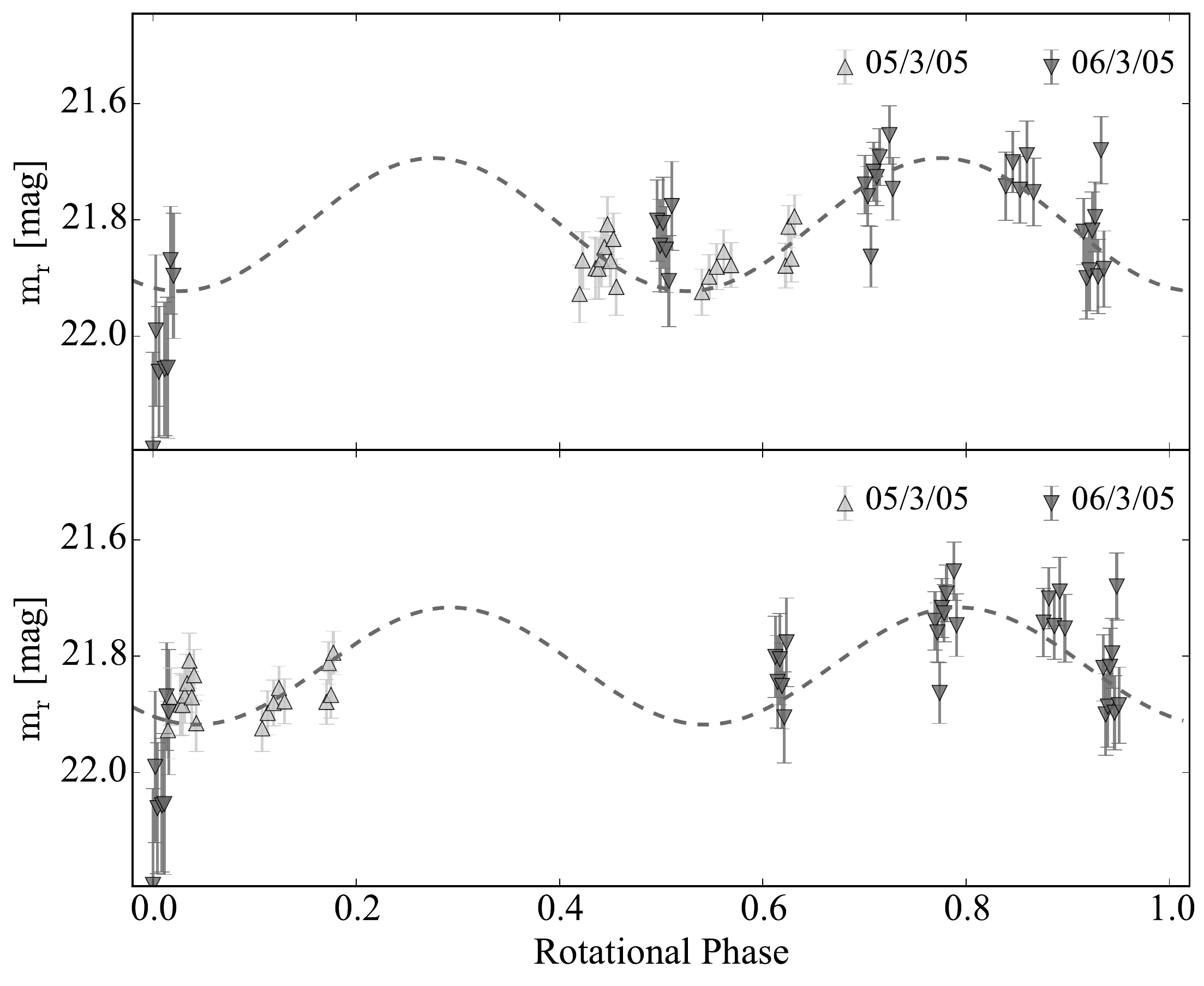}
   \caption{Rotational lightcurve of 47P with the data from 2005, folded with periods 10.8 hours (top) and 14.1 hours (bottom). It is impossible to select between these two periods.}
    \label{47P_2005}%
    \end{figure}

    We re-analysed the data from 2005 using our new absolute-photometry calibration method. The PS1 night-to-night calibration led to the identification of a smaller brightness variation and different possible periods than those in \cite{Snodgrass2006}. The two strongest peaks of our LS periodogram were at $P_{\mathrm{rot,1}}$ = 10.8 and $P_{\mathrm{rot,2}}$ = 14.1 hours (Fig. \ref{47P_2005_LS}), and it is impossible to choose between them unambiguously (Fig. \ref{47P_2005}). The brightness variation of the resulting lightcurve was $\Delta$$m_{\mathrm{r}}$ = 0.33 $\pm$ 0.06 mag suggesting axis ratio of $a/b$ $\geq$ 1.36 $\pm$ 0.07. 
    
\begin{figure}
    \centering
   \includegraphics[width=0.48\textwidth]{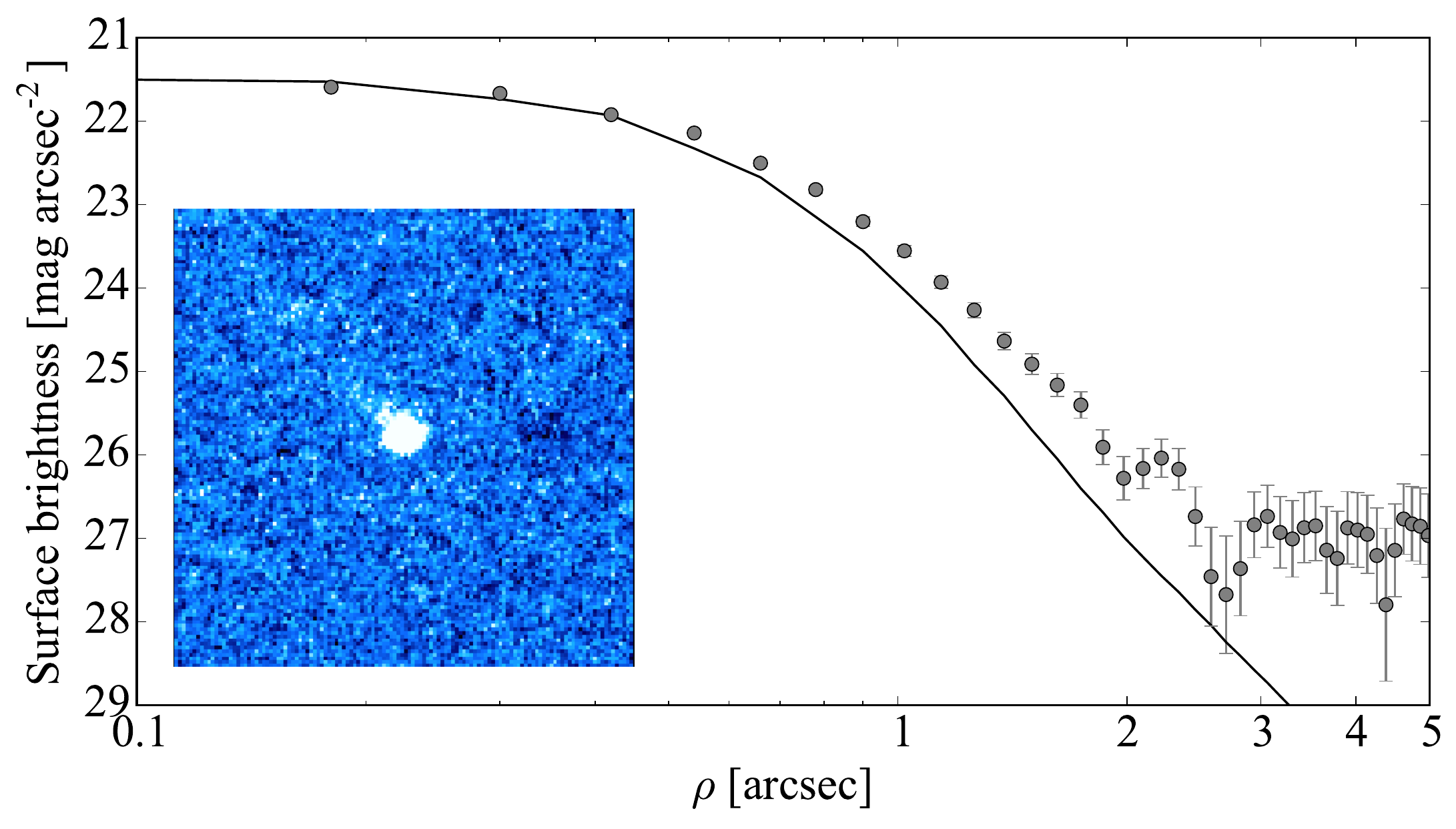}
   \caption{Same as Fig. \ref{14P_2004_PSF}, for 47P on 24 April 2015. The co-added composite image of 47P is made up of 26 $\times$ 80 s exposures. The comet appears to be slightly active with a tail detected to the north east.}
    \label{47P_2015_PSF}%
    \end{figure}

\begin{figure}
    \centering
   \includegraphics[width=0.48\textwidth]{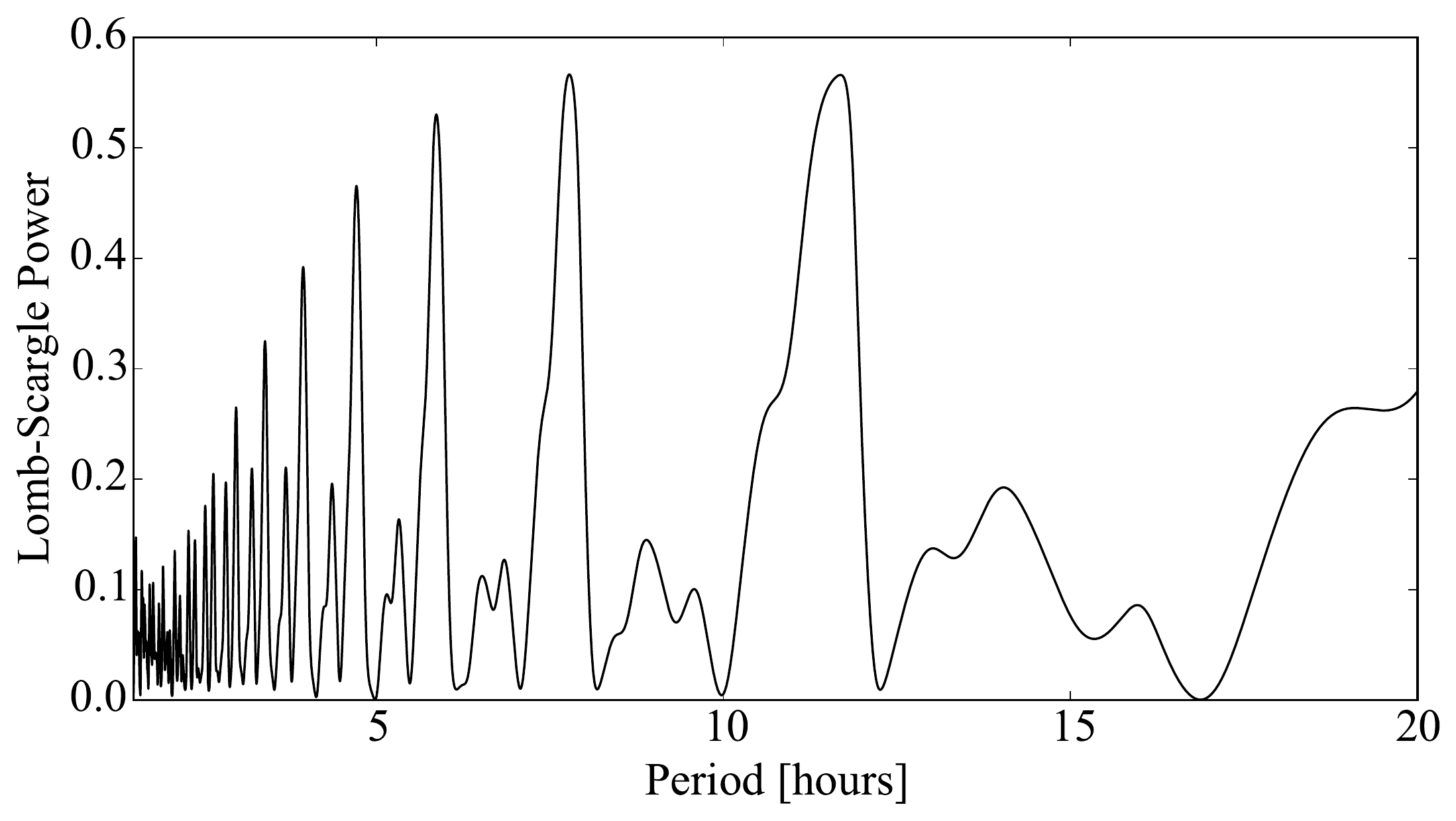}
   \caption{Lomb-Scargle periodogram of 47P with the data from 2015. The two highest peaks correspond to $P_{\mathrm{rot}}$ = 23.7 hours and $P_{\mathrm{rot}}$ = 15.6 hours, although the period of 23.7 is most likely a 24-hour alias.}
    \label{47P_2015_LS}%
    \end{figure}

    When 47P was observed again in 2015, it appeared to be slightly active (Fig. \ref{47P_2015_PSF}). Nevertheless, the new time series showed sufficient brightness variation to enable a rotation period determination. The two highest peaks on the LS periodogram of the 2015 dataset suggested $P_{\mathrm{rot,1}}$ = 15.6 hours or $P_{\mathrm{rot,2}}$ = 23.7 hours (Fig. \ref{47P_2015_LS}). However, we consider that $P_{\mathrm{rot,2}}$ = 23.7 hours is an alias due to the nightly sampling of the observations. Phasing the lightcurve of the comet with 23.7 hours produced a non-realistic noisy lightcurve, and confirmed that this period does not correspond to the rotation rate of 47P. 
    
\begin{figure}
    \centering
   \includegraphics[width=0.48\textwidth]{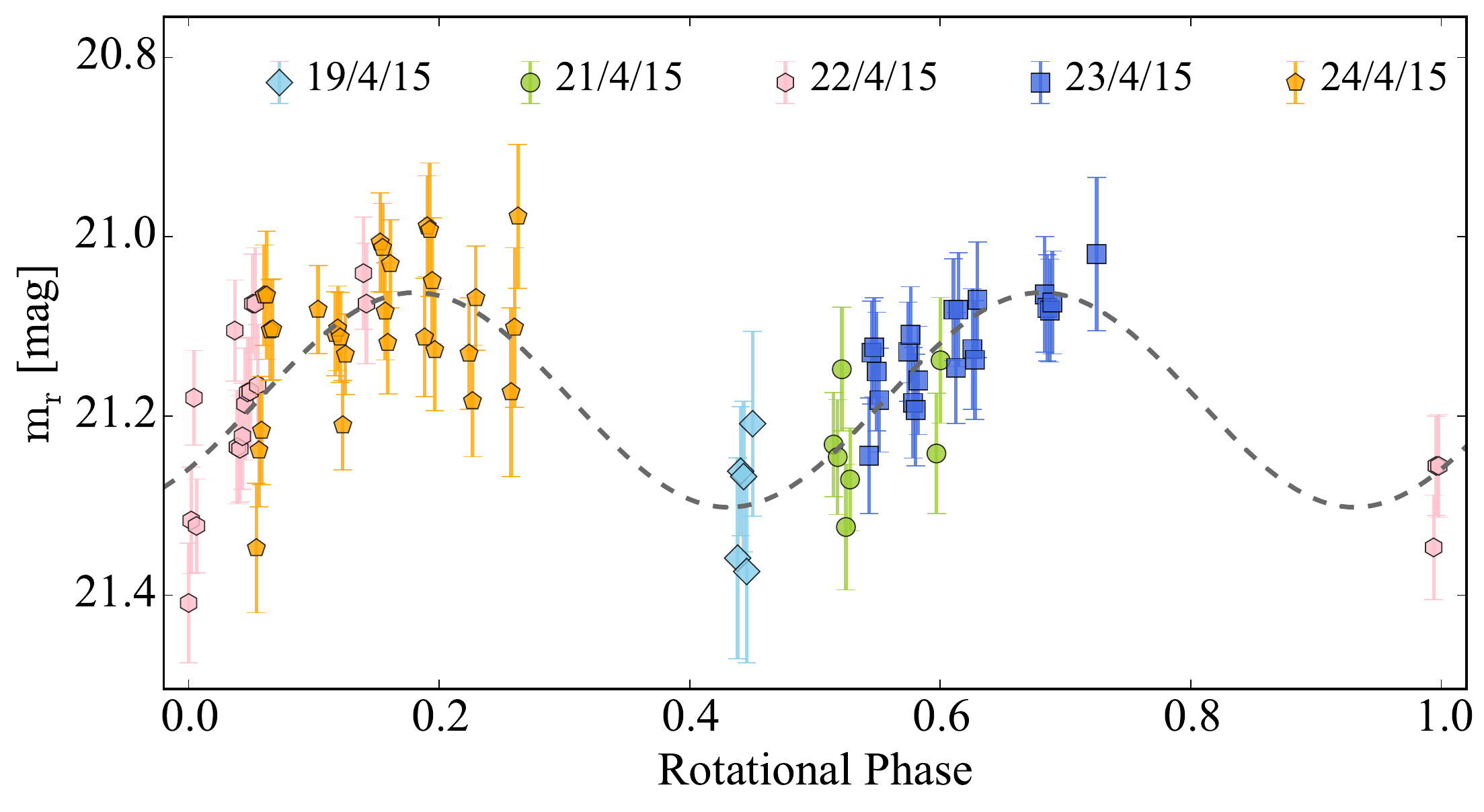}
   \caption{Rotational lightcurve of 47P with the data from 2015. The lightcurve is folded with the period of 15.6 hours derived from the MC method.}
    \label{47P_2015}%
    \end{figure}

    We ran the Monte Carlo simulation for periods between 3 and 23 hours (to avoid the 24-hour alias) and determined $P_{\mathrm{rot}}$ = 15.6 $\pm$ 0.1 hours. The resulting plots of the MC simulation here and for most objects below are not shown since they are similar to Fig. \ref{14P_BOTH_MC}, and do not provide additional information on the simulation outcomes. The brightness variation of the lightcurve (Fig. \ref{47P_2015}) was $\Delta$$m_{\mathrm{r}}$ = 0.24 $\pm$ 0.06 mag, suggesting $a/b$ $\geq$ 1.25 $\pm$ 0.07 and $D_{\mathrm{N}}$ $\geq$ 0.06 $\pm$ 0.02 $\mathrm{g \ cm^{-3}}$. 
    
    Besides deriving the lightcurve of the comet, one of the main aims of the new observations from 2015 was to constrain the phase function of 47P. To address this, we first considered the previous brightness measurements from \cite{Licandro2000}, \cite{Lamy2011} and \cite{Snodgrass2008}. Their magnitude measurements were converted to PS1 magnitudes using the colour indices of 47P (B-V) = 0.78 $\pm$ 0.08  and (V-R) = 0.40 $\pm$ 0.08 \citep{Lamy2011}, and the conversions from \cite{Tonry2012}. 
    
\begin{figure}
    \centering
   \includegraphics[width=0.48\textwidth]{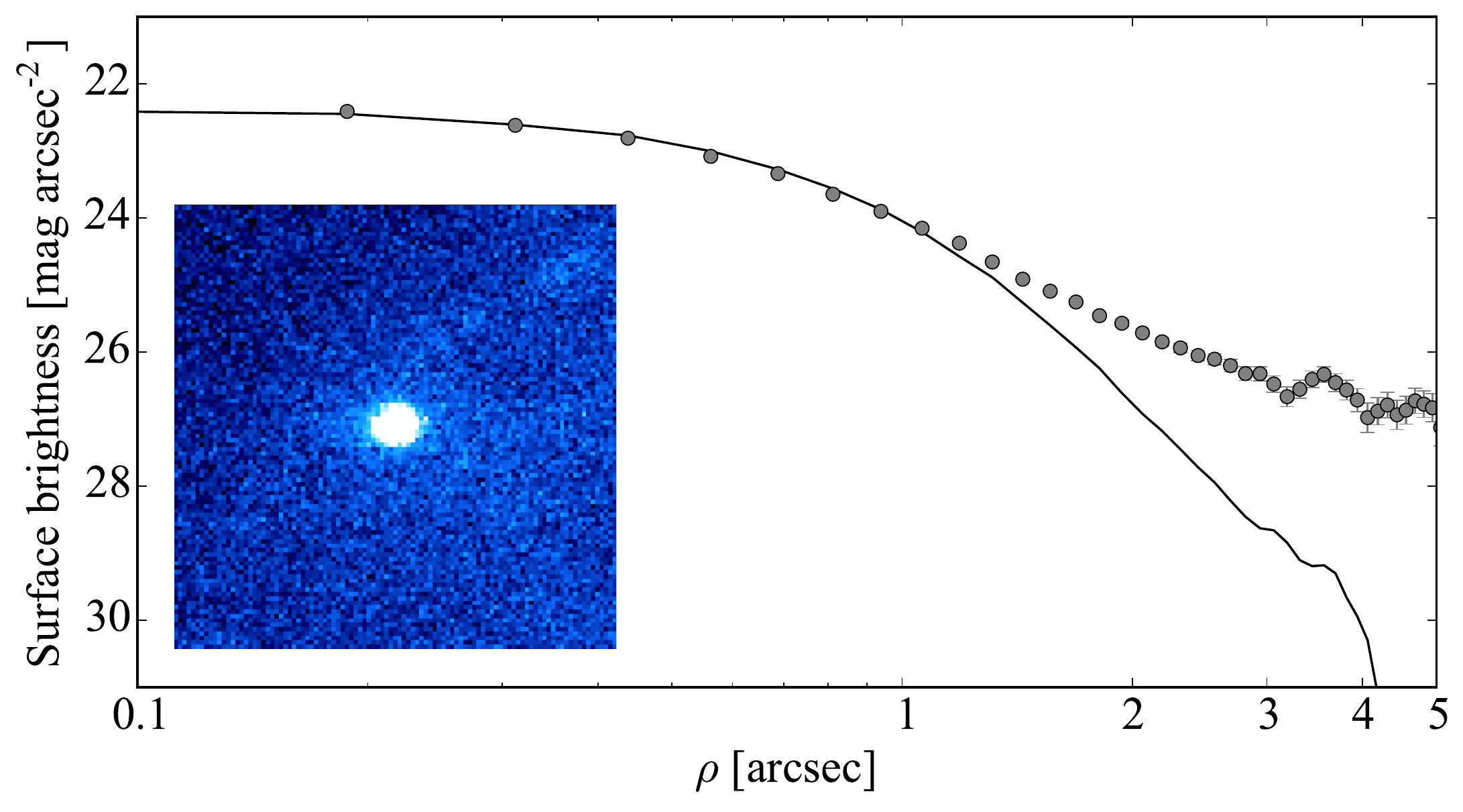}
   \caption{Same as Fig. \ref{14P_2004_PSF}, for 47P on 1 June 2006. The co-added composite image is made up of 4 $\times$ 300 s exposures. Due to the small number of frames, the composite image was made without subtraction of the average stellar background in order to avoid artefacts from the comet's slow position change. The comet appears active on the image, and its surface brightness profile deviates from the stellar PSF.}
    \label{47P_2006_PSF}%
    \end{figure}

   Additionally, we attempted to add an archival VLT data set from June 2006 when the comet was close to aphelion. However, these observations could not be used since the comet was clearly active on the frames (Fig. \ref{47P_2006_PSF}). Instead, these data complemented the data set from March 2006 \citep{Snodgrass2008}, and confirmed that the comet had an outburst around aphelion. 

\begin{figure}
    \centering
   \includegraphics[width=0.48\textwidth]{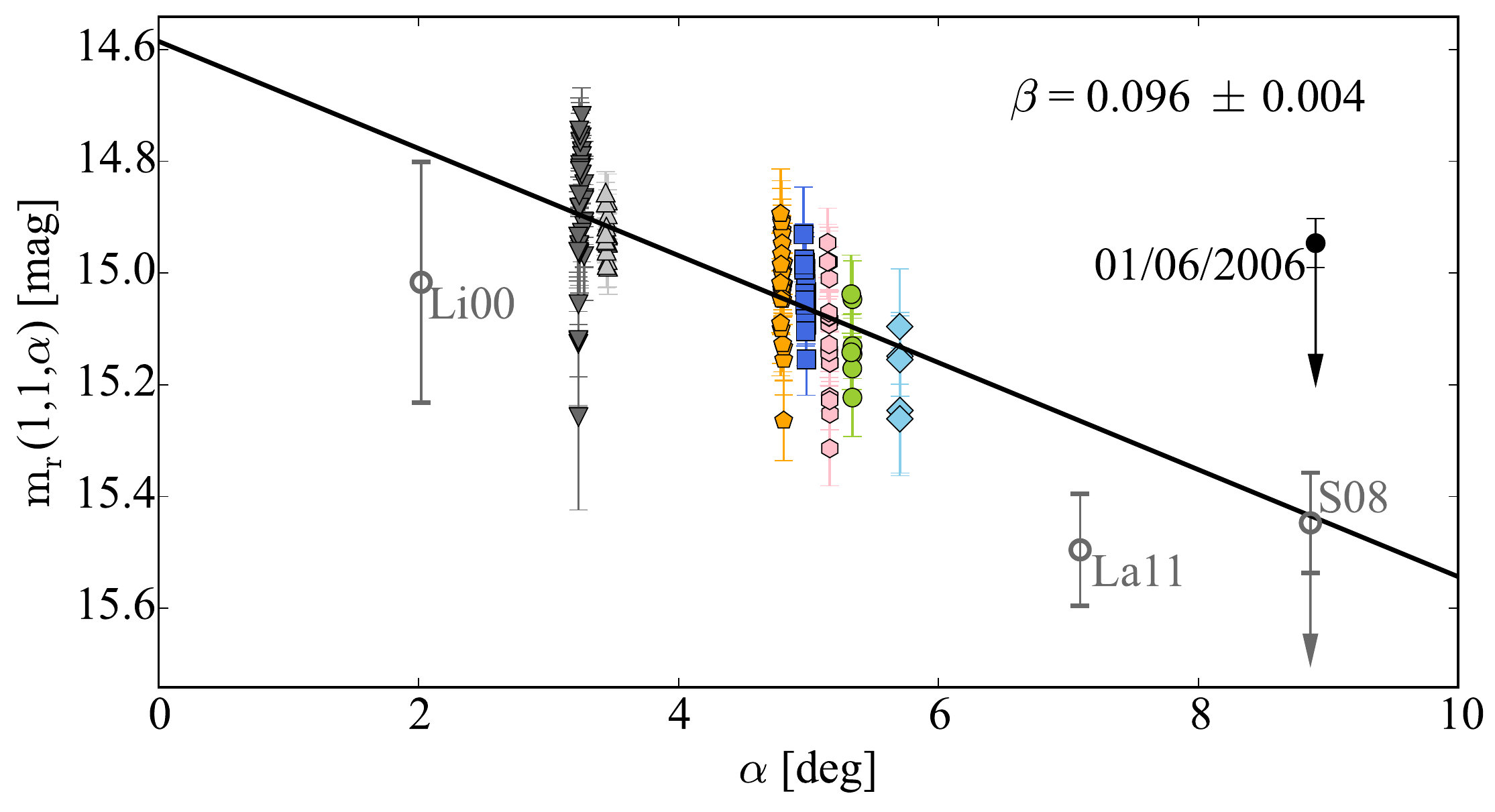}
   \caption{Phase function of comet 47P derived from the observing runs in 2005 and 2015. The symbols from 2005 and 2015 correspond to these used on Figs. \ref{47P_2005} and \ref{47P_2015}. The linear phase function slope $\beta$ determined with the MC method is 0.096 $\pm$ 0.004 mag deg\textsuperscript{-1}. Despite being unusually steep, the phase function is consistent with the previous observations of the comet from \protect\cite{Licandro2000,Snodgrass2008,Lamy2011}. However, since the comet was probably active in 2005 and 2015, the derived phase function slope is not conclusive. } 
    \label{47P_PHASE}%
    \end{figure}

    To derive the phase function coefficient $\beta$, we used the Monte Carlo approach considering only the long time-series from 2005 and 2015. We did not include the other observations where the comet was active, or where the photometric calibration had been done using different methods. The Monte Carlo method resulted in a coefficient $\beta$ = 0.096 $\pm$ 0.004 mag deg\textsuperscript{-1}. The derived phase function appears to be in good agreement with all previous observations (Fig.\ref{47P_PHASE}), although it is unusually steep compared to the typical phase function for JFCs (see Table \ref{tab_albedo_phase}).
    
    Using that value for $\beta$ to convert the observed magnitude, we calculated $H_{\mathrm{r}}$(1,1,0) = 14.59 $\pm$ 0.06 mag. Using the radius from SEPPCoN and Eq. \ref{eq:albedo}, we derived an albedo $A_{\mathrm{r}}$ = 5.0$\pm$0.7 \%. 
    
    We interpret these results with caution because of the the slight activity detected on the stacked frames from 2005 and 2015, as well as the unusually steep phase function. If the coma contribution was large and/or the actual nucleus phase function slope was shallower, we would expect the absolute magnitude of 47P to be fainter. In that case, the comet must also have a smaller albedo ($A_{\mathrm{r}}$ $\leq$ 5.0 \%). 
    
	   Similarly, the derived period  $P_{\mathrm{rot}}$ = 15.6 $\pm$ 0.1 hours must also be regarded as uncertain. The comet was found to be active at the time of the observations and therefore the nucleus signal was likely dampened by the present coma making the brightness variation more difficult to detect. Since the periods from both epochs were uncertain due to the limited sampling and the potential activity, we could not search for period changes occurring between 2005 and 2015.

\subsection{93P/Lovas}
\label{sec_res_93P}

\begin{figure}
    \centering
   \includegraphics[width=0.48\textwidth]{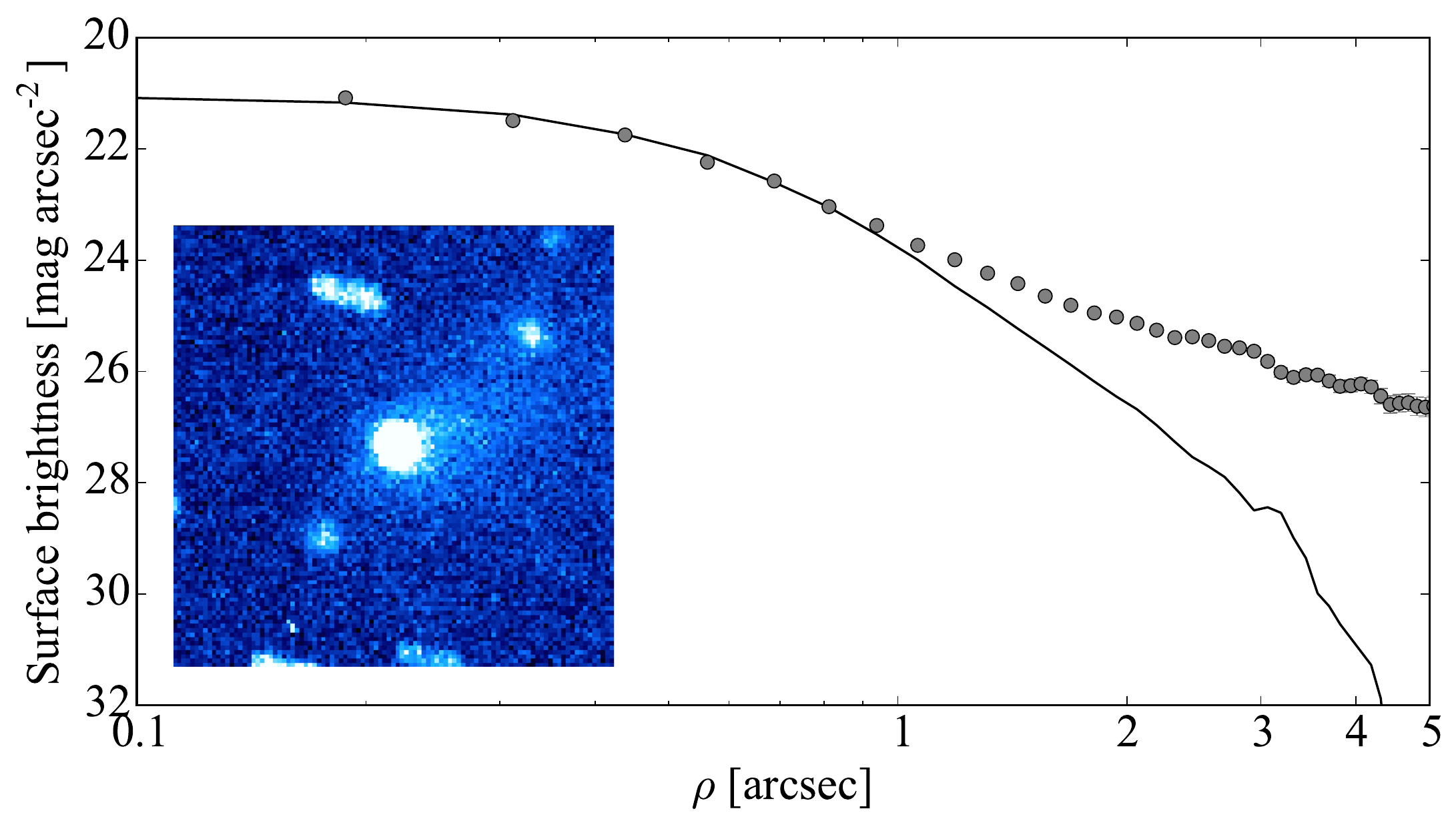}
   \caption{Same as Fig. \ref{14P_2004_PSF}, for the VLT observations of 93P on 24 January 2009. Due to the small number of frames, the composite image was made without subtraction of the average stellar background in order to avoid artefacts from the comet's slow position change. The co-added composite image is made up of 8 $\times$ 250 s exposures. A tail to the west can be clearly distinguished. The comet profile appears stellar close to the centre but deviates from that of the comparison star at larger radii.}
    \label{93P_PSF}%
    \end{figure}

	93P/Lovas was observed with three different instruments during six nights in January 2009 as part of SEPPCoN. The observations were taken at heliocentric distance of 3.8 au when 93P was outbound. The composite images of the comet from each night contained traces of activity, and a tail to the west could clearly be resolved on the VLT frames (Fig. \ref{93P_PSF}).
    
\begin{figure}
    \centering
   \includegraphics[width=0.48\textwidth]{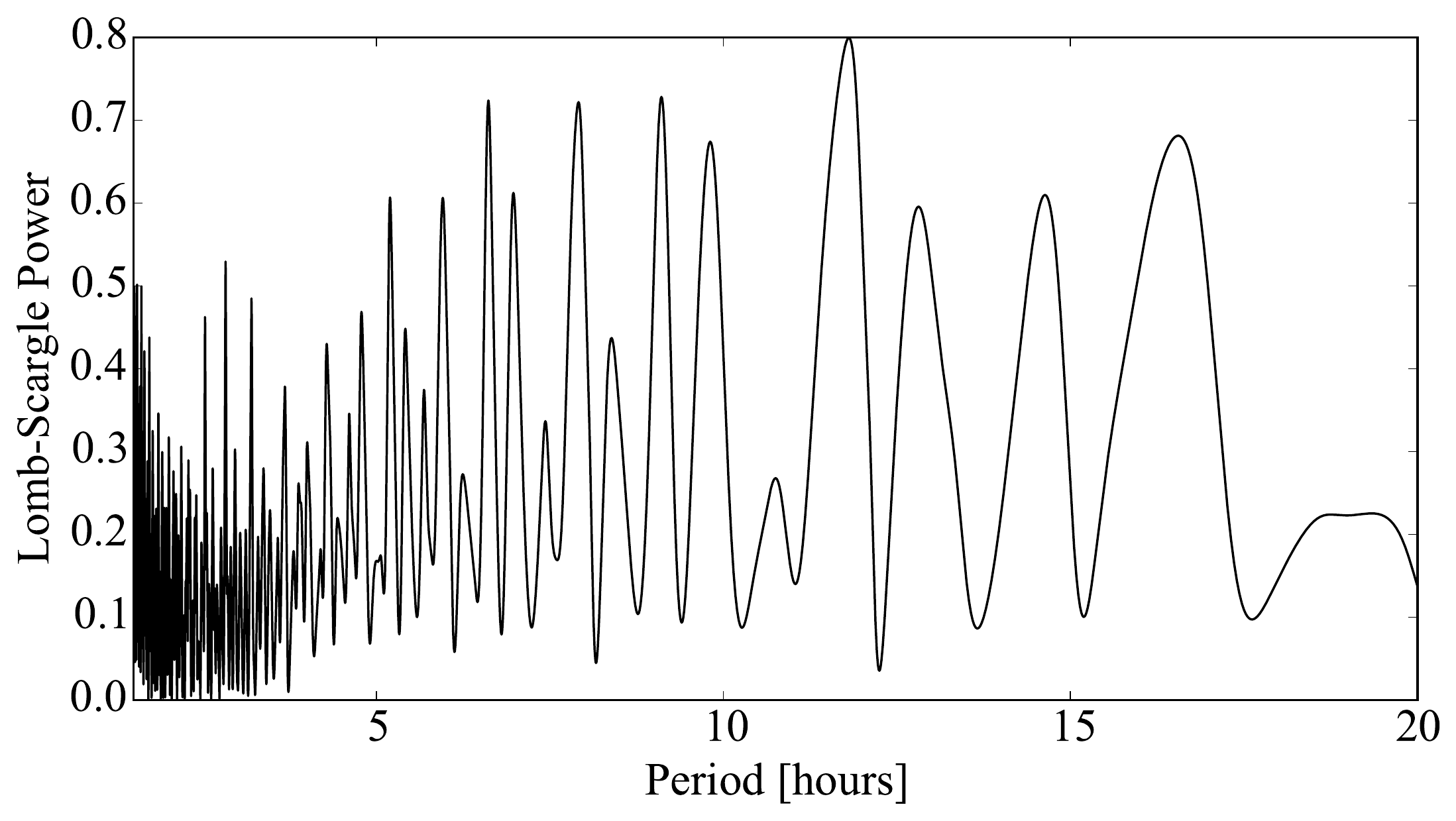}
   \caption{Lomb-Scargle periodogram of 93P showing the LS power versus period. The highest peak corresponds to a 24-hour alias. The next three peaks correspond to $P_{\mathrm{rot}}$ = 18.2, 13.2 and 15.8 hours.}
    \label{93P_LS}%
    \end{figure}

\begin{figure}
    \centering
   \includegraphics[width=0.48\textwidth]{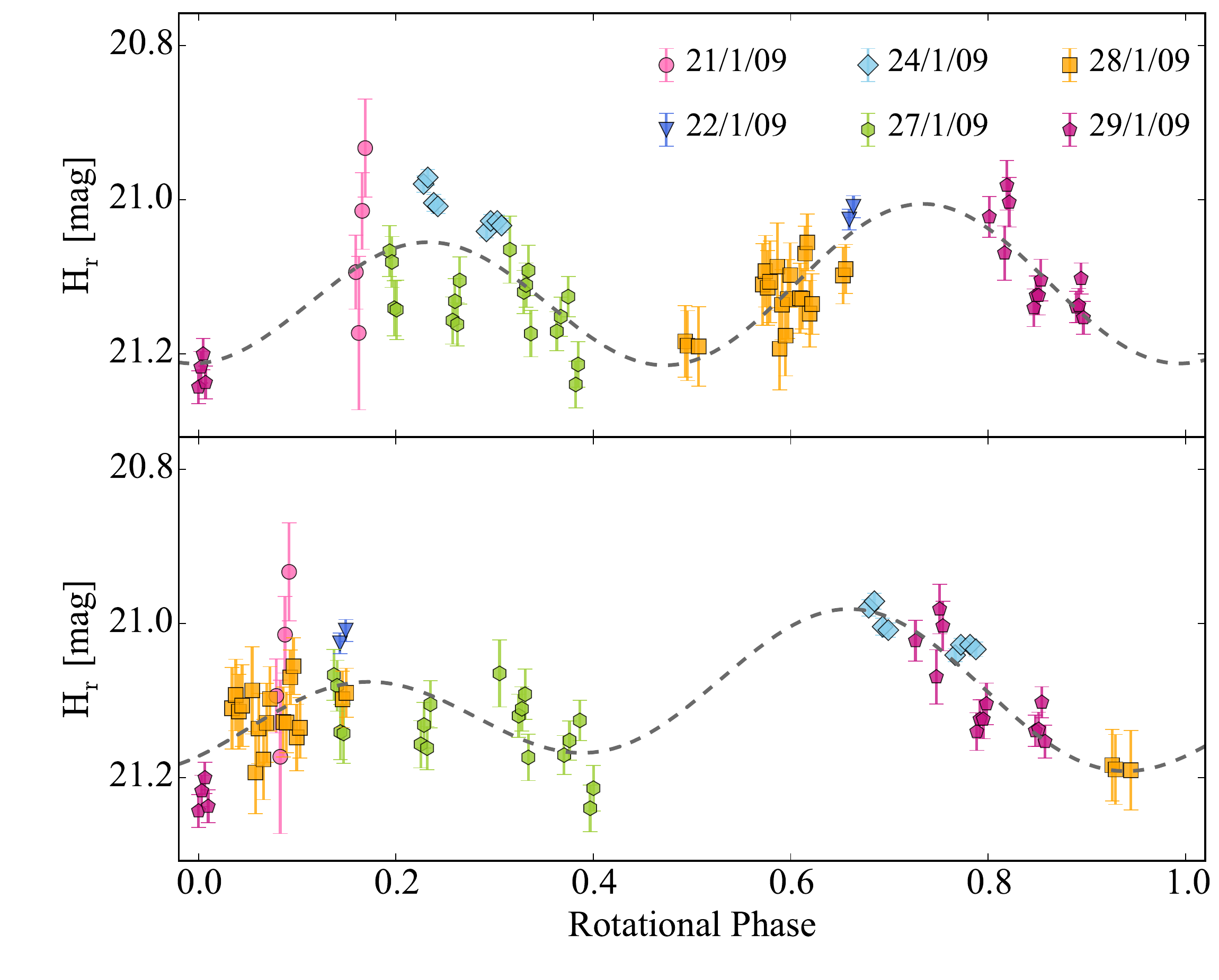}
   \caption{Rotational lightcurve of 93P folded with the two most likely periods 18.2 hours (top) and 13.2 hours (bottom). The dashed line corresponds to second-order Fourier series which aim to reproduce an asymmetric double-peaked lightcurve. The lightcurve phased with 13.2 hours shows less scatter, but the data are not sufficient to discriminate between the two periods.}
    \label{93P_2005}%
    \end{figure}

    Despite the weak activity, the brightness variation in the time series from each night suggested that the nucleus signal could still be detected. The LS periodogram of the combined dataset can be seen in Fig. \ref{93P_LS}. The strongest peak at $\sim$ 24 hours does not produce a typical lightcurve and corresponds to a 24-hour alias. From the remaining peaks, those at $P_{\mathrm{rot}}$ = 18.2 hours and $P_{\mathrm{rot}}$ = 13.2 hours result in possible lightcurves (Fig. \ref{93P_2005}). 
    
\begin{figure}
    \centering
   \includegraphics[width=0.48\textwidth]{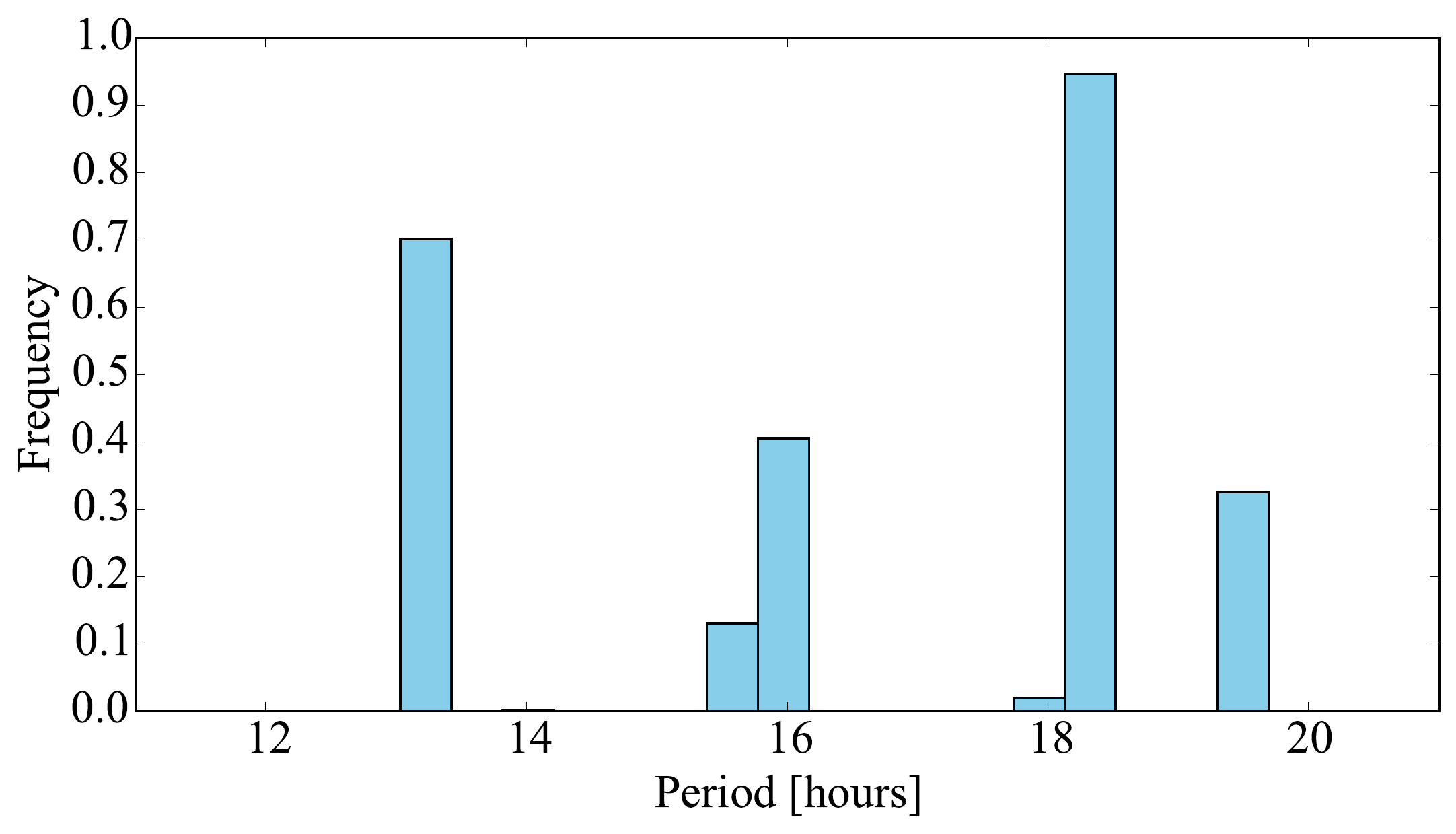}
   \caption{Monte Carlo simulation results for the rotation period of 93P. The most frequently preferred rotation period is 18.2 hours, but the large range of possible periods does not allow us to uniquely determine the rotation rate of the comet.}
    \label{93P_MC}%
    \end{figure}

    We used the MC method to look for the best period between 3 and 23 hours (to avoid the aliasing at 24 hours). The simulation resulted in possible periods between 13.1 and 19.7 with the most frequently preferred period of 18.2 hours (29\% of the iterations, Fig. \ref{93P_MC}). It is impossible to deduce the precise spin rate of 93P from these data, but the period can be constrained to the range $P_{\mathrm{rot}}=\mathrm{18.2^{+1.5}_{-5}}$ hours. 
    
     The brightness variation of 93P is $\Delta$$m_{\mathrm{r}}$ = 0.21 $\pm$ 0.05 mag and suggests an axis ratio $a/b$ $\geq$ 1.21 $\pm$ 0.06. The mean magnitude of the comet is $m_{\mathrm{r}}$ = 21.09 $\pm$ 0.05 mag which corresponds to $H_{\mathrm{r}}$(1,1,0) = 15.17 $\pm$ 0.05 mag, for a typical phase function $\beta$ = 0.04 mag/deg. Using Eq. \ref{eq:albedo} and the SEPPCoN radius from \cite{Fernandez2013}, we estimate that the albedo of 93P is $A_{\mathrm{r}}$ = 4.2$\pm$0.9 \%. 
     
   Since the comet showed signatures of activity during the time of the observations, the brightness and albedo values we have derived need to be treated as upper limits. If the coma contribution of the frames is significant, the absolute magnitude of the nucleus must be larger, and therefore the resulting albedo must be  smaller. In order to derive more certain estimates of the nucleus parameters, the comet needs to be observed at higher heliocentric distances where it is more likely to be inactive.

\subsection{94P/Russell}
\label{sec_res_94P}

	In the analysis described here, we attempted to determine the rotation rate of 94P/Russell after combining three data sets from 2005, 2007 and 2009. The observations were taken before and after the same aphelion passage in 2007. 
    
    The dataset from 2005 was previously used to determine a period of $\sim$ 33 hours \cite{Snodgrass2008}. We re-processed the data and used our method for absolute calibration to combine the observations from the four observing nights in 2005. 
    
\begin{figure}
    \centering
   \includegraphics[width=0.48\textwidth]{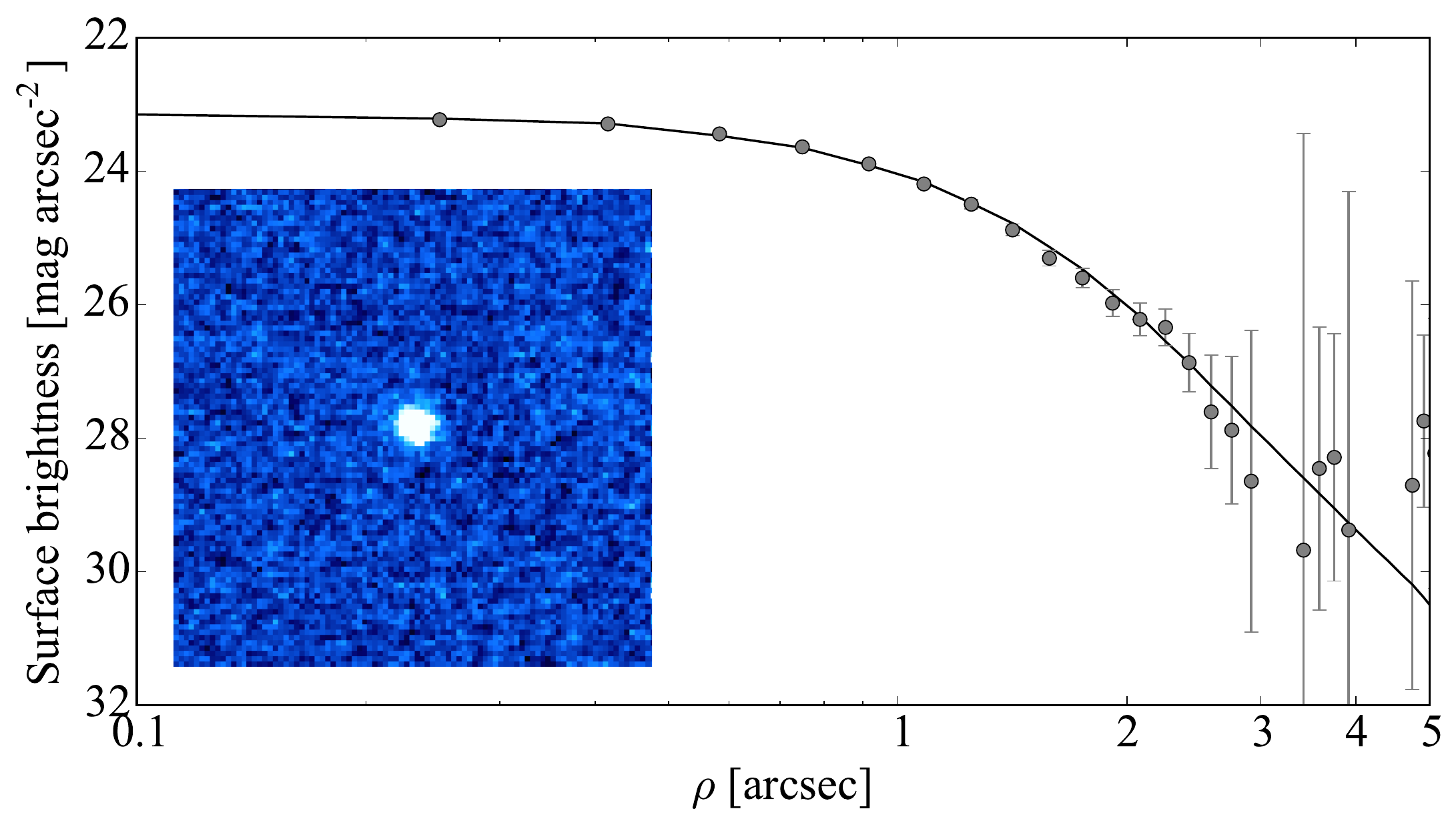}
   \caption{Same as Fig. \ref{14P_2004_PSF}, for the observations of 94P in 2005. The co-added composite image is made up of 15 $\times$ 75 s exposures taken on 7 July 2005.}
    \label{94P_2005_PSF}%
    \end{figure}

The surface brightness profile presented in \cite{Snodgrass2008} suggested that the comet could have been weakly active at the time of the observations. We performed a careful background subtraction of the comet composite images for each night, and concluded that 94P appeared stellar on each night of the run (see Fig. \ref{94P_2005_PSF}). 

\begin{figure}
    \centering
   \includegraphics[width=0.48\textwidth]{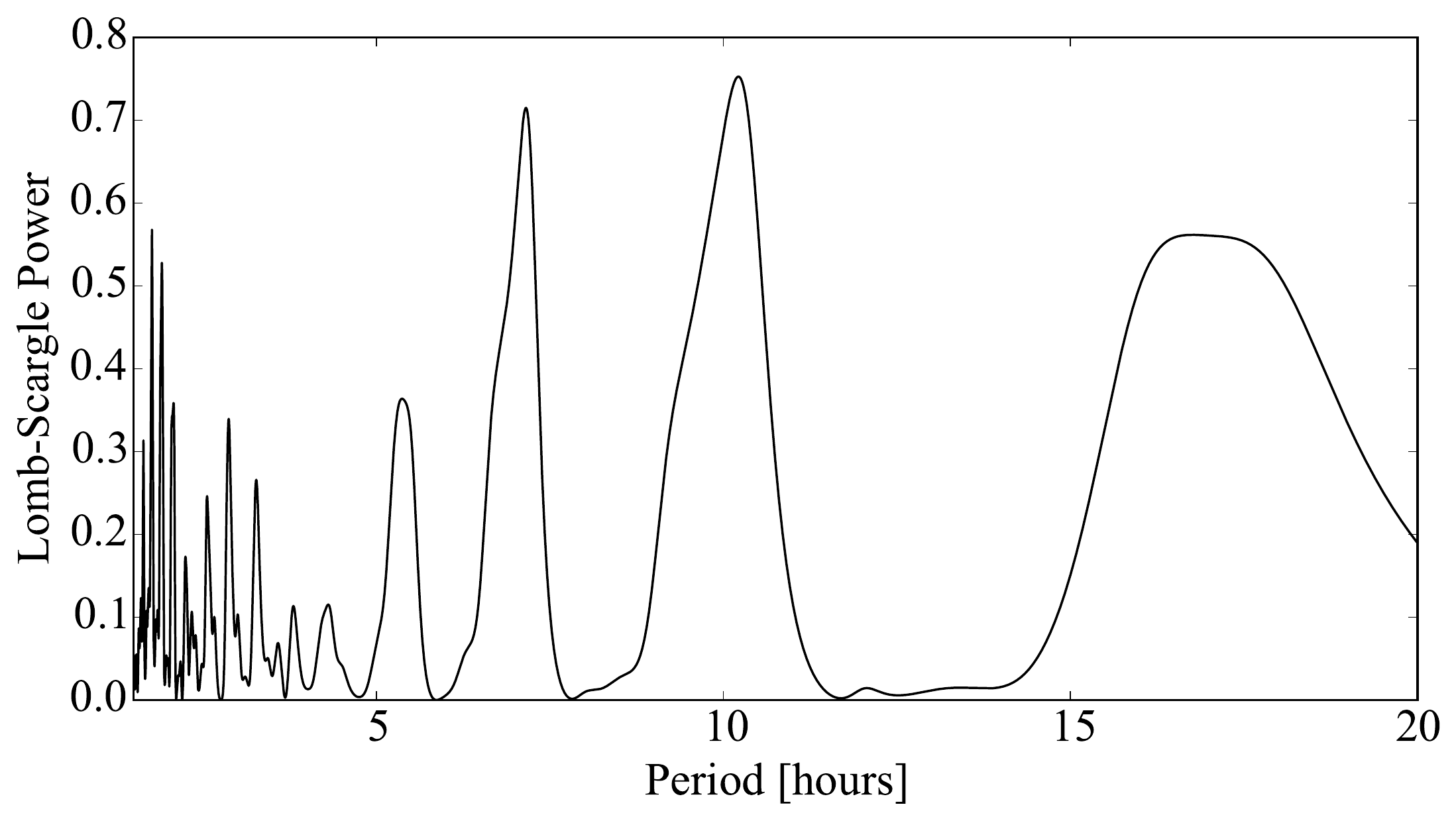}
   \caption{Lomb-Scargle periodogram of 94P from the dataset taken in 2005. The highest peaks correspond to the most likely periods $P_{\mathrm{rot,1}}$ = 20.43 hours and $P_{\mathrm{rot,2}}$ = 14.31 hours.}
    \label{94P_2005_LS}%
    \end{figure}

\begin{figure}
    \centering
   \includegraphics[width=0.48\textwidth]{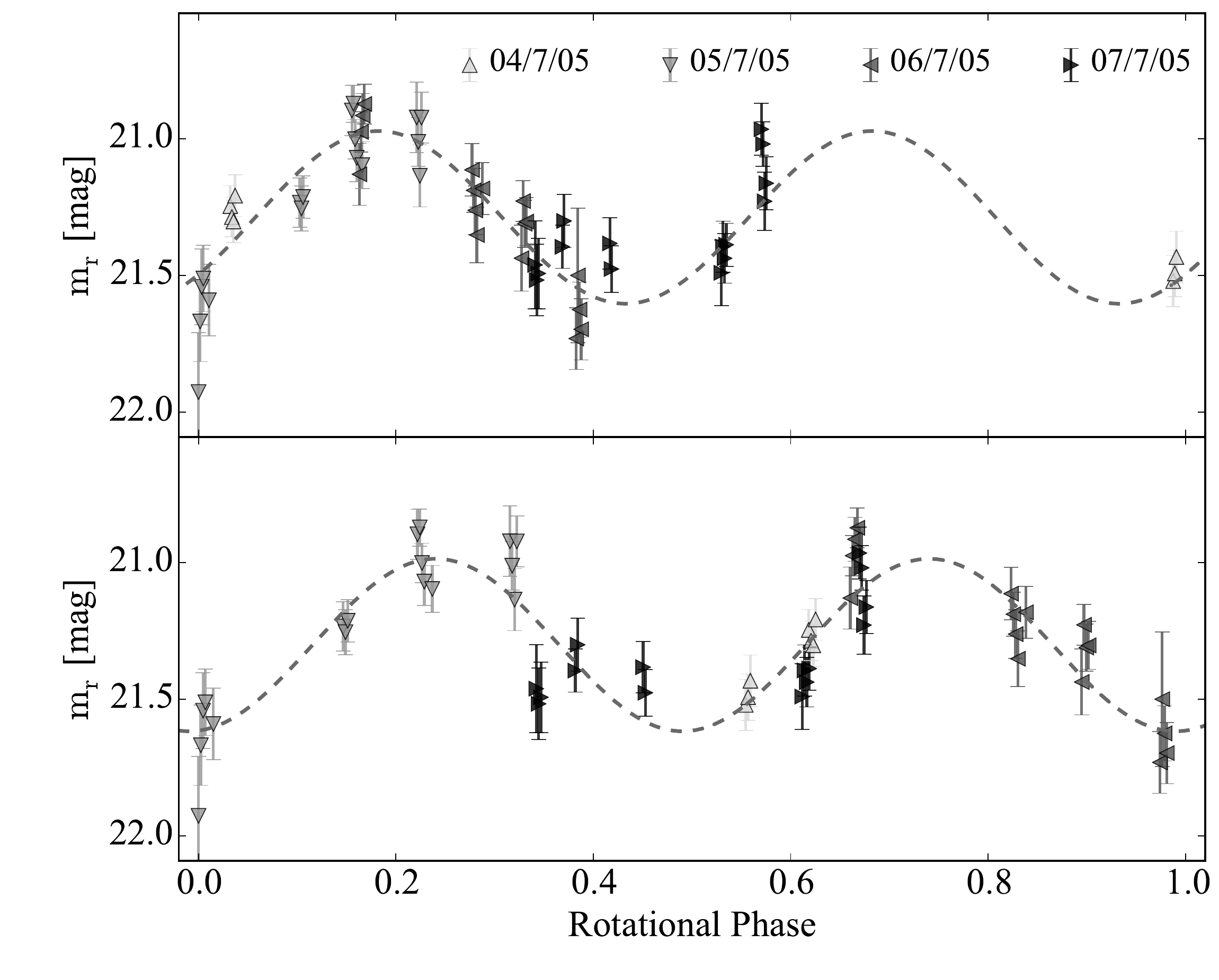}
   \caption{Rotational lightcurve of 94P from the data obtained in 2005. The lightcurve is folded with $P_{\mathrm{rot,1}}$ = 20.43 hours (top) and $P_{\mathrm{rot,2}}$ = 14.31 hours (bottom). We cannot choose between the two periods based on the appearance of the two lightcurves. However, $P_{\mathrm{rot,1}}$ = 20.43 hours is preferred by the MC method, and is therefore considered as more likely. }
    \label{94P_2005}%
    \end{figure}

	The Lomb-Scargle periodogram of the data taken in 2005 has two strong peaks corresponding to 20.43 and 14.31 hours (Fig. \ref{94P_2005_LS}). The lightcurves phased with these periods are plotted in Fig. \ref{94P_2005}. It is not possible to reject the second-best period based on the appearance of the lightcurve. However, in all iterations of the MC simulation the larger period was preferred and therefore the period was determined to be $P_{\mathrm{rot}}$ = 20.43 $\pm$ 0.05 hours. 
    
    The resulting lightcurve had a brightness variation $\Delta$$m_{\mathrm{r}}$ = 0.7 $\pm$ 0.1 mag. This corresponds to an axis ratio $a/b$ $\geq$ 1.9 $\pm$ 0.2 and density $D_{\mathrm{N}}$ $\geq$ 0.05 $\pm$ 0.01 $\mathrm{g \ cm^{-3}}$ .

\begin{figure}
    \centering
   \includegraphics[width=0.48\textwidth]{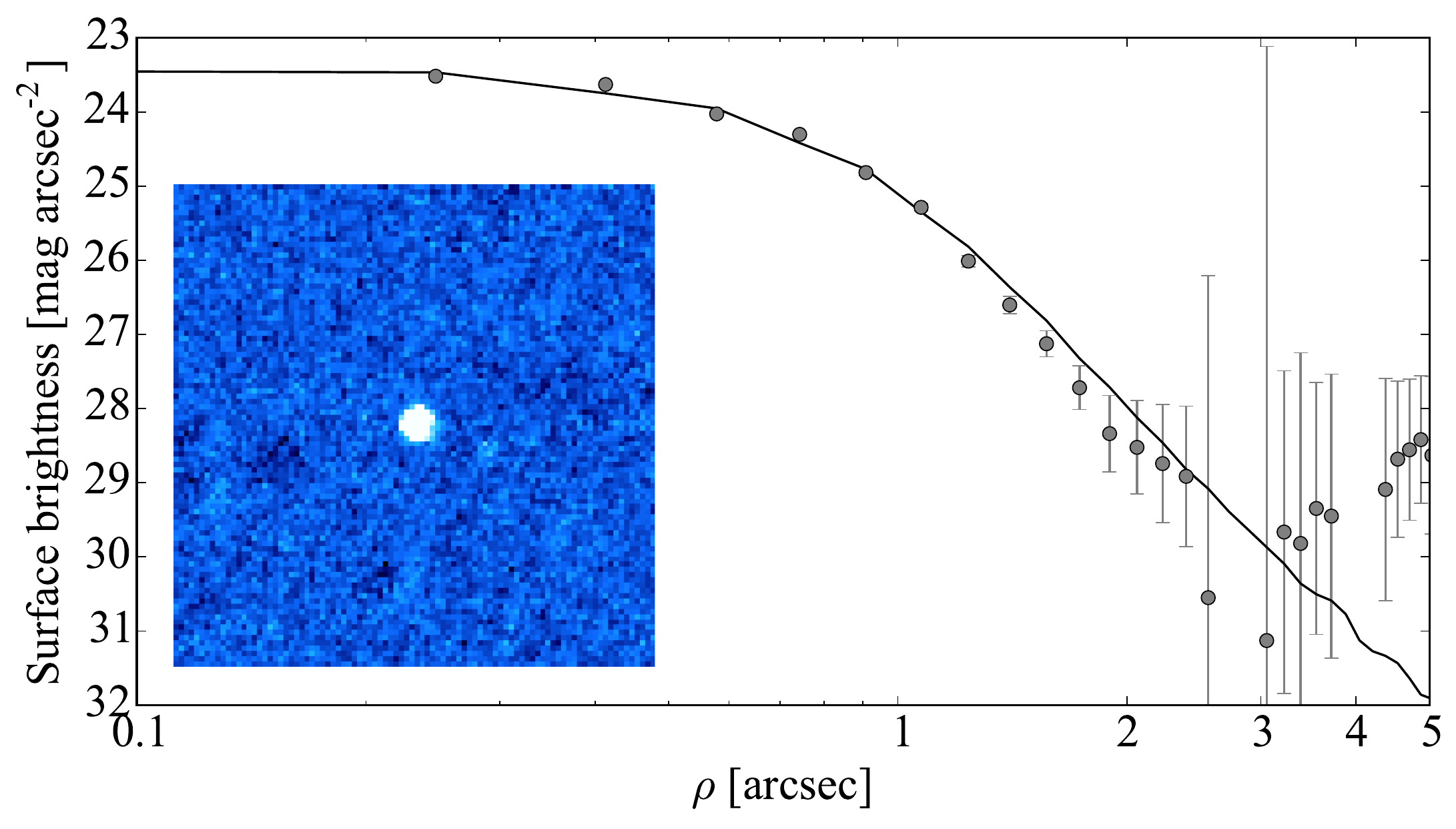}
   \caption{Same as Fig. \ref{14P_2004_PSF}, for the observations of 94P in 2007. The co-added composite image is made up of 8 $\times$ 400 s exposures taken on 20 July 2007.}
    \label{94P_2007_PSF}%
    \end{figure}

\begin{figure}
    \centering
   \includegraphics[width=0.48\textwidth]{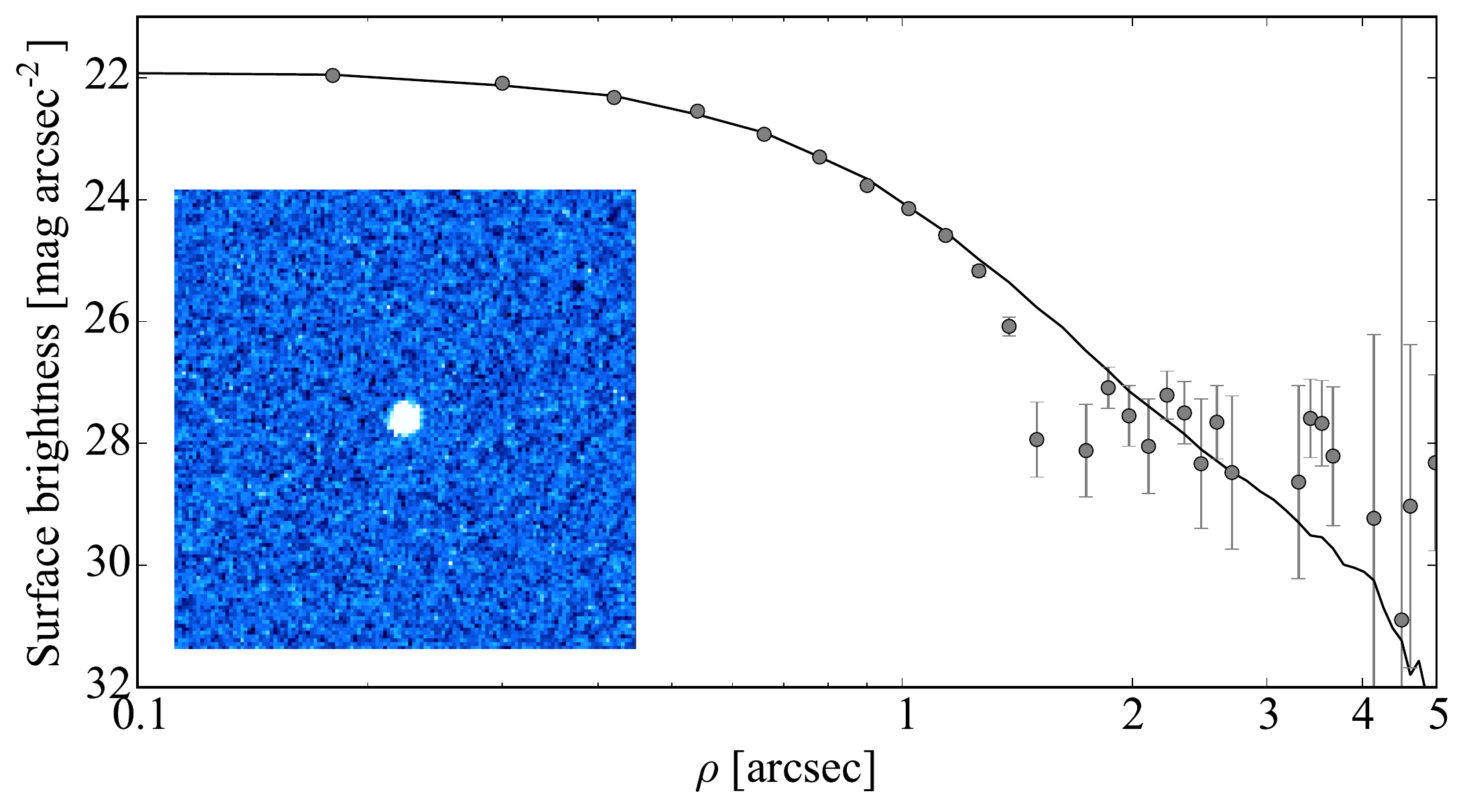}
   \caption{Same as Fig. \ref{14P_2004_PSF}, for the observations of 94P in 2009. The co-added composite image is made up of 8 $\times$ 100 s exposures taken on 28 January 2009.}
    \label{94P_2009_PSF}%
    \end{figure}

The data taken during the SEPPCoN runs in 2007 and 2009 were also checked for the presence of activity (Fig. \ref{94P_2007_PSF} and \ref{94P_2009_PSF}). Due to the faintness of the comet, in both cases its surface brightness profiles levelled out within 5 arcseconds from the nucleus. However, we can conclude that 94P was inactive in both epochs considering the good matches with the stellar PSF close to the centre, as well as the appearance of the composite images.

    Neither of the two datasets from 2007 and 2009 were sufficient to derive the  rotation rate of 94P independently. We therefore only used them to estimate the nucleus magnitude and the minimum brightness variation at each epoch. We measured $m_{\mathrm{r}}$ = 22.6 $\pm$ 0.2 and $\Delta$$m_{\mathrm{r}}$ = 1.0 $\pm$ 0.2 mag for 2007, and $m_{\mathrm{r}}$ = 21.30 $\pm$ 0.05 and $\Delta$$m_{\mathrm{r}}$ = 0.80 $\pm$ 0.05 mag for 2009. 
    
   We combined all three datasets to determine the precise rotation rate of the comet. The analysis of the joined datasets was done under the following assumptions: 1) the comet was inactive during all observations and the measured magnitudes had no coma contributions; 2) the rotation period remained constant during the entire aphelion passage, and 3) the changing viewing geometry between the different observations did not affect the lightcurve shape significantly.   
   
\begin{figure}
    \centering
   \includegraphics[width=0.48\textwidth]{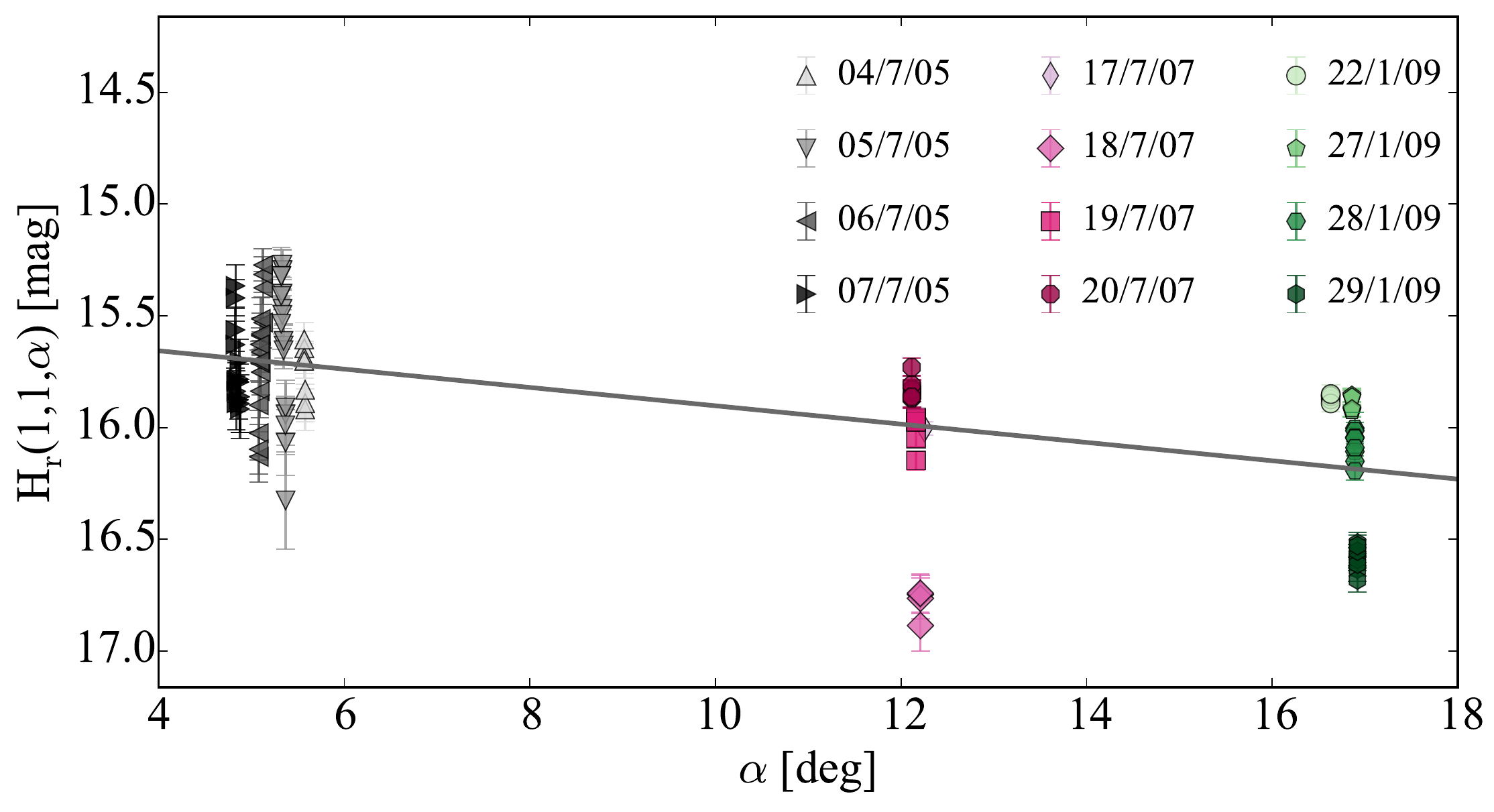}
   \caption{Phase function of comet 94P combining the datasets from 2005, 2007, and 2009. The linear phase function coefficient derived with the Monte Carlo method is $\beta$ = 0.039 $\pm$ 0.002 mag deg\textsuperscript{-1}. } 
    \label{94P_PHASE}%
    \end{figure}

\begin{figure}
    \centering
   \includegraphics[width=0.48\textwidth]{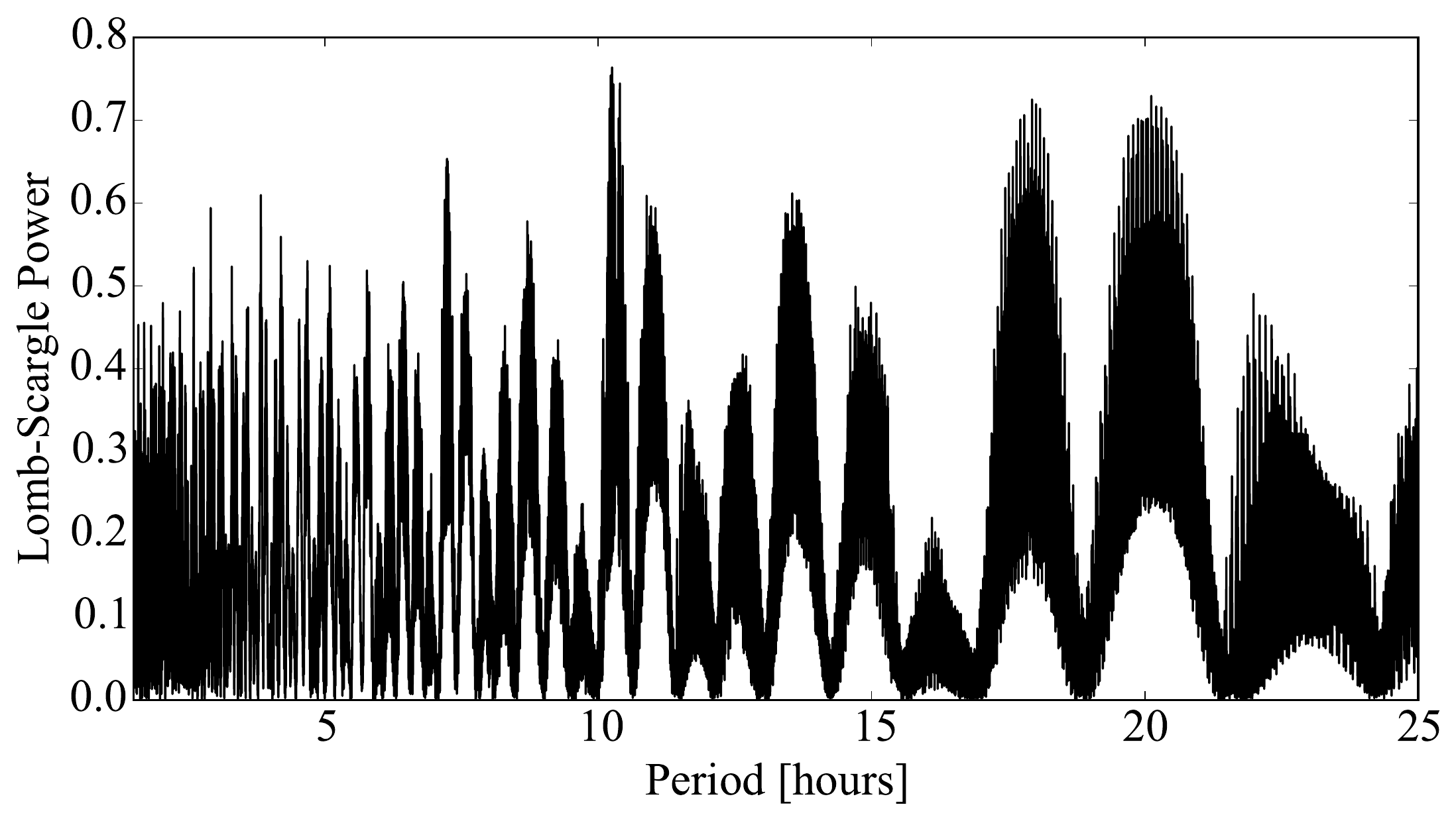}
   \caption{Lomb-Scargle periodogram of 94P with the datasets from 2005, 2007, and 2009 combined. The highest peak corresponds to the most likely period $P_{\mathrm{rot}}$ = 20.70 hours.}
    \label{94P_ALL_LS}%
    \end{figure}

   With these assumptions in mind, we used the MC method to derive a phase function with a slope $\beta$ = 0.039 $\pm$ 0.002 mag deg\textsuperscript{-1} (Fig. \ref{94P_PHASE}). The LS periodogram of the combined dataset on Fig. \ref{94P_ALL_LS} peaks at $P_{\mathrm{rot}}$ = 20.70 hours.  The period $P_{\mathrm{rot}}$ = 20.70 hours was also suggested by PDM and SLM. The other two peaks of the LS periodogram close to 38 and 40 hours were also inspected but their lightcurves were significantly noisier.
   
\begin{figure}
    \centering
   \includegraphics[width=0.48\textwidth]{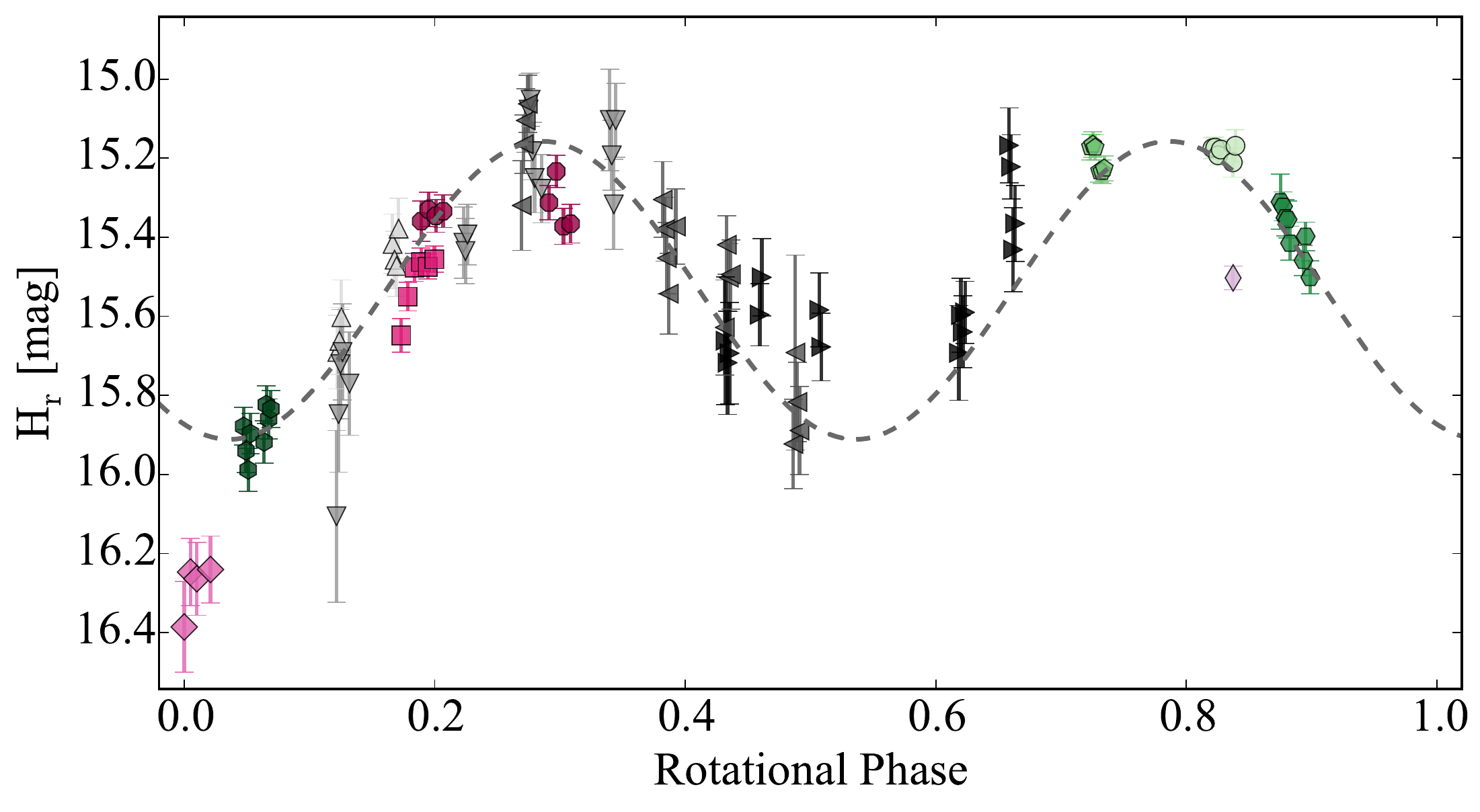}
   \caption{Rotational lightcurve of 94P with the combined datasets from 2005, 2007 and 2009. The symbols of each data set correspond to those used on Fig. \ref{94P_PHASE}. The lightcurve is folded with the best-fitting period $P_{\mathrm{rot}}$ = 20.70 hours. The fitted first-order Fourier series (dashed line) agree with all points except for the ones from 18 July 2007. These fainter points could be interpreted as signatures of an asymmetric lightcurve with one deep minimum, or alternatively as results from the changing viewing geometry between the three epochs. 
    \label{94P_ALL}}%
    \end{figure}

   The period of 20.70 hours was preferred in 86\% of the MC iterations, which allows us to set the rotation rate of 94P to $P_{\mathrm{rot}}$ = 20.70 $\pm$ 0.07 hours. The corresponding lightcurve plotted in Fig. \ref{94P_ALL} shows a very good agreement between the separate datasets. 
   
      The absolute magnitude of 94P from the combined dataset was $H_{\mathrm{r}}$(1,1,0) = 15.50 $\pm$ 0.09 mag. The albedo of 94P was determined with Eq.\ref{eq:albedo} to be $A_{\mathrm{r}}$ = 4.0$\pm$0.6 \%. 
      
    The only data sets which deviate from the first-order Fourier series in Fig. \ref{94P_ALL} are the ones from July 2007. These points are fainter than the comet magnitude from the rest of the nights. There were no indications of problems with the images or the photometric calibration during these nights. We can conclude that the lightcurve must be asymmetric, with one of the minima being sharper and deeper than the other one. Such a lightcurve would have $\Delta$$m_{\mathrm{r}}$ = 1.11 $\pm$ 0.09 mag which corresponds to $a/b$ $\geq$ 2.8 $\pm$ 0.2 and density $D_{\mathrm{N}}$ $\geq$ 0.07 $\pm$ 0.02 $\mathrm{g \ cm^{-3}}$. 
   
   Another effect which could produce the observed lightcurve is the change of viewing geometry. Comet 94P moved approximately 120\textsuperscript{$\circ$} along its orbit between 2005 and 2009, which could be sufficient to produce a noticeable variation in the total surface area of the nucleus for an observer on Earth. Alternatively, the shift in brightness might be caused by weak activity  in the 2005 and 2009 data when the comet was closer to the Sun. Such activity is not evident in the profiles on Figs.  \ref{94P_2005_PSF}, \ref{94P_2007_PSF} and \ref{94P_2009_PSF} but it is possible for some weak activity to be hidden within the seeing disc of distant comets \citep[e.g.][]{Snodgrass2016}. With the limited data here, we cannot determine whether the deep minimum in the lightcurve is a feature of the nucleus or if it is caused by other effects.

\subsection{110P/Hartley 3}
\label{sec_res_110P}
 
Comet 110P/Hartley 3 was observed with VLT-FORS2 and NTT-EFOSC2 during 8 nights between June and August 2011. The aim of the observations was to sample the comet's phase function in the phase angle range between 1\textsuperscript{$\circ$} and 10\textsuperscript{$\circ$}. The method for precise absolute photometric calibration with PS1 allowed us to combine these datasets and to derive the comet's phase function as well as to study its rotational lightcurve. 

\begin{figure}
    \centering
   \includegraphics[width=0.48\textwidth]{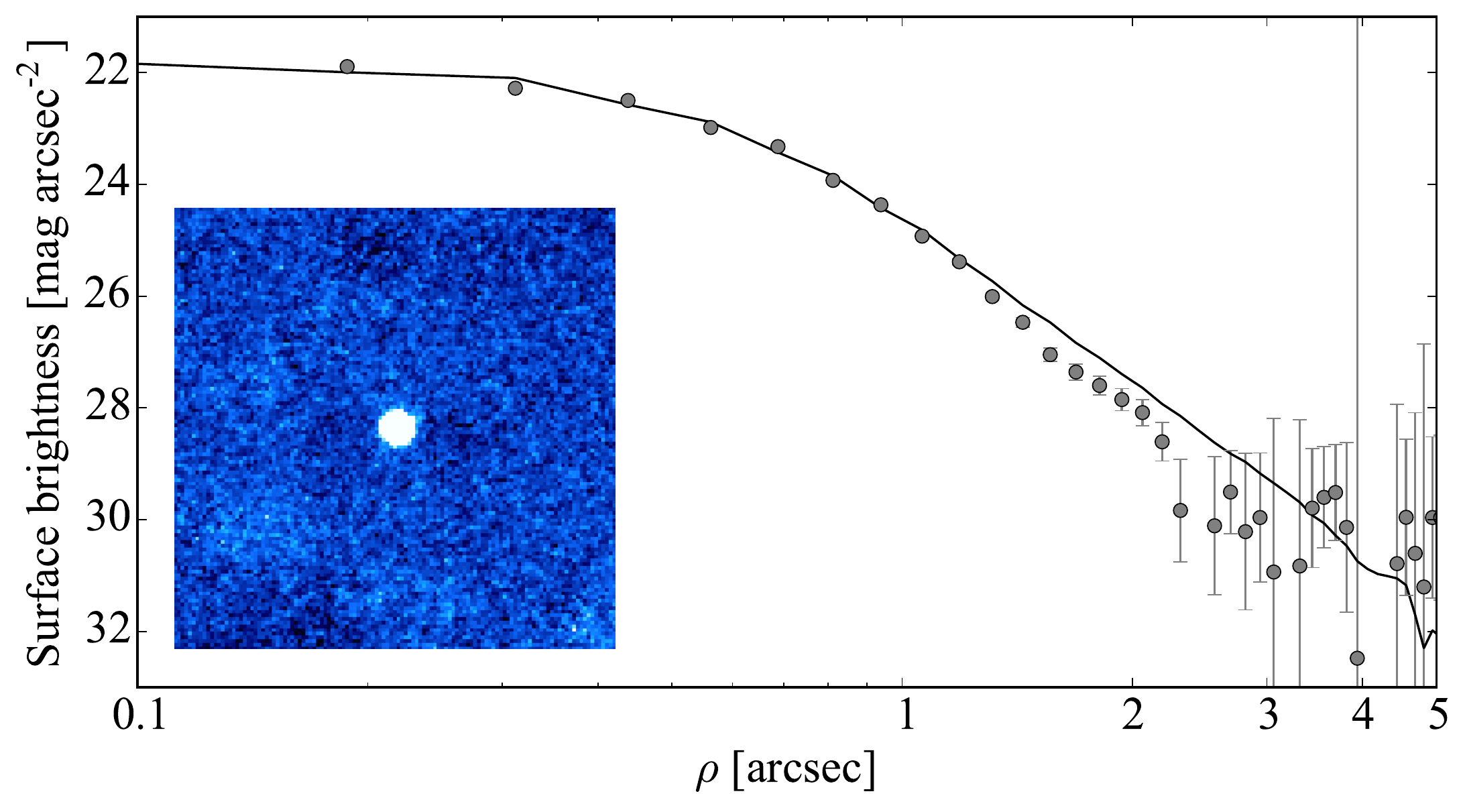}
   \caption{Same as Fig. \ref{14P_2004_PSF}, for 110P on 15 July 2012. The co-added composite image is made up of 18 $\times$ 70 s exposures. The comet appears inactive and its surface brightness profile follows that of the comparison star.}
    \label{110P_2012_PSF}%
    \end{figure}

We looked for signatures of activity on comet composite images for each individual night, and on Fig. \ref{110P_2012_PSF} we have presented an example for the middle of the observing period. The comet did not show any indication of coma presence throughout the observing period, and we assume that the derived photometry from each night contains only signal from the nucleus.  
        
\begin{figure}
    \centering
   \includegraphics[width=0.48\textwidth]{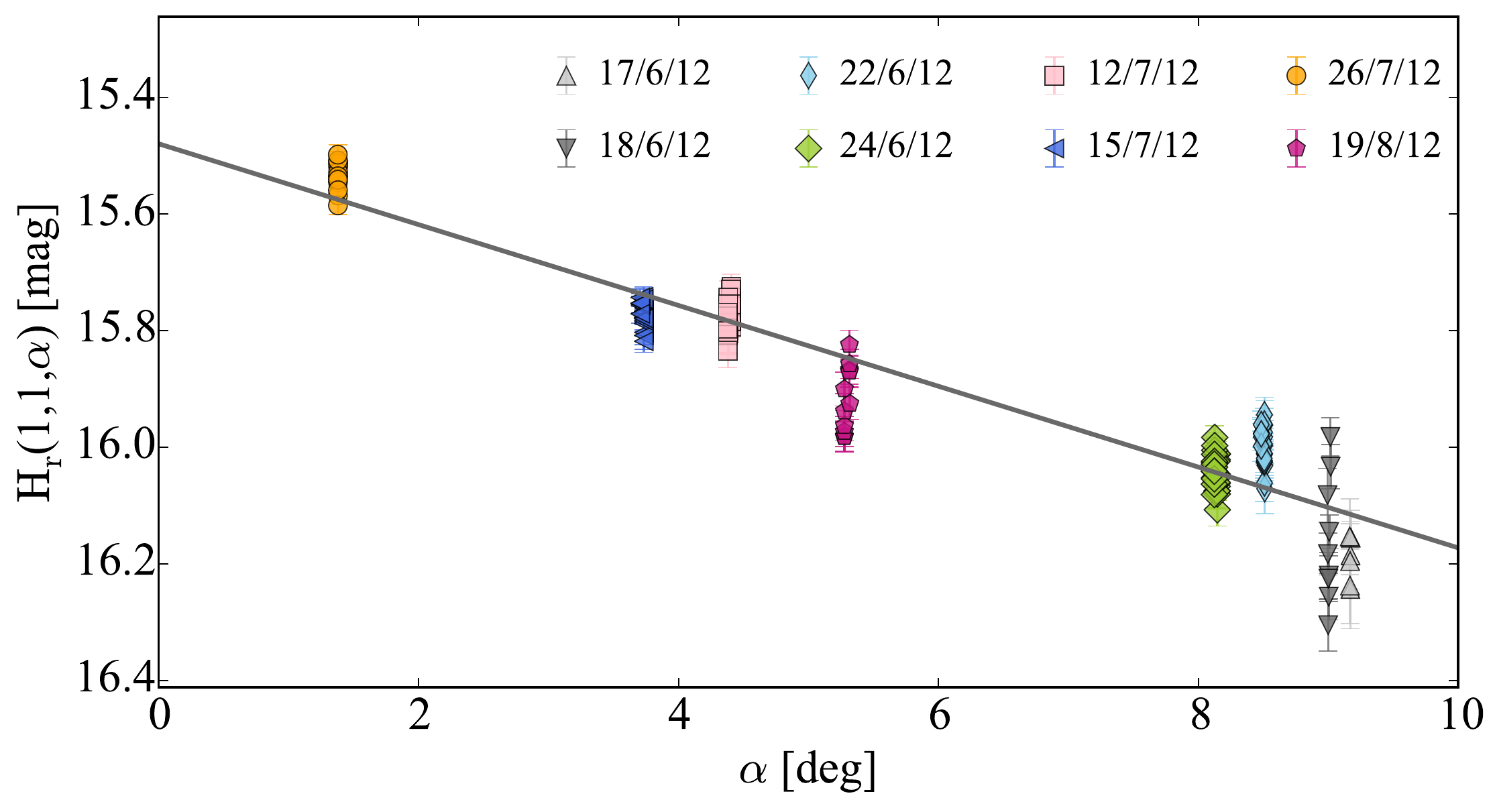}
   \caption{Phase function of comet 110P. The linear slope $\beta$ derived with the Monte Carlo method is 0.069 $\pm$ 0.002 mag deg\textsuperscript{-1}. The NTT-EFOSC2 points from 17 and 18 June 2012 were binned since the S/N of the individual points was low due to bad observing conditions.} 
    \label{110P_PHASE}%
    \end{figure}

We used the MC method to derive a phase function for 110P.
The determined phase function with linear slope $\beta$ = 0.069 $\pm$ 0.002 mag deg\textsuperscript{-1} is in excellent agreement with all individual datasets (Fig. \ref{110P_PHASE}). 

\begin{figure}
    \centering
   \includegraphics[width=0.48\textwidth]{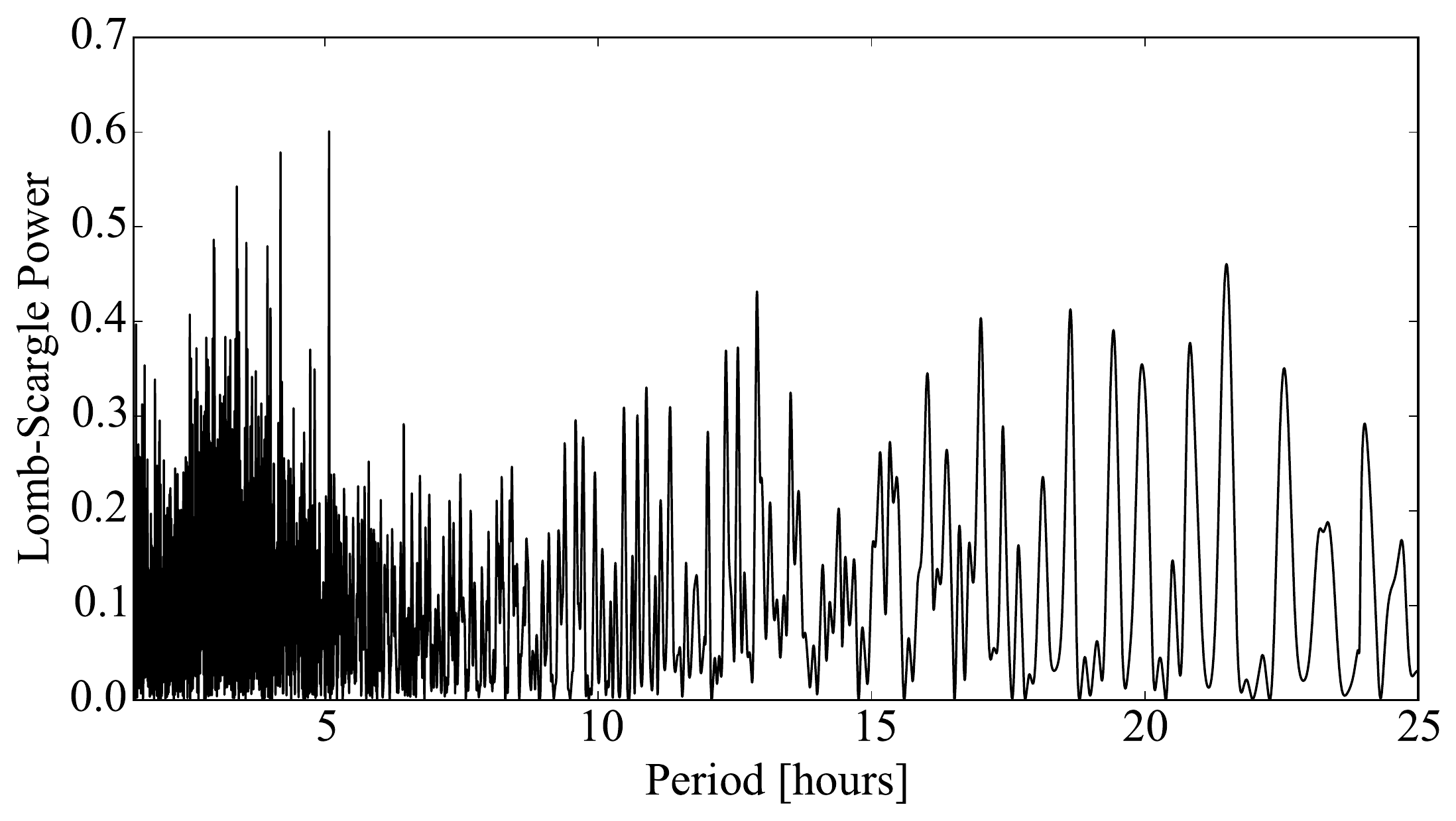}
   \caption{Lomb-Scargle periodogram of 110P for the combined dataset with all observations from 2012. The three highest peaks correspond to $P_{\mathrm{rot,1}}$ = 10.153 hours, $P_{\mathrm{rot,2}}$ = 8.375 hours and $P_{\mathrm{rot,3}}$ = 6.779 hours.}
    \label{110P_LS}%
    \end{figure}

\begin{figure}
    \centering
   \includegraphics[width=0.48\textwidth]{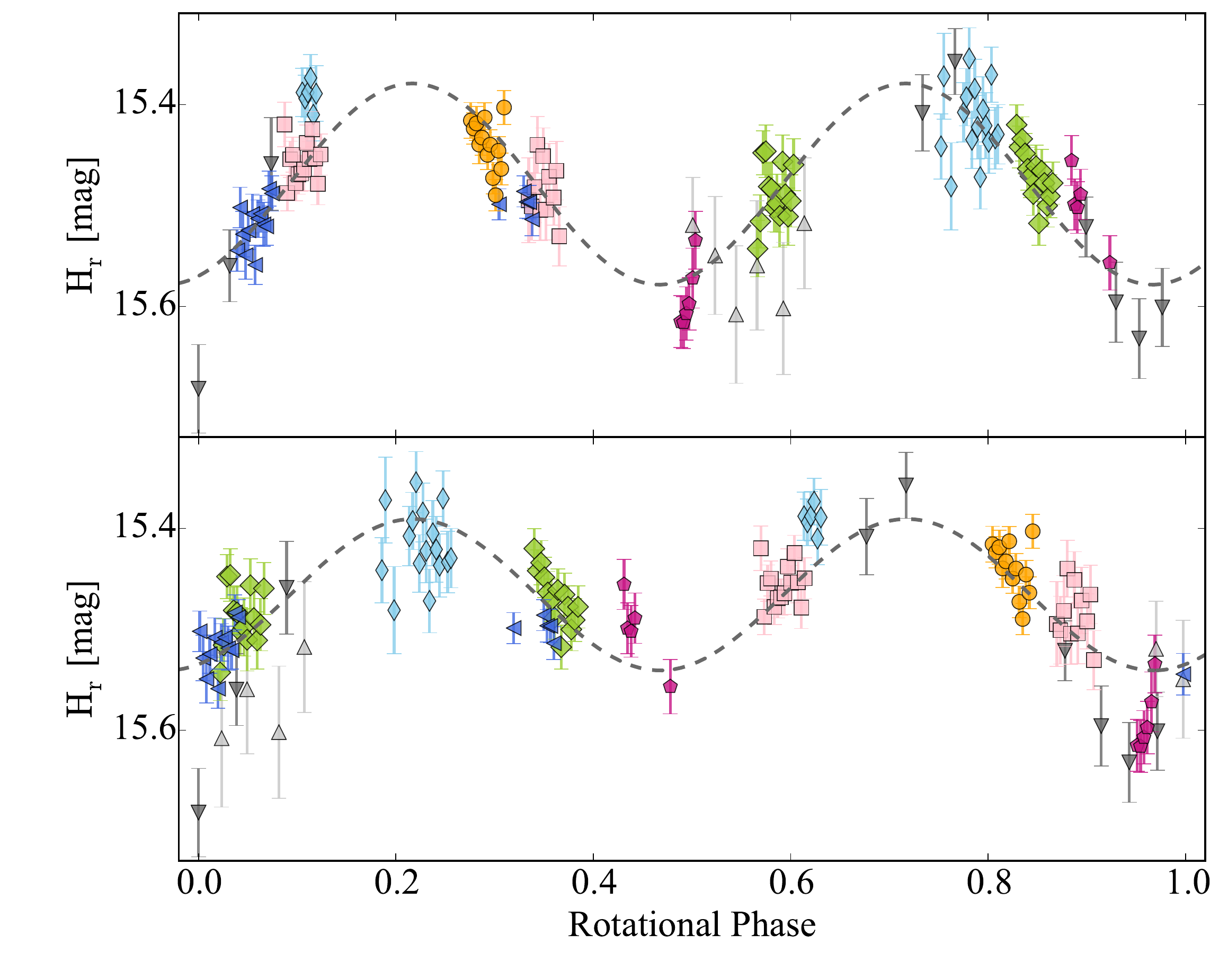}
   \caption{Rotational lightcurve of 110P with all of the data from 2012. The lightcurve is folded with the two most-likely periods 10.153 h (top) and 8.375 hours (bottom) derived from the MC method. The lightcurve with $P_{\mathrm{rot,1}}$ = 10.153 hours is preferred by the MC method (in 75\% of the iterations) and it is in better agreement with the brightness variation within the individual nights. The symbols are the same as in Fig. \ref{110P_PHASE}. The NTT-EFOSC2 points from 17 and 18 June 2012 were binned since the S/N of the individual points was low due to bad observing conditions.}
    \label{110P_2012}%
    \end{figure}

All datasets were used to derive the comet's lightcurve under the same assumptions as those described earlier for 14P, 47P and 94P. The LS periodogram in Fig.~\ref{110P_LS} has three pronounced peaks at $P_{\mathrm{rot,1}}$ = 10.153 hours, $P_{\mathrm{rot,2}}$ = 8.375 hours and $P_{\mathrm{rot,3}}$ = 6.779 hours. The MC method outlines $P_{\mathrm{rot,1}}$ = 10.153 $\pm$ 0.001 hours (75\% of the iterations) and $P_{\mathrm{rot,2}}$ = 8.375 $\pm$ 0.001 hours (17\% of the iterations) as most likely solutions (Fig. \ref{110P_2012}). Qualitatively, the lightcurve phased with $P_{\mathrm{rot,1}}$ = 10.153 $\pm$ 0.001 hours presents less scatter of the points and agrees with the trends in the individual observing blocks better. Since $P_{\mathrm{rot,1}}$ is also preferred by the MC method, we report 10.153 $\pm$ 0.001 hours as the most likely period of 110P. 

The brightness variation of the resulting lightcurve is $\Delta$$m_{\mathrm{r}}$ = 0.20 $\pm$ 0.03 which puts a lower limit on the comet axis ratio $a/b$ $\geq$ 1.20 $\pm$ 0.03. Using $P_{\mathrm{rot,1}}$, we can estimate the nucleus density $D_{\mathrm{N}}$ $\geq$ 0.13 $\pm$ 0.02 $\mathrm{g \ cm^{-3}}$. The mean absolute magnitude of the comet was $H_{\mathrm{r}}$(1,1,0) = 15.47 $\pm$ 0.03 mag, which corresponds to a nucleus radius $r_{\mathrm{N}}$ = 2.31 $\pm$ 0.03 km, assuming an albedo of 4\%.

Our results are in good agreement with those of \cite{Lamy2011} (see Section \ref{sec:rev_110P}). This validates our results and confirms that it is possible to constrain both the phase function and the lightcurve of the comet from sparse observations spread over months. Although the two observations were taken at different apparitions and a small period change could have occurred during the active phase of the comet, due to the large uncertainty in the period from \cite{Lamy2011}, we cannot search for period changes between the two epochs.

\subsection{123P/West-Hartley} 
\label{sec_res_123P}

\begin{figure}
    \centering
   \includegraphics[width=0.48\textwidth]{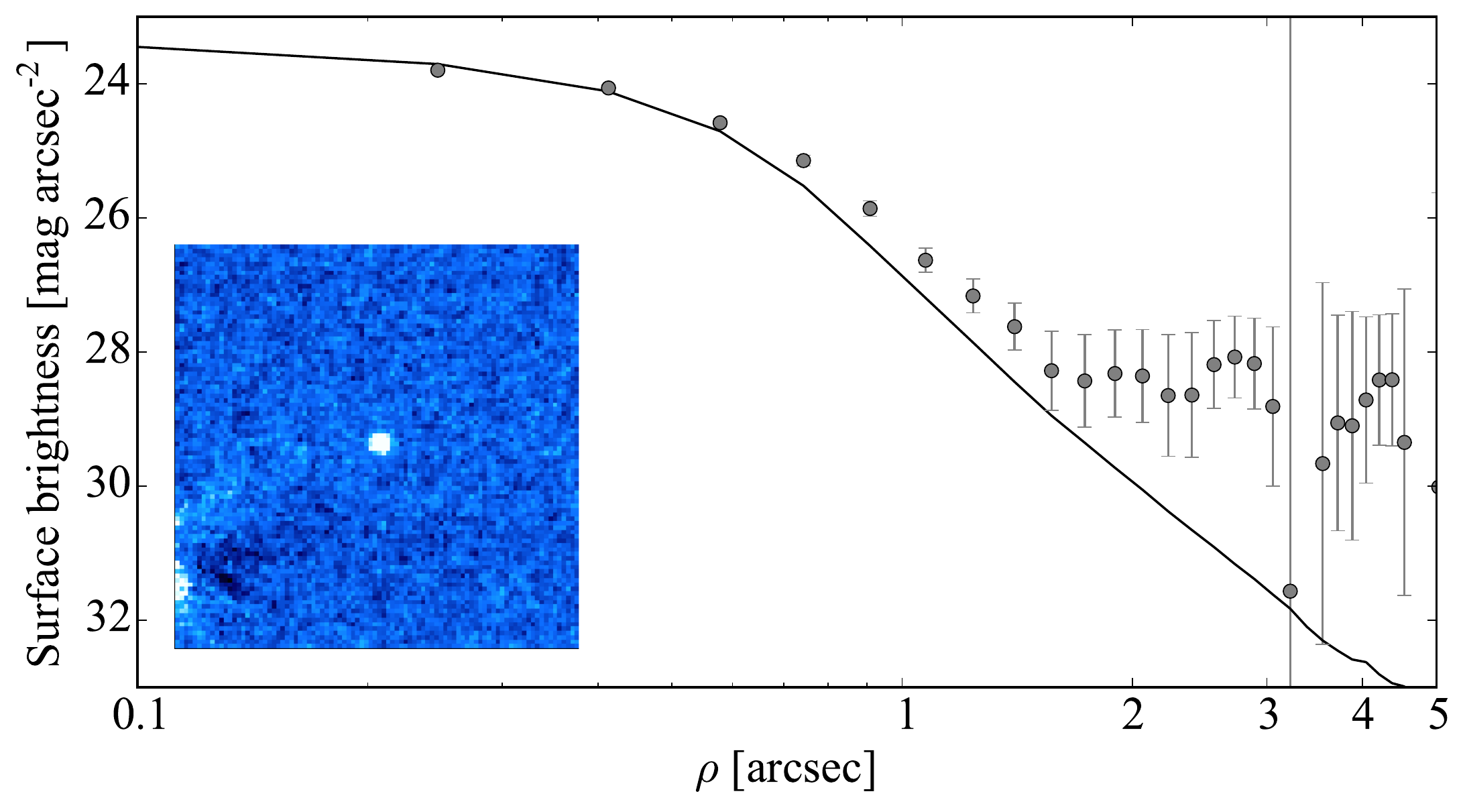}
   \caption{Same as Fig. \ref{14P_2004_PSF}, for 123P on 18 July 2007. The co-added composite image is made up of 23 $\times$ 110 s exposures. The comet appears stellar on the composite image, however its surface brightness profile deviates from that of the comparison star, which indicates that the comet was weakly active during the time of the observations. }
    \label{123P_PSF}%
    \end{figure}

This SEPPCoN target was observed on three consecutive nights in July 2007 while it was at heliocentric distance of 5.6 au. A careful examination of the images indicated that despite the large heliocentric distance at the time of the observations, 123P was weakly active (Fig. \ref{123P_PSF}). 

\begin{figure}
    \centering
   \includegraphics[width=0.48\textwidth]{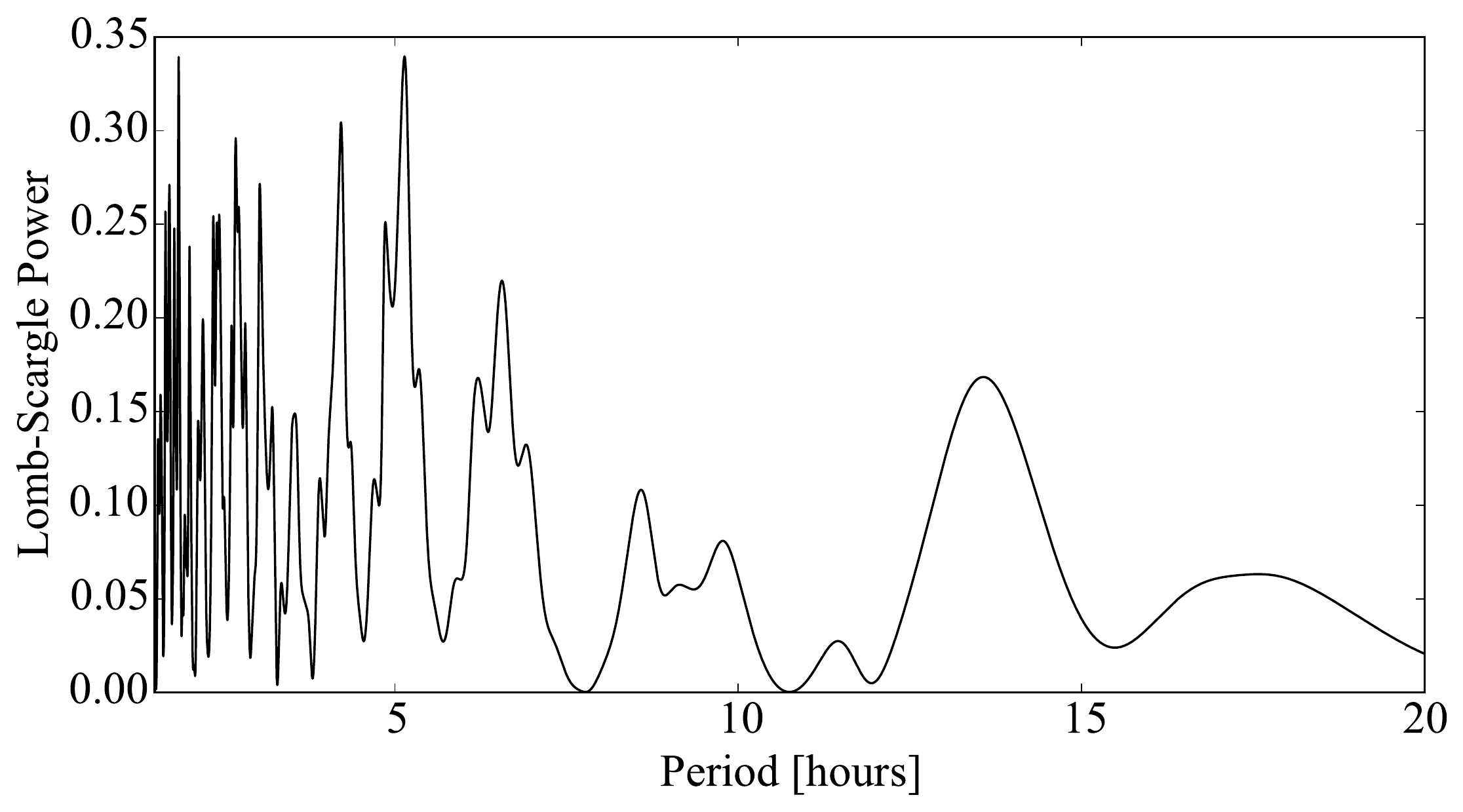}
   \caption{Lomb-Scargle periodogram of 123P. The two highest peaks correspond to $P_{\mathrm{rot,1}}$ = 3.7 hours and $P_{\mathrm{rot,2}}$ = 10.7.}
    \label{123P_LS}%
    \end{figure}

The observations from the individual nights clearly indicated a brightness variation of the nucleus. However, the LS periodogram of the data did not reveal any pronounced peaks with significant power (Fig. \ref{123P_LS}). The two highest peaks correspond to 3.7 and 10.3 hours. Those two periods were also preferred by the MC simulation, which picked $P_{\mathrm{rot}}$ = 3.70 $\pm$ 0.02 hours in 66\% of the iterations and $P_{\mathrm{rot}}$ = 10.27 $\pm$ 0.05 hours (34\%). 

\begin{figure}
    \centering
   \includegraphics[width=0.48\textwidth]{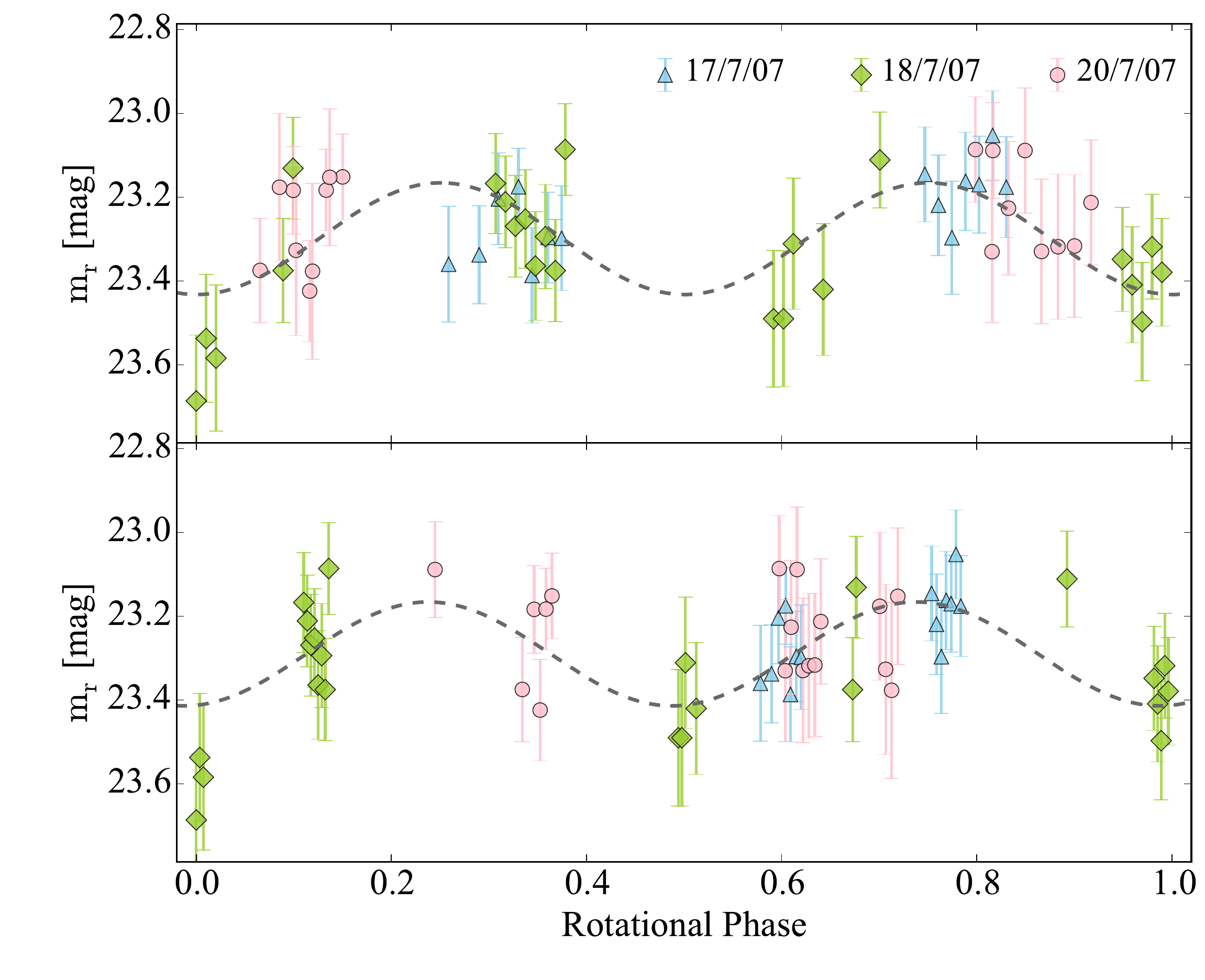}
   \caption{Rotational lightcurve of 123P with all of the data from 2007. The lightcurve is folded with the most-likely periods 3.7 h (top) and 10.7 hours (bottom).}
    \label{123P_2007}%
    \end{figure}

The lightcurves resulting from these two periods are plotted in Fig. \ref{123P_2007}. Both periods appear to be in agreement with the data, and it is not possible to choose between them. Moreover, the data phased with other periods selected by the periodogram produce lightcurves with similar quality. Therefore, we conclude that the collected data are not sufficient to determine the spin rate of 123P. 

We estimated a brightness variation $\Delta$$m_{\mathrm{r}}$ = 0.5 $\pm$ 0.1 mag which corresponds to an axis ratio $a/b$ $\geq$ 1.6 $\pm$ 0.1. The mean measured magnitude of 123P was $m_{\mathrm{r}}$ = 23.3 $\pm$ 0.1 mag which converts to $H_{\mathrm{r}}$(1,1,0) = 15.7 $\pm$ 0.1 mag if a phase function with $\beta$ = 0.04 mag deg\textsuperscript{-1} is used. Our absolute magnitude and the radius measured by \cite{Fernandez2013} convert to an albedo $A_\mathrm{r}$ = 3.6 $\pm$ 0.8\% (Eq. \ref{eq:albedo}). It is however important to note that the surface brightness profile of 123P indicated a weak activity, which implies that the absolute magnitude $H_{\mathrm{r}}$(1,1,0) of the nucleus could be fainter and the determined albedo must be treated as an upper limit.
        
\subsection{137P/Shoemaker-Levy 2} 
\label{sec_res_137P}
    
\begin{figure}
    \centering
   \includegraphics[width=0.48\textwidth]{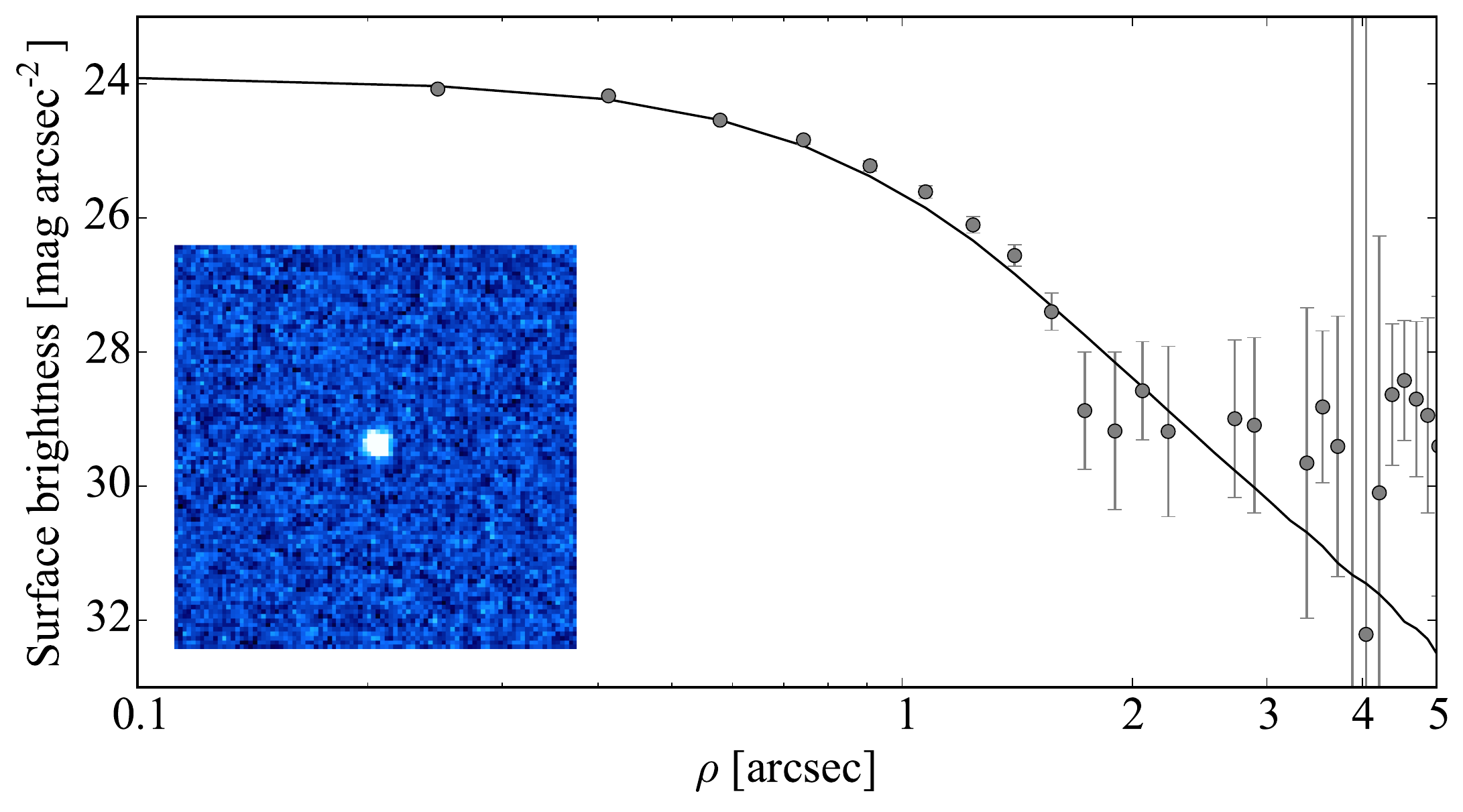}
   \caption{Same as Fig. \ref{14P_2004_PSF} for 137P on 6 March 2005. The co-added composite image is made up of 23 $\times$ 110 s exposures. The comet appears inactive and its surface brightness profile follows that of the comparison star close to the centre before it levels out at the background noise level. }
    \label{137P_2005_PSF}%
    \end{figure}

\begin{figure}
    \centering
   \includegraphics[width=0.48\textwidth]{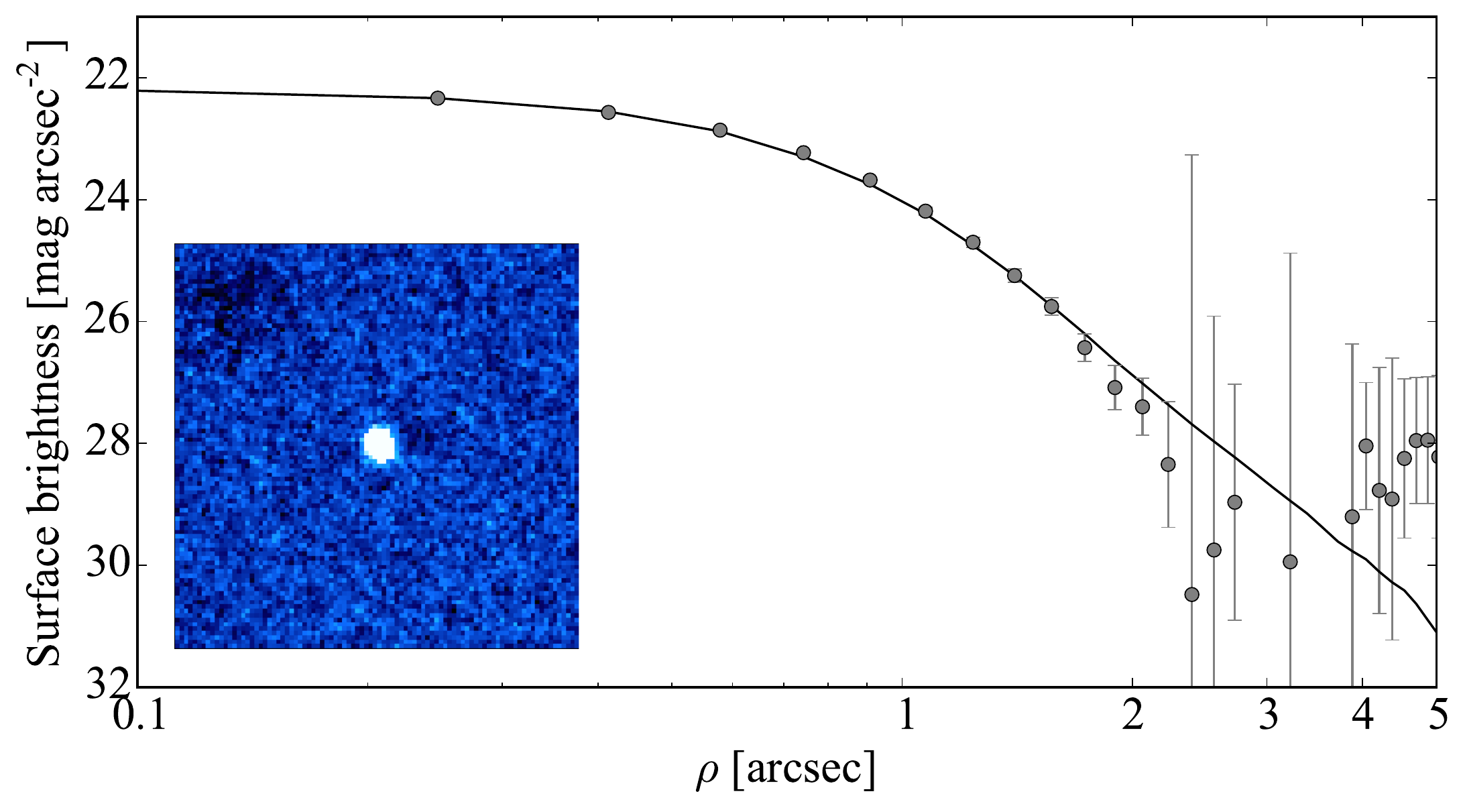}
   \caption{Same as Fig. \ref{14P_2004_PSF} for 137P on 13 July 2007. The co-added composite image is made up of 20 $\times$ 75 s exposures. The comet appears inactive and its surface brightness profile matches that of the comparison star.}
    \label{137P_2007_PSF}%
    \end{figure}

   Comet 137P was observed during one night in 2005 and two nights in 2007 as part of SEPPCoN. It appeared inactive during both observing epochs (Figs. \ref{137P_2005_PSF} and \ref{137P_2007_PSF}).
   
\begin{figure}
    \centering
   \includegraphics[width=0.48\textwidth]{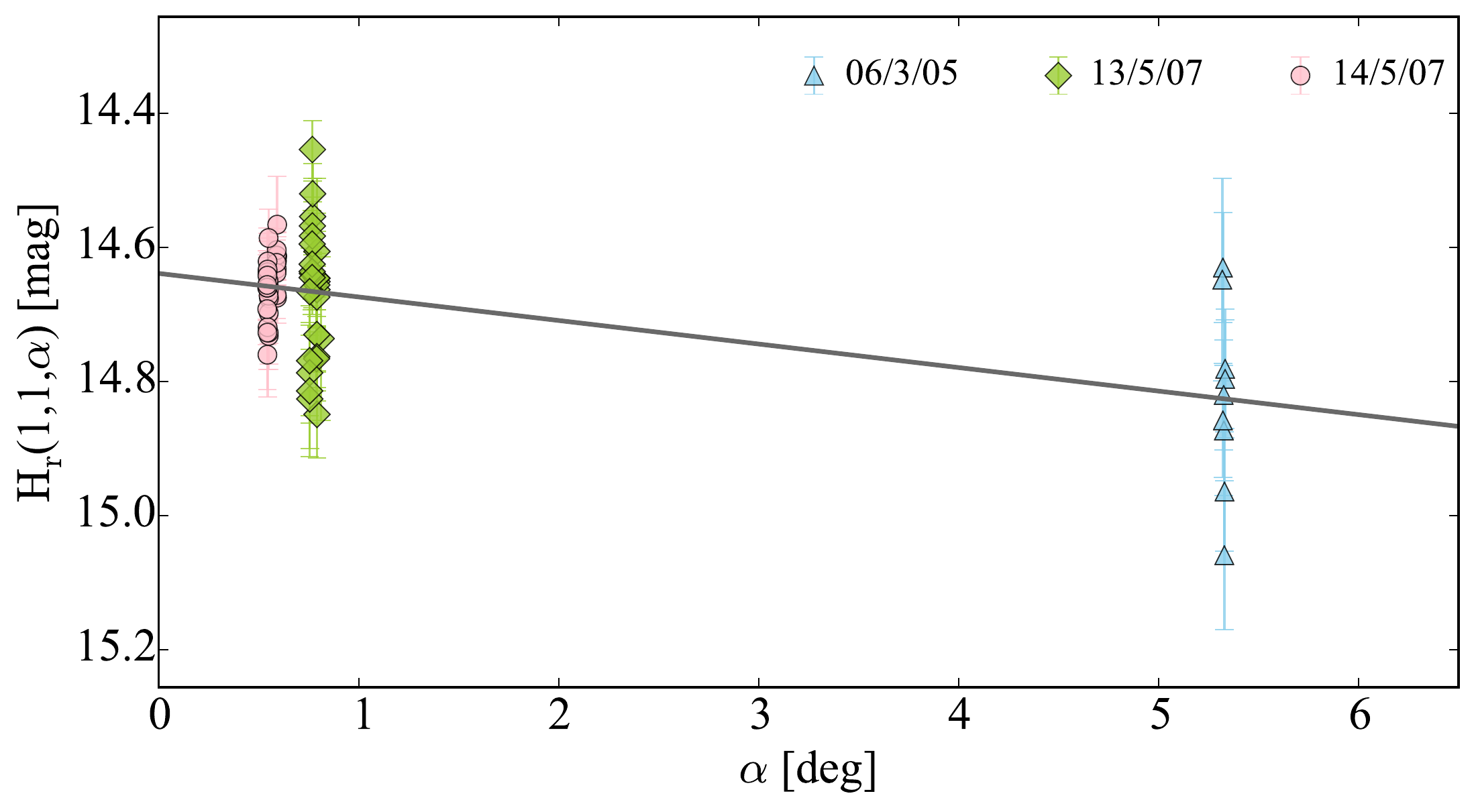}
   \caption{Phase function of comet 137P. The linear phase function coefficient derived from the Monte Carlo simulations is $\beta$ = 0.035 $\pm$ 0.004 mag deg\textsuperscript{-1}.} 
    \label{137P_PHASE}%
    \end{figure}

   We applied the MC method on the combined dataset from all three nights to determine the comet's phase function (Fig. \ref{137P_PHASE}). The derived phase function slope was $\beta$ = 0.035 $\pm$ 0.004 mag deg\textsuperscript{-1}. 
   
\begin{figure}
    \centering
   \includegraphics[width=0.48\textwidth]{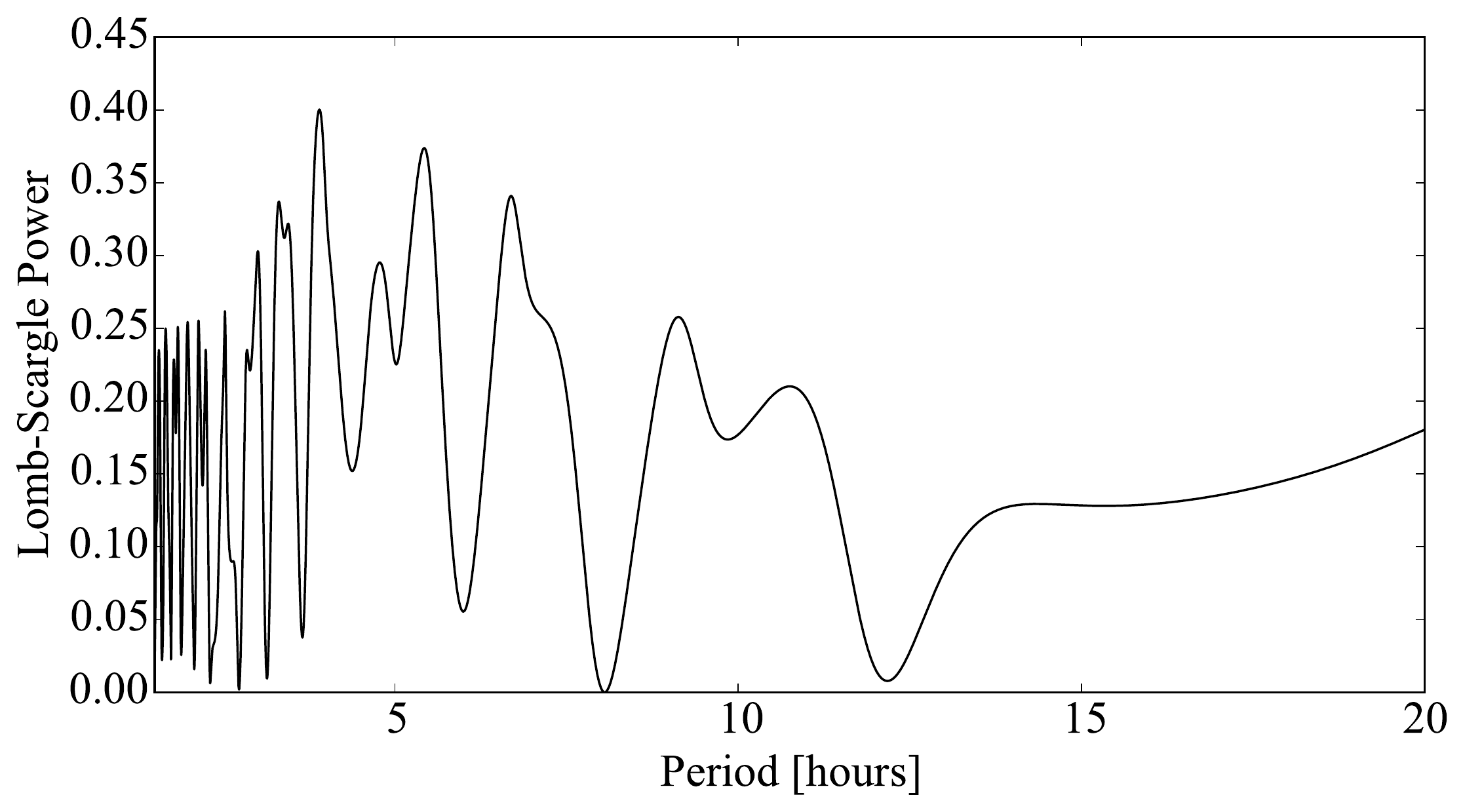}
   \caption{Lomb-Scargle periodogram of 137P from the 2007 dataset. The highest peak corresponds to a period of $P_{\mathrm{rot}}$ = 7.7 hours.}
    \label{137P_LS}%
    \end{figure}

   Next, we attempted to determine the lightcurve period from the data taken in 2005. The highest peak of the periodogram in Fig. \ref{137P_LS} corresponds to a rotation period of 7.7 hours. However, all peaks on the periodogram have low powers which are not sufficient to determine the rotation rate of 137P. 
   
\begin{figure}
    \centering
   \includegraphics[width=0.48\textwidth]{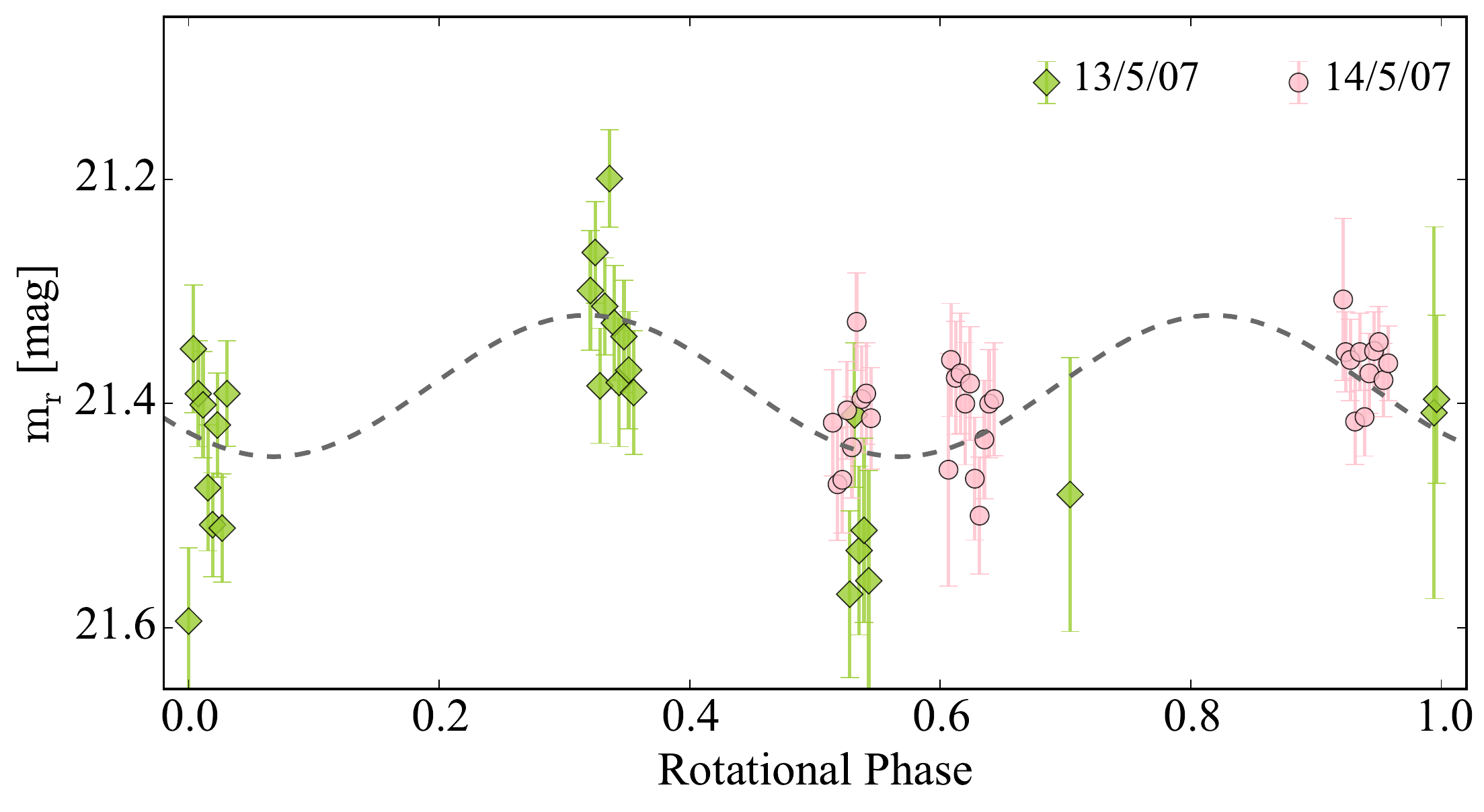}
   \caption{Rotational lightcurve of 137P with all of the data from 2007 folded with one of the possible periods, 7.7 hours. The uncertainty of the points is large in comparison to the brightness variation of the comet, which obstructs the period determination.}
    \label{137P_2007}%
    \end{figure}

   The lightcurve phased with a period of 7.7 hours is plotted in Fig. \ref{137P_2007}. Its brightness variation is $\Delta$$m_{\mathrm{r}}$ = 0.18 $\pm$ 0.05 mag, which converts to $a/b$ $\geq$ 1.18 $\pm$ 0.05. The uncertainties of the individual points are large in comparison with the detected brightness variation. Therefore, it is not possible to derive a precise rotation rate for the comet from this data set. We attempted to improve the period determination by combining all data from 2005 and 2007. However, the photometry from 2007 has even larger photometric uncertainties and does not lead to improvement of the period estimation. 

   The absolute magnitude of 137P is $H_{\mathrm{r}}$(1,1,0) = 14.63 $\pm$ 0.05 mag. Using Eq. \ref{eq:albedo} and the SEPPCoN radius from \cite{Fernandez2013}, we estimated and albedo $A_{\mathrm{r}}$ = 2.8$\pm$0.5\%.

\subsection{149P/Mueller 4}
\label{sec_res_149P}

\begin{figure}
    \centering
   \includegraphics[width=0.48\textwidth]{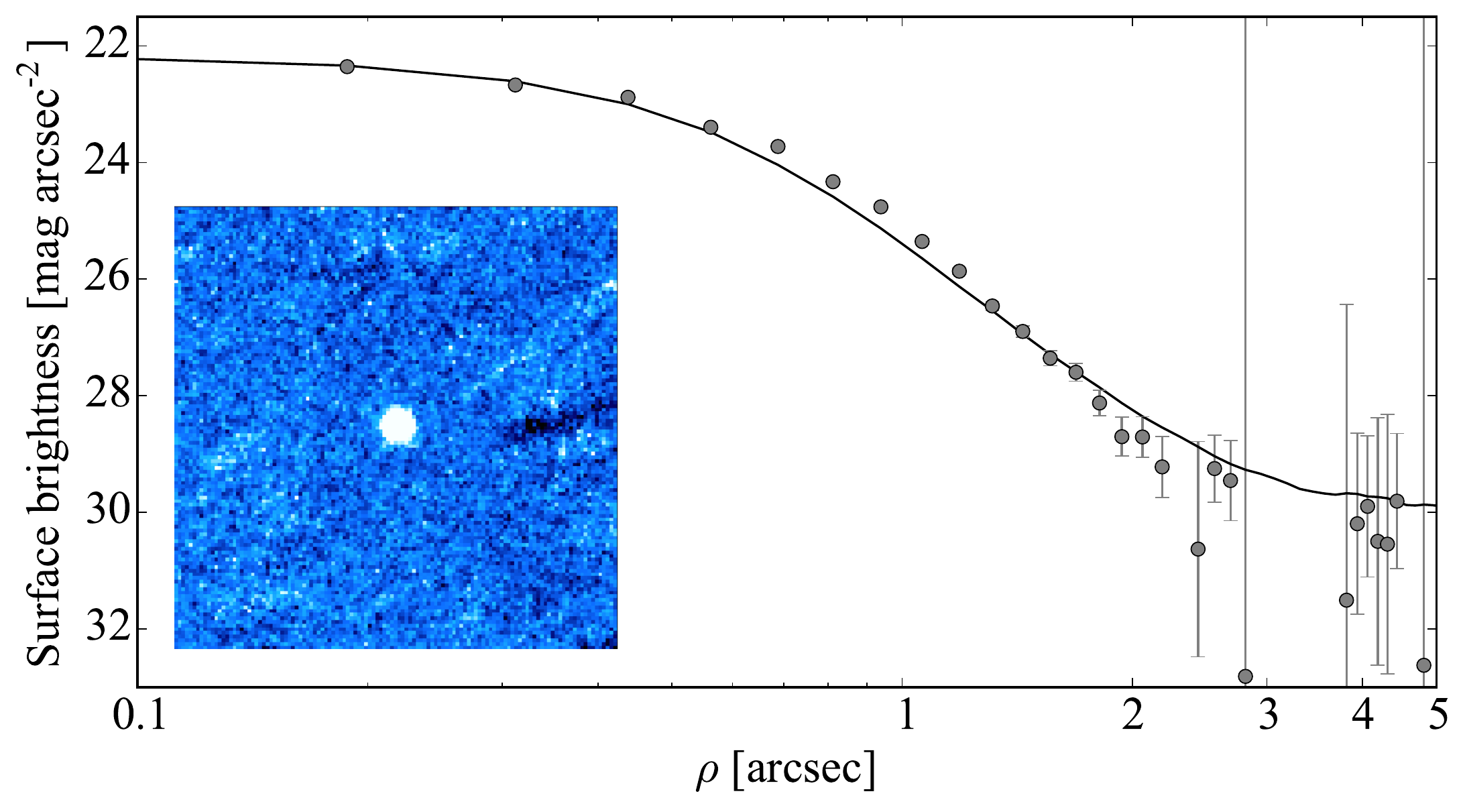}
   \caption{Same as Fig. \ref{14P_2004_PSF} for 149P on 23 January 2009. The co-added composite image is made up of 15 $\times$ 80 s exposures. The comet appears inactive and its surface brightness profile matches that of the comparison star.}
    \label{149P_PSF}%
    \end{figure}

Comet 149P was observed using NTT, WHT and VLT during  7 nights at the end of January 2009.  The surface brightness profiles of the comet for each night indicated that it was not active at  the time of the observations (see Fig. \ref{149P_PSF}).

The phase angle of 149P changed between 8.5 and 10 degrees  between the first and the last observing night. We used the MC method to constrain the phase function slope of the comet as $\beta$ = 0.03 $\pm$ 0.02 mag deg\textsuperscript{-1}. 

The periodogram of the time series corrected for geometric effects peaks at  $P_{\mathrm{rot}}$ = 11.9 hours. The period of 11.9 $\pm$ 0.1 is preferred by the MC simulation in 84\% of the iterations. However, the power of the peaks on the periodogram is too small and we cannot select the best period unambiguously. A rotation period near 12 hours would make this measurement for 149P difficult, and a clear determination of such a period using an Earth-based facility would require a longer photometric time sequence.

\begin{figure}
    \centering
   \includegraphics[width=0.48\textwidth]{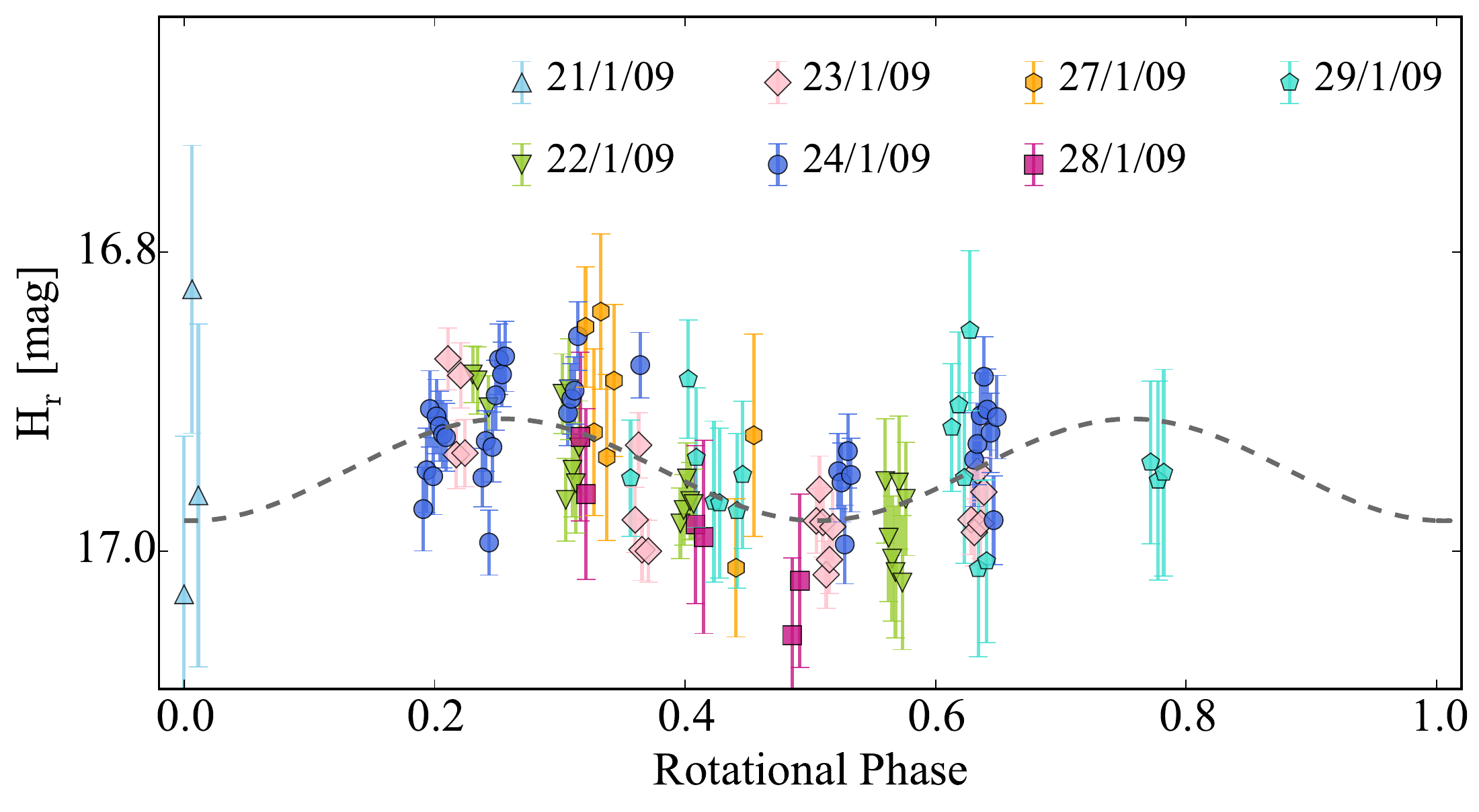}
   \caption{Rotational lightcurve of 149P with all of the data from 2009. The points from WHT and NTT were binned  The lightcurve is folded with the most-likely period of 11.88 hours.}
    \label{149P_2007}%
    \end{figure}

Figure  \ref{149P_2007} shows the lightcurve of 149P with the best fit from the MC method. The  photometric uncertainty of the individual points is large with respect to the total brightness variation  of the lightcurve, which confirms that the derived lightcurve is uncertain. 

The brightness variation of the comet is  $\Delta m_{\mathrm{r}}$ = 0.11 $\pm$ 0.04 mag which converts to  $a/b$ $\geq$ 1.11 $\pm$ 0.04. The observed mean magnitude of 149P was $m_{\mathrm{r}}$ = 22.14 $\pm$ 0.04 mag which corresponds to $H_{\mathrm{r}}$(1,1,0) = 16.93 $\pm$ 0.04 if the derived phase function with $\beta$ = 0.03 $\pm$ 0.02 mag deg\textsuperscript{-1} is used. Using  Eq. \ref{eq:albedo}, we can calculate that the albedo of 149P is $A_{\mathrm{r}}$ = 2.8 $\pm$ 0.4\%. 
   
     \begin{figure}
    \centering
   \includegraphics[width=0.48\textwidth]{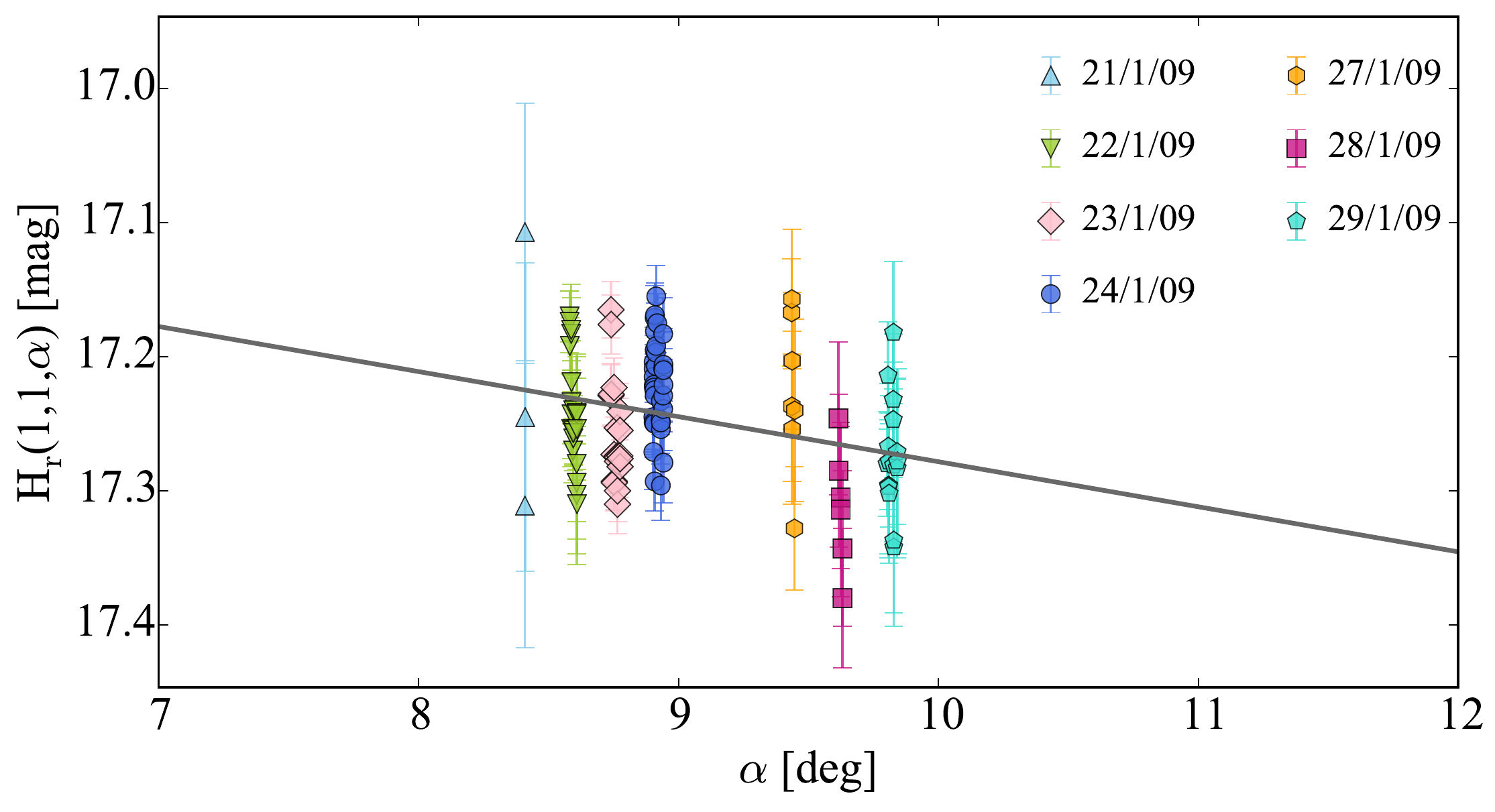}
   \caption{Phase function of comet 149P. The linear phase function coefficient derived from the Monte Carlo simulations is $\beta$ = 0.03 $\pm$ 0.02 mag deg\textsuperscript{-1}.} 
    \label{149P_PHASE}%
    \end{figure}
   
   \begin{figure}
    \centering
   \includegraphics[width=0.48\textwidth]{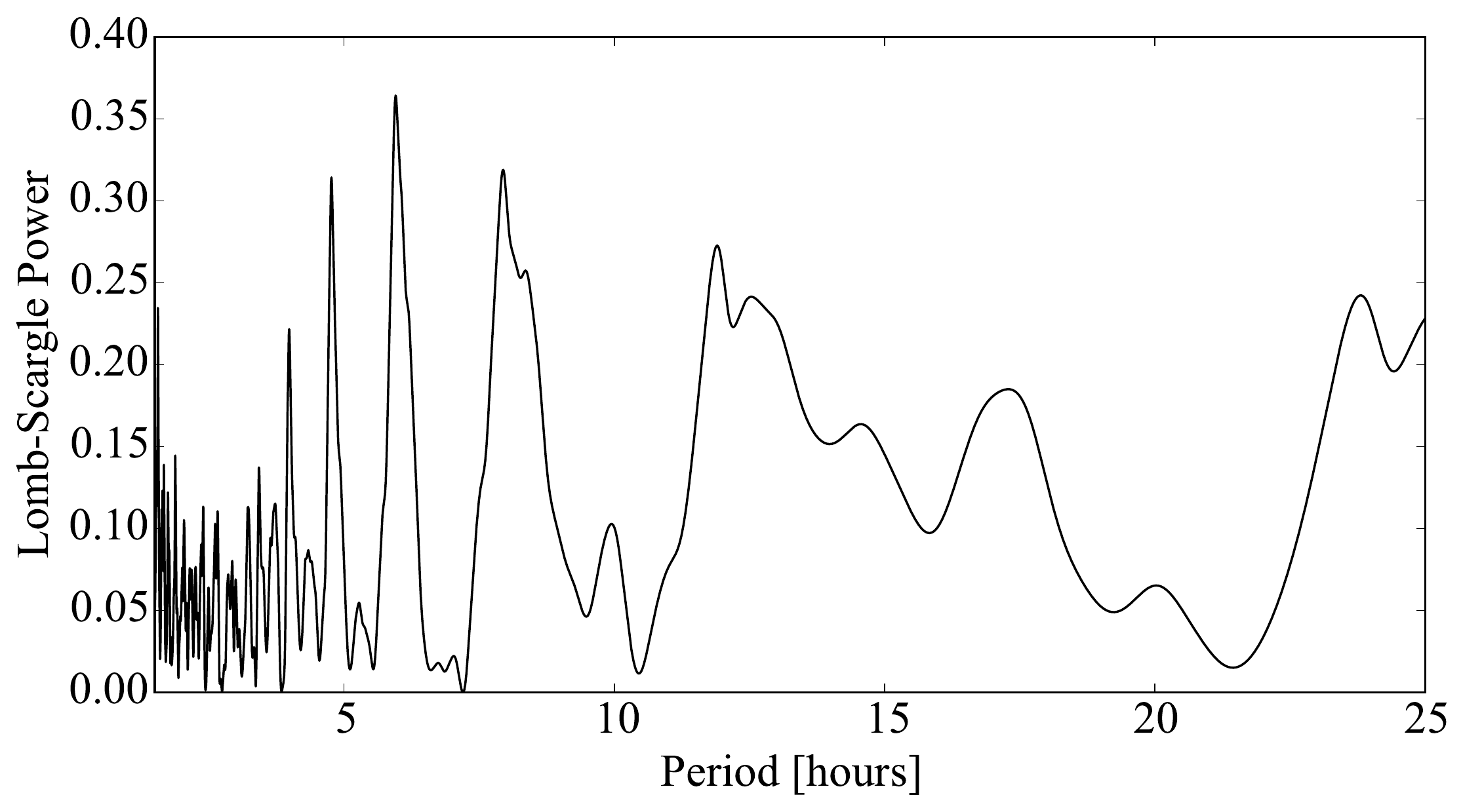}
   \caption{Lomb-Scargle periodogram of the combined datasets for 149P showing the LS power versus period. The highest peak corresponds to the most likely period $P_{\mathrm{rot}}$ = 11.88 hours. Since all peaks have low power, the spin period of the comet cannot be determined unambiguously.}
    \label{149P_LS}%
    \end{figure}
    
\subsection{162P/Siding Spring}
\label{sec_res_162P}

Comet 162P was observed in 2007 around its aphelion, and again in 2012 close to its next aphelion passage. The first set of observations aimed to determine the comet's lightcurve, while the second data set focused on its phase function. 

\begin{figure}
    \centering
   \includegraphics[width=0.48\textwidth]{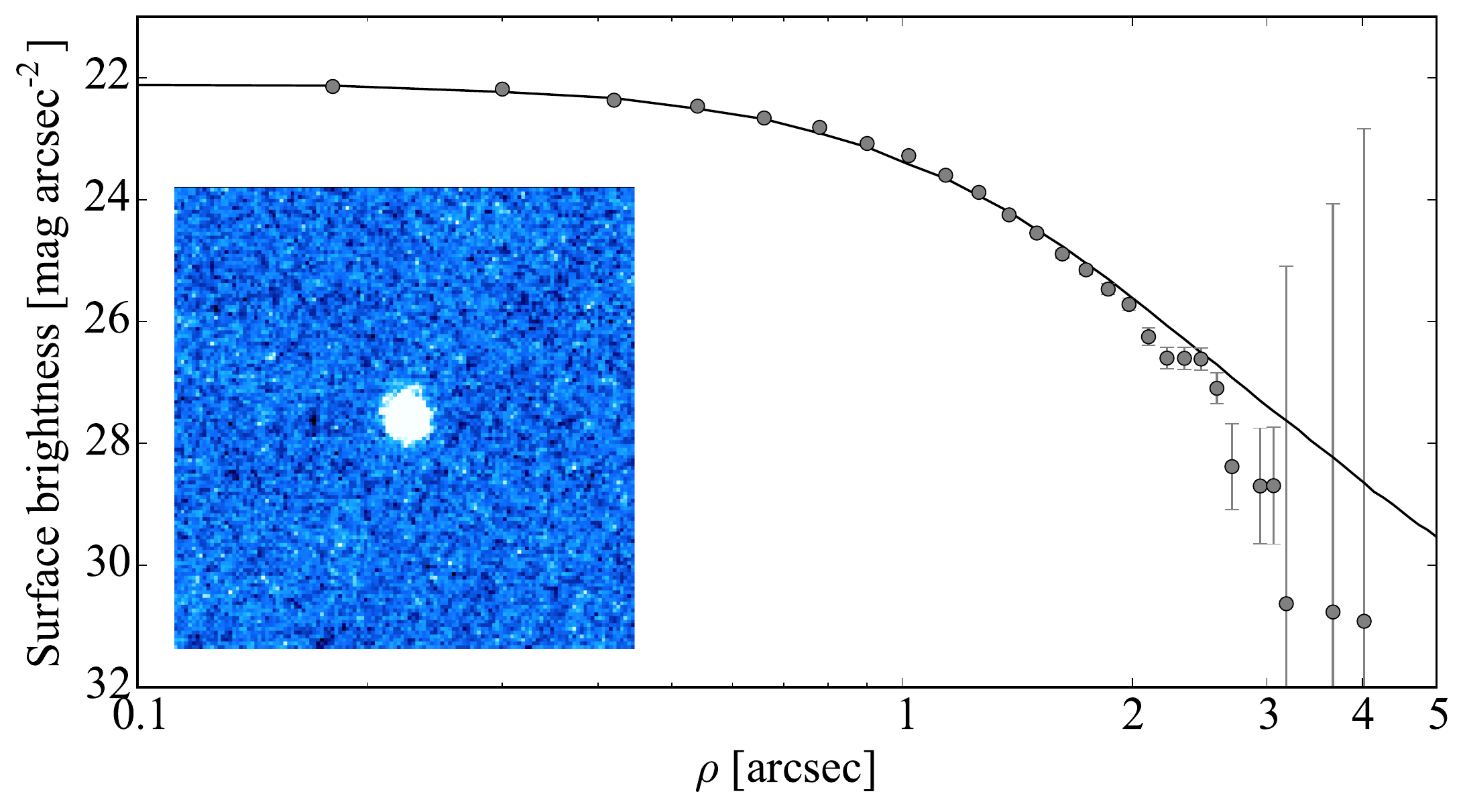}
   \caption{Same as Fig. \ref{14P_2004_PSF}, for 162P on 18 May 2007. The co-added composite image is made up of 10 $\times$ 110 s exposures. The comet appears inactive and its surface brightness profile agrees with that of the comparison star.}
    \label{162P_2007_PSF}%
    \end{figure}

\begin{figure}
    \centering
   \includegraphics[width=0.48\textwidth]{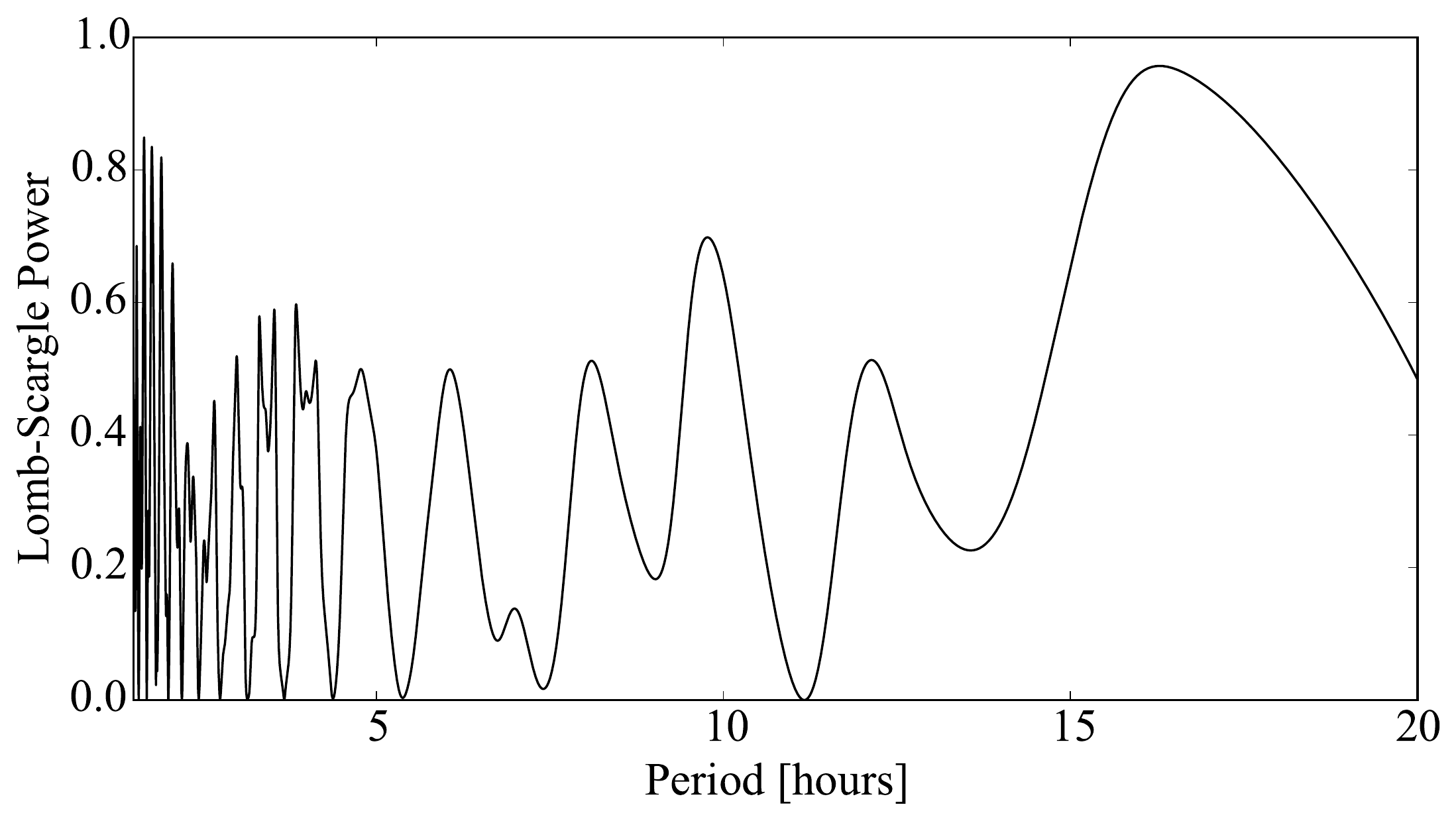}
   \caption{Lomb-Scargle periodogram of the 2007 dataset for 162P showing the LS power versus period. The highest peak corresponds to the most likely period $P_{\mathrm{rot}}$ = 32.6 hours.}
    \label{162P_2007_LS}%
    \end{figure}

\begin{figure}
    \centering
   \includegraphics[width=0.48\textwidth]{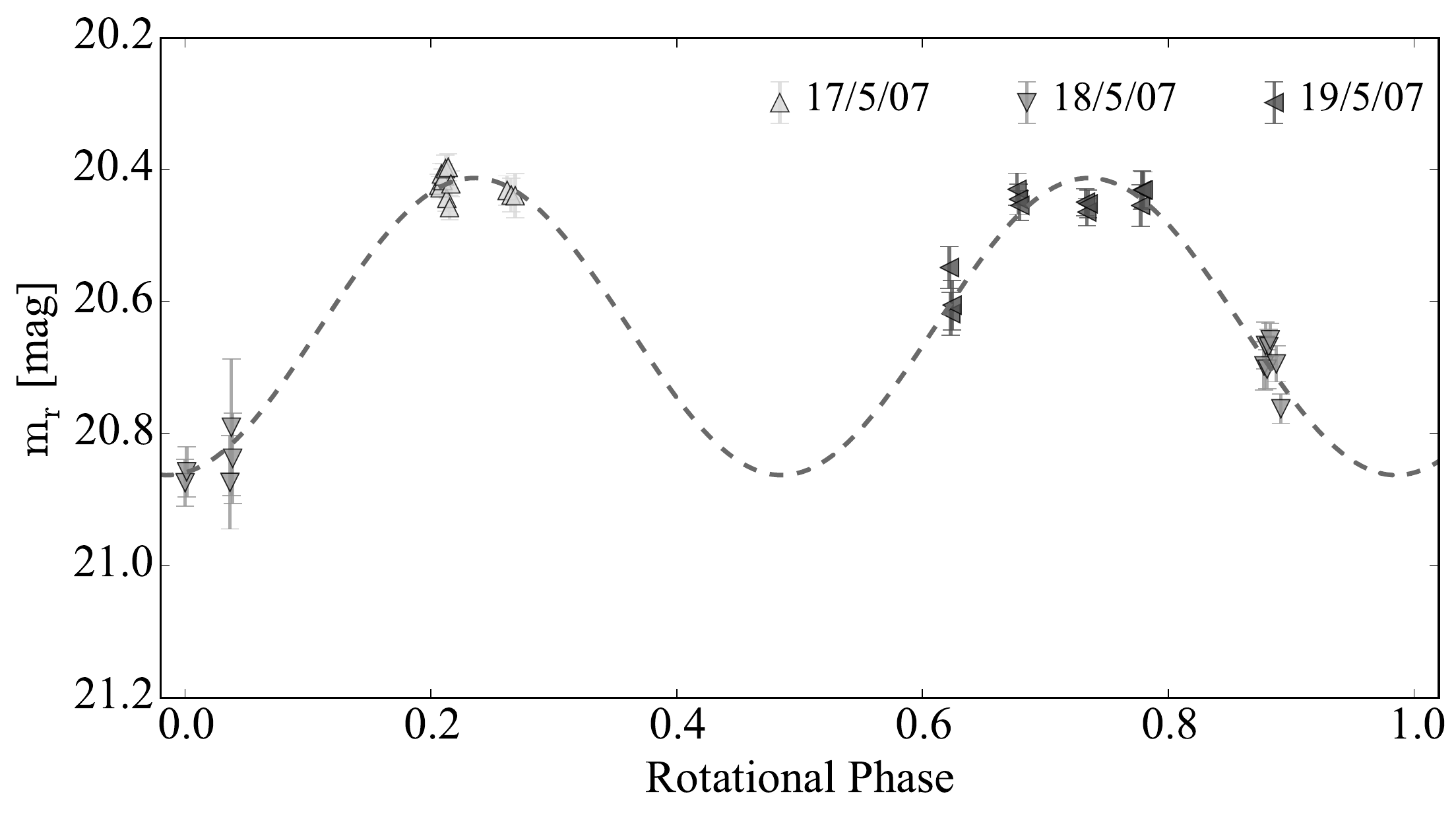}
   \caption{Rotational lightcurve of 162P with the data from 2007. The lightcurve is folded with period 32.6 hours.}
    \label{162P_2007}%
    \end{figure}

The comet had a stellar profile and appeared to be inactive in 2007 (Fig. \ref{162P_2007_PSF}). The LS periodogram of the data from the three observing nights in 2007 is shown in Fig. \ref{162P_2007_LS}. The most pronounced peak in the periodogram corresponds to  $P_{\mathrm{rot}}$ = 32.6 hours, and the lightcurve phased with that period can be seen in Fig. \ref{162P_2007}. Using the MC method without phase function correction, we determined the rotation period of the comet to be $P_{\mathrm{rot}}$ = 32.6 $\pm$ 1 hours. This period is in good agreement with the value of $\sim$ 33 hours determined by the team of La Ca\~{n}ada observatory (see Section \ref{sec:rev_162P}). 

From the observations in 2007, we measured the mean magnitude of 162P to be $m_{\mathrm{r}}$ = 20.63 $\pm$ 0.05 mag. The brightness variation of the comet was $\Delta$$m_{\mathrm{r}}$ = 0.45 $\pm$ 0.05 mag, which corresponds to $a/b$ $\geq$ 1.51 $\pm$ 0.07.

\begin{figure}
    \centering
   \includegraphics[width=0.48\textwidth]{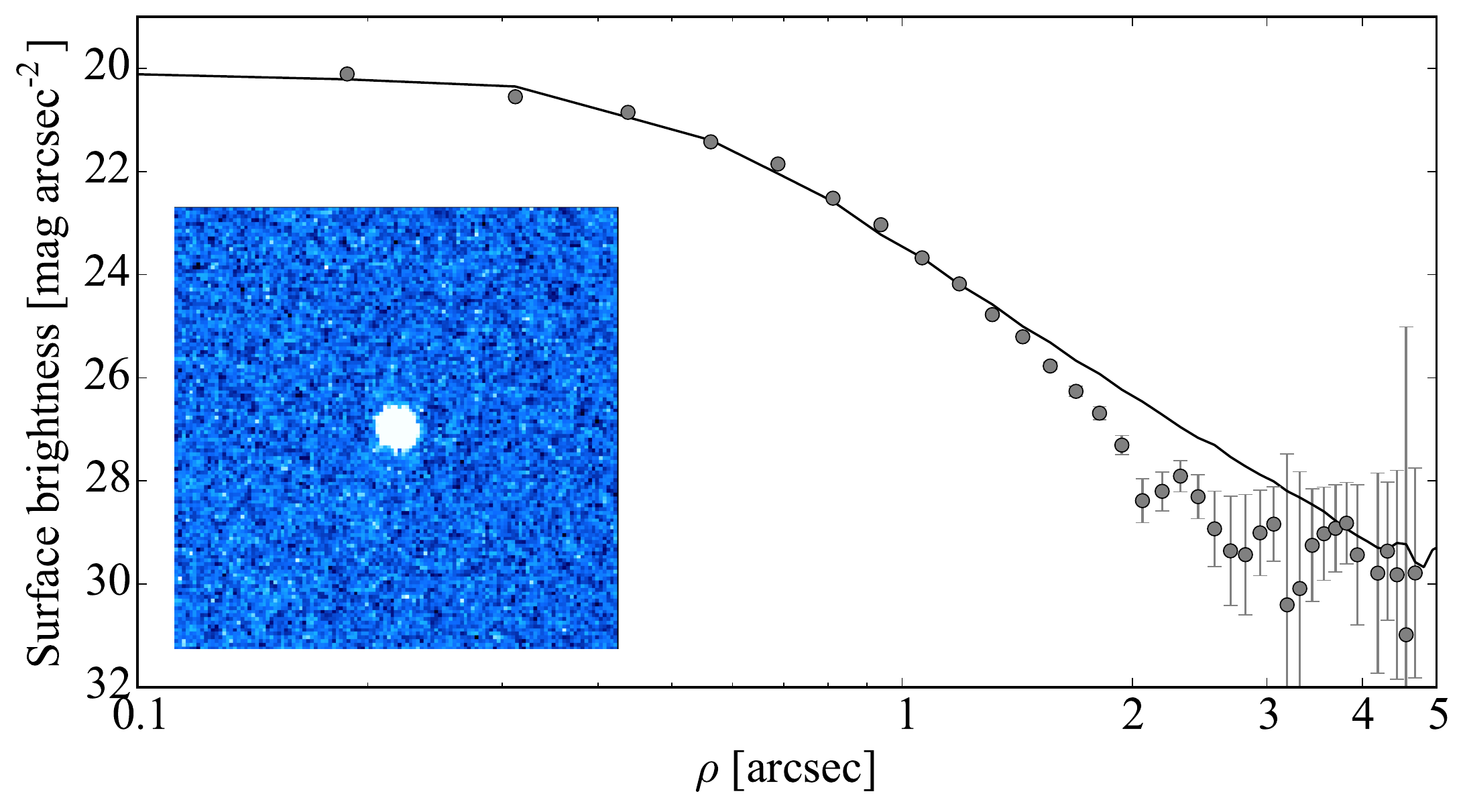}
   \caption{Same as Fig. \ref{14P_2004_PSF}, for 162P on 23 April 2012. The co-added composite image is made up of 5 $\times$ 60 s exposures. The comet appears inactive and its surface brightness profile generally agrees with that of the comparison star. The narrower profile of the comet is most likely an artefact of the position uncertainty of the comet on the frames.}
    \label{162P_2012_PSF}%
    \end{figure}

Comet 162P was also inactive during all observations in 2012, which is demonstrated by the surface brightness plot in Fig. \ref{162P_2012_PSF}. Since the observations were taken at a large phase angle range (4-12\textsuperscript{$\circ$}), we could only combine the data after deriving the comet's phase function. The MC method determined a phase function coefficient $\beta$ = 0.039 $\pm$ 0.002 mag deg\textsuperscript{-1}. 

\begin{figure}
    \centering
   \includegraphics[width=0.48\textwidth]{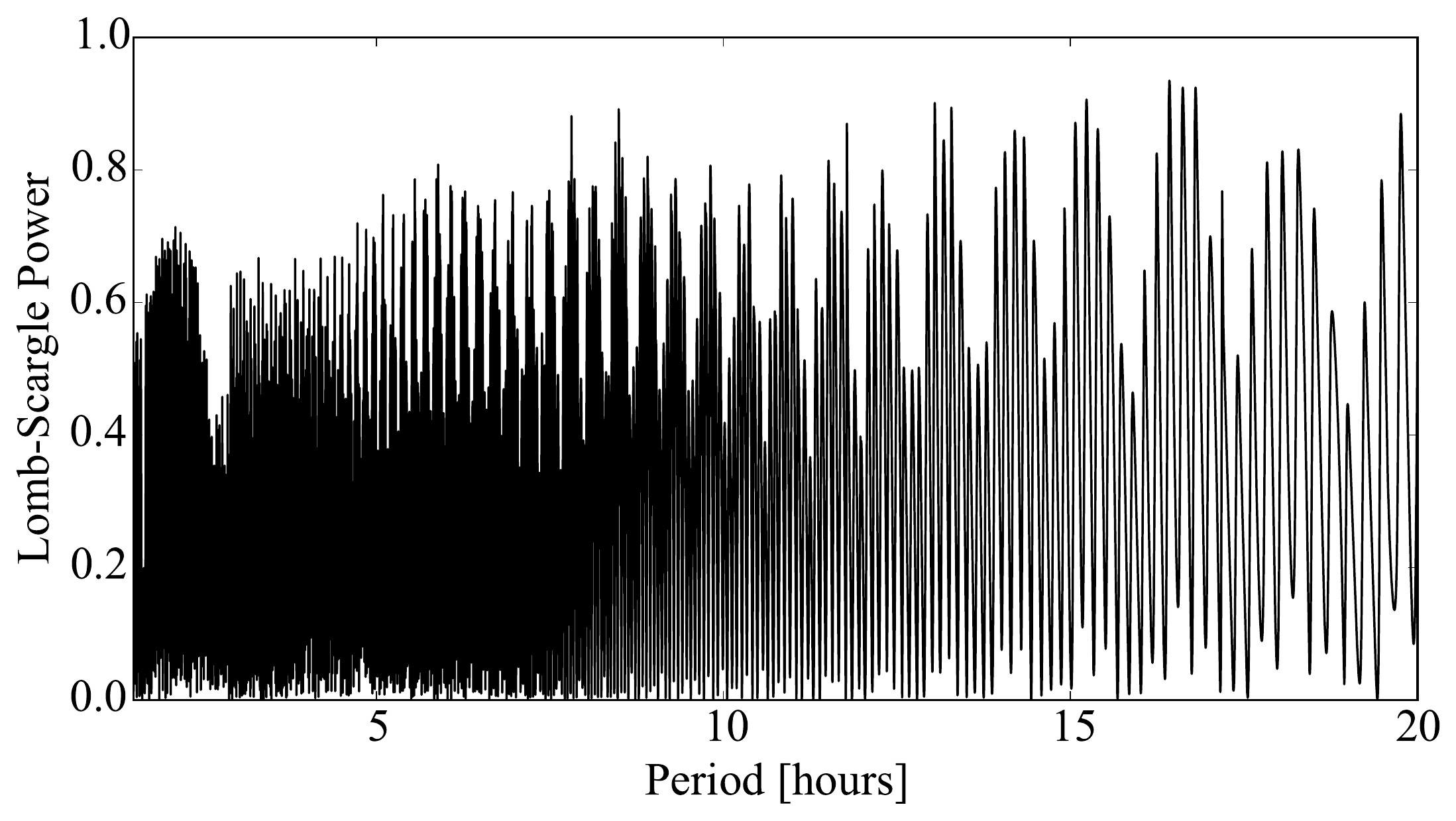}
   \caption{Lomb-Scargle periodogram of the 2012 dataset for 162P showing the LS power versus period. There are a number of possible periods as well as secondary peaks caused by aliasing. The highest peaks correspond to rotation periods of 32.852 hours and 33.237 hours.  }
    \label{162P_2012_LS}%
    \end{figure}

\begin{figure}
    \centering
   \includegraphics[width=0.48\textwidth]{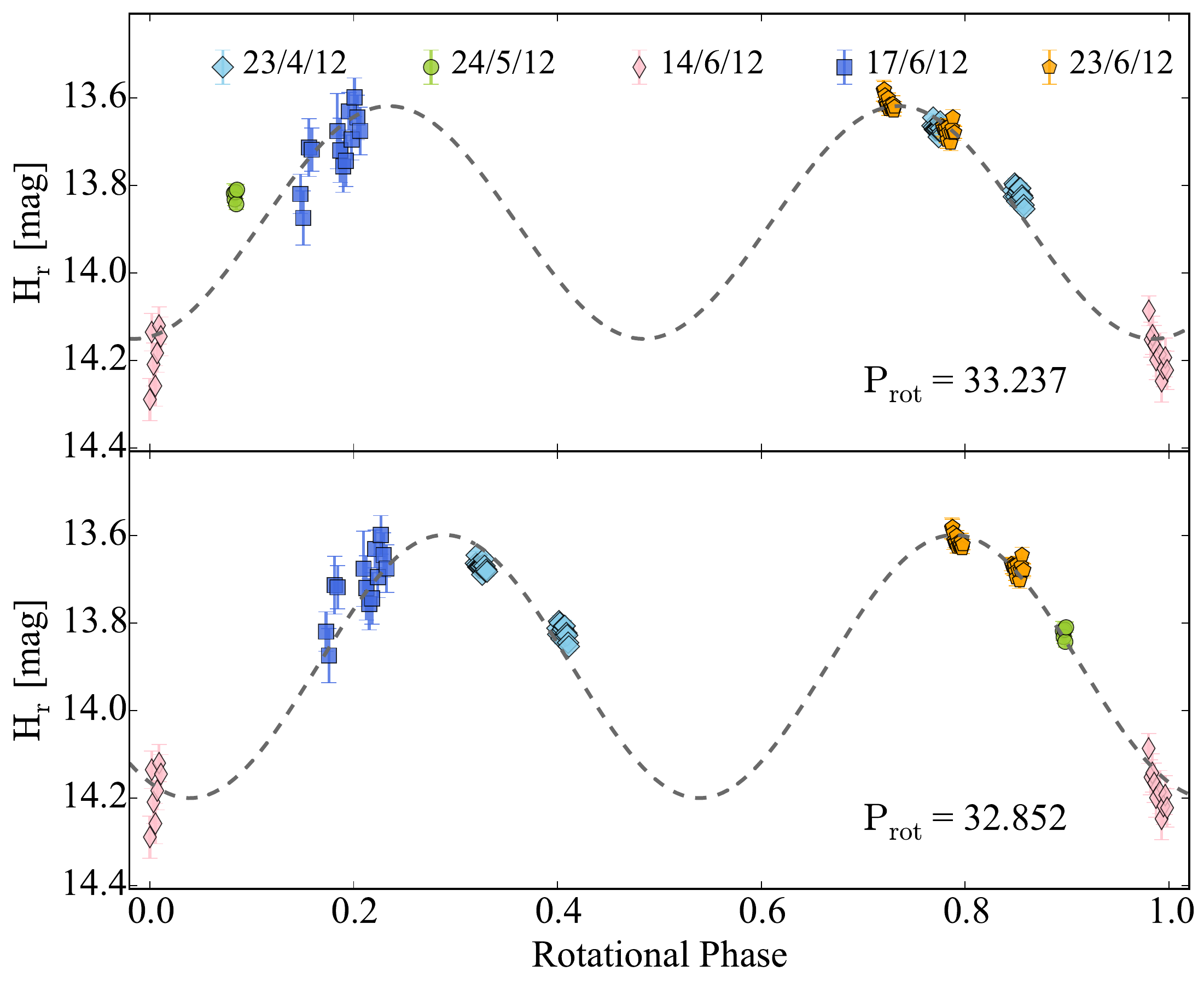}
   \caption{Rotational lightcurve of 162P with the data from 2012. The lightcurve is folded with $P_{\mathrm{rot,1}}$ = 33.237 hours (top) and $P_{\mathrm{rot,2}}$ = 32.852  hours (bottom). It is not possible to choose between the two periods from the data set collected in 2012.}
    \label{162P_2012}%
    \end{figure}

The LS periodogram of the combined data set from 2012 suggested multiple possible rotation periods for 162P (Fig. \ref{162P_2012_LS}). The MC method preferred $P_{\mathrm{rot,1}}$ = 33.237 $\pm$ 0.008 hours in 62\% of the iterations and $P_{\mathrm{rot,2}}$ = 32.852 $\pm$ 0.003 hours in 35\% of the iterations. The lightcurves in Fig. \ref{162P_2012} confirm that due to the limited sampling of the lightcurve, it is impossible to choose between these two possibilities, although it is worth noting that the points from 24 May 2012 agree better with $P_{\mathrm{rot,2}}$ = 32.852. 


The brightness variation in the 2012 observations was $\Delta m_{\mathrm{r}} = 0.59 \pm 0.04$ mag, which corresponds to $a/b \geq 1.72 \pm 0.06$. The absolute magnitude of 162P from the 2012 dataset was $H_{\mathrm{r}}(1,1,0) = 13.91 \pm 0.04$ mag. If we use Eq. \ref{eq:albedo}, we can estimate the albedo of 162P to be $A_\mathrm{r} = 1.8 \pm 0.3$\%. This result makes comet 162P the JFC with the lowest known albedo (see Section \ref{sec:disc_surface}).

As a final step in the analysis of the data for 162P, we combined the two datasets from 2007 and 2012 in order to attempt constraining the comet's lightcurve and phase function better.  It is possible that the period of 162P slightly changed between 2007 and 2012 while the comet was active close to perihelion. Besides, it is not excluded that since the two observations were done at different geometries, the resulting lightcurves can appear different. Nevertheless, it is worth attempting to combine the two data sets as the increased number of observations can provide a better understanding of the nucleus' properties. 

\begin{figure}
    \centering
   \includegraphics[width=0.48\textwidth]{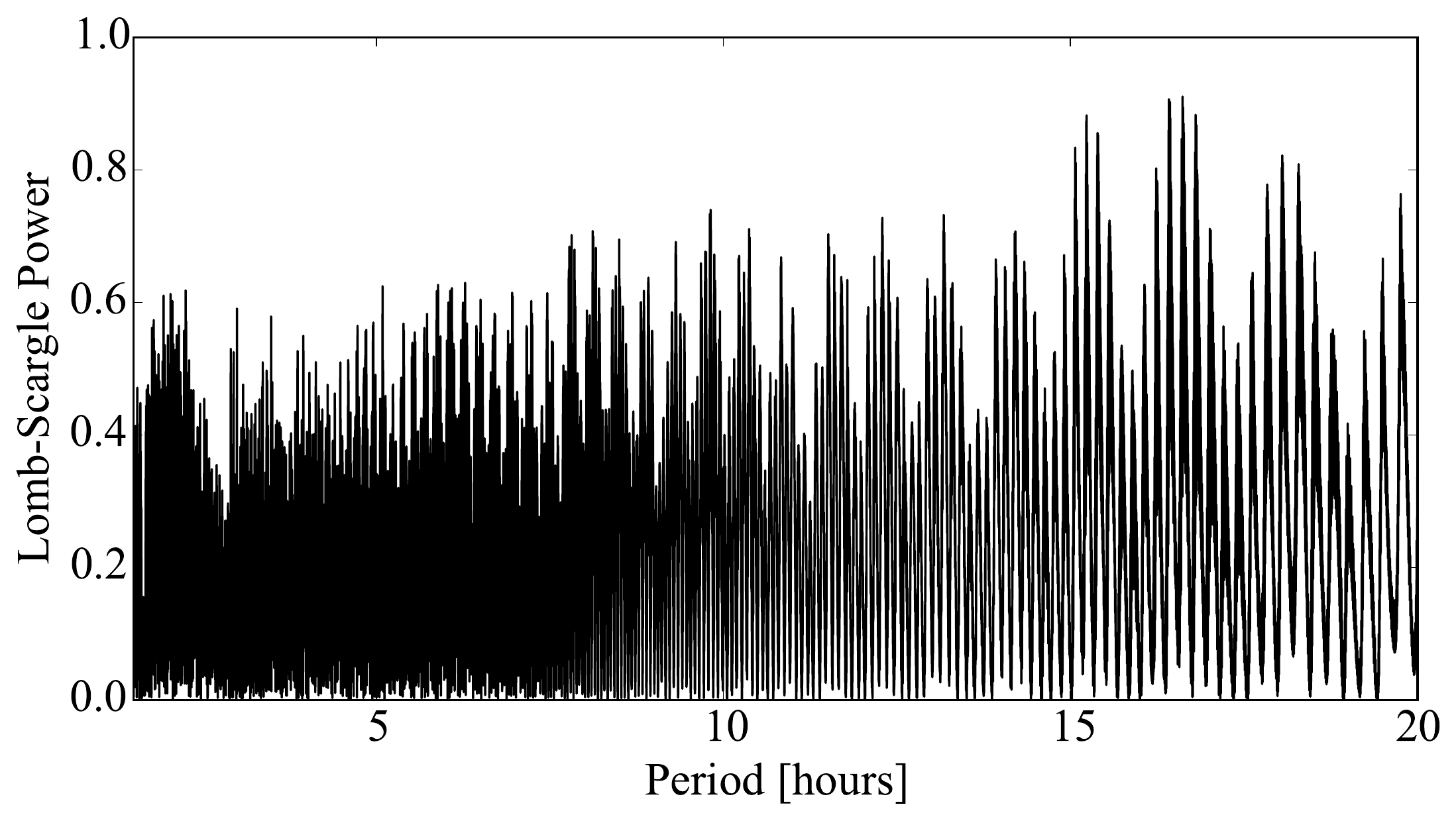}
   \caption{Lomb-Scargle periodogram of the combined datasets of 162P from 2007 and 2012 showing the LS power versus period. The highest peak corresponds to $P_{\mathrm{rot}}$ = 32.853 hours.}
    \label{162P_BOTH_LS}%
    \end{figure}

With these caveats in mind, we proceeded to analyse the combined data from 2007 and 2012. The MC method suggested a phase function with a slope $\beta = 0.038 \pm 0.002$ mag deg\textsuperscript{-1} and a lightcurve with period $P_{\mathrm{rot}} = 32.853 \pm 0.002$ hours. This period corresponds to the highest peak of the LS periodogram in Fig. \ref{162P_BOTH_LS}. 

The derived parameters from the combined data set are very close to those of the 2012 data set alone (See. Table \ref{tab:resultsl}). However since they were derived using data from two different apparitions, we consider the values from just the 2012 data set to be less uncertain.

    \begin{figure}
    \centering
   \includegraphics[width=0.48\textwidth]{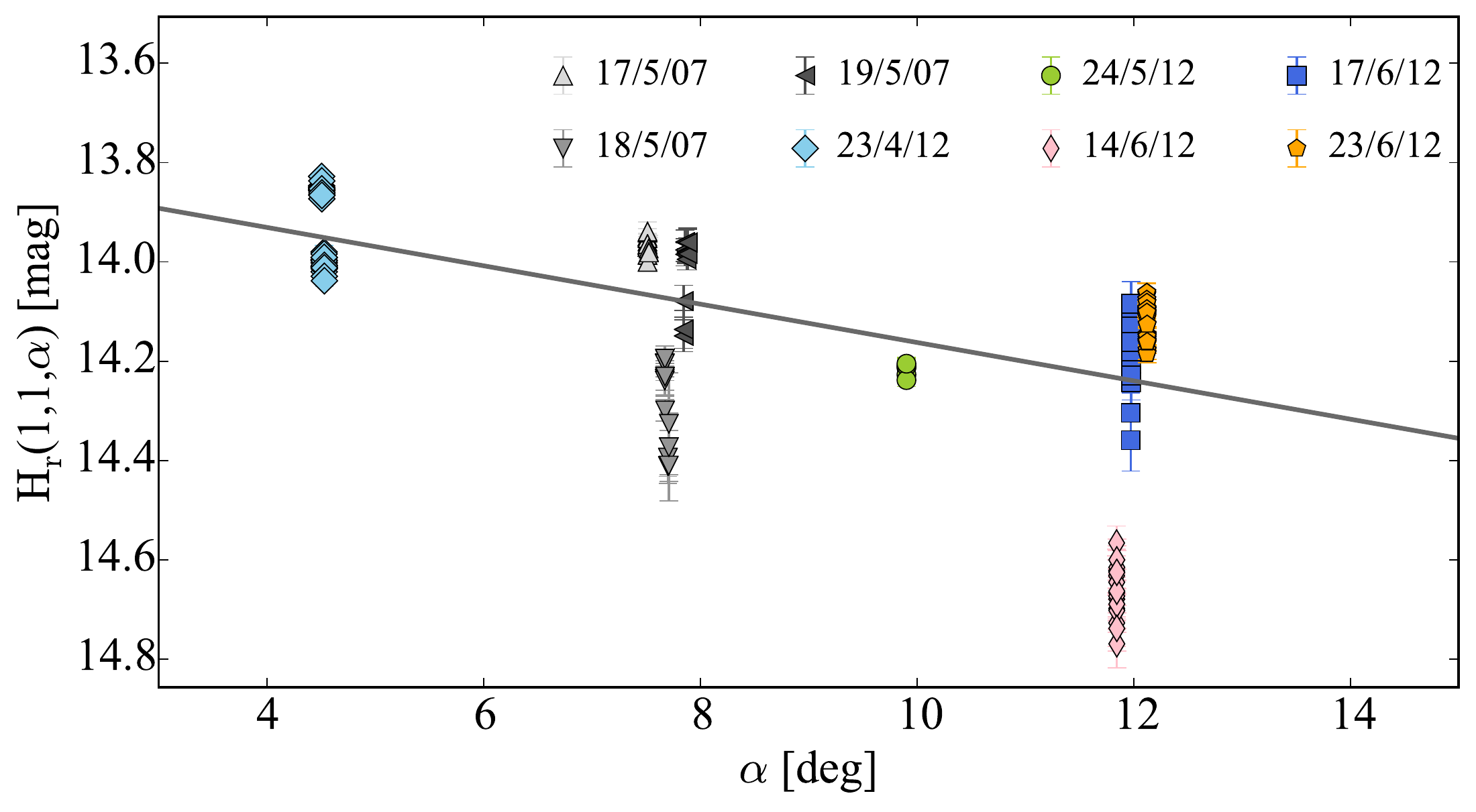}
   \caption{Phase function of comet 162P. The linear phase function slope derived from the Monte Carlo simulations is $\beta = 0.038 \pm 0.002$ mag deg\textsuperscript{-1}.} 
    \label{162P_PHASE}%
    \end{figure}
    
    \begin{figure}
    \centering
   \includegraphics[width=0.48\textwidth]{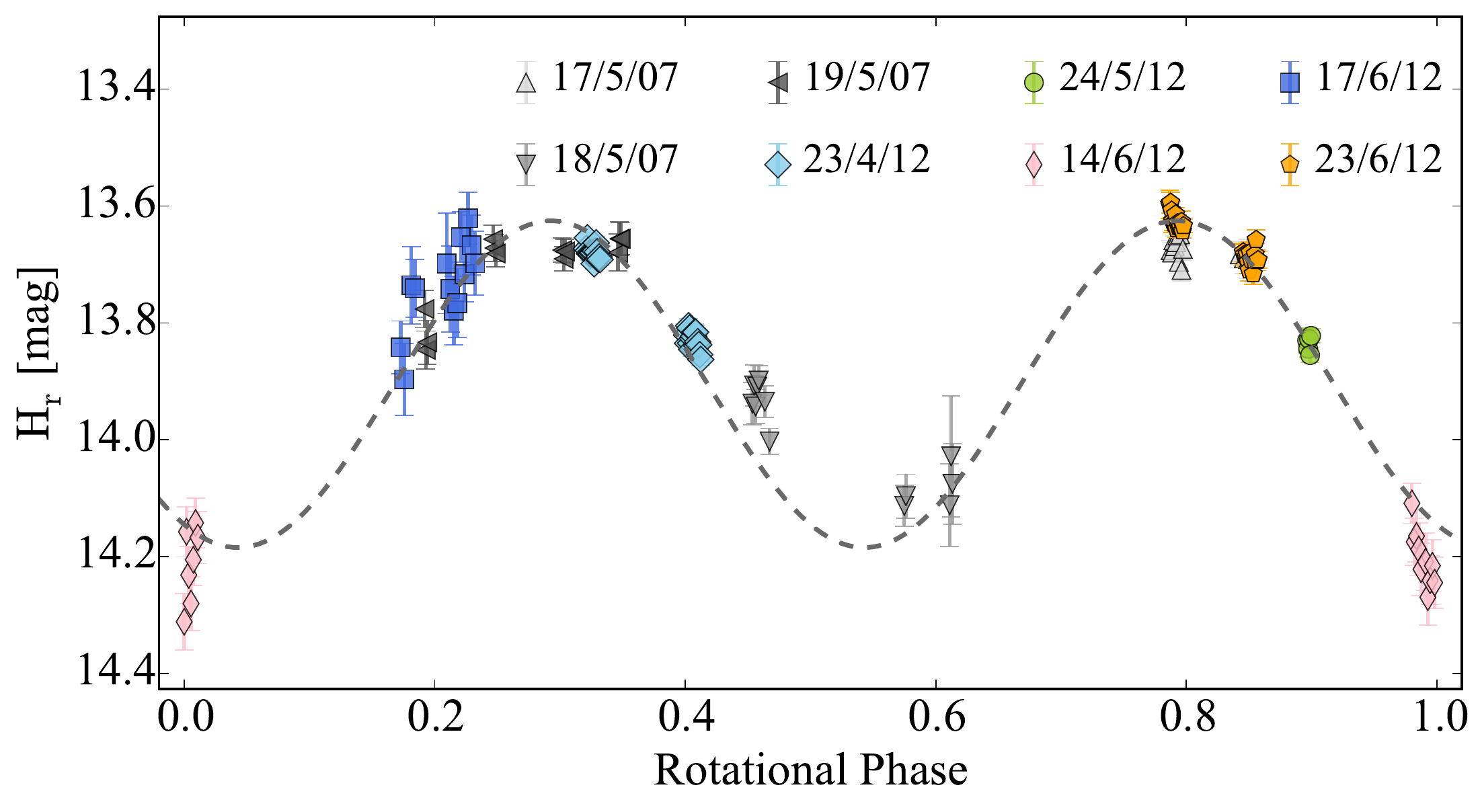}
   \caption{Rotational lightcurve of 162P with the data from 2007 and 2012. The lightcurve is folded with the most likely period of 32.853 hours.}
    \label{162P_BOTH}%
    \end{figure}

\begin{table*}
\centering
\caption{Derived physical parameters for all observed comets.}
\label{tab:resultsl}
\setlength\tabcolsep{3.5pt}
\begin{tabular}{lllllllllll}
\hline
Comet & Epoch & $m_{\mathrm{r}}$\textsuperscript{1} & $H_{\mathrm{r}}$(1,1,0)\textsuperscript{1} & $P_{\mathrm{rot}}$ [h]\textsuperscript{2} & $\beta$  [mag/deg]\textsuperscript{3} & $r_{\mathrm{N}}$ [km]\textsuperscript{4} & $A_{\mathrm{r}}$ [\%]\textsuperscript{5} & $\Delta$$m_{\mathrm{r}}$ & $a/b$ & $D_{\mathrm{N}}$ [$\mathrm{g \ cm^{-3}}$] \\
\hline
14P 	& 2004 			& 22.58$\pm$0.05 		& - 								& 8.93$\pm$0.04 				& - 				& - 			& - 								& 0.36$\pm$0.05	& 1.39$\pm$0.06	& 0.19$\pm$0.04\\
 		& 2007 			& 21.06$\pm$0.05 		& - 								& 9.02$\pm$0.04 				& - 				& - 			& - 								& 0.39$\pm$0.05	& 1.43$\pm$0.07	& 0.19$\pm$0.04\\
 		& Combined 		& - 					& 14.87$\pm$0.05 					& 9.02$\pm$0.01 				& 0.060$\pm$0.005	& - 			& 4.3$\pm$0.6 						& 0.37$\pm$0.05	& 1.41$\pm$0.06	& 0.19$\pm$0.03\\
47P 	& 2005* 			& 21.83$\pm$0.06 		& - 								& 10.8/14.1 					& - 				& - 			& - 								& 0.33$\pm$0.06	& 1.36$\pm$0.07	& -	\\
 		& 2006* 		& 21.55$\pm$0.04 		& - 								& - 							& - 				& - 			& - 								& -		 		& -				& -	\\
 		& 2015* 		& 21.11$\pm$0.06 		& 14.58$\pm$0.06\textsuperscript{a} & 15.6$\pm$0.1 					& - 			& - 			& - 								& 0.24$\pm$0.06	& 1.25$\pm$0.07	& 0.06$\pm$0.02\\
 		& 2005 + 2015** & - 					& 14.59$\pm$0.06 					& - 							& 0.096$\pm$0.004 	& - 			& 5.0$\pm$0.7\textsuperscript{c} 	& -			 	& -				& -	\\
93P 	& 2009* 		& 21.09$\pm$0.05 		& 15.17$\pm$0.05\textsuperscript{b} & $\mathrm{18.2^{+1.5}_{-15}}$ 	& - 				& - 			& 4.2$\pm$0.9\textsuperscript{c} 	& 0.21$\pm$0.05	& 1.21$\pm$0.06	& -	\\
94P 	& 2005 			& 21.3$\pm$0.1 			& - 								& 20.43$\pm$0.05 				& - 				& - 			& - 								& 0.7$\pm$0.1	& 1.9$\pm$0.2	& 0.05$\pm$0.01\\
 		& 2007 			& 22.6$\pm$0.2 			& - 								& - 							& - 				& - 			& - 								& 1$\pm$0.2		& 2.5$\pm$0.5	& -	\\
 		& 2009 			& 21.30$\pm$0.05 		& - 								& - 							& - 				& - 			& - 								& 0.80$\pm$0.05	& 2.09$\pm$0.10	& -	\\
 		& Combined 		& - 					& 15.50$\pm$0.09 					& 20.70$\pm$0.07 				& 0.039$\pm$0.002 	& - 			& 4.0$\pm$0.6 						& 1.11$\pm$0.09	& 2.8$\pm$0.2	& 0.07$\pm$0.02\\
110P 	& 2012 			& - 					& 15.47$\pm$0.03 					& 10.153$\pm$0.001 				& 0.069$\pm$0.002 	& 2.31$\pm$0.03 & - 								& 0.20$\pm$0.03	& 1.20$\pm$0.03	& 0.13$\pm$0.02\\
123P 	& 2007* 		& 23.3$\pm$0.1 			& 15.7$\pm$0.1\textsuperscript{b} 	& - 							& - 				& - 			& 3.6$\pm$0.8\textsuperscript{c} 	& 0.5$\pm$0.1	& 1.6$\pm$0.1	& -	\\
137P 	& 2007 			& 21.39$\pm$0.05 		& - 								& - 							& - 				& - 			& - 								& 0.18$\pm$0.05	& 1.18$\pm$0.05	& -	\\
 		& 2005 + 2007 	& - 					& 14.63$\pm$0.05 					& - 							& 0.035$\pm$0.004 	& - 			& 2.8$\pm$0.5 						& -			 	& -				& -	\\
149P 	& 2009 			& 22.14$\pm$0.04 		& 16.93$\pm$0.04 					& - 							& 0.03$\pm$0.02 	& - 			& 2.8$\pm$0.4 						& 0.11$\pm$0.04	& 1.11$\pm$0.04	& -	\\
162P 	& 2007 			& 20.63$\pm$0.05 		& - 								& 32.6$\pm$1			 		& - 				& - 			& - 								& 0.45$\pm$0.05	& 1.51$\pm$0.07	& -	\\
 		& 2012 			& - 					& 13.91$\pm$0.04 					& 33.237/32.852 				& 0.039$\pm$0.002 	& - 			& 1.8$\pm$0.3 						& 0.59$\pm$0.04	& 1.72$\pm$0.06	& 0.017$\pm$0.003\\
 		& Combined** 	& - 					& 13.90$\pm$0.05 					& 32.853$\pm$0.002 				& 0.038$\pm$0.002 	& - 			& 1.8$\pm$0.3 						& 0.62$\pm$0.05	& 1.77$\pm$0.08	& 0.018$\pm$0.003\\

\hline
\end{tabular}
\begin{minipage}{\textwidth}
	\textsuperscript{1} Magnitudes in PS1 system. \\
     \textsuperscript{2} The synodic rotation periods and their uncertainties were derived from the mean and standard deviation from the MC method (see Section \ref{sec:MC}). \\
     \textsuperscript{3} The linear phase function coefficients and their uncertainties were derived from the mean and standard deviation from the MC method (see Section \ref{sec:MC}). \\
	\textsuperscript{4} Calculated from $H_{\mathrm{r}}$(1,1,0) assuming an albedo A=4\%. \\
    \textsuperscript{5} Calculated using Eq. \ref{eq:albedo} from $H_{\mathrm{r}}$(1,1,0) and the effective radius $R_{\mathrm{eff}}$ from \cite{Fernandez2013} (see Tab. \ref{tab:review}). \\
	\textsuperscript{*} The comet was weakly active. The results do not include corrections for the presence of a near-nucleus coma. \\
     \textsuperscript{**} The data are from different apparitions. \\
    \textsuperscript{a} The $\beta$ value for the $H_{\mathrm{r}}$(1,1,0) was taken from the phase function fit of the combined 2005 and 2015 datasets.\\
    \textsuperscript{b} Calculated for $\beta$ = 0.04 mag deg\textsuperscript{-1}.\\
    \textsuperscript{c} The comet was weakly active at the time of the observation. The albedo estimates are therefore upper limits. \\

\end{minipage}
\end{table*}

\section{Discussion}
\label{sec:discussion}
	In Table \ref{tab:review}, we summarised the physical characteristics of all JFCs with known rotation rates. With the newly analysed lightcurves in Section \ref{sec:results}, we have added six additional lightcurves, seven phase functions and eight albedo estimates. Here, we compare our newly obtained results with the overall JFC characteristics and use the expanded sample to draw conclusions about the collective population properties.

    \subsection{Spin rate distribution}

\begin{figure}
    \centering
   \includegraphics[width=0.48\textwidth]{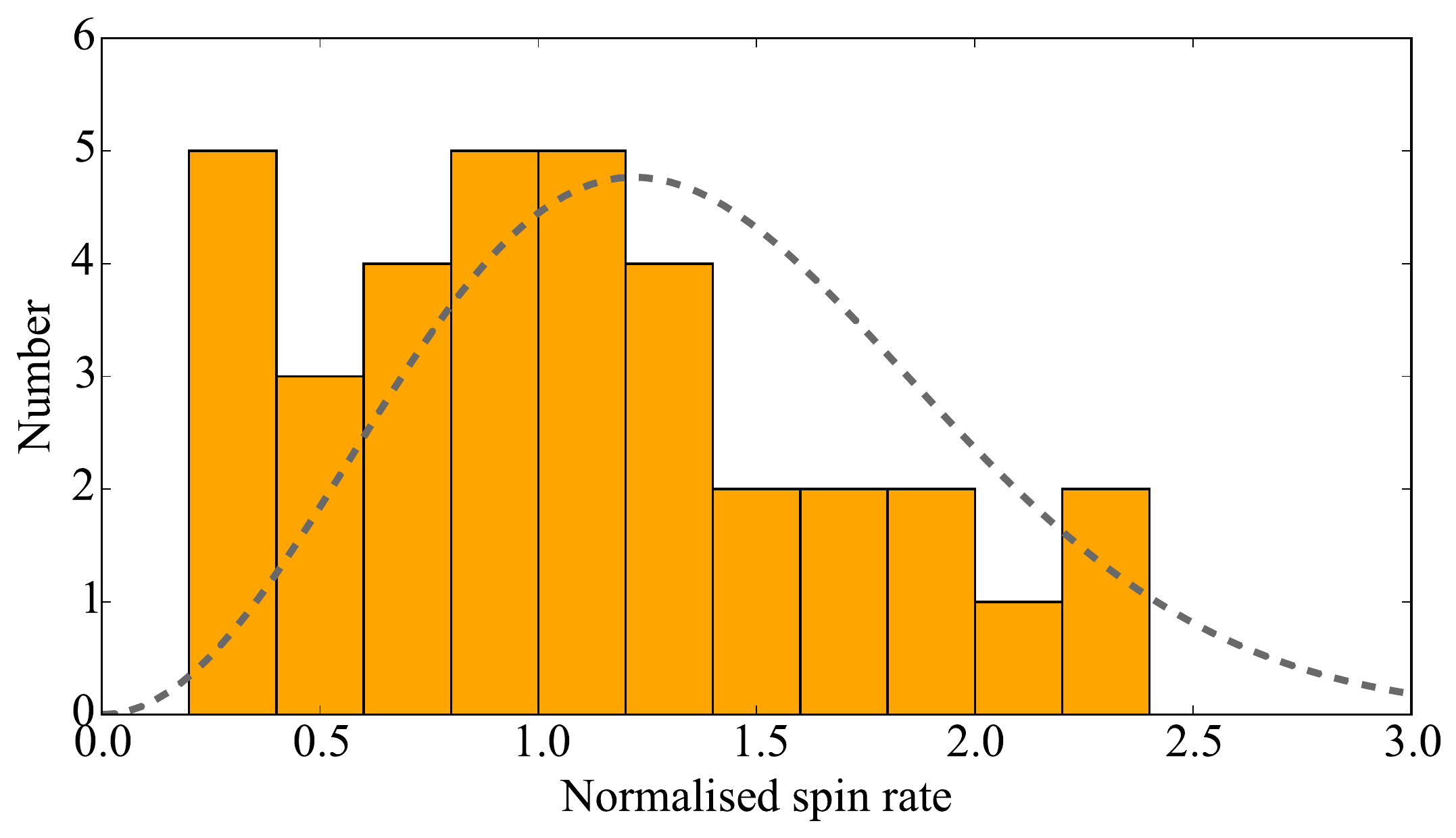}
   \caption{Histogram of the normalised rotation rates of 37 JFCs. The normalised spin rate is calculated as $f/\langle f \rangle$ where $f$ = 1 / $P_\mathrm{rot}$ and $\langle f \rangle$ is the geometric mean of $f$. The dashed line corresponds to the best-fitting Maxwellian distribution.}
    \label{res-hist_periods}%
    \end{figure}

   The distribution of the rotation rates of comets can be used to study their collisional history. Fig. \ref{res-hist_periods} shows a histogram of all known  spin rates of JFCs.  We have plotted the rotation frequency $f = 1/P_\mathrm{rot}$ which was normalised using the geometric mean $\langle f \rangle$ of the whole sample. Similar plots for asteroids have shown that the distribution of asteroid spin rates is Maxwellian which has suggested that asteroids are a collisionally evolved population \citep{Harris1996,Pravec2002}. 
    
    The best-fitting Maxwell distribution in Fig. \ref{res-hist_periods} does not show good agreement with the measured spin rates. We performed Kolmogorov-Smirnov tests comparing the normalised frequency distribution in Fig. \ref{res-hist_periods} to Maxwell distribution and flat distribution with the same mean and standard deviation. The resulting $D$ statistics were 0.20 ($p$ = 0.09) and 0.13 ($p$ = 0.44) for the uniform and Maxwell distributions respectively. We cannot reject the null hypothesis in either of the cases, and therefore both distributions can possibly describe the data. 
    
The cumulative size distribution (CSD) of JFCs was found to be  very close to the one expected for a collisionally relaxed population of strengthless bodies \rk{\citep[][and references therein]{Lamy2004,Snodgrass2011,Fernandez2013}}. However, this result has a large uncertainty and cannot be used as a proof that JFCs originate from disrupted larger bodies (e.g KBOs). In turn, it suggests that due to the continuous mass loss of JFCs their size distribution can be shaped by a complex combination of collisional processes in the past and activity in the present epoch \citep{Snodgrass2011}. 

Similarly, our results for the spin distribution of comets suggest that their rotation can be determined by the ongoing activity. The mass lost through activity jets is able to exert a torque on the nucleus, which in turn changes the spin rate of the comet on orbital timescales \citep[e.g.][]{Samarasinha2004}. This mechanism can be responsible for reshaping the original distribution of the spin rates, and could explain the current spin rate distribution of JFC. However, it is important to know that Fig. \ref{res-hist_periods} includes data from just 37 comets, many of which have lightcurve periods with large uncertainties. This highlights the need to increase the sample of JFCs with known rotational properties in order to enable the understanding of the population history. 

It is worth noting that evidence from Rosetta, such as the low density/high porosity, and presence of hypervolatiles like O$_2$ and N$_2$, suggests that 67P is not a collisional fragment \citep[see][and references therein]{Davidsson2016}. The apparent coincidence of sizes and spin rates of JFC nuclei being consistent with collisional evolution, while in situ measurements of their bulk properties suggest otherwise, is surprising. This may instead support the hypothesis by \cite{Jutzi2016} that JFCs have undergone significant collisional evolution, but the distributions presented here do not yet allow a definitive conclusion.

    \subsection{Shapes}
    
\begin{figure}
    \centering
   \includegraphics[width=0.48\textwidth]{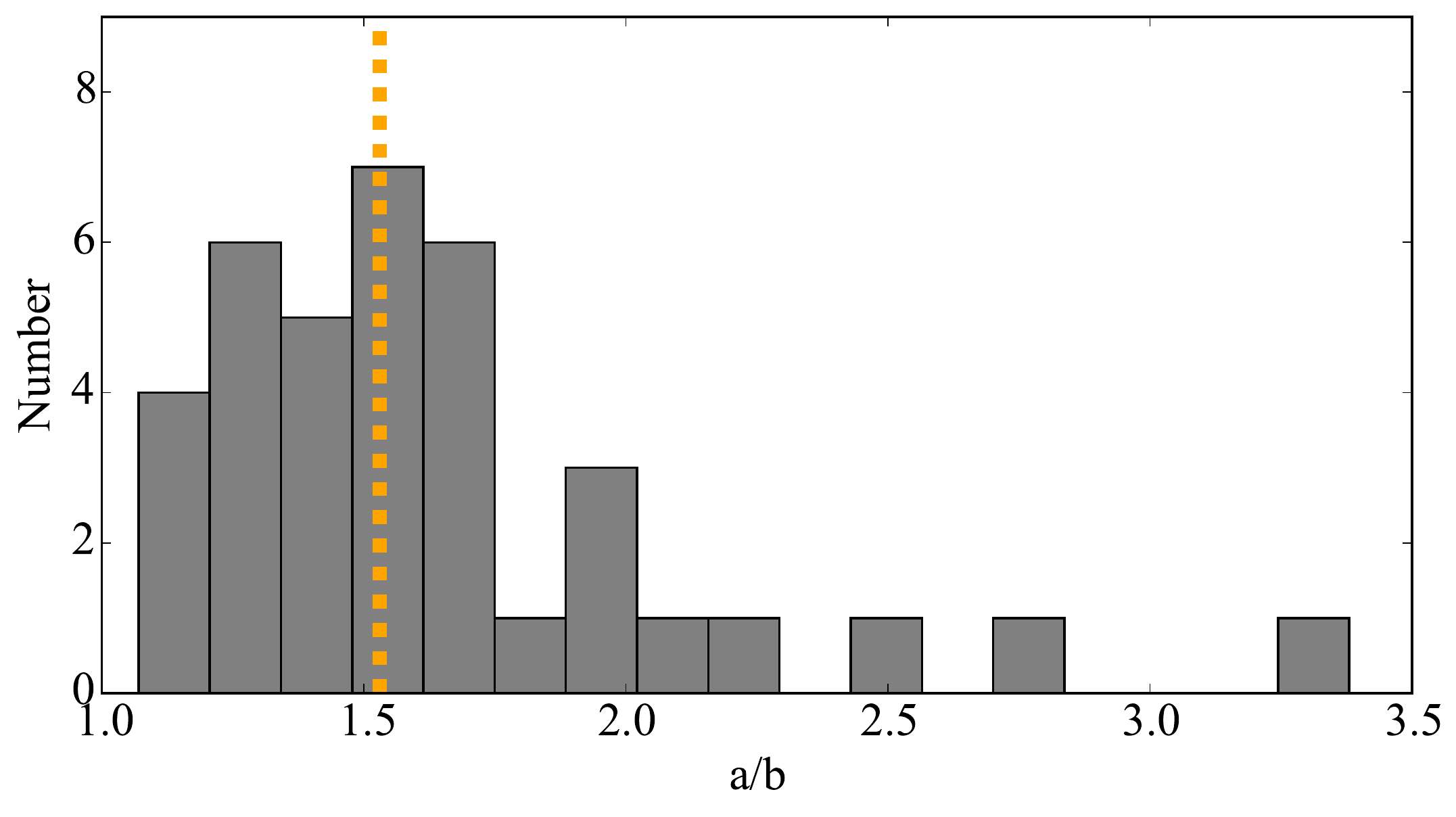}
   \caption{Distribution of the axis ratios $a/b$ of JFCs. The vertical line corresponds to the median value of $a/b$ = 1.5. For all comets (except 9P, 19P,  67P, 81P, 103P), the given axis ratio is obtained from ground- and space-based telescope and is therefore just a lower limit of the elongation. }
    \label{res-hist_elong}%
    \end{figure}

    Fig. \ref{res-hist_elong} shows the distribution of the axis ratios of all comets. Most $a/b$ values are smaller than $a/b$ = 2 and the median of the distribution is at $a/b$ = 1.5. However, all comets with shape models obtained from in situ observations (9P, 19P,  67P, 81P, 103P) have significantly higher axis ratios (see Table \ref{tab:review}). For all other objects the axis ratio is a lower limit since it was calculated from the lightcurve brightness variation. It is therefore possible that the typical elongation of JFCs is higher than the one we estimated from the current distribution, suggesting that bilobate shapes (like those seen by spacecraft at 67P and 103P) may be common, in agreement with recent formation models \citep{Davidsson2016}.
    
    \subsection{Bulk densities and stability against rotational splitting}
    \label{sec:disc-density}
    
    We attempted to use the expanded sample of JFCs with estimated rotation rates and elongations to constrain the comet density and tensile strength. As we discussed in Section \ref{sec:method-equations}, it is commonly assumed that comets have negligible tensile strengths. Under this assumption, it is possible to set a lower limit on the density necessary to keep JFCs stable against rotational instabilities \citep[Eq. \ref{eq:density}; ][]{Pravec2002}. 
    
\begin{figure*}
   \centering
   \includegraphics[width=\textwidth]{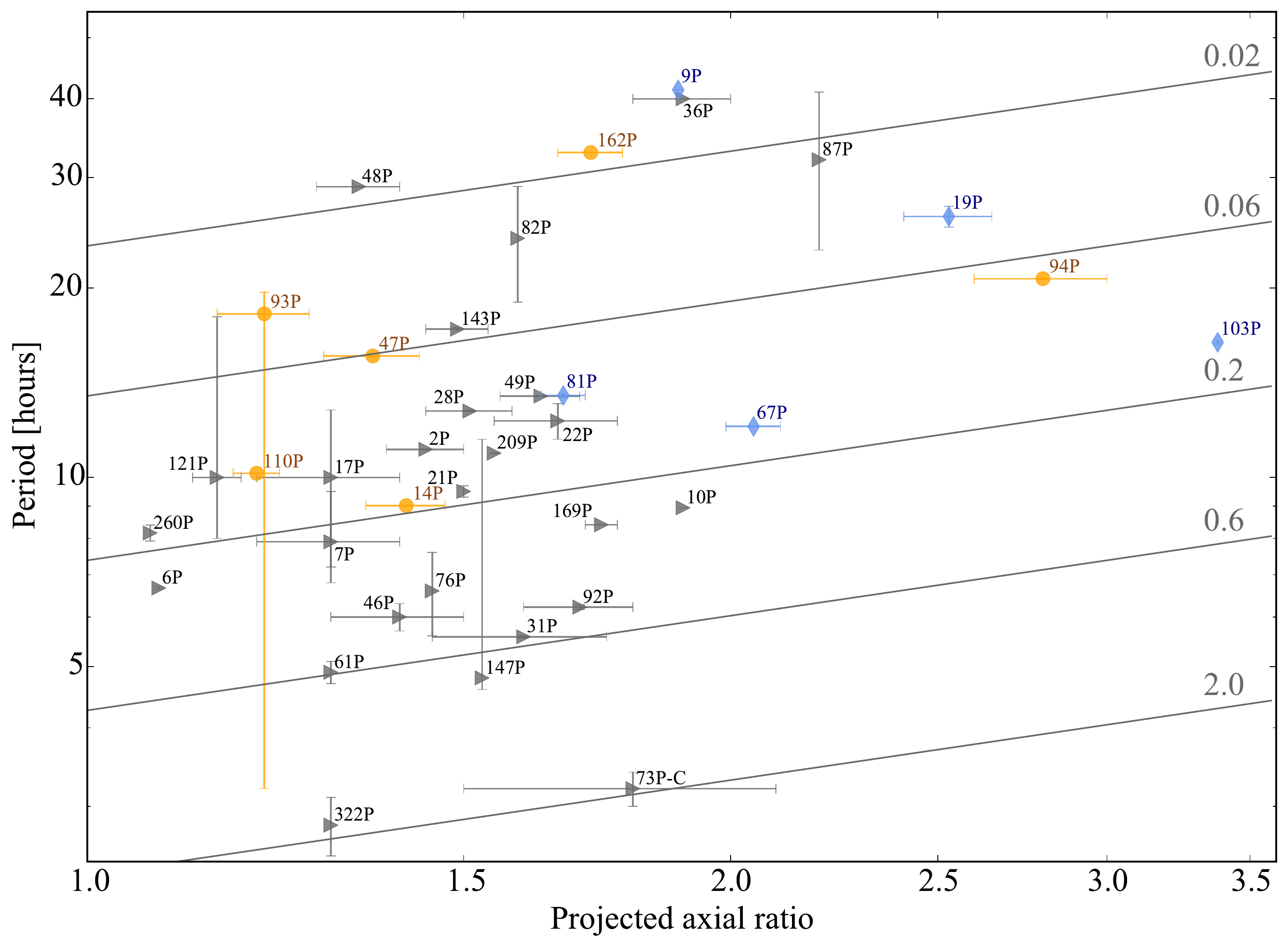}
   \caption{Rotation period against projected axis ratio for JFC nuclei. The grey triangles denote comets with parameters determined from lightcurve or radar measurements. The orange circles are the comets from this work. For these points, the axis ratio is a lower limit and the uncertainties are plotted when they were stated by the authors. The blue diamonds correspond to comets visited by spacecraft with precise shape models. The diagonal lines indicate the minimum density (denoted in g cm\textsuperscript{-3} to the right), which a strengthless body of the given axis ratio and spin period requires to remain intact. Apart from the unusual cases of 323P and 73P, which are discussed in the text, no comet requires a density greater than $\sim$ 0.6 g cm\textsuperscript{-3} to remain stable against rotational splitting.}
   \label{res-fig-density}
    \end{figure*}

    In Fig. \ref{res-fig-density} we plot the rotation versus projected axis ratio for all comets in the expanded sample. 
Using a similar plot, \cite{Lowry2003} discovered that comets do not require densities higher than approximately 0.6 g cm\textsuperscript{-3} in order to be stable against rotational instabilities. Here we confirm this result for all objects except for 322P, 73P-C and 147P. 

As we discussed in section \ref{sec:rev_322P}, according to \cite{Knight2016} it is not clear whether 322P has asteroidal or cometary origin. Therefore, the fact that it requires higher density can be interpreted as evidence in favour of the hypothesis that it is an asteroid. Comet 147P lies very close to the limit of 0.6 g cm\textsuperscript{-3} and has a large period uncertainty. Therefore, we do not consider it as an outlier.  \rk{Additionally, 147P belongs to the class of quasi-Hilda comets and might have asteroidal origin \citep{Ohtsuka2008}.} Comet 73P-C on the other hand clearly has a JFC origin and therefore should be similar to the other objects in our sample. However, since it seems to be continuously disintegrating (see section \ref{sec:rev_73P}), it cannot be used to study the stability criterion.  \rk{It is also possible that the breakup of the comet exposed the innermost part of the pre-breakup nucleus which could have a larger tensile strength \citep[see][and references therein]{Gundlach2016}}

If we exclude these three comets, our expanded sample confirms the density limit of 0.6 g cm\textsuperscript{-3} discovered by \cite{Lowry2003}. By analogy with the clear cut off in rotation rates of asteroids at 2.2 g cm\textsuperscript{-3} \citep{Pravec2002}, we interpret the cut-off for comets as an indication that 0.6 g cm\textsuperscript{-3} is a typical density for JFCs. This agrees with the density estimates from recent spacecraft measurements \citep{Richardson2007,Jorda2016}.

\begin{figure*}
   \centering
   \includegraphics[width=\textwidth]{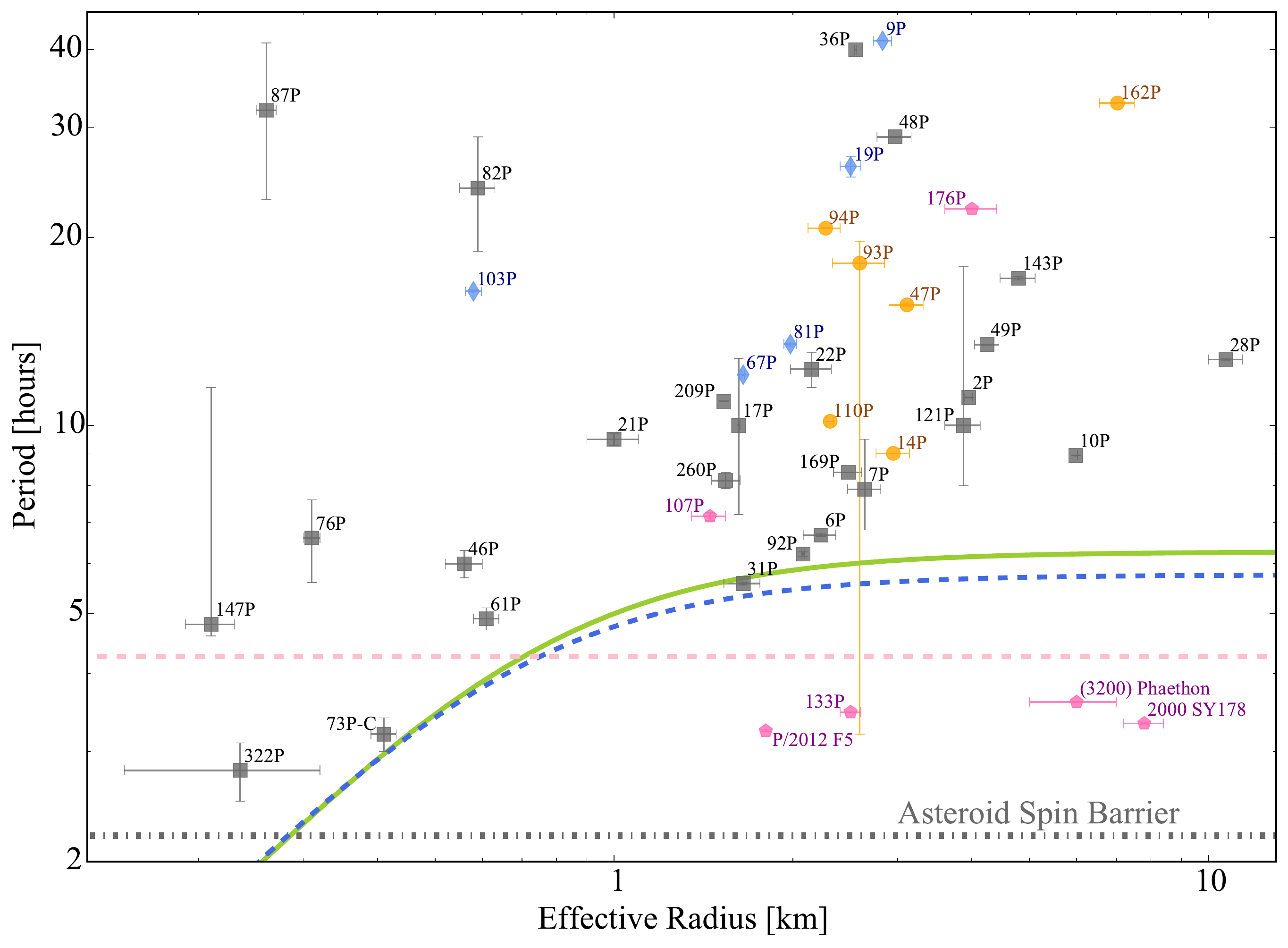}
   \caption{Rotation period against effective radius of the JFC nuclei. The blue diamonds are comets visited by spacecraft; the grey squares are comets observed from ground and the orange circles are the comets added in this work. For comparison we plotted active asteroids with known rotation rates (pink pentagons). The lower horizontal dotted line corresponds to the asteroid spin barrier \protect\citep{Harris1996,Pravec2002}. The upper dashed pink line shows the minimum possible rotation rate for strengthless spherical bodies with density $\rho$ = 0.6 g cm\textsuperscript{-3}. The curves are derived from the model for prolate ellipsoids stable against rotational instability by \protect\citet{Davidsson2001}. The solid green line is the model for density $\rho$ = 532 kg m\textsuperscript{-3}, axis ratio $a/b$ = 2 and tensile strength $T$ = 15 Pa, which corresponds to the parameters measured for 67P from Rosetta \protect\citep{Jorda2016,Groussin2015}. The dashed blue curve is for the same density but  $a/b$ = 1.6 (the value for 31P) and $T$ = 10 Pa. By varying the model parameters, we can conclude that for typical densities and axis ratios ($a/b$ $\leq$ 2.0), none of the observed comets requires a tensile strength larger than 25 Pa to remain stable against rotational splitting.}
   \label{res-rot-vs-rad}%
    \end{figure*}

Further insights into the material properties of JFCs can be determined from comparing their rotation rates and sizes. \rk{In previous studies, \cite{Davidsson1999,Davidsson2001} and \cite{Toth2006} already explored the location of comets and other primitive minor bodies in the radius-rotation period plane.} In Fig. \ref{res-rot-vs-rad} we plot the distribution of rotation rates with radius for all comets. A key feature of the distribution of comets in the plot is that the domain in the lower right corner is not populated. 

In order to interpret this observation, we employ the recent results from the Rosetta mission. The in situ measurements of comet 67P provide precise estimates of the nucleus bulk parameters. It has density of 0.532 $\pm$ 0.007 g cm\textsuperscript{-3} \citep{Jorda2016}, axis ratio $a/b$ = 2.05 $\pm$ 0.06 \citep[calculated from the axis estimates in][]{Jorda2016}, and tensile strength of 3-15 Pa with an upper limit of 150 Pa \citep{Groussin2015}. If we assume that 67P is a representative example for JFCs, we can use these values to study the properties of the whole population. 

In Fig. \ref{res-rot-vs-rad}, we have plotted the asteroid spin barrier \citep{Pravec2002} which corresponds to the minimum rotation period of a strengthless body with density $\sim$ 3 g cm\textsuperscript{-3}. For a comparison, we have also plotted the rotation limit for a spherical object with density of 0.6 g cm\textsuperscript{-3}. The position of the limit for comets will change for different elongations and densities since less dense and more elongated objects are easier to disrupt. 

So far in the analysis, we have treated comets as strengthless, however the measurements of the tensile strength of 67P allow us to explore more complicated models which take the material strength of comets into account. We have used the analytical models developed by \cite{Davidsson1999,Davidsson2001} to determine the maximum rotation rate of prolate ellipsoids which are stable against rotational instabilities using the density, axis ratio and tensile strength of 67P  (Fig. \ref{res-rot-vs-rad}, solid green curve). This curve agrees very well with the observed data and puts 73P-C right at the limit of stability, which agrees with its frequent fragmentation events. Although comet 31P lies below the stability line, its projected  axis ratio is lower than that of 67P. 

We have therefore investigated the stability limit for objects with density of 0.5 g cm\textsuperscript{-3} and a typical axis ratio of $a/b$ = 1.6 (equal to the elongation of 31P). We determine that under these assumptions none of the comets requires tensile strength higher than $\sim$ 10 Pa to remain stable against rotational instabilities (Fig. \ref{res-rot-vs-rad}, dashed blue curve). We varied the axis-ratio parameter of the model for ratios $a/b$ $\leq$ 2.0 and concluded that none of the observed comets requires a tensile strength larger than 25 Pa to remain stable against rotational splitting. This confirms the low-tensile strength estimates of 67P by \cite{Groussin2015} and of Shoemaker-Levy 9 \citep{Asphaug1996}. 

An interesting test of this model would come from future observations of the rotation rate of 31P. The comet's period was previously very well determined by \cite{Luu1992}. If new observations of its lightcurve show that the nucleus is spinning up, this comet would be a strong candidate for future rotational splitting.

Despite the small number of nuclei with radii larger than 3 km in the sample, it is noticeable that all of them lie far above the stability limit. The simplest explanation for this effect could be deduced from the understanding of activity-induced rotational changes. According to the relations derived in \cite{Samarasinha2013}, the rotation changes induced by outgassing are proportional to the square of the rotation period and inversely proportional to the square of the radius. In this scenario, if a large nucleus is spinning up due to reaction torques, the faster it gets, the less it can spin up with every orbit. Therefore, it is very hard to spin up the large nuclei which already rotate with relatively short periods.

At this stage, we cannot evaluate this hypothesis further since spin changes are poorly investigated and to this date only 5 comets have confirmed period changes \citep[see][and references therein]{Samarasinha2013}. To improve the understanding of the rotation of large comets, we need to measure the rotation rates of more large nuclei and to increase the number of comets with period determinations at multiple apparitions.

Finally, in Fig. \ref{res-rot-vs-rad}, we have also plotted all active asteroids with known periods and radii \citep{Jewitt2015b}. Most of them lie in the lower right domain of the plot where no JFCs can be found. However, it is particularly interesting to note that 107P fulfils the stability criteria for comets too. This object has sparked a long-standing debate on whether it is a comet or an active asteroid \citep[see][and references therein]{Jewitt2015b} Since 107P is above the stability limit for typical JFCs, we cannot reject the possibility that it has a cometary origin.

\subsection{Surface Properties}
\label{sec:disc_surface}

Prior to this work, there were only nine comets for which both the albedo and the phase function were known \citep{Snodgrass2011}. We have significantly increased this number by updating the values for one comet and adding the measurements for five additional comets from this work. We have summarized the albedos and phase function coefficients for 24 comets in Table \ref{tab_albedo_phase}. The median of all known linear phase function slopes is \rk{0.046 mag/deg and the standard deviation is 0.017 mag/deg}. The median of all albedos is 4.2\% and the standard deviation is 1.3\%.

\begin{figure}
    \centering
   \includegraphics[width=0.48\textwidth]{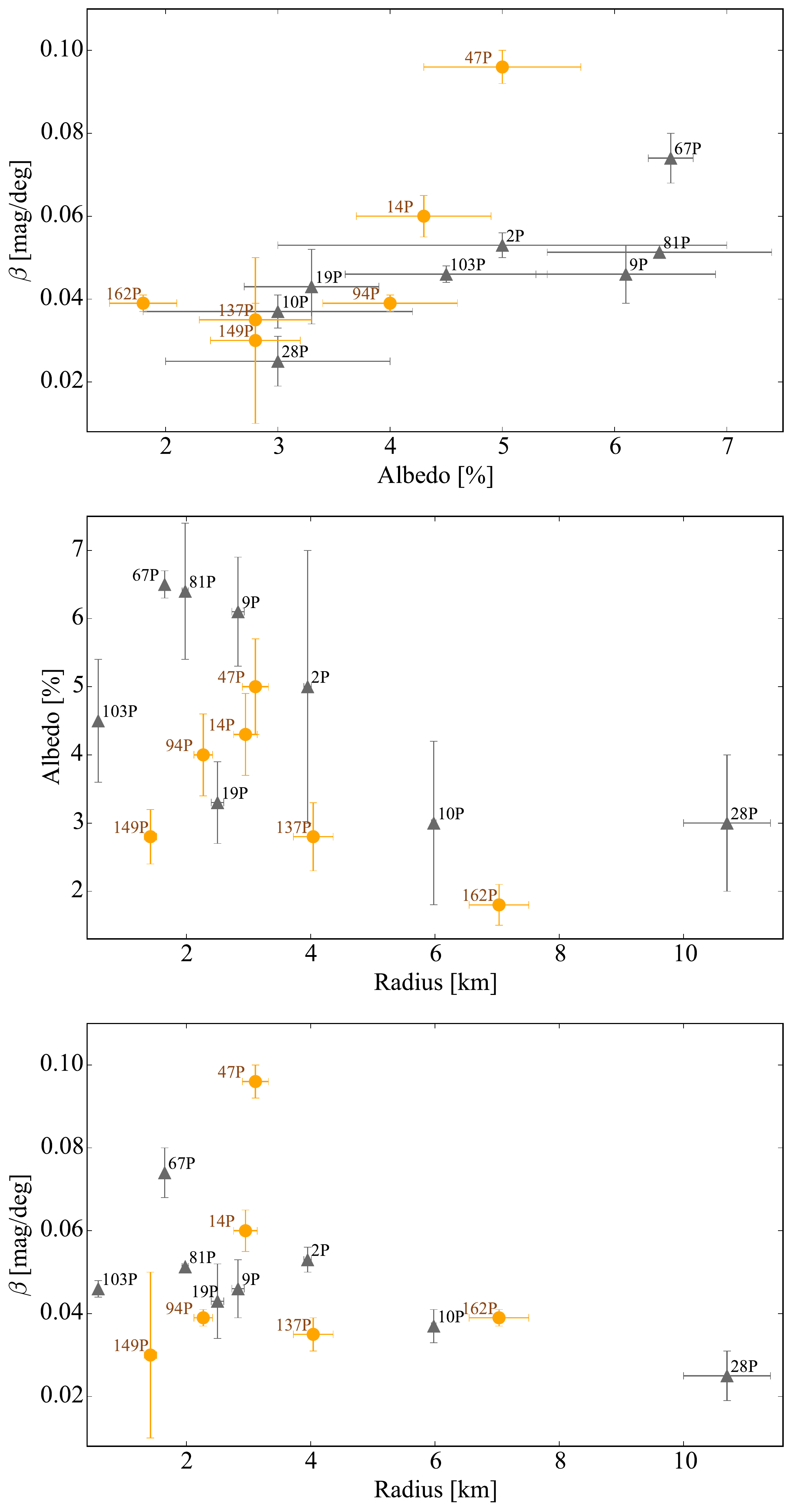}
   \caption{Surface properties of all JFCs with known radius, albedo and phase function. The orange circles correspond to comets with properties derived in this work. Top: Phase function slope versus albedo. There is a trend of increasing phase function slope with increasing albedo. Comet 47P was active at the time of the observations, so in reality its phase function coefficient might be smaller and its albedo might be higher. Middle: albedo versus radius. Bottom: phase function coefficient versus radius. }
    \label{fig:phase_vs_albedo}%
    \end{figure}

We have looked for possible correlations between the surface properties of the comets and their sizes. In Fig. \ref{fig:phase_vs_albedo} it can be seen that large JFCs tend to have low albedos and small phase function coefficients. The albedo distribution with size agrees with the one presented by \cite{Fernandez2016}, which consisted of a larger sample of approximately 50 comets with albedos derived within the SEPPCoN program.

We note a possible correlation between the phase function coefficient and the albedo in Fig. \ref{fig:phase_vs_albedo} (top panel). It is well established that similar correlations exist between albedo or spectral type and phase functions for asteroids \citep[e.g.][]{Oszkiewicz2012}. 

Comet 47P was determined to be active at the time of the observations which were used to determine its albedo and phase function. Under these conditions, it is possible that we have overestimated the nucleus brightness and therefore underestimated its albedo. Additionally, the activity possibly led us to determine an inaccurate phase function. Due to these concerns, we prefer to exclude it from the analysis.

We performed a Spearman rank correlation test between the phase function coefficient and the albedo of all comets (excluding 47P). The test produced rank $\rho$ of 0.82 and $p$-value of 0.0005 which suggests a possible correlation between the phase function coefficients and albedos. In order to confirm this possible correlation and to be able to interpret it, we need to increase the number of JFCs with well-determined surface properties.

 \begin{table*}
\centering
\caption{Albedo and phase function measurements for JFCs.}
\label{tab_albedo_phase}
\begin{tabular}{llllll}
\hline
Comet & A [\%] & Reference & $\beta$ [mag/deg] & $\alpha$ Range [deg] & Reference. \\
\hline
2P & 5.0 $\pm$ 2.0 & \cite{Fernandez2000} & 0.053 $\pm$ 0.003 & - & Weighted mean \\
'' & - & - & 0.060 $\pm$ 0.005 & 0-110 & \cite{Fernandez2000} \\
'' & - & - & 0.060 $\pm$ 0.005 & 4-28 & \cite{Boehnhardt2008} \\
9P & 6.1 $\pm$ 0.8 & Weighted mean & 0.046 $\pm$ 0.007 & 4-117 & \cite{Li2007b} \\
'' & 6.4 $\pm$ 1.3 & \cite{Li2007b} & - & - & - \\
'' & 4.6 $\pm$ 1.5 & \cite{Lisse2005} & - & - & - \\
'' & 7.2 $\pm$ 1.6 & \cite{Fernandez2003} & - & - & - \\
10P & 3.0 $\pm$ 1.2 & \cite{AHearn1989} & 0.037 $\pm$ 0.004 & 9-28 & \cite{Sekanina1991} \\
14P & 4.3 $\pm$ 0.6 & This work & 0.060 $\pm$ 0.005 & 5-9 & This work \\
19P & 3.3 $\pm$ 0.6 & Weighted mean & 0.043 $\pm$ 0.009 & 13-80 & \cite{Li2007a} \\
'' & 2.9 $\pm$ 0.6 & \cite{Buratti2004} & - & - & - \\
'' & 7.2 $\pm$ 2.0 & \cite{Li2007a} & - & - & - \\
22P & 4.8 $\pm$ 1.0 & \cite{Lamy2002} & - & - & - \\
28P & 3.0 $\pm$ 1.0 & \cite{Jewitt1988} & 0.025 $\pm$ 0.006 & 0-15 & \cite{Delahodde2001} \\
36P & - & - & 0.060 $\pm$ 0.019 & 1-11 & \cite{Snodgrass2008} \\
45P & -  & - & $\sim$0.06  & 88-93 & \cite{Lamy2004} \\

47P & $\leq$ 5.0 $\pm$ 0.7 & This work & 0.096 $\pm$ 0.004 & 3-9 & This work \\
'' & - & - & 0.083 $\pm$ 0.006 & 2-9 & \cite{Snodgrass2008} \\
48P & - & - & 0.059 $\pm$ 0.002 & 5-16 & \cite{Jewitt2004} \\
49P & 4.5 $\pm$ 1.9 & \cite{Campins1995} & - & - & - \\
67P & 6.5 $\pm$ 0.2 & \cite{Fornasier2015a} & 0.074 $\pm$ 0.006 & 1-10 & \cite{Fornasier2015a} \\
'' & 5.4 $\pm$ 0.6 & \cite{Kelley2009} & 0.076 $\pm$ 0.003 & 0-11 & \cite{Tubiana2008} \\
81P & 6.4 $\pm$ 1.0 & \cite{Li2009} & 0.0513 $\pm$ 0.0002 & 0-100 & \cite{Li2009} \\
93P & 4.2 $\pm$ 0.9 & This work & - & - & - \\
94P & 4.0 $\pm$ 0.6 & This work & 0.039 $\pm$ 0.002 & 5-17 & This work \\
103P & 4.5 $\pm$ 0.9 & \cite{Li2013} & 0.046 $\pm$ 0.002 & 79-95 & \cite{Li2013}\\
110P & - & - & 0.069 $\pm$ 0.002 &  1-9 & This work \\
123P & 3.6 $\pm$ 0.8 & This work & - & - & - \\
137P & 2.8 $\pm$ 0.5 & This work & 0.035 $\pm$ 0.004 & 0.5-6 & This work \\
143P & - & - & 0.043 $\pm$ 0.001 & 5-13 & \cite{Jewitt2003} \\
149P & 2.8 $\pm$ 0.4 & This work & 0.03 $\pm$ 0.02 & 8-10 & This work \\
162P & 1.8 $\pm$ 0.3 & This work & 0.039 $\pm$ 0.002 & 4-12 & This work \\
'' & 3.7 $\pm$ 1.4 & \cite{Fernandez2006} & - & - & -\\
169P & 3.0 $\pm$ 1.0 & \cite{DeMeo2008} & - & - & - \\

\hline
\end{tabular}
\end{table*}

\section{Summary}
\label{sec:conclusions}

We have developed a method for precise absolute calibration of photometric time series using Pan-STARRS DR1 stars. With this technique we achieved photometric calibration with uncertainty as low as 0.02 mag. Thus we were able to study the rotation, shapes, and surfaces of nine Jupiter family comets, most of which were observed at multiple epochs using different instruments. We have collected an up-to-date sample of JFCs with published rotational properties and expanded it with the measurements from this work. We used the extended sample to characterise the bulk properties of JFCs. The results are as follows:

\begin{enumerate}

\item{We have used time-series photometry of nine JFCs taken in the period 2004-2015 to study their lightcurves. We have derived the rotation rates of six comets (14P, 47P, 93P, 94P, 110P, and 162P). For comets 123P, 137P and 149P the collected data were insufficient to derive unambiguous rotation periods. To our knowledge, for comets 93P, 94P and 162P these are the first published rotation rates. Comets 14P, 47P and 110P had previous lightcurves but our results significantly improved the period estimates.}

\item{Lower limits on the axis ratios of all observed comets have been derived from the brightness variation of the time series. Three of the comets, 47P, 93P and 123P, were most likely active at the times of the observations and therefore their brightness variation was most likely underestimated. }

\item{We have determined the linear phase function coefficients of seven of the observed comets - 14P, 47P, 94P, 110P, 137P, 149P, and 162P. To our knowledge, for all comets except 47P, this is the first phase function determination. Our results have increased the number of comets with well-constrained phase function coefficients from 13 to 19. }

\item{The derived phase function coefficients have been used in the calculation of the absolute magnitudes $H_{\mathrm{r}}$(1,1,0). For comets 93P and 123P, we used a phase function coefficient $\beta$=0.04 mag/deg. All comets except for 110P were part of SEPPCoN and had radius measurements derived from Spitzer infrared observations \citep{Fernandez2013}. Using these radii and the derived absolute magnitudes, we have estimated the albedos of all eight comets. }

\item{We derived a geometric albedo of 1.8 $\pm$ 0.3 \% for comet 162P. This makes 162P the JFC with lowest measured geometric albedo to date.}

\item{Prior to this work, there were nine comets for which both the albedo and the phase function coefficient were known \citep[see][]{Snodgrass2011}. We have updated the values for 47P and added five more comets (14P, 94P, 137P, 149P, 162P) to this sample. 

The increased number of comets has allowed us to look for correlations between the surface properties of JFCs. Large nuclei ($R_\mathrm{eff}$ $\geq$ 5 km) appear to have low albedos ($\leq$ 3 \%) and low phase function coefficients ($\leq$ 0.04 mag/deg). However, since only three comets in that size range have been observed, this needs to be confirmed with future observations. We have discovered a possible correlation between the phase function coefficient and the albedo, where comets with larger albedos have steeper phase functions. In order to confirm as well as to interpret this result, we would require further phase function observations.   }

\item{\rk{In Table \ref{tab_albedo_phase} we have collected the known albedos and phase functions of 24 JFCs. The distribution of the linear phase function slopes has a median of 0.046 mag/deg and standard deviation of 0.017 mag/deg. The known albedos have a median of 4.2\% and standard deviation of 1.3\%}}

\item{We have reviewed the properties of all JFCs which (to the extent of our knowledge) had published rotation rates. After adding the six comets from this work, the total size of the sample has reached 37 comets. }

\item{We have attempted to use the distribution of spin rates to improve the understanding of JFC evolution. The employed Kolmogorov-Smirnov tests determined that the normalised spin rates of comets is consistent both with a Maxwell distribution and a flat distribution. Therefore, we cannot distinguish  between the possibilities that JFCs are a collisionally-dominated population like asteroids or their spin rate distribution is dominated by other processes, such as activity-driven spin changes. }

\item{The distribution of the axis ratios shows that the majority of comets have projected axis ratios smaller than 2. The median of the whole JFC sample is 1.5. However, ground observations only give a lower limit to the axis ratio. All five comets with shape models determined from in situ space craft observations have axis ratios larger than 1.6.}

\item{Under the assumption that JFCs have negligible tensile strengths, we have used their axis ratios and periods to constrain their bulk densities. We have  confirmed the result from \cite{Lowry2003} that a density of 0.6 g $\mathrm{cm^{-3}}$ is sufficient to keep all of the studied nuclei stable against rotational instabilities.}

\item{If we instead model JFCs as prolate ellipsoids with non-negligible tensile strengths using the model from \cite{Davidsson2001}, we conclude that none of the observed comets requires tensile strength higher than 10-25 Pa in order to be stable against rotational splitting. Comet 73P-C lies very close to the stability limit we derived, which suggests that its ongoing splitting might be due to rotational instabilities. An interesting outcome of this analysis is that comet 31P also lies very close to the stability limit which makes it a candidate for potential future splitting.}

\end{enumerate}

\section*{Acknowledgements}
\rk{We would like to thank the referee, Imre Toth, for the helpful suggestions that improved the manuscript.}
This work is based on observations collected at the European Organisation for Astronomical Research in the Southern Hemisphere under ESO programmes 072.C-0233(A), 074.C-0125(A), 077.C-0609(B), 079.C-0297(A), 079.C-0297(B), 082.C-0517(A), 082.C-0517(B), 089.C-0372(A), 089.C-0372(B), and 194.C-0207(C). We thank the observatory staff who helped us obtain the various data sets in this paper, especially those collected in service mode at the VLT. Also based on observations made with the WHT and INT operated on the island of La Palma by the Isaac Newton Group of Telescopes in the Spanish Observatorio del Roque de los Muchachos of the Instituto de Astrofisica de Canarias, under UK PATT programmes I/2005A/11, W/2007A/20 and W/2008B/23.

CS is funded by a STFC Ernest Rutherford fellowship. AF acknowledges support from STFC grant ST/L000709/1. 

The Pan-STARRS1 Surveys (PS1) and the PS1 public science archive have been made possible through contributions by the Institute for Astronomy, the University of Hawaii, the Pan-STARRS Project Office, the Max-Planck Society and its participating institutes, the Max Planck Institute for Astronomy, Heidelberg and the Max Planck Institute for Extraterrestrial Physics, Garching, The Johns Hopkins University, Durham University, the University of Edinburgh, the Queen's University Belfast, the Harvard-Smithsonian Center for Astrophysics, the Las Cumbres Observatory Global Telescope Network Incorporated, the National Central University of Taiwan, the Space Telescope Science Institute, the National Aeronautics and Space Administration under Grant No. NNX08AR22G issued through the Planetary Science Division of the NASA Science Mission Directorate, the National Science Foundation Grant No. AST-1238877, the University of Maryland, Eotvos Lorand University (ELTE), the Los Alamos National Laboratory, and the Gordon and Betty Moore Foundation.




\bibliographystyle{mnras}
\bibliography{Comets_bib.bib} 

\begin{thebibliography}{}
\makeatletter
\relax
\def\mn@urlcharsother{\let\do\@makeother \do\$\do\&\do\#\do\^\do\_\do\%\do\~}
\def\mn@doi{\begingroup\mn@urlcharsother \@ifnextchar [ {\mn@doi@}
  {\mn@doi@[]}}
\def\mn@doi@[#1]#2{\def\@tempa{#1}\ifx\@tempa\@empty \href
  {http://dx.doi.org/#2} {doi:#2}\else \href {http://dx.doi.org/#2} {#1}\fi
  \endgroup}
\def\mn@eprint#1#2{\mn@eprint@#1:#2::\@nil}
\def\mn@eprint@arXiv#1{\href {http://arxiv.org/abs/#1} {{\tt arXiv:#1}}}
\def\mn@eprint@dblp#1{\href {http://dblp.uni-trier.de/rec/bibtex/#1.xml}
  {dblp:#1}}
\def\mn@eprint@#1:#2:#3:#4\@nil{\def\@tempa {#1}\def\@tempb {#2}\def\@tempc
  {#3}\ifx \@tempc \@empty \let \@tempc \@tempb \let \@tempb \@tempa \fi \ifx
  \@tempb \@empty \def\@tempb {arXiv}\fi \@ifundefined
  {mn@eprint@\@tempb}{\@tempb:\@tempc}{\expandafter \expandafter \csname
  mn@eprint@\@tempb\endcsname \expandafter{\@tempc}}}

\bibitem[\protect\citeauthoryear{A'Hearn, Campins, Schleicher  \&
  Millis}{A'Hearn et~al.}{1989}]{AHearn1989}
A'Hearn M.~F.,  Campins H.,  Schleicher D.~G.,   Millis R.~L.,  1989, \mn@doi
  [The Astrophysical Journal] {10.1086/168204}, 347, 1155

\bibitem[\protect\citeauthoryear{A'Hearn et~al.,}{A'Hearn
  et~al.}{2005}]{AHearn2005}
A'Hearn M.~F.,  et~al., 2005, \mn@doi [Science (New York, N.Y.)]
  {10.1126/science.1118923}, 310, 258

\bibitem[\protect\citeauthoryear{A'Hearn et~al.,}{A'Hearn
  et~al.}{2011}]{AHearn2011a}
A'Hearn M.~F.,  et~al., 2011, \mn@doi [Science] {10.1126/science.1204054}, 332,
  1396

\bibitem[\protect\citeauthoryear{Appenzeller et~al.,}{Appenzeller
  et~al.}{1998}]{Appenzeller1998}
Appenzeller I.,  et~al., 1998, The Messenger, vol. 94, p. 1-6, 94, 1

\bibitem[\protect\citeauthoryear{Asphaug \& Benz}{Asphaug \&
  Benz}{1996}]{Asphaug1996}
Asphaug E.,  Benz W.,  1996, \mn@doi [Icarus] {10.1006/icar.1996.0083}, 121,
  225

\bibitem[\protect\citeauthoryear{Belskaya, Levasseur-Regourd, Shkuratov  \&
  Muinonen}{Belskaya et~al.}{2008}]{Belskaya2008}
Belskaya I.~N.,  Levasseur-Regourd A.-C.,  Shkuratov Y.~G.,   Muinonen K.,
  2008, The Solar System Beyond Neptune, M. A. Barucci, H. Boehnhardt, D. P.
  Cruikshank, and A. Morbidelli (eds.), University of Arizona Press, Tucson,
  592 pp., p.115-127, pp 115--127

\bibitem[\protect\citeauthoryear{Belton, Samarasinha, Fernandez  \&
  Meech}{Belton et~al.}{2005}]{Belton2005}
Belton M.,  Samarasinha N.,  Fernandez Y.,   Meech K.,  2005, \mn@doi [Icarus]
  {10.1016/j.icarus.2004.10.029}, 175, 181

\bibitem[\protect\citeauthoryear{Belton et~al.,}{Belton
  et~al.}{2011}]{Belton2011}
Belton M.~J.,  et~al., 2011, \mn@doi [Icarus] {10.1016/j.icarus.2011.01.006},
  213, 345

\bibitem[\protect\citeauthoryear{Belton et~al.,}{Belton
  et~al.}{2013}]{Belton2013}
Belton M.~J.,  et~al., 2013, \mn@doi [Icarus] {10.1016/j.icarus.2012.06.037},
  222, 595

\bibitem[\protect\citeauthoryear{Boehnhardt et~al.,}{Boehnhardt
  et~al.}{2002}]{Boehnhardt2002}
Boehnhardt H.,  et~al., 2002, \mn@doi [Astronomy and Astrophysics]
  {10.1051/0004-6361:20020494}, 387, 1107

\bibitem[\protect\citeauthoryear{Boehnhardt, Tozzi, Bagnulo, Muinonen, Nathues
  \& Kolokolova}{Boehnhardt et~al.}{2008}]{Boehnhardt2008}
Boehnhardt H.,  Tozzi G.~P.,  Bagnulo S.,  Muinonen K.,  Nathues A.,
  Kolokolova L.,  2008, \mn@doi [Astronomy and Astrophysics, Volume 489, Issue
  3, 2008, pp.1337-1343] {10.1051/0004-6361:200809922}, 489, 1337

\bibitem[\protect\citeauthoryear{Bohnhardt, Kaufl, Keen, Camilleri, Carvajal
  \& Hale}{Bohnhardt et~al.}{1995}]{Bohnhardt1995}
Bohnhardt H.,  Kaufl H.~U.,  Keen R.,  Camilleri P.,  Carvajal J.,   Hale A.,
  1995, IAU Circ., No. 6274, {\#}1 (1995). Edited by Green, D. W. E., 6274

\bibitem[\protect\citeauthoryear{Buratti, Hicks, Soderblom, Britt, Oberst  \&
  Hillier}{Buratti et~al.}{2004}]{Buratti2004}
Buratti B.,  Hicks M.,  Soderblom L.,  Britt D.,  Oberst J.,   Hillier J.,
  2004, \mn@doi [Icarus] {10.1016/j.icarus.2003.05.002}, 167, 16

\bibitem[\protect\citeauthoryear{Buzzoni et~al.,}{Buzzoni
  et~al.}{1984}]{Buzzoni1984}
Buzzoni B.,  et~al., 1984, ESO Messenger (ISSN 0722-6691), Dec. 1984, p. 9-13.,
  38, 9

\bibitem[\protect\citeauthoryear{Campins}{Campins}{1988}]{Campins1988}
Campins H.,  1988, \mn@doi [Icarus] {10.1016/0019-1035(88)90060-7}, 73, 508

\bibitem[\protect\citeauthoryear{Campins, Osip, Rieke  \& Rieke}{Campins
  et~al.}{1995}]{Campins1995}
Campins H.,  Osip D.~J.,  Rieke G.,   Rieke M.,  1995, \mn@doi [Planetary and
  Space Science] {10.1016/0032-0633(94)E0074-Z}, 43, 733

\bibitem[\protect\citeauthoryear{Campins, Ziffer, Licandro, Pinilla-Alonso,
  Fern{\'{a}}ndez, de Le{\'{o}}n, Moth{\'{e}}-Diniz  \& Binzel}{Campins
  et~al.}{2006}]{Campins2006}
Campins H.,  Ziffer J.,  Licandro J.,  Pinilla-Alonso N.,  Fern{\'{a}}ndez Y.,
  de Le{\'{o}}n J.,  Moth{\'{e}}-Diniz T.,   Binzel R.~P.,  2006, \mn@doi [The
  Astronomical Journal] {10.1086/506253}, 132, 1346

\bibitem[\protect\citeauthoryear{Chambers et~al.,}{Chambers
  et~al.}{2016}]{Chambers2016}
Chambers K.~C.,  et~al., 2016, http://arxiv.org/abs/1612.05560

\bibitem[\protect\citeauthoryear{Chesley et~al.,}{Chesley
  et~al.}{2013}]{Chesley2013}
Chesley S.,  et~al., 2013, \mn@doi [Icarus] {10.1016/j.icarus.2012.03.022},
  222, 516

\bibitem[\protect\citeauthoryear{Ciarniello et~al.,}{Ciarniello
  et~al.}{2015}]{Ciarniello2015}
Ciarniello M.,  et~al., 2015, \mn@doi [Astronomy {\&} Astrophysics]
  {10.1051/0004-6361/201526307}, 583, A31

\bibitem[\protect\citeauthoryear{Crovisier, Biver, Bockelee-Morvan, Colom,
  Gerard, Jorda  \& Rauer}{Crovisier et~al.}{1995}]{Crovisier1995}
Crovisier J.,  Biver N.,  Bockelee-Morvan D.,  Colom P.,  Gerard E.,  Jorda L.,
    Rauer H.,  1995, IAU Circ., No. 6227, {\#}1 (1995). Edited by Green, D. W.
  E., 6227

\bibitem[\protect\citeauthoryear{Davidsson}{Davidsson}{1999}]{Davidsson1999}
Davidsson B. J.~R.,  1999, \mn@doi [Icarus] {10.1006/icar.1999.6214}, 142, 525

\bibitem[\protect\citeauthoryear{Davidsson}{Davidsson}{2001}]{Davidsson2001}
Davidsson B. J.~R.,  2001, \mn@doi [Icarus] {10.1006/icar.2000.6540}, 149, 375

\bibitem[\protect\citeauthoryear{Davidsson, Sierks, Guttler, Marzari, Pajola
  \& Rickman}{Davidsson et~al.}{2016}]{Davidsson2016}
Davidsson B. J.~R.,  Sierks H.,  Guttler C.,  Marzari F.,  Pajola M.,   Rickman
  H.,  2016, \mn@doi [Astronomy {\&} Astrophysics]
  {10.1051/0004-6361/201526968}

\bibitem[\protect\citeauthoryear{Davis}{Davis}{1999}]{Davis1999}
Davis L.~E.,  1999, in Craine E.~R.,  Tucker R.~A.,   Barnes J.~V.,  eds,  Vol.
  189, Precision CCD Photometry. ASP Conference Series, p.~35

\bibitem[\protect\citeauthoryear{DeMeo \& Binzel}{DeMeo \&
  Binzel}{2008}]{DeMeo2008}
DeMeo F.,  Binzel R.~P.,  2008, \mn@doi [Icarus]
  {10.1016/j.icarus.2007.10.011}, 194, 436

\bibitem[\protect\citeauthoryear{Delahodde, Meech, Hainaut  \& Dotto}{Delahodde
  et~al.}{2001}]{Delahodde2001}
Delahodde C.~E.,  Meech K.~J.,  Hainaut O.~R.,   Dotto E.,  2001, \mn@doi
  [Astronomy and Astrophysics] {10.1051/0004-6361:20011028}, 376, 672

\bibitem[\protect\citeauthoryear{Drahus, K{\"{u}}ppers, Jarchow, Paganini,
  Hartogh  \& Villanueva}{Drahus et~al.}{2010}]{Drahus2010}
Drahus M.,  K{\"{u}}ppers M.,  Jarchow C.,  Paganini L.,  Hartogh P.,
  Villanueva G.~L.,  2010, \mn@doi [Astronomy and Astrophysics]
  {10.1051/0004-6361/20078882}, 510, A55

\bibitem[\protect\citeauthoryear{Drahus et~al.,}{Drahus
  et~al.}{2011}]{Drahus2011a}
Drahus M.,  et~al., 2011, \mn@doi [The Astrophysical Journal Letters, Volume
  734, Issue 1, article id. L4, 6 pp. (2011).] {10.1088/2041-8205/734/1/L4},
  734

\bibitem[\protect\citeauthoryear{Duncan, Levison  \& Dones}{Duncan
  et~al.}{2004}]{Duncan2004}
Duncan M.,  Levison H.,   Dones L.,  2004, Comets II, M. C. Festou, H. U.
  Keller, and H. A. Weaver (eds.), University of Arizona Press, Tucson, 745
  pp., p.193-204, pp 193--204

\bibitem[\protect\citeauthoryear{Duxbury, Newburn  \& Brownlee}{Duxbury
  et~al.}{2004}]{Duxbury2004}
Duxbury T.~C.,  Newburn R.~L.,   Brownlee D.~E.,  2004, \mn@doi [Journal of
  Geophysical Research] {10.1029/2004JE002316}, 109, E12S02

\bibitem[\protect\citeauthoryear{Dworetsky}{Dworetsky}{1983}]{Dworetsky1983}
Dworetsky M.~M.,  1983, Monthly Notices of the Royal Astronomical Society, 203,
  917

\bibitem[\protect\citeauthoryear{Farnham}{Farnham}{2001}]{Farnham2001}
Farnham T.~L.,  2001, American Astronomical Society, DPS Meeting {\#}33,
  id.12.10; Bulletin of the American Astronomical Society, Vol. 33, p.1047, 33,
  1047

\bibitem[\protect\citeauthoryear{Fern{\'{a}}ndez}{Fern{\'{a}}ndez}{2000}]{Fernandez2000}
Fern{\'{a}}ndez Y.,  2000, \mn@doi [Icarus] {10.1006/icar.2000.6431}, 147, 145

\bibitem[\protect\citeauthoryear{Fern{\'{a}}ndez, Meech, Lisse, A'Hearn,
  Pittichov{\'{a}}  \& Belton}{Fern{\'{a}}ndez et~al.}{2003}]{Fernandez2003}
Fern{\'{a}}ndez Y.,  Meech K.,  Lisse C.,  A'Hearn M.,  Pittichov{\'{a}} J.,
  Belton M.,  2003, \mn@doi [Icarus] {10.1016/S0019-1035(03)00142-8}, 164, 481

\bibitem[\protect\citeauthoryear{Fern{\'{a}}ndez, Lowry, Weissman, Mueller,
  Samarasinha, Belton  \& Meech}{Fern{\'{a}}ndez et~al.}{2005}]{Fernandez2005}
Fern{\'{a}}ndez Y.,  Lowry S.,  Weissman P.,  Mueller B.,  Samarasinha N.,
  Belton M.,   Meech K.,  2005, \mn@doi [Icarus]
  {10.1016/j.icarus.2004.10.019}, 175, 194

\bibitem[\protect\citeauthoryear{Fernandez, Campins, Kassis, Hergenrother,
  Binzel, Licandro, Hora  \& Adams}{Fernandez et~al.}{2006}]{Fernandez2006}
Fernandez Y.~R.,  Campins H.,  Kassis M.,  Hergenrother C.~W.,  Binzel R.~P.,
  Licandro J.,  Hora J.~L.,   Adams J.~D.,  2006, \mn@doi [The Astronomical
  Journal, Volume 132, Issue 3, pp. 1354-1360.] {10.1086/506252}, 132, 1354

\bibitem[\protect\citeauthoryear{Fern{\'{a}}ndez et~al.,}{Fern{\'{a}}ndez
  et~al.}{2013}]{Fernandez2013}
Fern{\'{a}}ndez Y.,  et~al., 2013, \mn@doi [Icarus]
  {10.1016/j.icarus.2013.07.021}, 226, 1138

\bibitem[\protect\citeauthoryear{Fernandez, Weaver, Lisse, Meech, Lowry, Bauer,
  Fitzsimmons  \& Snodgrass}{Fernandez et~al.}{2016}]{Fernandez2016}
Fernandez Y.~R.,  Weaver H.~A.,  Lisse C.~M.,  Meech K.~J.,  Lowry S.~C.,
  Bauer J.~M.,  Fitzsimmons A.,   Snodgrass C.,  2016, American Astronomical
  Society, AAS Meeting {\#}227, id.141.22, 227

\bibitem[\protect\citeauthoryear{Fornasier et~al.,}{Fornasier
  et~al.}{2015}]{Fornasier2015a}
Fornasier S.,  et~al., 2015, \mn@doi [Astronomy {\&} Astrophysics]
  {10.1051/0004-6361/201525901}, 583, A30

\bibitem[\protect\citeauthoryear{Glassmeier, Boehnhardt, Koschny, K{\"{u}}hrt
  \& Richter}{Glassmeier et~al.}{2007}]{Glassmeier2007}
Glassmeier K.-H.,  Boehnhardt H.,  Koschny D.,  K{\"{u}}hrt E.,   Richter I.,
  2007, \mn@doi [Space Science Reviews] {10.1007/s11214-006-9140-8}, 128, 1

\bibitem[\protect\citeauthoryear{Groussin, Lamy, Jorda  \& Toth}{Groussin
  et~al.}{2004}]{Groussin2004}
Groussin O.,  Lamy P.,  Jorda L.,   Toth I.,  2004, \mn@doi [Astronomy and
  Astrophysics] {10.1051/0004-6361:20040073}, 419, 375

\bibitem[\protect\citeauthoryear{Groussin et~al.,}{Groussin
  et~al.}{2015}]{Groussin2015}
Groussin O.,  et~al., 2015, \mn@doi [Astronomy {\&} Astrophysics, Volume 583,
  id.A32, 12 pp.] {10.1051/0004-6361/201526379}, 583

\bibitem[\protect\citeauthoryear{Gundlach, Blum  \& Blum}{Gundlach
  et~al.}{2016}]{Gundlach2016}
Gundlach B.,  Blum J.,   Blum J.,  2016, \mn@doi [Astronomy {\&} Astrophysics]
  {10.1051/0004-6361/201527260}, 589, A111

\bibitem[\protect\citeauthoryear{Gutierrez, de Leon, Jorda, Licandro, Lara  \&
  Lamy}{Gutierrez et~al.}{2003}]{Gutierrez2003}
Gutierrez P.~J.,  de Leon J.,  Jorda L.,  Licandro J.,  Lara L.~M.,   Lamy P.,
  2003, \mn@doi [Astronomy and Astrophysics] {10.1051/0004-6361:20031066}, 407,
  L37

\bibitem[\protect\citeauthoryear{Harmon \& Nolan}{Harmon \&
  Nolan}{2005}]{Harmon2005}
Harmon J.,  Nolan M.,  2005, \mn@doi [Icarus] {10.1016/j.icarus.2005.01.012},
  176, 175

\bibitem[\protect\citeauthoryear{Harmon, Nolan, Howell, Giorgini  \&
  Taylor}{Harmon et~al.}{2011}]{Harmon2011}
Harmon J.~K.,  Nolan M.~C.,  Howell E.~S.,  Giorgini J.~D.,   Taylor P.~A.,
  2011, \mn@doi [The Astrophysical Journal] {10.1088/2041-8205/734/1/L2}, 734,
  L2

\bibitem[\protect\citeauthoryear{Harris}{Harris}{1996}]{Harris1996}
Harris A.~W.,  1996, Lunar and Planetary Science, volume 27, page 493, 27

\bibitem[\protect\citeauthoryear{Hergenrother}{Hergenrother}{2014}]{Hergenrother2014}
Hergenrother C.,  2014, IAU CBET, 3870

\bibitem[\protect\citeauthoryear{Hoenig}{Hoenig}{2005}]{Hoenig2005}
Hoenig S.~F.,  2005, \mn@doi [Astronomy and Astrophysics, Volume 445, Issue 2,
  January II 2006, pp.759-763] {10.1051/0004-6361:20053991}, 445, 759

\bibitem[\protect\citeauthoryear{Howell}{Howell}{1989}]{Howell1989}
Howell S.~B.,  1989, \mn@doi [Publications of the Astronomical Society of the
  Pacific] {10.1086/132477}, 101, 616

\bibitem[\protect\citeauthoryear{Howell et~al.,}{Howell
  et~al.}{2014}]{Howell2014}
Howell E.~S.,  et~al., 2014, American Astronomical Society, DPS meeting {\#}46,
  id.209.24, 46

\bibitem[\protect\citeauthoryear{Jehin, Manfroid, Hutsemekers, Gillon  \&
  Magain}{Jehin et~al.}{2010}]{Jehin2010}
Jehin E.,  Manfroid J.,  Hutsemekers D.,  Gillon M.,   Magain P.,  2010,
  Central Bureau Electronic Telegrams, No. 2589, {\#}1 (2010). Edited by Green,
  D. W. E., 2589

\bibitem[\protect\citeauthoryear{Jester et~al.,}{Jester
  et~al.}{2005}]{Jester2005}
Jester S.,  et~al., 2005, \mn@doi [The Astronomical Journal, Volume 130, Issue
  3, pp. 873-895.] {10.1086/432466}, 130, 873

\bibitem[\protect\citeauthoryear{Jewitt}{Jewitt}{2009}]{Jewitt2009a}
Jewitt D.,  2009, \mn@doi [The Astronomical Journal, Volume 137, Issue 5, pp.
  4296-4312 (2009).] {10.1088/0004-6256/137/5/4296}, 137, 4296

\bibitem[\protect\citeauthoryear{Jewitt \& Luu}{Jewitt \&
  Luu}{1989}]{Jewitt1989}
Jewitt D.,  Luu J.,  1989, \mn@doi [The Astronomical Journal] {10.1086/115118},
  97, 1766

\bibitem[\protect\citeauthoryear{Jewitt \& Meech}{Jewitt \&
  Meech}{1987}]{Jewitt1987}
Jewitt D.,  Meech K.,  1987, \mn@doi [The Astronomical Journal]
  {10.1086/114437}, 93, 1542

\bibitem[\protect\citeauthoryear{Jewitt \& Meech}{Jewitt \&
  Meech}{1988}]{Jewitt1988}
Jewitt D.~C.,  Meech K.~J.,  1988, \mn@doi [The Astrophysical Journal]
  {10.1086/166351}, 328, 974

\bibitem[\protect\citeauthoryear{Jewitt \& Sheppard}{Jewitt \&
  Sheppard}{2004}]{Jewitt2004}
Jewitt D.,  Sheppard S.,  2004, \mn@doi [The Astronomical Journal]
  {10.1086/382097}, 127, 1784

\bibitem[\protect\citeauthoryear{Jewitt, Sheppard  \& Fernndez}{Jewitt
  et~al.}{2003}]{Jewitt2003}
Jewitt D.,  Sheppard S.,   Fernndez Y.,  2003, \mn@doi [The Astronomical
  Journal] {10.1086/374947}, 125, 3366

\bibitem[\protect\citeauthoryear{Jewitt, Hsieh  \& Agarwal}{Jewitt
  et~al.}{2015}]{Jewitt2015b}
Jewitt D.,  Hsieh H.,   Agarwal J.,  2015, \mn@doi [Asteroids IV, Patrick
  Michel, Francesca E. DeMeo, and William F. Bottke (eds.), University of
  Arizona Press, Tucson, 895 pp. ISBN: 978-0-816-53213-1, 2015., p.221-241]
  {10.2458/azu_uapress_9780816532131-ch012}, pp 221--241

\bibitem[\protect\citeauthoryear{Jorda, Lamy, Groussin, Toth, A'Hearn  \&
  Peschke}{Jorda et~al.}{2000}]{Jorda2000}
Jorda L.,  Lamy P.,  Groussin O.,  Toth I.,  A'Hearn M.~F.,   Peschke S.,
  2000, ISO Beyond Point Sources: Studies of Extended Infrared Emission,
  September 14-17, 1999, ISO Data Centre, Villafranca del Castillo, Madrid,
  Spain. Edited by R. J. Laureijs, K. Leech and M. F. Kessler, ESA-SP 455,
  2000. p. 61., 455, 61

\bibitem[\protect\citeauthoryear{Jorda et~al.,}{Jorda et~al.}{2016}]{Jorda2016}
Jorda L.,  et~al., 2016, \mn@doi [Icarus] {10.1016/j.icarus.2016.05.002}, 277,
  257

\bibitem[\protect\citeauthoryear{Jutzi, Benz, Toliou, Morbidelli  \&
  Brasser}{Jutzi et~al.}{2017}]{Jutzi2016}
Jutzi M.,  Benz W.,  Toliou A.,  Morbidelli A.,   Brasser R.,  2017, \mn@doi
  [Astronomy {\&} Astrophysics, Volume 597, id.A61, 13 pp.]
  {10.1051/0004-6361/201628963}, 597

\bibitem[\protect\citeauthoryear{Kaiser et~al.,}{Kaiser
  et~al.}{2002}]{Kaiser2002}
Kaiser N.,  et~al., 2002, in Tyson J.~A.,  Wolff S.,  eds,  Vol. 4836, Survey
  and Other Telescope Technologies and Discoveries. Edited by Tyson, J.
  Anthony; Wolff, Sidney. Proceedings of the SPIE, Volume 4836, pp. 154-164
  (2002).. p.~154, \mn@doi{10.1117/12.457365}, \url
  {http://proceedings.spiedigitallibrary.org/proceeding.aspx?doi=10.1117/12.457365}

\bibitem[\protect\citeauthoryear{Kaiser et~al.,}{Kaiser
  et~al.}{2010}]{Kaiser2010}
Kaiser N.,  et~al., 2010, in Stepp L.~M.,  Gilmozzi R.,   Hall H.~J.,  eds,
  Vol. 7733, Ground-based and Airborne Telescopes III. Edited by Stepp, Larry
  M.; Gilmozzi, Roberto; Hall, Helen J. Proceedings of the SPIE, Volume 7733,
  article id. 77330E, 14 pp. (2010).. p. 77330E, \mn@doi{10.1117/12.859188},
  \url
  {http://proceedings.spiedigitallibrary.org/proceeding.aspx?doi=10.1117/12.859188}

\bibitem[\protect\citeauthoryear{Kamoun, Campbell, Ostro, Pettengill  \&
  Shapiro}{Kamoun et~al.}{1982}]{Kamoun1982}
Kamoun P.~G.,  Campbell D.~B.,  Ostro S.~J.,  Pettengill G.~H.,   Shapiro
  I.~I.,  1982, \mn@doi [Science] {10.1126/science.216.4543.293}, 216, 293

\bibitem[\protect\citeauthoryear{Kasuga, Balam  \& Wiegert}{Kasuga
  et~al.}{2010}]{Kasuga2010}
Kasuga T.,  Balam D.~D.,   Wiegert P.~A.,  2010, \mn@doi [The Astronomical
  Journal] {10.1088/0004-6256/140/6/1806}, 140, 1806

\bibitem[\protect\citeauthoryear{Keller, Mottola, Skorov  \& Jorda}{Keller
  et~al.}{2015}]{Keller2015}
Keller H.~U.,  Mottola S.,  Skorov Y.,   Jorda L.,  2015, \mn@doi [Astronomy
  {\&} Astrophysics] {10.1051/0004-6361/201526421}, 579, L5

\bibitem[\protect\citeauthoryear{Kelley, Wooden, Tubiana, Boehnhardt, Woodward
  \& Harker}{Kelley et~al.}{2009}]{Kelley2009}
Kelley M.~S.,  Wooden D.~H.,  Tubiana C.,  Boehnhardt H.,  Woodward C.~E.,
  Harker D.~E.,  2009, \mn@doi [The Astronomical Journal, Volume 137, Issue 6,
  pp. 4633-4642 (2009).] {10.1088/0004-6256/137/6/4633}, 137, 4633

\bibitem[\protect\citeauthoryear{Knight, Farnham, Schleicher  \&
  Schwieterman}{Knight et~al.}{2011}]{Knight2011}
Knight M.~M.,  Farnham T.~L.,  Schleicher D.~G.,   Schwieterman E.~W.,  2011,
  \mn@doi [The Astronomical Journal] {10.1088/0004-6256/141/1/2}, 141, 2

\bibitem[\protect\citeauthoryear{Knight, Schleicher, Farnham, Schwieterman  \&
  Christensen}{Knight et~al.}{2012}]{Knight2012}
Knight M.~M.,  Schleicher D.~G.,  Farnham T.~L.,  Schwieterman E.~W.,
  Christensen S.~R.,  2012, \mn@doi [The Astronomical Journal]
  {10.1088/0004-6256/144/5/153}, 144, 153

\bibitem[\protect\citeauthoryear{Knight, Mueller, Samarasinha  \&
  Schleicher}{Knight et~al.}{2015}]{Knight2015b}
Knight M.~M.,  Mueller B. E.~A.,  Samarasinha N.~H.,   Schleicher D.~G.,  2015,
  \mn@doi [The Astronomical Journal, Volume 150, Issue 1, article id. 22, 14
  pp. (2015).] {10.1088/0004-6256/150/1/22}, 150

\bibitem[\protect\citeauthoryear{Knight, Fitzsimmons, Kelley  \&
  Snodgrass}{Knight et~al.}{2016}]{Knight2016}
Knight M.~M.,  Fitzsimmons A.,  Kelley M. S.~P.,   Snodgrass C.,  2016, \mn@doi
  [The Astrophysical Journal Letters, Volume 823, Issue 1, article id. L6, 6
  pp. (2016).] {10.3847/2041-8205/823/1/L6}, 823

\bibitem[\protect\citeauthoryear{Lamy, Toth, Jorda, Weaver  \& A'Hearn}{Lamy
  et~al.}{1998a}]{Lamy1998a}
Lamy P.~L.,  Toth I.,  Jorda L.,  Weaver H.~A.,   A'Hearn M.,  1998a, Astronomy
  and Astrophysics, v.335, p.L25-L29 (1998), 335, L25

\bibitem[\protect\citeauthoryear{Lamy, Toth  \& Weaver}{Lamy
  et~al.}{1998b}]{Lamy1998}
Lamy P.~L.,  Toth I.,   Weaver H.~A.,  1998b, Astronomy and Astrophysics,
  v.337, p.945-954 (1998), 337, 945

\bibitem[\protect\citeauthoryear{Lamy, Toth, A'Hearn, Weaver  \& Weissman}{Lamy
  et~al.}{2001}]{Lamy2001}
Lamy P.,  Toth I.,  A'Hearn M.~F.,  Weaver H.~A.,   Weissman P.~R.,  2001,
  \mn@doi [Icarus] {10.1006/icar.2001.6705}, 154, 337

\bibitem[\protect\citeauthoryear{Lamy, Toth, Jorda, Groussin, A'Hearn  \&
  Weaver}{Lamy et~al.}{2002}]{Lamy2002}
Lamy P.,  Toth I.,  Jorda L.,  Groussin O.,  A'Hearn M.~F.,   Weaver H.~A.,
  2002, \mn@doi [Icarus] {10.1006/icar.2001.6785}, 156, 442

\bibitem[\protect\citeauthoryear{Lamy, Toth, Fernandez  \& Weaver}{Lamy
  et~al.}{2004}]{Lamy2004}
Lamy P.~L.,  Toth I.,  Fernandez Y.~R.,   Weaver H.~A.,  2004, Comets II

\bibitem[\protect\citeauthoryear{Lamy, Toth, Weaver, Jorda, Kaasalainen  \&
  Guti{\'{e}}rrez}{Lamy et~al.}{2006}]{Lamy2006}
Lamy P.~L.,  Toth I.,  Weaver H.~A.,  Jorda L.,  Kaasalainen M.,
  Guti{\'{e}}rrez P.~J.,  2006, \mn@doi [Astronomy and Astrophysics]
  {10.1051/0004-6361:20065253}, 458, 669

\bibitem[\protect\citeauthoryear{Lamy, Toth, Weaver, A'Hearn  \& Jorda}{Lamy
  et~al.}{2009}]{Lamy2009}
Lamy P.~L.,  Toth I.,  Weaver H.~A.,  A'Hearn M.~F.,   Jorda L.,  2009, \mn@doi
  [Astronomy and Astrophysics] {10.1051/0004-6361/200811462}, 508, 1045

\bibitem[\protect\citeauthoryear{Lamy, Toth, Weaver, A'Hearn  \& Jorda}{Lamy
  et~al.}{2011}]{Lamy2011}
Lamy P.~L.,  Toth I.,  Weaver H.~A.,  A'Hearn M.~F.,   Jorda L.,  2011, \mn@doi
  [Monthly Notices of the Royal Astronomical Society]
  {10.1111/j.1365-2966.2010.17934.x}, 412, 1573

\bibitem[\protect\citeauthoryear{Lamy, Faury, Llebaria, Knight, A'Hearn  \&
  Battams}{Lamy et~al.}{2013}]{Lamy2013}
Lamy P.,  Faury G.,  Llebaria A.,  Knight M.,  A'Hearn M.,   Battams K.,  2013,
  \mn@doi [Icarus] {10.1016/j.icarus.2013.07.035}, 226, 1350

\bibitem[\protect\citeauthoryear{Leibowitz \& Brosch}{Leibowitz \&
  Brosch}{1986}]{Leibowitz1986}
Leibowitz E.~M.,  Brosch N.,  1986, \mn@doi [Icarus]
  {10.1016/0019-1035(86)90049-7}, 68, 430

\bibitem[\protect\citeauthoryear{Levison}{Levison}{1996}]{Levison1996}
Levison H.~F.,  1996, in Completing the Inventory of the Solar System,
  Astronomical Society of the Pacific Conference Proceedings, volume 107, T.W.
  Rettig and J.M. Hahn, Eds., pp. 173-191.. Astronomical Society of the Pacific
  (ASP), pp 173--191, \url {http://adsabs.harvard.edu/abs/1996ASPC..107..173L}

\bibitem[\protect\citeauthoryear{Levison, Terrell, Wiegert, Dones  \&
  Duncan}{Levison et~al.}{2006}]{Levison2006}
Levison H.~F.,  Terrell D.,  Wiegert P.~A.,  Dones L.,   Duncan M.~J.,  2006,
  \mn@doi [Icarus] {10.1016/j.icarus.2005.12.016}, 182, 161

\bibitem[\protect\citeauthoryear{Li et~al.,}{Li et~al.}{2007a}]{Li2007b}
Li J.-Y.,  et~al., 2007a, \mn@doi [Icarus] {10.1016/j.icarus.2006.09.018}, 187,
  41

\bibitem[\protect\citeauthoryear{Li, A'Hearn, McFadden  \& Belton}{Li
  et~al.}{2007b}]{Li2007a}
Li J.,  A'Hearn M.,  McFadden L.,   Belton M.,  2007b, \mn@doi [Icarus]
  {10.1016/j.icarus.2006.11.015}, 188, 195

\bibitem[\protect\citeauthoryear{Li, A'Hearn, Farnham  \& McFadden}{Li
  et~al.}{2009}]{Li2009}
Li J.-Y.,  A'Hearn M.~F.,  Farnham T.~L.,   McFadden L.~A.,  2009, \mn@doi
  [Icarus] {10.1016/j.icarus.2009.06.002}, 204, 209

\bibitem[\protect\citeauthoryear{Li et~al.,}{Li et~al.}{2013}]{Li2013}
Li J.-Y.,  et~al., 2013, \mn@doi [Icarus] {10.1016/j.icarus.2012.11.001}, 222,
  559

\bibitem[\protect\citeauthoryear{Licandro, Tancredi, Lindgren, Rickman  \&
  Hutton}{Licandro et~al.}{2000}]{Licandro2000}
Licandro J.,  Tancredi G.,  Lindgren M.,  Rickman H.,   Hutton R.~G.,  2000,
  \mn@doi [Icarus] {10.1006/icar.2000.6442}, 147, 161

\bibitem[\protect\citeauthoryear{Lisse et~al.,}{Lisse et~al.}{2005}]{Lisse2005}
Lisse C.~M.,  et~al., 2005, \mn@doi [The Astrophysical Journal]
  {10.1086/431238}, 625, L139

\bibitem[\protect\citeauthoryear{Lisse et~al.,}{Lisse et~al.}{2009}]{Lisse2009}
Lisse C.~M.,  et~al., 2009, \mn@doi [Publications of the Astronomical Society
  of Pacific, Volume 121, Issue 883, pp. 968-975 (2009).] {10.1086/605546},
  121, 968

\bibitem[\protect\citeauthoryear{Lomb}{Lomb}{1976}]{Lomb1976}
Lomb N.~R.,  1976, \mn@doi [Astrophysics and Space Science]
  {10.1007/BF00648343}, 39, 447

\bibitem[\protect\citeauthoryear{Lowry \& Fitzsimmons}{Lowry \&
  Fitzsimmons}{2001}]{Lowry2001a}
Lowry S.~C.,  Fitzsimmons A.,  2001, \mn@doi [Astronomy and Astrophysics]
  {10.1051/0004-6361:20000180}, 365, 204

\bibitem[\protect\citeauthoryear{Lowry \& Weissman}{Lowry \&
  Weissman}{2003}]{Lowry2003}
Lowry S.~C.,  Weissman P.~R.,  2003, \mn@doi [Icarus]
  {10.1016/S0019-1035(03)00129-5}, 164, 492

\bibitem[\protect\citeauthoryear{Lowry \& Weissman}{Lowry \&
  Weissman}{2007}]{Lowry2007}
Lowry S.~C.,  Weissman P.~R.,  2007, \mn@doi [Icarus]
  {10.1016/j.icarus.2006.11.014}, 188, 212

\bibitem[\protect\citeauthoryear{Lowry, Fitzsimmons, Cartwright  \&
  Williams}{Lowry et~al.}{1999}]{Lowry1999}
Lowry S.~C.,  Fitzsimmons A.,  Cartwright I.~M.,   Williams I.~P.,  1999,
  Astronomy and Astrophysics, v.349, p.649-659 (1999), 349, 649

\bibitem[\protect\citeauthoryear{Lowry, Fitzsimmons  \& Collander-Brown}{Lowry
  et~al.}{2003}]{Lowry2003b}
Lowry S.~C.,  Fitzsimmons A.,   Collander-Brown S.,  2003, \mn@doi [Astronomy
  and Astrophysics] {10.1051/0004-6361:20021486}, 397, 329

\bibitem[\protect\citeauthoryear{Lowry, Fitzsimmons, Jorda, Kaasalainen, Lamy
  \& Toth}{Lowry et~al.}{2006}]{Lowry2006}
Lowry S.~C.,  Fitzsimmons A.,  Jorda L.,  Kaasalainen M.,  Lamy P.,   Toth I.,
  2006, American Astronomical Society, 38

\bibitem[\protect\citeauthoryear{Lowry, Fitzsimmons, Lamy  \& Weissman}{Lowry
  et~al.}{2008}]{Lowry2008}
Lowry S.,  Fitzsimmons A.,  Lamy P.,   Weissman P.,  2008, in , The Solar
  System Beyond Neptune, M. A. Barucci, H. Boehnhardt, D. P. Cruikshank, and A.
  Morbidelli (eds.), University of Arizona Press, Tucson, 592 pp., p.397-410.
pp 397--410, \url {http://adsabs.harvard.edu/abs/2008ssbn.book..397L}

\bibitem[\protect\citeauthoryear{Lowry, Duddy, Rozitis, Green, Fitzsimmons,
  Snodgrass, Hsieh  \& Hainaut}{Lowry et~al.}{2012}]{Lowry2012a}
Lowry S.,  Duddy S.~R.,  Rozitis B.,  Green S.~F.,  Fitzsimmons A.,  Snodgrass
  C.,  Hsieh H.~H.,   Hainaut O.,  2012, \mn@doi [Astronomy {\&} Astrophysics]
  {10.1051/0004-6361/201220116}, 548, A12

\bibitem[\protect\citeauthoryear{Luu \& Jewitt}{Luu \& Jewitt}{1990}]{Luu1990a}
Luu J.,  Jewitt D.,  1990, \mn@doi [Icarus] {10.1016/0019-1035(90)90199-J}, 86,
  69

\bibitem[\protect\citeauthoryear{Luu \& Jewitt}{Luu \& Jewitt}{1992}]{Luu1992}
Luu J.~X.,  Jewitt D.~C.,  1992, \mn@doi [The Astronomical Journal]
  {10.1086/116400}, 104, 2243

\bibitem[\protect\citeauthoryear{Manzini, Oldani, Crippa, Borrero, Bryssink,
  Mobberley  \& Nicolas}{Manzini et~al.}{2014}]{Manzini2014}
Manzini F.,  Oldani V.,  Crippa R.,  Borrero J.,  Bryssink E.,  Mobberley M.,
  Nicolas J.,  2014, \mn@doi [Astrophysics and Space Science]
  {10.1007/s10509-014-1854-6}, 351, 435

\bibitem[\protect\citeauthoryear{Masoumzadeh et~al.,}{Masoumzadeh
  et~al.}{2017}]{Masoumzadeh2017}
Masoumzadeh N.,  et~al., 2017, \mn@doi [Astronomy {\&} Astrophysics]
  {10.1051/0004-6361/201629734}, 599, A11

\bibitem[\protect\citeauthoryear{Meech, Hainaut  \& Marsden}{Meech
  et~al.}{2004}]{Meech2004a}
Meech K.,  Hainaut O.,   Marsden B.,  2004, \mn@doi [Icarus]
  {10.1016/j.icarus.2004.03.014}, 170, 463

\bibitem[\protect\citeauthoryear{Meech et~al.,}{Meech et~al.}{2005}]{Meech2005}
Meech et~al., 2005, \mn@doi [Science] {10.1126/science.1118978}, 310, 265

\bibitem[\protect\citeauthoryear{Meech et~al.,}{Meech et~al.}{2009}]{Meech2009}
Meech K.~J.,  et~al., 2009, American Astronomical Society, DPS meeting {\#}41,
  id.20.07, 41

\bibitem[\protect\citeauthoryear{Meech et~al.,}{Meech
  et~al.}{2011a}]{Meech2011a}
Meech K.,  et~al., 2011a, \mn@doi [Icarus] {10.1016/j.icarus.2011.02.016}, 213,
  323

\bibitem[\protect\citeauthoryear{Meech et~al.,}{Meech
  et~al.}{2011b}]{Meech2011}
Meech K.~J.,  et~al., 2011b, \mn@doi [The Astrophysical Journal]
  {10.1088/2041-8205/734/1/L1}, 734, L1

\bibitem[\protect\citeauthoryear{Millis, A'Hearn  \& Campins}{Millis
  et~al.}{1988}]{Millis1988}
Millis R.~L.,  A'Hearn M.~F.,   Campins H.,  1988, \mn@doi [The Astrophysical
  Journal] {10.1086/165974}, 324, 1194

\bibitem[\protect\citeauthoryear{Mottola et~al.,}{Mottola
  et~al.}{2014}]{Mottola2014}
Mottola S.,  et~al., 2014, \mn@doi [Astronomy {\&} Astrophysics]
  {10.1051/0004-6361/201424590}, 569, L2

\bibitem[\protect\citeauthoryear{Mueller}{Mueller}{1992}]{Mueller1992}
Mueller B. E.~A.,  1992, In Lunar and Planetary Inst., Asteroids, Comets,
  Meteors 1991 p 425-428 (SEE N93-19113 06-90)

\bibitem[\protect\citeauthoryear{Mueller \& Ferrin}{Mueller \&
  Ferrin}{1996}]{Mueller1996}
Mueller B.~E.,  Ferrin I.,  1996, \mn@doi [Icarus] {10.1006/icar.1996.0172},
  123, 463

\bibitem[\protect\citeauthoryear{Mueller \& Samarasinha}{Mueller \&
  Samarasinha}{2002}]{Mueller2002}
Mueller B. E.~A.,  Samarasinha N.~H.,  2002, Earth, Moon, and Planets, v. 90,
  Issue 1, p. 463-471 (2002)., 90, 463

\bibitem[\protect\citeauthoryear{Mueller \& Samarasinha}{Mueller \&
  Samarasinha}{2015}]{Mueller2015}
Mueller B. E.~A.,  Samarasinha N.~H.,  2015, American Astronomical Society

\bibitem[\protect\citeauthoryear{Mueller, Samarasinha  \& Fernandez}{Mueller
  et~al.}{2008}]{Mueller2008}
Mueller B. E.~A.,  Samarasinha N.~H.,   Fernandez Y.~R.,  2008, American
  Astronomical Society, 40

\bibitem[\protect\citeauthoryear{Mueller, Farnham, Samarasinha  \&
  A'Hearn}{Mueller et~al.}{2010a}]{Mueller2010a}
Mueller B. E.~A.,  Farnham T.~L.,  Samarasinha N.~H.,   A'Hearn M.~F.,  2010a,
  American Astronomical Society, DPS meeting {\#}42, id.28.31; Bulletin of the
  American Astronomical Society, Vol. 42, p.966, 42, 966

\bibitem[\protect\citeauthoryear{Mueller, Samarasinha, Rauer  \&
  Helbert}{Mueller et~al.}{2010b}]{Mueller2010}
Mueller B. E.~A.,  Samarasinha N.~H.,  Rauer H.,   Helbert J.,  2010b, \mn@doi
  [Icarus] {10.1016/j.icarus.2010.05.004}, 209, 745

\bibitem[\protect\citeauthoryear{Nolan, Harmon, Howell, Benner, Giorgini,
  Ostro, Campbell  \& Margot}{Nolan et~al.}{2006}]{Nolan2006}
Nolan M.~C.,  Harmon J.~K.,  Howell E.~S.,  Benner L.~A.,  Giorgini J.~D.,
  Ostro S.~J.,  Campbell D.~B.,   Margot J.~L.,  2006, American Astronomical
  Society, DPS meeting {\#}38, id.12.06; Bulletin of the American Astronomical
  Society, Vol. 38, p.504, 38, 504

\bibitem[\protect\citeauthoryear{Ohtsuka, Ito, Yoshikawa, Asher  \&
  Arakida}{Ohtsuka et~al.}{2008}]{Ohtsuka2008}
Ohtsuka K.,  Ito T.,  Yoshikawa M.,  Asher D.~J.,   Arakida H.,  2008, \mn@doi
  [Astronomy and Astrophysics, Volume 489, Issue 3, 2008, pp.1355-1362]
  {10.1051/0004-6361:200810321}, 489, 1355

\bibitem[\protect\citeauthoryear{Oszkiewicz, Bowell, Wasserman, Muinonen,
  Penttil{\"{a}}, Pieniluoma, Trilling  \& Thomas}{Oszkiewicz
  et~al.}{2012}]{Oszkiewicz2012}
Oszkiewicz D.~A.,  Bowell E.,  Wasserman L.~H.,  Muinonen K.,  Penttil{\"{a}}
  A.,  Pieniluoma T.,  Trilling D.~E.,   Thomas C.~A.,  2012, \mn@doi [Icarus,
  Volume 219, Issue 1, p. 283-296.] {10.1016/j.icarus.2012.02.028}, 219, 283

\bibitem[\protect\citeauthoryear{Pravec \& Harris}{Pravec \&
  Harris}{2000}]{Pravec2000}
Pravec P.,  Harris A.~W.,  2000, \mn@doi [Icarus] {10.1006/icar.2000.6482},
  148, 12

\bibitem[\protect\citeauthoryear{Pravec, Sarounova  \& Wolf}{Pravec
  et~al.}{1996}]{Pravec1996}
Pravec P.,  Sarounova L.,   Wolf M.,  1996, \mn@doi [Icarus]
  {10.1006/icar.1996.0223}, 124, 471

\bibitem[\protect\citeauthoryear{Pravec, Harris  \& Michalowski}{Pravec
  et~al.}{2002}]{Pravec2002}
Pravec P.,  Harris A.~W.,   Michalowski T.,  2002, Asteroids III, W. F. Bottke
  Jr., A. Cellino, P. Paolicchi, and R. P. Binzel (eds), University of Arizona
  Press, Tucson, p.113-122, pp 113--122

\bibitem[\protect\citeauthoryear{Reyniers, Degroote, Bodewits, Cuypers  \&
  Waelkens}{Reyniers et~al.}{2009}]{Reyniers2009}
Reyniers M.,  Degroote P.,  Bodewits D.,  Cuypers J.,   Waelkens C.,  2009,
  \mn@doi [Astronomy and Astrophysics, Volume 494, Issue 1, 2009, pp.379-389]
  {10.1051/0004-6361:20079225}, 494, 379

\bibitem[\protect\citeauthoryear{Richardson, Melosh, Lisse  \&
  Carcich}{Richardson et~al.}{2007}]{Richardson2007}
Richardson J.~E.,  Melosh H.~J.,  Lisse C.~M.,   Carcich B.,  2007, \mn@doi
  [Icarus] {10.1016/j.icarus.2007.08.001}, 190, 357

\bibitem[\protect\citeauthoryear{Samarasinha \& Mueller}{Samarasinha \&
  Mueller}{2013}]{Samarasinha2013}
Samarasinha N.~H.,  Mueller B. E.~A.,  2013, \mn@doi [The Astrophysical
  Journal] {10.1088/2041-8205/775/1/L10}, 775, L10

\bibitem[\protect\citeauthoryear{Samarasinha, Mueller, Belton  \&
  Jorda}{Samarasinha et~al.}{2004}]{Samarasinha2004}
Samarasinha N.~H.,  Mueller B. E.~A.,  Belton M. J.~S.,   Jorda L.,  2004,
  Comets II

\bibitem[\protect\citeauthoryear{Samarasinha, Mueller, A'Hearn  \&
  Farnham}{Samarasinha et~al.}{2010}]{Samarasinha2010}
Samarasinha N.~H.,  Mueller B. E.~A.,  A'Hearn M.~F.,   Farnham T.~L.,  2010,
  IAU Circ., No. 9178, {\#}1 (2010). Edited by Green, D. W. E., 9178

\bibitem[\protect\citeauthoryear{Samarasinha, Mueller, A'Hearn, Farnham  \&
  Gersch}{Samarasinha et~al.}{2011}]{Samarasinha2011}
Samarasinha N.~H.,  Mueller B. E.~A.,  A'Hearn M.~F.,  Farnham T.~L.,   Gersch
  A.,  2011, \mn@doi [The Astrophysical Journal] {10.1088/2041-8205/734/1/L3},
  734, L3

\bibitem[\protect\citeauthoryear{Samarasinha et~al.,}{Samarasinha
  et~al.}{2012}]{Samarasinha2012}
Samarasinha N.~H.,  et~al., 2012, American Astronomical Society, DPS meeting
  {\#}44, id.506.03, 44

\bibitem[\protect\citeauthoryear{Scargle}{Scargle}{1982}]{Scargle1982}
Scargle J.~D.,  1982, \mn@doi [The Astrophysical Journal] {10.1086/160554},
  263, 835

\bibitem[\protect\citeauthoryear{Schleicher \& Knight}{Schleicher \&
  Knight}{2016}]{Schleicher2016}
Schleicher D.~G.,  Knight M.~M.,  2016, eprint arXiv:1605.01705

\bibitem[\protect\citeauthoryear{Schleicher, Knight  \& Levine}{Schleicher
  et~al.}{2013}]{Schleicher2013}
Schleicher D.~G.,  Knight M.~M.,   Levine S.~E.,  2013, \mn@doi [The
  Astronomical Journal, Volume 146, Issue 5, article id. 137, 8 pp. (2013).]
  {10.1088/0004-6256/146/5/137}, 146

\bibitem[\protect\citeauthoryear{Scotti et~al.,}{Scotti
  et~al.}{1996}]{Scotti1996}
Scotti J.~V.,  et~al., 1996, IAU Circ., No. 6301, {\#}1 (1996). Edited by
  Marsden, B. G., 6301

\bibitem[\protect\citeauthoryear{Sekanina}{Sekanina}{1987}]{Sekanina1987}
Sekanina Z.,  1987, In ESA, 278

\bibitem[\protect\citeauthoryear{Sekanina \& Zdenek}{Sekanina \&
  Zdenek}{1991}]{Sekanina1991}
Sekanina Z.,  Zdenek 1991, \mn@doi [The Astronomical Journal] {10.1086/115881},
  102, 350

\bibitem[\protect\citeauthoryear{Sekanina, Brownlee, Economou, Tuzzolino  \&
  Green}{Sekanina et~al.}{2004}]{Sekanina2004}
Sekanina Z.,  Brownlee D.~E.,  Economou T.~E.,  Tuzzolino A.~J.,   Green S.~F.,
   2004, \mn@doi [Science] {10.1126/science.1098388}, 304, 1769

\bibitem[\protect\citeauthoryear{Sierks et~al.,}{Sierks
  et~al.}{2015}]{Sierks2015}
Sierks H.,  et~al., 2015, \mn@doi [Science (New York, N.Y.)]
  {10.1126/science.aaa1044}, 347, aaa1044

\bibitem[\protect\citeauthoryear{Snodgrass \& Carry}{Snodgrass \&
  Carry}{2013}]{Snodgrass.2013}
Snodgrass C.,  Carry B.,  2013, {Automatic Removal of Fringes from EFOSC
  Images}.
 Vol. 152, European Southern Observatory, \url
  {http://adsabs.harvard.edu/abs/2013Msngr.152...14S}

\bibitem[\protect\citeauthoryear{Snodgrass, Fitzsimmons  \& Lowry}{Snodgrass
  et~al.}{2005}]{Snodgrass2005}
Snodgrass C.,  Fitzsimmons A.,   Lowry S.~C.,  2005, \mn@doi [Astronomy and
  Astrophysics] {10.1051/0004-6361:20053237}, 444, 287

\bibitem[\protect\citeauthoryear{Snodgrass, Lowry  \& Fitzsimmons}{Snodgrass
  et~al.}{2006}]{Snodgrass2006}
Snodgrass C.,  Lowry S.~C.,   Fitzsimmons A.,  2006, \mn@doi [Monthly Notices
  of the Royal Astronomical Society] {10.1111/j.1365-2966.2006.11121.x}, 373,
  1590

\bibitem[\protect\citeauthoryear{Snodgrass, Saviane, Monaco  \&
  Sinclaire}{Snodgrass et~al.}{2008a}]{Snodgrass2008a}
Snodgrass C.,  Saviane I.,  Monaco L.,   Sinclaire P.,  2008a, The Messenger,
  vol. 132, p. 18-19, 132, 18

\bibitem[\protect\citeauthoryear{Snodgrass, Lowry  \& Fitzsimmons}{Snodgrass
  et~al.}{2008b}]{Snodgrass2008}
Snodgrass C.,  Lowry S.~C.,   Fitzsimmons A.,  2008b, \mn@doi [Monthly Notices
  of the Royal Astronomical Society] {10.1111/j.1365-2966.2008.12900.x}, 385,
  737

\bibitem[\protect\citeauthoryear{Snodgrass, Fitzsimmons, Lowry  \&
  Weissman}{Snodgrass et~al.}{2011}]{Snodgrass2011}
Snodgrass C.,  Fitzsimmons A.,  Lowry S.~C.,   Weissman P.,  2011, \mn@doi
  [Monthly Notices of the Royal Astronomical Society]
  {10.1111/j.1365-2966.2011.18406.x}, 414, 458

\bibitem[\protect\citeauthoryear{Snodgrass et~al.,}{Snodgrass
  et~al.}{2016}]{Snodgrass2016}
Snodgrass C.,  et~al., 2016, \mn@doi [Astronomy {\&} Astrophysics, Volume 588,
  id.A80, 12 pp.] {10.1051/0004-6361/201527834}, 588

\bibitem[\protect\citeauthoryear{Soderblom et~al.,}{Soderblom
  et~al.}{2002}]{Soderblom2002}
Soderblom L.~A.,  et~al., 2002, \mn@doi [Science] {10.1126/science.1069527},
  296, 1087

\bibitem[\protect\citeauthoryear{Stellingwerf}{Stellingwerf}{1978}]{Stellingwerf1978}
Stellingwerf R.~F.,  1978, \mn@doi [The Astrophysical Journal]
  {10.1086/156444}, 224, 953

\bibitem[\protect\citeauthoryear{Storm et~al.,}{Storm et~al.}{2006}]{Storm2006}
Storm S.~P.,  et~al., 2006, American Astronomical Society, DPS meeting {\#}38,
  id.12.08; Bulletin of the American Astronomical Society, Vol. 38, p.504, 38,
  504

\bibitem[\protect\citeauthoryear{Tancredi, Fern{\'{a}}ndez, Rickman  \&
  Licandro}{Tancredi et~al.}{2000}]{Tancredi2000}
Tancredi G.,  Fern{\'{a}}ndez J.~A.,  Rickman H.,   Licandro J.,  2000, \mn@doi
  [Astronomy and Astrophysics Supplement Series] {10.1051/aas:2000263}, 146, 73

\bibitem[\protect\citeauthoryear{Thomas et~al.,}{Thomas
  et~al.}{2013a}]{Thomas2013}
Thomas P.,  et~al., 2013a, \mn@doi [Icarus] {10.1016/j.icarus.2012.02.037},
  222, 453

\bibitem[\protect\citeauthoryear{Thomas et~al.,}{Thomas
  et~al.}{2013b}]{Thomas2013a}
Thomas P.~C.,  et~al., 2013b, \mn@doi [Icarus] {10.1016/j.icarus.2012.05.034},
  222, 550

\bibitem[\protect\citeauthoryear{Tody}{Tody}{1986}]{Tody1986}
Tody D.,  1986, in Crawford D.~L.,  ed.,  Vol. 627, IN: Instrumentation in
  astronomy VI; Proceedings of the Meeting, Tucson, AZ, Mar. 4-8, 1986. Part 2
  (A87-36376 15-35). Bellingham, WA, Society of Photo-Optical Instrumentation
  Engineers, 1986, p. 733.. pp 733--748, \mn@doi{10.1117/12.968154}, \url
  {http://proceedings.spiedigitallibrary.org/proceeding.aspx?articleid=1242189}

\bibitem[\protect\citeauthoryear{Tody}{Tody}{1993}]{Tody1993}
Tody D.,  1993, Astronomical Data Analysis Software and Systems II, A.S.P.
  Conference Series, Vol. 52, 1993, R. J. Hanisch, R. J. V. Brissenden, and
  Jeannette Barnes, eds., p. 173., 52, 173

\bibitem[\protect\citeauthoryear{Tonry et~al.,}{Tonry et~al.}{2012}]{Tonry2012}
Tonry J.~L.,  et~al., 2012, \mn@doi [The Astrophysical Journal]
  {10.1088/0004-637X/750/2/99}, 750, 99

\bibitem[\protect\citeauthoryear{Toth \& Lisse}{Toth \& Lisse}{2006}]{Toth2006}
Toth I.,  Lisse C.,  2006, \mn@doi [Icarus] {10.1016/j.icarus.2005.10.012},
  181, 162

\bibitem[\protect\citeauthoryear{Toth, Lamy  \& Weaver}{Toth
  et~al.}{2005}]{Toth2005}
Toth I.,  Lamy P.,   Weaver H.,  2005, \mn@doi [Icarus]
  {10.1016/j.icarus.2005.04.013}, 178, 235

\bibitem[\protect\citeauthoryear{Tubiana, Barrera, Drahus  \&
  Boehnhardt}{Tubiana et~al.}{2008}]{Tubiana2008}
Tubiana C.,  Barrera L.,  Drahus M.,   Boehnhardt H.,  2008, \mn@doi [Astronomy
  and Astrophysics] {10.1051/0004-6361:20078792}, 490, 377

\bibitem[\protect\citeauthoryear{Tubiana, B{\"{o}}hnhardt, Agarwal, Drahus,
  Barrera  \& Ortiz}{Tubiana et~al.}{2011}]{Tubiana2011}
Tubiana C.,  B{\"{o}}hnhardt H.,  Agarwal J.,  Drahus M.,  Barrera L.,   Ortiz
  J.~L.,  2011, \mn@doi [Astronomy {\&} Astrophysics]
  {10.1051/0004-6361/201016027}, 527, A113

\bibitem[\protect\citeauthoryear{Tubiana, Snodgrass, Michelsen, Haack,
  Boehnhardt, Fitzsimmons  \& Williams}{Tubiana et~al.}{2015}]{Tubiana2015}
Tubiana C.,  Snodgrass C.,  Michelsen R.,  Haack H.,  Boehnhardt H.,
  Fitzsimmons A.,   Williams I.~P.,  2015, \mn@doi [Astronomy {\&}
  Astrophysics, Volume 584, id.A97, 10 pp.] {10.1051/0004-6361/201425512}, 584

\bibitem[\protect\citeauthoryear{VanderPlas \& Ivezic}{VanderPlas \&
  Ivezic}{2015}]{VanderPlas2015}
VanderPlas J.~T.,  Ivezic Z.,  2015, \mn@doi [The Astrophysical Journal, Volume
  812, Issue 1, article id. 18, 15 pp. (2015).] {10.1088/0004-637X/812/1/18},
  812

\bibitem[\protect\citeauthoryear{Volk \& Malhotra}{Volk \&
  Malhotra}{2008}]{Volk2008}
Volk K.,  Malhotra R.,  2008, \mn@doi [The Astrophysical Journal]
  {10.1086/591839}, 687, 714

\bibitem[\protect\citeauthoryear{Warner}{Warner}{2006}]{Warner2006}
Warner B.~D.,  2006, {The Minor planet bulletin bulletin of the Minor Planets
  Section of the Association of Lunar and Planetary Observers : MPB.}.
 Vol. 33, Univ, \url {http://adsabs.harvard.edu/abs/2006MPBu...33...35W}

\bibitem[\protect\citeauthoryear{Warner \& Fitzsimmons}{Warner \&
  Fitzsimmons}{2005}]{Warner2005}
Warner B.~D.,  Fitzsimmons A.,  2005, IAU Circ., No. 8578, {\#}1 (2005). Edited
  by Green, D. W. E., 8578

\bibitem[\protect\citeauthoryear{Weaver, Stern  \& Parker}{Weaver
  et~al.}{2003}]{Weaver2003}
Weaver H.~A.,  Stern S.~A.,   Parker J.~W.,  2003, \mn@doi [The Astronomical
  Journal] {10.1086/375752}, 126, 444

\bibitem[\protect\citeauthoryear{Weissman, Doressoundiram, Hicks, Chamberlin,
  Sykes, Larson  \& Hergenrother}{Weissman et~al.}{1999}]{Weissman1999}
Weissman P.~R.,  Doressoundiram A.,  Hicks M.~D.,  Chamberlin A.,  Sykes M.~V.,
   Larson S.,   Hergenrother C.,  1999, Bulletin of the Astronomical Society,
  Vol. 31, No. 4, p. 1121, id.30.03, 31, 1121

\bibitem[\protect\citeauthoryear{Williams}{Williams}{2017}]{Williams2017}
Williams G.~V.,  2017, Central Bureau Electronic Telegrams, 4359, 1 (2017).
  Edited by Green, D. W. E., 4359

\makeatother
\end{thebibliography}

\bsp	
\label{lastpage}
\end{document}